\documentclass[11pt,superscriptaddress,amsmath,amssymb,nofootinbib,longbibliography,preprint]{revtex4-1}

\usepackage{graphicx} 
\usepackage{textcomp} 
\usepackage{dcolumn} 
\usepackage{bm}
\usepackage{amssymb}
\usepackage{amsmath} 
\usepackage{color}
\usepackage[normalem]{ulem}
\usepackage{hyperref}

\newcommand{\tb}{\tan\beta}
\newcommand{\sba}{s_{\beta-\alpha}}
\newcommand{\cba}{c_{\beta-\alpha}}
\newcommand{\hpm}{H^{\pm}}

\newcommand{\ken}[1]{{\color{black}#1}}

\allowdisplaybreaks[1]

\begin{document} 

\title{
Probing extended Higgs sectors by the 
synergy between direct searches at the LHC and precision tests at future lepton colliders
}

\preprint{OU-HET 1075}
\preprint{KA-TP-15-2020}

\author{Masashi Aiko}
\email{m-aikou@het.phys.sci.osaka-u.ac.jp}
\affiliation{Department of Physics, Osaka University, Toyonaka, Osaka 560-0043, Japan}

\author{Shinya Kanemura}
\email{kanemu@het.phys.sci.osaka-u.ac.jp}
\affiliation{Department of Physics, Osaka University, Toyonaka, Osaka 560-0043, Japan}

\author{Mariko Kikuchi}
\email{kikuchi@kct.ac.jp}
\affiliation{National Institute of Technology, Kitakyushu College, Kitakyushu, Fukuoka 802-0985, Japan}

\author{Kentarou Mawatari}
\email{mawatari@iwate-u.ac.jp}
\affiliation{Faculty of Education, Iwate University, Morioka, Iwate 020-8550, Japan}

\author{Kodai Sakurai}
\email{kodai.sakurai@kit.edu}
\affiliation{Institute for Theoretical Physics, Karlsruhe Institute of Technology, 76131 Karlsruhe, Germany}

\author{Kei Yagyu}
\email{yagyu@het.phys.sci.osaka-u.ac.jp}
\affiliation{Department of Physics, Osaka University, Toyonaka, Osaka 560-0043, Japan}

\begin{abstract}

We discuss a possibility that the parameter space of the two Higgs doublet model 
is significantly narrowed down by considering the synergy between direct searches for additional Higgs bosons 
at the LHC and its luminosity upgraded operation and precision measurements of the Higgs boson properties 
at future electron-positron colliders such as the International Linear Collider. 
We show that, in the case where the coupling constants of the discovered Higgs boson are slightly different from the predicted values in the standard model, 
most of the parameter space is explored by the direct searches of extra Higgs bosons, in particular for the decays of the extra Higgs bosons into the discovered Higgs boson,   
and also by the theoretical arguments such as perturbative unitarity and vacuum stability. 
This can be done because there appears an upper limit on the mass of the extra Higgs bosons as long as the deviation exists in the Higgs boson coupling. 
We also show that in the alignment limit where all the Higgs boson couplings take the standard model like values most of the parameter space cannot be excluded because 
most of the Higgs to Higgs decays are suppressed and also there is no upper limit on the masses from the theoretical arguments. 

\end{abstract}
\maketitle
\tableofcontents
\newpage
\section{Introduction}
\noindent

The current observations at the LHC experiments indicate that properties of the discovered Higgs boson with the mass of 125 GeV 
coincide with those predicted in the standard model (SM)~\cite{Aad:2019mbh,Sirunyan:2018koj}.
This, however, does not mean that the Higgs sector in the SM, which plays an essential role in the electroweak (EW) symmetry breaking, is verified. 
While the minimal Higgs sector is composed of one Higgs doublet field in the SM, there is no principle to determine the structure of the Higgs sector. 
In fact, it is possible to consider a variety of non-minimal Higgs sectors. 
Extended Higgs sectors are often introduced in new physics models
which can explain observed phenomena beyond the SM, such as neutrino oscillations,  
dark matter and baryon asymmetry of the Universe.  
In addition, they also appear in some of the new paradigms motivated from a theoretical problem in the SM; e.g., the hierarchy problem.
Therefore, new physics beyond the SM can be revealed by thoroughly testing the Higgs sector. 

A definitive probe of extended Higgs sectors would be direct detection of new scalar particles. 
At the LHC, especially after the discovery of the 125~GeV Higgs boson,
direct searches for additional Higgs bosons have been conducted exhaustively in a wide variety of the search channels~\cite{Aaboud:2017sjh,Aaboud:2017sjh,Aad:2019zwb,Aaboud:2018mjh,Aaboud:2018knk,Aaboud:2017gsl,Aaboud:2017rel,Aaboud:2017cxo,Aaboud:2017cxo,Aaboud:2018cwk,Aaboud:2018gjj,Khachatryan:2016qkc,Sirunyan:2018taj,Sirunyan:2017uhk,Sirunyan:2017isc,Khachatryan:2016cfx,Sirunyan:2019xls}.
Observations of such new particles have not been reported yet,
leading to constraints on parameters of extended Higgs models such as masses and coupling constants.
The direct searches are still one of the key programs at the LHC as well as at the high-luminosity LHC (HL-LHC)~\cite{ApollinariG.:2017ojx}. 

In addition to the direct searches, extended Higgs sectors can be explored by measuring various properties of the discovered Higgs boson such as cross sections, the width, branching ratios and coupling constants. 
If deviations from the SM are observed, we can extract upper limits on the mass scale of the second Higgs boson by taking into account theoretical consistencies. 
Furthermore, by looking at the pattern of the deviation we can extract the structure of the Higgs sector; e.g., 
the representation of the weak isospin, the number of Higgs fields, and symmetries. 
To this end, precision measurements of the Higgs boson couplings are most important. 
Although the current accuracy of the measurements is not enough, typically order 10 (20) percent level for the Higgs 
boson coupling to weak bosons (third generation fermions)~\cite{Aad:2019mbh,Sirunyan:2018koj}, it is expected to be improved 
at the HL-LHC~\cite{ApollinariG.:2017ojx} and further significantly at future lepton colliders; e.g., the International Linear Collider (ILC)~\cite{Baer:2013cma,Fujii:2017vwa,Asai:2017pwp,Fujii:2019zll}, 
the Future Circular Collider (FCC-ee)~\cite{Gomez-Ceballos:2013zzn} and the Circular Electron Positron Collider (CEPC)~\cite{CEPC-SPPCStudyGroup:2015csa}. 

It goes without saying that accurate calculations of the Higgs boson couplings are inevitable in order to 
compare theory predictions with the future precision measurements. 
It has been well known that QCD corrections to Higgs boson couplings with quarks or gluons can be quite large. 
For example, QCD corrections to the decay rate of the Higgs boson into gluons at the next-to-leading order (NLO) is about $70\%$ level~\cite{Dawson:1990zj,Djouadi:1991tka,Spira:1995rr}. 
Thus, QCD corrections must be included for calculations, by which we can discuss the deviation from the SM prediction. 
On the other hand, EW corrections are typically much smaller than QCD ones, but 
they have a sensitivity to the structure of the Higgs sector, particularly non-decoupling nature of extra scalar fields. 
So far, EW corrections to Higgs boson couplings and/or decays have been investigated in models with extended Higgs sectors
such as those with extra 
singlets~\cite{Kanemura:2015fra,Kanemura:2017wtm,Kanemura:2016lkz,He:2016sqr,Kanemura:2018yai,Kanemura:2019kjg}, 
doublets~\cite{Arhrib:2003ph,Arhrib:2016snv,Kanemura:2004mg,Kanemura:2014dja,Kanemura:2015mxa,Kanemura:2017wtm,Kanemura:2018yai,Kanemura:2019kjg,Gu:2017ckc,Chen:2018shg,Han:2020lta,LopezVal:2010vk,Castilla-Valdez:2015sng,Xie:2018yiv, Altenkamp:2017ldc,Altenkamp:2017kxk,Altenkamp:2018bcs,Kanemura:2016sos,Arhrib:2015hoa} and triplets~\cite{Kanemura:2012rs,Aoki:2012yt,Aoki:2012jj,Chiang:2017vvo,Chiang:2018xpl}. 
Therefore, calculations with both QCD and EW corrections are quite important for the precision measurements in near future, and 
several numerical tools have been available; e.g., {\tt H-COUP}~\cite{Kanemura:2017gbi,Kanemura:2019slf}, {\tt 2HDECAY}~\cite{Krause:2018wmo} and {\tt Prophecy4f}~\cite{Denner:2019fcr}. 

In this paper, we investigate the impact of the combined study of direct searches for new particles at hadron colliders and precision measurements of Higgs boson couplings at future lepton colliders. 
We perform such study including higher-order QCD corrections. 
We consider two Higgs doublet models (THDMs) as a representative extended Higgs model.
The models are one of the well-motivated extensions of the SM, and some of new physics models contain two Higgs doublets, 
such as the minimal supersymmetric extension of the SM~\cite{Haber:1984rc,Gunion:1989we,Djouadi:2005gj}, models for electroweak baryogenesis~\cite{Bochkarev:1990fx,McLerran:1990zh,Turok:1990zg,Turok:1991uc,Funakubo:1993jg,Trodden:1998ym,Basler:2016obg,Basler:2017uxn}, and those for radiative neutrino mass generation~\cite{Zee:1980ai,Ma:2006km,Aoki:2008av,Aoki:2009vf} and so on.
\ken{The parameter regions in the THDMs have been explored by direct
searches for the additional Higgs bosons at the
LEP~\cite{Abdallah:2004wy,Schael:2006cr,Abbiendi:2013hk} and the
LHC~\cite{Aaboud:2017sjh,Aaboud:2017sjh,Aad:2019zwb,Aaboud:2018mjh,Aaboud:2018knk,Aaboud:2017gsl,Aaboud:2017rel,Aaboud:2017cxo,Aaboud:2017cxo,Aaboud:2018cwk,Aaboud:2018gjj,Celis:2013ixa,Dumont:2014wha,Bernon:2014nxa,Craig:2015jba,Bernon:2015qea,Bernon:2015wef,Chowdhury:2017aav,Su:2019dsf,Kling:2020hmi}. 
The prospect at the HL-LHC and the ILC has been studied in Ref.~\cite{Kanemura:2014dea}. 
Furthermore, there are studies which discuss the observed data
for the discovered Higgs boson at the LHC in the THDMs
\cite{Kanemura:2014bqa,Bernon:2015qea,Bernon:2015wef,Chowdhury:2017aav,Haller:2018nnx}.
The signatures of the additional Higgs bosons at the future lepton
colliders have been examined in
Refs.~\cite{Gunion:1988tf,Djouadi:1996ah,Kanemura:2000cw,Moretti:2002pa,Kanemura:2014dea}.}

The observed Higgs boson couplings are consistent with those in the SM under current experimental and theoretical uncertainties~\cite{Aad:2019mbh,Sirunyan:2018koj}, so that 
this fact gives a strong motivation to investigate the  {\it alignment} scenario where the Higgs boson couplings are nearly or exactly SM like.
In the near alignment region, the decays of the extra Higgs bosons into the discovered Higgs boson such as $A\rightarrow Zh$ and $H\rightarrow hh$
can be dominant, and at the same time the discovered Higgs boson couplings can deviate from the SM predictions. 
These decay modes of extra Higgs bosons can be well tested at the HL-LHC~\cite{Cepeda:2019klc}, by which we can set a lower limit on the masses of extra Higgs bosons. 
In addition, we can impose an upper limit on the masses~\cite{Kanemura:2014bqa,Kanemura:2015ska,Blasi:2017zel} 
when deviations of the Higgs boson couplings are found at future lepton colliders.


We show that by utilizing the synergy between the direct search for additional Higgs bosons and the precision measurement of the Higgs boson couplings  
a large portion of the parameter space can be explored in the near alignment region. 
We also show that in the alignment limit; i.e., all the Higgs boson couplings are exactly same as the SM values, plenty of the parameter space still remains even if the mass of the additional Higgs bosons are around the EW scale. This is because most of the Higgs to Higgs decays are prohibited and also there is no upper limit on masses of additional Higgs bosons. 

This paper is organized as follows. 
In Sec.~\ref{sec:model}, we define the THDMs and give the Higgs potential, the kinetic terms and the Yukawa interactions. 
Theoretical constraints from perturbative unitarity and vacuum stability are also discussed. 
Constraints from flavor physics and previous colliders are summarized. 
Sec.~\ref{sec:decay} is devoted to the discussion for decays of the Higgs bosons. 
We first give the analytic expressions of the decay rates with higher-order QCD corrections 
and then numerically show total widths and branching ratios of the Higgs bosons. 
In Sec.~\ref{sec:direct}, we show the excluded region of the parameter space from the direct searches at the LHC Run-II experiments. 
In Sec.~\ref{sec:synergy}, we discuss how the parameter space is widely explored by combining direct searches at the HL-LHC and 
precision measurements of the Higgs boson couplings at future lepton colliders. 
Conclusions are given in Sec.~\ref{sec:conclusion}. 
In Appendix, we present the analytic expressions for the perturbative unitarity and the vacuum stability conditions (Appendix~\ref{sec:app0}) and 
the decay rates of the Higgs bosons at the leading order (LO) (Appendix~\ref{sec:app1}). 

\section{Model}\label{sec:model}

We discuss the THDM, whose Higgs sector is composed of two isospin doublet scalar fields $\Phi_1$ and $\Phi_2$. 
In order to avoid flavor changing neutral currents (FCNCs) at tree level, we impose the $Z_2$ symmetry~\cite{Glashow:1976nt} ($\Phi_1 \to +\Phi_1$, $\Phi_2 \to -\Phi_2$) which can be softly-broken 
by a dimensionful parameter in the Higgs potential.  
The most general Higgs potential under the $Z_2$ symmetry is given by 
\begin{align}
V & =  m_1^2|\Phi_1|^2+m_2^2|\Phi_2|^2 - (m_3^2\Phi_1^\dagger \Phi_2 +\text{h.c.})\notag\\
& +\frac{\lambda_1}{2}|\Phi_1|^4+\frac{\lambda_2}{2}|\Phi_2|^4+\lambda_3|\Phi_1|^2|\Phi_2|^2+\lambda_4|\Phi_1^\dagger\Phi_2|^2
+\left[\frac{\lambda_5}{2}(\Phi_1^\dagger\Phi_2)^2+\text{h.c.}\right],  \label{eq:pot-thdm}
\end{align}
where $m_3^2$ is a soft-breaking parameter of the $Z_2$ symmetry. 
Throughout this paper, we assume CP-conservation in the Higgs sector, so that the $m_3^2$ and $\lambda_5$ parameters are taken to be real. 
It is convenient to define the Higgs basis~\cite{Georgi:1978ri,Donoghue:1978cj,Gunion:2002zf} as 
\begin{align}\label{eq:HiggsBasis}
\left(\begin{array}{c}
\Phi_1\\
\Phi_2
\end{array}\right)&=R(\beta)
\left(\begin{array}{c}
\Phi\\
\Phi'
\end{array}\right),\quad 
R(\theta) = 
\begin{pmatrix}
c_\theta & -s_\theta \\
s_\theta &  c_\theta
\end{pmatrix},
\end{align}
where 
\begin{align}
\Phi = \left(\begin{array}{c}
G^+\\
\frac{v + h_1' + iG^0}{\sqrt{2}}
\end{array}\right),\quad 
\Phi' = \left(\begin{array}{c}
H^+\\
\frac{h_2' + iA}{\sqrt{2}}
\end{array}\right),
  \label{eq:doublets}
\end{align} 
with the rotation angle $\beta$ being determined by $\tan\beta = v_2/v_1$ ($v_i \equiv \sqrt{2}\langle \Phi_i^0\rangle$, $i=1,2$). 
We introduced short-hand notation for trigonometric functions $s_\theta \equiv \sin\theta$ and $c_\theta \equiv \cos\theta$. 
In Eq.~(\ref{eq:doublets}), $\Phi$ contains the vacuum expectation value (VEV) $v \equiv \sqrt{v_1^2 + v_2^2} = (\sqrt{2}G_F)^{-1/2}$ with $G_F$ being the Fermi constant 
and the Nambu-Goldstone bosons $G^\pm$ and $G^0$ which are absorbed into the longitudinal component of the $W^\pm$ and $Z$ boson, respectively. 
On the other hand, $\Phi'$ contains the physical charged Higgs boson $H^\pm$ and the CP-odd Higgs boson $A$. 
Remaining two states $h_1'$ and $h_2'$ are related to the mass eigenstates of the CP-even Higgs bosons via 
\begin{align}
\left(\begin{array}{c}
h_1'\\
h_2'
\end{array}\right)=R(\alpha-\beta)
\left(\begin{array}{c}
H\\
h
\end{array}\right), \label{mixing}
\end{align}
where $h$ can be identified as the discovered Higgs boson with the mass of 125 GeV. 

The squared masses of the physical Higgs bosons are expressed as follows: 
\begin{align}
m_{H^\pm}^2 & = M^2  - \frac{v^2}{2}(\lambda_4+\lambda_5), \label{eq:masses1}\\
m_A^2      & = M^2 - v^2\lambda_5,  \\
m_H^2      & = M_{11}^2c^2_{\beta-\alpha} +  M_{22}^2s^2_{\beta-\alpha} -  M_{12}^2s_{2(\beta-\alpha)},\\ 
m_h^2      & =  M_{11}^2s^2_{\beta-\alpha} +  M_{22}^2c^2_{\beta-\alpha} + M_{12}^2s_{2(\beta-\alpha)},  
\end{align}
where $M^2 \equiv m_3^2/(s_\beta c_\beta)$ and $M_{ij}^2$ are the elements of the squared mass matrix in the basis of ($h_1',h_2'$) given by
\begin{align}
M_{11}^2&=v^2(\lambda_1c^4_\beta+\lambda_2s^4_\beta +2\lambda_{345}s^2_{\beta}c^2_{\beta}),   \\
M_{22}^2&=M^2 + \frac{v^2}{4}s^2_{2\beta}(\lambda_1+\lambda_2-2\lambda_{345}), \label{m22}  \\
M_{12}^2&=\frac{v^2}{2} s_{2\beta}( -\lambda_1c^2_\beta +  \lambda_2s^2_\beta  + \lambda_{345}c_{2\beta}), 
\end{align}
with $\lambda_{345}\equiv \lambda_3+\lambda_4+\lambda_5$. 
The mixing angle $\beta-\alpha$ can also be expressed by these matrix elements as 
\begin{align}
&\tan 2(\beta-\alpha)= -\frac{2M_{12}^2}{M_{11}^2-M_{22}^2}.  \label{eq:b-a}
\end{align} 
We can choose the following six variables as the free parameters: 
\begin{align}
m_H^{},~~m_A^{},~~m_{H^\pm}^{},~~M^2,~~\tan\beta,~~s_{\beta-\alpha},  \label{input:thdm}
\end{align} 
where we define $0 < \beta < \pi/2$ and $0 <\beta-\alpha < \pi$ such that $\tan\beta > 0$ and $0 < s_{\beta-\alpha} \leq 1$. 

\begin{table}[t]
\begin{center}
\begin{tabular}{l||ccccc|ccc}\hline\hline
&$Q_L$&$L_L$&$u_R$&$d_R$&$e_R$&$\zeta_u$ &$\zeta_d$&$\zeta_e$ \\\hline\hline
Type-I &$+$&$+$&
$-$&$-$&$-$&$\cot\beta$&$\cot\beta$&$\cot\beta$ \\\hline
Type-II&$+$&$+$&
$-$
&$+$&$+$& $\cot\beta$&$-\tan\beta$&$-\tan\beta$ \\\hline
Type-X (lepton specific)&$+$&$+$&
$-$
&$-$&$+$&$\cot\beta$&$\cot\beta$&$-\tan\beta$ \\\hline
Type-Y (flipped) &$+$&$+$&
$-$
&$+$&$-$& $\cot\beta$&$-\tan\beta$&$\cot\beta$ \\\hline\hline
\end{tabular}
\caption{$Z_2$ charge assignments in four types of the Yukawa interactions 
and the $\zeta_f$ ($f=u,d,e$) factors appearing in Eq.~(\ref{eq:yuk-thdm}). 
}
\label{tab:z2}
\end{center}
\end{table}

The kinetic terms for the Higgs doublets are written in the Higgs basis as 
\begin{align}
{\cal L}_{\text{kin}} =  |D_\mu \Phi_{1}|^2 + |D_\mu \Phi_{2}|^2
= |D_\mu \Phi|^2 + |D_\mu \Phi'|^2. \label{eq:kin}
\end{align} 
The covariant derivative is defined by $D_{\mu}=\partial_{\mu} -ig I^{a}W_{\mu}^{a}-ig^{\prime}YB_{\mu}$,
with the ${\rm SU(2)_{L}}$ generator $I^{a}$ $(a=1$-$3)$ and the hypercharge $Y$, 
from which electric charge $Q$ is derived by $Q=I^{3}+Y$. 
In the expression of $D_{\mu}$, 
$W_{\mu}^{a}\ (g)$ and $B_{\mu} \ (g^{\prime})$ denote the ${\rm SU(2)_{L} }$ and ${\rm U(1)_{Y} }$ gauge bosons (coupling), respectively. 
The $W^{\pm}$ bosons and the neutral gauge bosons are then identified as $W_{\mu}^{\pm}=(W^{1}_{\mu}\mp iW^{2}_{\mu})/\sqrt{2}$ and $(Z_{\mu},~A_{\mu})^{T}=R(\theta_{W})(W_{\mu}^{3},B_{\mu})^{T}$, respectively, with $\theta_{W}$ being the Weinberg angle. 
It is clear from Eq.~(\ref{eq:doublets}) that the masses of the $W^\pm$ and $Z$ bosons are given only from the term $|D_\mu \Phi|^2$, which also includes the gauge-gauge-scalar type interactions. 
On the other hand, the term $|D_\mu \Phi^{'}|^2$, contains the scalar-scalar-gauge type interaction terms such as $AZh$. 

Under the $Z_2$ symmetry, the Yukawa interaction terms are expressed as
\begin{align}
{\mathcal L}_Y =
&-Y_{u}{\bar Q}_L\tilde{\Phi}_uu_R^{}
-Y_{d}{\bar Q}_L\Phi_dd_R^{}
-Y_{e}{\bar L}_L\Phi_e e_R^{}+\text{h.c.},
\end{align}
where $\tilde{\Phi}^{}_{u}= i\sigma_2 \Phi^{*}_{u}$ with $\sigma_2$ being the second Pauli matrix, 
and $\Phi_{u,d,e}$ denote $\Phi_1$ or $\Phi_2$. 
We here do not explicitly show the flavor indices. 
Using the Higgs basis Eq.~\eqref{eq:HiggsBasis}, they can be rewritten  as
\begin{align}
{\cal L}_Y
&= -Y_u\bar{Q}_L (\tilde{\Phi} +\zeta_u \tilde{\Phi}') \,u_R^{} - Y_d\bar{Q}_L (\Phi +\zeta_d\Phi')\, d_R^{} - Y_e\bar{L}_L (\Phi +\zeta_e\Phi') \, e_R^{} + \text{h.c.}, \label{eq:yuk-thdm}
\end{align}
with $\tilde{\Phi}^{(\prime)}= i\sigma_2 \Phi^{(\prime)*}$. 
The Yukawa matrices $Y_f$ are related to the mass matrices for fermions by $M_f = Y_fv/\sqrt{2}$ which are diagonalized by unitary transformations of the left and the right handed fermions.
Thanks to the $Z_2$ symmetry, both $\Phi$ and $\Phi'$ are coupled with a common Yukawa matrix $Y_f$, so that FCNCs mediated by the neutral Higgs bosons do not appear at tree level. 
The $\zeta_f$ parameters are fixed by specifying the charges of the $Z_2$ symmetry for fermions, in which  
there are four independent choices of the charge assignments~\cite{Barger:1989fj,Grossman:1994jb,Aoki:2009ha}, the so-called Type-I, Type-II, Type-X and Type-Y as shown in Table~\ref{tab:z2}. 

For the later convenience, we introduce the scaling factors $\kappa_X^\phi$ which are defined by the ratio of the Higgs boson couplings at tree level: 
\begin{align}
\kappa_X^\phi \equiv \frac{g_{\phi X\bar{X}}}{g_{h_{\rm SM} X\bar{X}}},\quad \phi = h,H,A, 
\end{align}
where $h_{\rm SM}$ is the Higgs boson in the SM. 
From the above Lagrangians, the scaling factors can be extracted as follows: 
\begin{align}
\kappa_V^h &= s_{\beta-\alpha},  \quad \kappa_V^H = c_{\beta-\alpha}, \quad \kappa_V^A = 0, \\
\kappa_f^h &= s_{\beta-\alpha} + c_{\beta-\alpha}\zeta_f, \quad \kappa_f^H = c_{\beta-\alpha} - s_{\beta-\alpha}\zeta_f, \quad \kappa_f^A = -2iI_f\zeta_f,
\label{kappa_f}
\end{align}
where $V$ represents $W$ and $Z$, and $I^{3}_f = 1/2~(-1/2)$ for $f = u~(d,e)$. 
For the loop induced couplings $\phi\gamma\gamma$, $\phi Z\gamma$ and $\phi gg$, we define 
$\kappa_{XY}^\phi \equiv \sqrt{\Gamma(\phi \to XY)/\Gamma(\phi \to XY)_{\rm SM}}$ with $\Gamma(\phi \to XY)$ 
being the decay rate of $\phi \to XY$ and $XY = \gamma\gamma$, $Z\gamma$ and $gg$. 
For the charged Higgs bosons $H^\pm$, their Yukawa couplings are expressed as 
\begin{align}
{\cal L}_Y^{H^\pm} = \frac{\sqrt{2}}{v}\left[ \overline{u}
(m_{u} V_{ud} \zeta_u P_L - V_{ud} m_{d} \zeta_d P_R)dH^+
 - \zeta_e\overline{\nu}m_{e}   P_R e H^+
+\text{h.c.} \right], 
    \end{align}
where $P_L$ $(P_R)$ is the projection operator for left- (right-) handed fermions and $V_{ud}$ is the Cabibbo-Kobayashi-Maskawa (CKM) matrix element. 
Here, we also give the scalar trilinear couplings $\lambda_{\phi\phi^{'}\phi^{''}}$ defined by the coefficient of the corresponding Lagrangian term, 
which are relevant to the decay rates discussed in Sec.~\ref{sec:decay}: 
\begin{align}
\lambda_{H^+H^-h }&=\frac{1}{v}\left[(2M^2-2m_{H^\pm}^2-m_h^2)s_{\beta-\alpha}+2(M^2-m_h^2)\cot2\beta c_{\beta-\alpha}  \right],\\
\lambda_{H^+H^-H}&=-\frac{1}{v}\Big[2(M^2-m_H^2)\cot2\beta s_{\beta-\alpha}+(2m_{H^\pm}^2+m_H^2-2M^2)c_{\beta-\alpha}\Big], \\
\lambda_{H^+H^-A}&= 0, \label{lam_HpHmA} \\
\lambda_{Hhh}&=-\frac{c_{\beta-\alpha}}{2v}\Big\{4M^2-2m_h^2-m_H^2 \notag\\
& +(2m_h^2+m_H^2-3M^2)[2c^2_{\beta-\alpha}+s_{\beta-\alpha}c_{\beta-\alpha}(\tan\beta-\cot\beta)]\Big\}, \label{lam_Hhh}
\end{align}
where Eq.~\eqref{lam_HpHmA} is followed from the CP-invariance. 

Let us discuss the important limits of the parameters in the THDM, the decoupling limit and the alignment limit. 
First, the decoupling limit is realized by taking $M \to \infty$, by which all the masses of the additional Higgs bosons 
become infinity, and only $h$ remains at the EW scale\footnote{From Eq.~(\ref{eq:b-a}), $\tan2(\beta-\alpha)$ goes to zero at $M \to \infty$ which corresponds to $\beta-\alpha \to \pi/2$ in our convention. 
In this case, $m_H^2$ and $m_h^2$ are determined only by $M_{22}^2$ and $M_{11}^2$, respectively. }. 
In this limit, new physics effects on low energy observables disappear due to the decoupling theorem~\cite{Appelquist:1974tg,Gunion:2002zf}. 
Second, on the other hand, the alignment limit can be defined by taking $s_{\beta-\alpha} \to 1$, in which the 
$h_1'$ state in $\Phi$ coincides with the mass eigenstate $h$, and $\kappa_V^h = \kappa_f^h = 1$ is satisfied at tree level. 
We note that this limit is automatically realized in the decoupling limit. 
Here, the important thing is that if $\kappa_V^h \neq 1$ and/or $\kappa_f^h \neq 1$ are found at future collider experiments, we cannot take the decoupling limit.
This provides us a {\it new no-loose theorem}~\cite{Kanemura:2014bqa,Kanemura:2015fra,Blasi:2017zel}, where we can extract the upper bound on the mass scale of the second Higgs boson\footnote{An original no-loose theorem was discussed for the SM in Ref.~\cite{Lee:1977eg}.}. 
Quantitatively, such a bound is given by imposing the constraints from perturbative unitarity and vacuum stability as we will discuss them below. 
Notice here that the inverse of the above statement does not hold in general, namely, {\it alignment without decoupling} can be considered. 
Such scenario is well motivated by; e.g., the successful EW baryogenesis~\cite{Bochkarev:1990fx,McLerran:1990zh,Turok:1990zg,Turok:1991uc,Funakubo:1993jg,Trodden:1998ym,Basler:2016obg,Basler:2017uxn}. 

As mentioned above, we take into account the perturbative unitarity and the vacuum stability bounds. 
For the unitarity bound, we impose $|a_i^{}| \leq 1/2$, where 
$a_i$ are independent eigenvalues of the $s$-wave amplitude matrix for two-body to two-body scattering processes in the high-energy limit~\cite{Kanemura:1993hm,Akeroyd:2000wc,Ginzburg:2005dt,Kanemura:2015ska}. 
The analytic expressions for $a_i$ are given in Appendix~\ref{sec:app0}. 
In this limit, due to the equivalence theorem~\cite{Cornwall:1974km}, only the contact scalar interaction terms contribute to the $s$-wave amplitude, which can be written in terms of the scalar quartic couplings. 
Thus, the unitarity bound gives constraints on the masses of additional Higgs bosons and the mixing angle through the relations given in Eqs.~(\ref{eq:masses1})--(\ref{eq:b-a}), see also Eqs.~\eqref{eq:lam1}--\eqref{eq:lam45}. 
On the other hand, the vacuum stability is the requirement that the Higgs potential is bounded from below in any direction with large field values. 
The sufficient and necessary conditions are given in Refs.~\cite{Deshpande:1977rw,Klimenko:1984qx,Sher:1988mj,Nie:1998yn}. 
We give a comment on the true vacuum condition. 
The Higgs potential can have several extrema besides the EW true vacuum. 
In such a case, we need to ensure that the true vacuum is the deepest vacuum than all the other ones. 
In Ref.~\cite{Barroso:2013awa}, it has been shown that most of the parameter regions with $M^2 < 0$ are excluded by the true vacuum condition. 
Thus, throughout the paper we simply assume $M^2$ to be a positive value in order to satisfy the true vacuum condition. 

Before closing this section, we briefly mention constraints from various flavor observables which particularly sensitive to the 
mass of the charged Higgs bosons.
The comprehensive studies of these constraints in the $Z_{2}$ symmetric THDMs have been carried out in Refs.~\cite{Enomoto:2015wbn,Haller:2018nnx}.
In Type-II, the $B\rightarrow X_{s}\gamma$ process gives the lower bound of $m_{H^{\pm}}\gtrsim 800~{\rm GeV}$ at 95\% confidence level (CL) almost independently of the value of $\tan{\beta}$ for $\tan{\beta}\gtrsim 2$~\cite{Misiak:2020vlo}.
On the other hand in Type-I, the severe constraint on $m_{H^{\pm}}$ is given particularly for smaller $\tan\beta$; e.g., 
$m_{H^{\pm}}\gtrsim 450~{\rm GeV}$ for $\tan{\beta}=1$~\cite{Misiak:2017bgg}. 
However, above $\tan{\beta}\simeq 2$, the bound becomes weaker than the lower bound from the direct search at LEP; i.e., $m_{H^{\pm}}\gtrsim 80~{\rm GeV}$ \cite{Abbiendi:2013hk}.
Because the lepton Yukawa couplings are irrelevant to the $B\rightarrow X_{s}\gamma$ process, similar bounds given in Type-I and Type-II can be obtained in Type-X and Type-Y, respectively.
In Type-II, $B\rightarrow \tau\nu$ and $B_{s}\rightarrow \mu\bar\mu$ processes 
give an upper limit on $\tan\beta$; e.g., $\tan\beta\gtrsim20$ for $m_{H^{\pm}}=800~{\rm GeV}$~\cite{Haller:2018nnx}. 
In Type-X, constraint by $\tau \rightarrow \mu\nu\bar{\nu}$ becomes important for large $\tan{\beta}$ \cite{Aoki:2009ha,Krawczyk:2004na,Abe:2015oca}.
In the small $\tan{\beta}$ region, the neutral meson mixing processes $B^{0}-\bar{B}^{0}$ give a stronger bounds for $m_{H^{\pm}}$ compared to the bound from $B\rightarrow X_{s}\gamma$, and these exclude the wide region in all the types of THDMs.

\newpage
\section{Decays of the Higgs bosons}\label{sec:decay}

In this section, we give the analytic expressions for the decay rates of the Higgs bosons including higher-order corrections in QCD. 
In addition, some numerical results for the decays of the Higgs bosons are shown. 

\subsection{Running parameters}

We give the expressions for the running strong coupling $\alpha_s(\mu)$ and the running quark masses $\overline{m}_q(\mu)$ at the scale $\mu$ in the $\overline{\text{MS}}$ scheme. 
In order to compute these variables, we need the coefficients of the $\beta$ function for $\alpha_s(\mu)$ and those of the anomalous dimension for $\overline{m}_q(\mu)$.
Their formulae at the three-loop level are given by~\cite{Djouadi:2005gi,Gorishnii:1991zr} 
\begin{align}
&\beta_0 = 11- \frac{2}{3}N_{f},\quad 
\beta_1 = 51- \frac{19}{3}N_{f},\quad 
\beta_2 = 2857- \frac{5033}{9}N_{f} + \frac{325}{27}N_{f}^2,\\
&\gamma_0 = 4,\quad 
\gamma_1 = \frac{101}{3} - \frac{10}{9}N_{f},\quad 
\gamma_2 = 2498 - \left[\frac{4432}{27} + \frac{320}{3}\zeta(3)\right]N_{f} - \frac{280}{81}N_{f}^2, 
\end{align}
with $N_{f}$ being the number of active flavors and with $\zeta (n)$ indicating the Riemann zeta function.  
The running strong coupling $\alpha_{s}$ at the scale $\mu$ is expressed as 
\begin{align}
\alpha_{s}(\mu) = \frac{4\pi}{\beta_0\ell_\mu}\left\{ 1 - \frac{2\beta_1}{\beta_0^2}\frac{\ln \ell_\mu}{\ell_\mu} + \frac{4\beta_1^2}{\beta_0^4 \ell_\mu^2}\left[\left(\ln \ell_\mu -\frac{1}{2} \right)^2 
+ \frac{\beta_2\beta_0}{8\beta_1^2} -\frac{5}{4} \right] \right\}, 
\end{align}
 where $\ell_\mu = \ln(\mu^2/\Lambda_{\rm QCD}^2)$, and $\Lambda_{\rm QCD}$ is the asymptotic scale parameter \cite{Chetyrkin:1997sg}.
The running quark mass at the scale of the pole mass $m_{q}$ is given by \cite{Gray:1990yh, Chetyrkin:1999ys, Chetyrkin:1999qi, Melnikov:2000qh} 
\begin{align}
\overline{m}_q(m_q) = m_q&\Bigg[1 -\frac{4}{3}\frac{\alpha_s(m_q)}{\pi} + (1.0414N_{f} -14.3323)\frac{\alpha_s^2(m_q)}{\pi^2} \notag\\
&+(-0.65269 N_{f}^2 + 26.9239N_{f} -198.7068)\frac{\alpha_s^3(m_q)}{\pi^3} \Bigg]. 
\end{align}
The running quark mass at the scale $\mu$ is expressed as 
\begin{align} \label{eq:rmq}
\overline{m}_q(\mu) = \overline{m}_q(m_q)\times\frac{c[\alpha_s(\mu)/\pi]}{c[\alpha_s(m_q)/\pi]},
\end{align}
where the function $c(x)$ is given by~\cite{Chetyrkin:1997dh, Vermaseren:1997fq}
\begin{align}
c(x) = x^{\bar{\gamma}_0} \left\{1 + (\bar{\gamma}_1 - \bar{\beta}_1\bar{\gamma}_0)x + \frac{1}{2}[(\bar{\gamma}_1 -\bar{\beta}_1\bar{\gamma}_0)^2 + \bar{\gamma}_2 + \bar{\beta}_1^2 \bar{\gamma}_0 - \bar{\beta}_1 \bar{\gamma}_1 - \bar{\beta}_2\bar{\gamma}_0]x^2\right\}, 
\end{align}
with $\bar{\beta}_i = \beta_i/\beta_0$ and $\bar{\gamma}_i = \gamma_i/\beta_0$. 

If the renormalization group (RG) evolution crosses the flavor threshold, we need to take into account the matching condition of the running strong coupling \cite{Chetyrkin:1997un} and the running quark mass \cite{Bernreuther:1981sg, Bernreuther:1983zp}
\begin{align}
\alpha_{s}^{(N_{f}-1)}(\mu) &= \zeta_{g}^{2}\alpha_{s}^{(N_{f})}(\mu), \\
\overline{m}_{q}^{(N_{f}-1)}(\mu) &= \zeta_{m}\overline{m}_{q}^{(N_{f})}(\mu),
\end{align}
where $\zeta_{g}$ and $\zeta_{m}$ are the matching coefficients. We note that the matching coefficients are unity up to NLO, and we use $\zeta_{g}=\zeta_{m}=1$ in the following. 
For example, the running charm quark mass at $m_h$ can be evaluated as
\begin{align}
\overline{m}_{c}(m_h) = \frac{c^{(5)}[\alpha_s(m_h)/\pi]}{c^{(5)}[\alpha_s(m_{b})/\pi]}\times \frac{c^{(4)}[\alpha_s(m_{b})/\pi]}{c^{(4)}[\alpha_s(m_{c})/\pi]}\times \overline{m}_{c}(m_{c}). 
\end{align}


\subsection{QCD corrections to the neutral Higgs decays}

In the following, we describe how to include QCD corrections for processes of the neutral Higgs bosons $\phi~(=h,H,A)$ in our calculations. 
For the decay rates of $h$, we adopt the formulae of incorporating those QCD corrections in {\tt H-COUP~v2}~\cite{Kanemura:2019slf}. 

The decay rate into a pair of light quarks ($q \neq t$) including next-to-next-to-leading order (NNLO) QCD corrections in the $\overline{\rm MS}$ scheme 
is given by~\cite{Mihaila:2015lwa, Gorishnii:1990zu,Gorishnii:1991zr,Chetyrkin:1995pd,Larin:1995sq} 
\begin{align}
\Gamma(\phi \to q\bar{q}) &=  \Gamma_0(\phi \to q\bar{q})(1 + \Delta_q^\phi), 
\end{align}
where
\begin{align}\label{eq:deltaq}
\Delta_q^\phi = \frac{\alpha_s(\mu)}{\pi}C_F\left(\frac{17}{4} + \frac{3}{2}\ln \frac{\mu^2}{m_\phi^2}\right)
+ \left(\frac{\alpha_s(\mu)}{\pi}\right)^2(35.94 - 1.36N_f^{})
+ \Delta_{\textrm{t-loop}}^{\phi},   
\end{align}
with the color factor $C_F = 4/3$. 
The last term $\Delta_\textrm{t-loop}^{\phi}$ indicates top-quark loop contributions, which calculated in the case with $m_t^{} \gg m_\phi^{}$ and $\mu=m_\phi^{}$ as 
\begin{align}
 &\Delta_{\textrm{t-loop}}^{H} = \frac{\kappa_t^{H}}{\kappa_q^{H}}\left(\frac{\alpha_s(\mu)}{\pi}\right)^2\left( 1.57 -\frac{2}{3}\ln\frac{m_H^2}{m_t^2} + \frac{1}{9}\ln^2\frac{\overline{m}_q^2(\mu)}{m_H^2} \right), \\
 &\Delta_{\textrm{t-loop}}^{A} = \frac{\kappa_t^{A}}{\kappa_q^{A}}\left(\frac{\alpha_s(\mu)}{\pi}\right)^2\left( 3.83 - \ln\frac{m_A^2}{m_t^2} + \frac{1}{6}\ln^2\frac{\overline{m}_q^2(\mu)}{m_A^2} \right).   
\end{align}
In the LO decay rate $\Gamma_{0}$, mass parameters arising from Yukawa couplings are replaced by the running masses $\overline{m}_q(\mu)$.
Thereby, large logarithmic corrections  induced by the light quark masses are resummed~\cite{Braaten:1980yq}.

For the top pair, the QCD correction factor $\Delta_t^{\phi}$ depends on the CP property of the Higgs boson. 
We obtain the decay rate at the NLO in the on-shell scheme as 
\begin{align}
\Gamma(\phi \to t\bar{t}) &=  \Gamma_0(\phi \to t\bar{t})(1 + \Delta_t^\phi), 
\end{align}
where~\cite{Drees:1989du,Djouadi:2005gj} 
\begin{align}
\Delta_t^{H} &=  \frac{\alpha_s(\mu)}{\pi}C_F \left[\frac{L(\beta_t)}{\beta_t} - \frac{1}{16\beta_t^3}(3+34\beta_t^2 -13\beta_t^4)\ln \rho_t + \frac{3}{8\beta_t^2}(7\beta_t^2-1) \right], \\
\Delta_t^A &= \frac{\alpha_s(\mu)}{\pi}C_F \left[\frac{L(\beta_t)}{\beta_t} - \frac{1}{16\beta_t^3}(19+ 2\beta_t^2 + 3\beta_t^4)\ln \rho_t + \frac{3}{8}(7-\beta_t^2)\right],  
\end{align}
with $\beta_t = \lambda^{1/2}(m_t^2/m_\phi^2,m_t^2/m_\phi^2)$ and $\rho_t = (1 - \beta_t)/(1 + \beta_t)$, 
where the function $\lambda$ is defined in Appendix B.  
The function $L(\beta_t)$ is given by 
\begin{align}
L(\beta_t) &= (1 + \beta_t^2) \left[4\text{Li}_2(\rho_t) + 2\text{Li}_2(-\rho_t) + 3\ln \rho_t \ln \frac{2}{1 + \beta_t} + 2\ln \rho_t \ln \beta_t \right] \notag\\
 &- 3\beta_t \ln \frac{4}{1-\beta_t^2} -4\beta_t\ln \beta_t, 
\end{align}
where $\text{Li}_2$ is the dilog function. 
In the chiral limit $\beta_t \to 1$, we obtain 
\begin{align}
\Delta_t^{\phi} &= \frac{\alpha_s(\mu)}{\pi}C_F\left(\frac{9}{4} + \frac{3}{2}\ln \frac{m_t^2}{m_\phi^2} \right). 
\end{align}
Contributions of the top quark mass in the NLO QCD corrections are significant near the threshold region. 
On the other hands, dominant contributions in  $m_{\phi}\gg m_{t}$ can be the logarithmic contribution, $\ln (m_t^2/m_\phi^2)$, which appears in the QCD corrections in the $\overline{\rm MS}$ scheme. 
In order to take into account both of the effects, we use interpolation for the corrections to $\phi \to t\bar{t}$ as discussed in Ref.~\cite{Djouadi:1997yw}. 

For the decays into an off-shell gauge boson $\phi \to VV^*$ and $\phi \to \phi V^*$ ($V = W,Z$), the QCD correction can enter in the $V^* \to q\bar{q}$ part. This effect can be included by~\cite{Albert:1979ix} 
\begin{align}
\Gamma(\phi \to VV^*\to Vq\bar{q})      &= \Gamma_0(\phi \to VV^*\to Vq\bar{q}) (1 + \Delta_{\rm QCD}), \\
\Gamma(\phi \to \phi' V^* \to \phi' q\bar{q}) &= \Gamma_0(\phi \to \phi' V^*\to \phi' q\bar{q})(1 + \Delta_{\rm QCD}), 
\end{align}
where 
\begin{align}\label{eq:deltaQCD}
\Delta_{\rm QCD} = C_F\frac{3\alpha_s(\mu)}{4\pi}. 
\end{align}

The fermion loop contribution to the decay rate of $\phi \to \gamma\gamma$ receives QCD corrections. 
At the NLO, the QCD correction can be implemented by the following replacement of the quark loop function $I_F^\phi(\tau_q^{})$ in the $\overline{\rm MS}$ scheme~\cite{Dawson:1993qf,Spira:1995rr}
\begin{align}
 I_F^\phi(\tau_q) \to I_F^\phi(\tau_q) \left[ 1+\frac{\alpha_s(\mu)}{\pi}C_\phi^{}\right], \label{eq:QCD_gamgam1}
\end{align}
where $I_F^\phi(\tau_q)$ is defined in Appendix B, and the factor $C_\phi$ is determined by the scale $\mu$ and the mass ratio $\tau_q \equiv m_\phi^2/(4m_q^2)$. 
In our computation, we adopt the analytic expression of $C_\phi^{}$ given in Ref.~\cite{Harlander:2005rq}, in which 
$C_\phi^{}$ is written in terms of the polylog functions, up to the $\text{Li}_4$ function.  
It has been known that the factor $C_\phi$ becomes the simple form in the large top mass limit, $\tau_t \to 0$, as~\cite{Dawson:1993qf,Spira:1995rr,Steinhauser:1996wy}
\begin{align}
&C_{H}^{} = -1,\quad C_{A}^{} = 0 \label{eq:QCD_gamgam2}. 
\end{align}
On the other hand, in the large Higgs mass limit or equivalently the massless fermion limit, the factor $C_\phi$ is common to the case for the CP-even and CP-odd Higgs boson~\cite{Spira:1995rr}: 
\begin{align}
\text{Re}I_F^\phi C_{\phi}^{} = -\frac{1}{18}\left[\ln^2(4\tau_q )-\pi^2 -\frac{2}{3}\ln (4\tau_q) + 2\ln \frac{\mu^2}{m_q^2}\right],\ 
\text{Im}I_F^\phi C_{\phi}^{}  = \frac{\pi}{3}\left[\frac{1}{3}\ln(4\tau_q) + 2\right].
\end{align}
For $H/A \rightarrow Z\gamma$ decays, we calculate them at the LO.

For the $\phi \to gg$ decays, we take into account the decay rate corrected up to NNLO expressed as, 
\begin{align}
\Gamma(\phi \to gg) = \Gamma_0(\phi \to gg)\left[1 + \frac{\alpha_s(\mu)}{\pi}  E_\phi^{(1)} 
 + \left(\frac{\alpha_s(\mu)}{\pi}\right)^2 E_\phi^{(2)} \right].  \label{eq:H->gg}
\end{align}
For the NLO QCD corrections to the $\phi \to gg$ decays, 
there are contributions from virtual gluon loops and those from real emissions of a gluon ($\phi \to ggg$) and a gluon splitting into quark pair ($\phi \to gq\bar{q}$). 
$E_\phi^{(1)}$ in Eq.~(\ref{eq:H->gg}) can be decomposed as~\cite{Spira:1995rr},
\begin{align}
E_\phi^{(1)} & = E_{\phi}^{\rm virt}({m_t \to \infty}) + E_{\phi}^{\rm real}(m_t \to \infty )  + \Delta E_{\phi}. 
\end{align}
The first and second terms respectively denote the contribution from virtual gluon loops and that from real gluon emissions in the large top-quark mass limit. 
These are expressed by 
\begin{align}
E_{H}^{\rm virt}({m_t \to \infty}) & = \frac{11}{2}+ \frac{33-2N_f}{6}\ln\frac{\mu^2}{m_{H}^2},\\
E_{A}^{\rm virt}({m_t \to \infty})      &= 6 + \frac{33-2N_f}{6}\ln\frac{\mu^2}{m_A^2}, \\
E_{H}^{\rm real}({m_t \to \infty}) &= E_A^{\rm real}({m_t \to \infty}) = \frac{73}{4} - \frac{7}{6}N_f. \label{eq:ephi}
\end{align}
The last term $\Delta E_{\phi}$ vanishes in the large top-quark mass limit, which can be decomposed into 
the following three parts: 
\begin{align}
\Delta E_{\phi} = \Delta E_{\phi}^{\rm virt} + \Delta E_{\phi}^{ggg} + N_f \Delta E_{\phi}^{gq\bar{q}}. 
\end{align}
Similar to the $\phi \to \gamma\gamma$ decays, we adopt the analytic expression for the virtual correction $\Delta E_{\phi}^{\rm virt}$ given in Ref.~\cite{Harlander:2005rq}. 
Those for the real emissions $\Delta E_{\phi}^{ggg}$ and $\Delta E_{\phi}^{gq\bar{q}}$ are given in Ref.~\cite{Spira:1995rr}, which 
are expressed in the form with a double integral with respect to phase space variables. 
According to Ref.~\cite{Spira:1995rr}, the factor $\Delta E_{\phi}$ is dominantly determined by the contribution from the virtual gluon loop $\Delta E_{\phi}^{\rm virt}$, 
so that in our computation we neglect the contributions from $\Delta E_{\phi}^{ggg}$ and $\Delta E_{\phi}^{gq\bar{q}}$. 
From Eq.~(\ref{eq:ephi}), $E_\phi$ is given to be about 18 at $\mu = m_\phi $ and $N_f = 5$, and it gives sizable correction to the decay rate; e.g., $\sim 70\%$ for $m_\phi = 100$ GeV. 
For NNLO contributions; i.e., $E_\phi^{(2)}$, we incorporate those in the limit with $m_t \gg m_\phi^{}$ and setting as $\mu=m_\phi^{}$, which are expressed as \cite{Chetyrkin:1997iv, Chetyrkin:1998mw}
\begin{align}
 E_{H}^{(2)} &= \frac{149533}{288} -\frac{363}{8}\zeta (2) -\frac{495}{8}\zeta (3) -\frac{19}{8}\ln\frac{m_t^2}{m_{H}^2} \notag\\
 &+N_f^{}\left(-\frac{4157}{72} +\frac{11}{2}\zeta (2)+\frac{5}{4}\zeta (3) -\frac{2}{3}\ln\frac{m_t^2}{m_{H}^2} \right)
    +N_f^2\left(\frac{127}{108} -\frac{1}{6}\zeta (2)  \right) ,\label{eq:E_H^2}\\
 E_A^{(2)} &=\frac{51959}{96} -\frac{363}{8}\zeta (2) -\frac{495}{8}\zeta (3) 
               +N_f^{}\left(-\frac{473}{8} +\frac{11}{2}\zeta (2) +\frac{5}{4}\zeta (3) -\ln\frac{m_t^2}{m_A^2} \right) \notag\\
        &+ N_f^2\left( \frac{251}{216} -\frac{\zeta (2)}{6} \right). 
\end{align}

\subsection{QCD corrections to the charged Higgs decays}
The QCD corrections to charged Higgs decays into light quarks are presented  in the $\overline{\rm MS}$ scheme. 
The expression can be written in the same way with the neutral Higgs boson decays as
\begin{align}
\Gamma(H^{\pm}\to qq^{\prime})=\Gamma_{0}(H^{\pm}\to qq^{\prime})(1+\Delta^{H^{\pm}}_{q}),
\end{align}
where the $\Delta^{H^{\pm}}_{q}$  is given by Eq.~\eqref{eq:deltaq} but without the last term $\Delta_{\textrm{t-loop}}^{\phi}$.
For the the decays into quarks including the top quark, we apply the QCD correction in the on-shell scheme. 
It is given in~\cite{Djouadi:1994gf,Djouadi:2005gj}
\begin{align}
\Gamma(H^{\pm}\to qq^{\prime})&=
\frac{ 3G_{F}m_{H^{\pm}} }{4\sqrt{2}\pi }|V_{qq^{\prime}}|^{2}
\lambda_{qq{'}}^{1/2} 
 \Bigg[ \left(1-\mu_{q}-\mu_{q^{\prime}} \right)
\Bigg\{ m_{q}^{2}\zeta_{q}^{2} \left(1+C_{F}\frac{\alpha_{s}(\mu)}{\pi}\Delta^{+}_{qq{'}}\right) \notag \\
&+m_{q^{\prime}}^{2}\zeta_{q^{\prime}}^{2} \left(1+C_{F}\frac{\alpha_{s}(\mu)}{\pi}\Delta^{-}_{qq{'}}\right)\Bigg\} 
-4\sqrt{\mu_{q} \mu_{q^{\prime}} } m_{q}m_{q^{\prime}}\zeta_{q}\zeta_{q^{\prime}}\Bigg],
\end{align}
where $\mu_{q}=m_{q}^{2}/m^{2}_{H^{\pm}}$, $\lambda_{qq^{\prime}}=(1-\mu_{q}-\mu_{q^{\prime}})^{2}-2\mu_{q}\mu_{q^{\prime}}$. 
The QCD corrections $\Delta^{+}_{qq^{'}}$ and $\Delta^{-}_{qq^{'}}$ are expressed by 
 \begin{align}
\Delta_{{qq^\prime}}^{+}&=\frac{9}{4}+\frac{3-2\mu_q+2\mu_{q^\prime}}{4}\ln\frac{\mu_q}{\mu_{q^\prime}}+\frac{(\frac{3}{2}-\mu_q-\mu_{q^\prime})\lambda_{qq^\prime}+5\mu_q\mu_{q^\prime}}{2\lambda_{qq^\prime}^{1/2}(1-\mu_q-\mu_{q^\prime})}\ln x_{q}x_{q^{\prime}}+B_{qq^\prime}, \\
\Delta_{{qq^\prime}}^{-}&=3+\frac{\mu_{q^\prime}-\mu_q}{2}\ln\frac{\mu_q}{\mu_{q^\prime}}+\frac{\lambda_{qq^\prime}+2(1-\mu_q-\mu_{q^\prime})}{2\lambda_{qq^\prime}^{1/2}}\ln x_{q}x_{q^{\prime}}+B_{qq^\prime},
 \end{align}
where  $x_{q}=2\mu_{q}/(1-\mu_{q}-\mu_{q^\prime}+\lambda_{qq^{\prime}}^{1/2})$. 
A function $B_{qq^{\prime}}$ is given in Ref.~\cite{Djouadi:2005gj}. 
In these expressions quark pole masses are used. 
Similar to $\phi \to t\bar{t}$, we incorporate the corrections with interpolation to consider the effect of the top quark mass and the logarithmic corrections due to  light down-type quark masses. 

For the off-shell decays into a neutral Higgs boson and  a W boson, $H^{\pm}\to \phi W^{\ast}$, the QCD correction can be applied as similar to $\phi \to \phi' V^{\ast}$.
It can be written as
\begin{align}
\Gamma(H^{\pm}\to \phi W^{\pm  \ast} \to \phi qq^{\prime} ) = 
\Gamma_{0}(H^{\pm}\to \phi W^{\pm \ast} \to \phi qq^{\prime})(1+\Delta_{\rm QCD}), 
\end{align}
where the QCD correction factor is given in~Eq.~\eqref{eq:deltaQCD}. 
For loop induced decay processes of the charged Higgs bosons, $H^{\pm}\to W^{\pm }V\ (V=Z,\gamma)$, which have been studied in Refs.~\cite{CapdequiPeyranere:1990qk,Kanemura:1997ej,DiazCruz:2001tn,HernandezSanchez:2004tq,Arhrib:2006wd,Abbas:2018pfp}, we calculate them at the LO.

\subsection{Total decay widths and decay branching ratios}
\label{Total widths and branching ratios}
\begin{figure}[t]
\includegraphics[width=\linewidth]{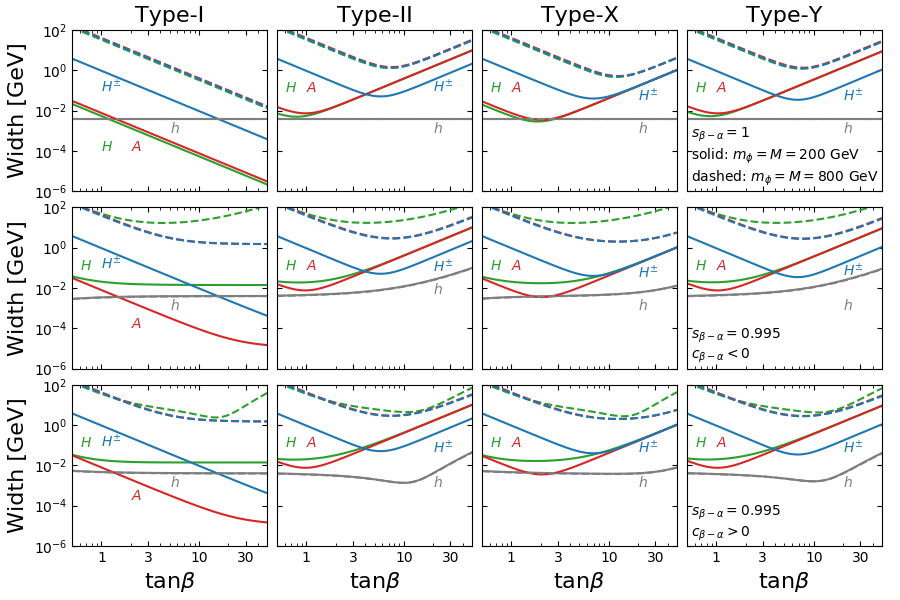}
\caption{
Total widths of $h$, $H$, $A$ and $H^\pm$ as a function of $\tan\beta$ in Type-I, Type-II, Type-X and Type-Y of the THDM from the left panels to the right panels. 
Solid lines and dashed lines show results  of $m_{\Phi}=M=200$ GeV and  $m_{\Phi}=M=800$ GeV, respectively. 
In the top panels, $\sba$ is set to be 1.
In the middle and bottom panels, $\sba$ is set to be 0.995 with $\cba<0$ and $\cba>0$, respectively. 
}
\label{FIG:width}
\end{figure}
\begin{figure}[t]
\includegraphics[width=\linewidth]{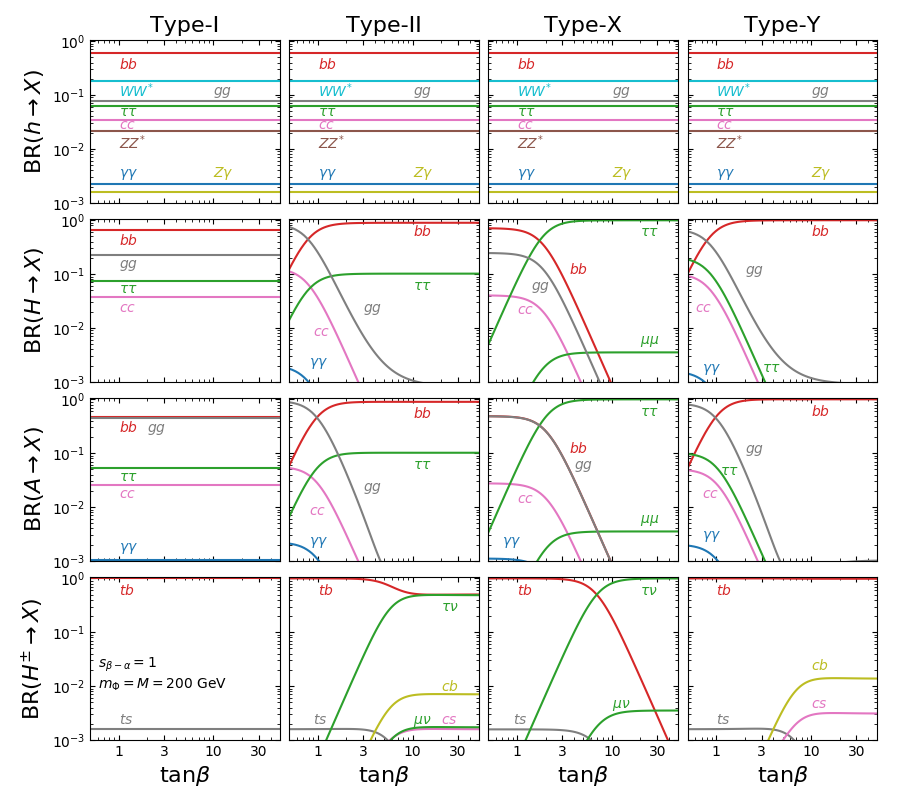}
\caption{
Decay branching ratios for $h$, $H$, $A$ and $H^\pm$ as a function of $\tan\beta$ in the case of $m_{\Phi}=M=200$ GeV and $\sba=1$. 
Results for Type-I, Type-II, Type-X and Type-Y of the THDM are shown from the left panels to the right panels. 
}
\label{FIG:BR1}
\end{figure}
\begin{figure}[t]
\includegraphics[width=\linewidth]{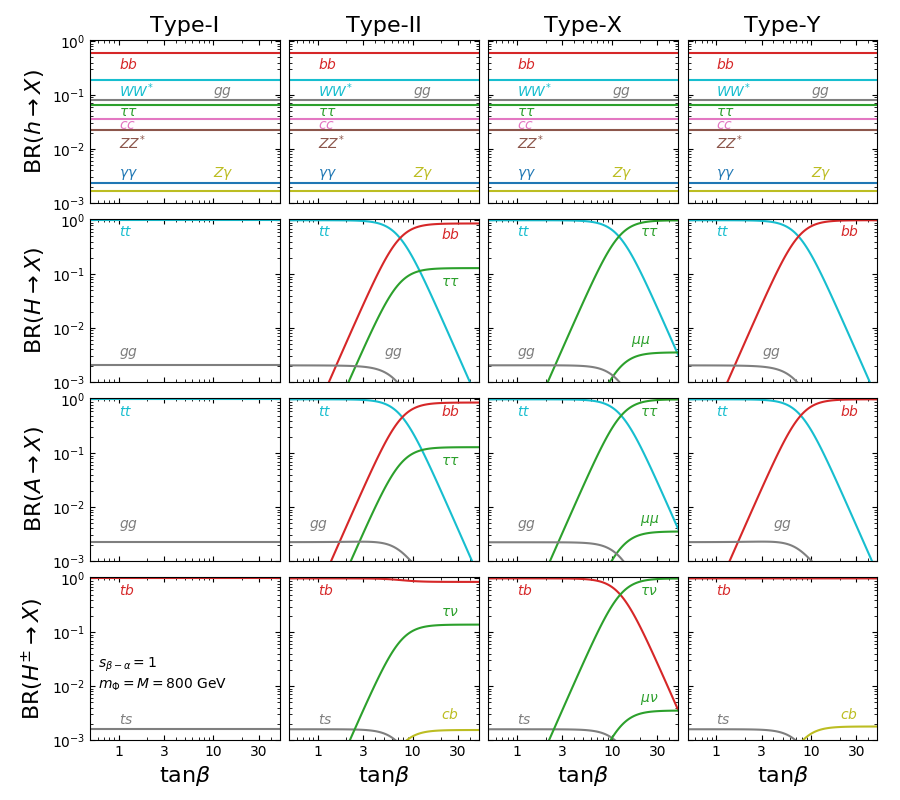}
\caption{
Decay branching ratios for $h$, $H$, $A$ and $H^\pm$ as a function of $\tan\beta$ in the case of $m_{\Phi}=M=800$ GeV and $\sba=1$. 
Results for Type-I, Type-II, Type-X and Type-Y of the THDM are shown from the left panels to the right panels. 
}
\label{FIG:BR2}
\end{figure}
\begin{figure}[t]
\includegraphics[width=\linewidth]{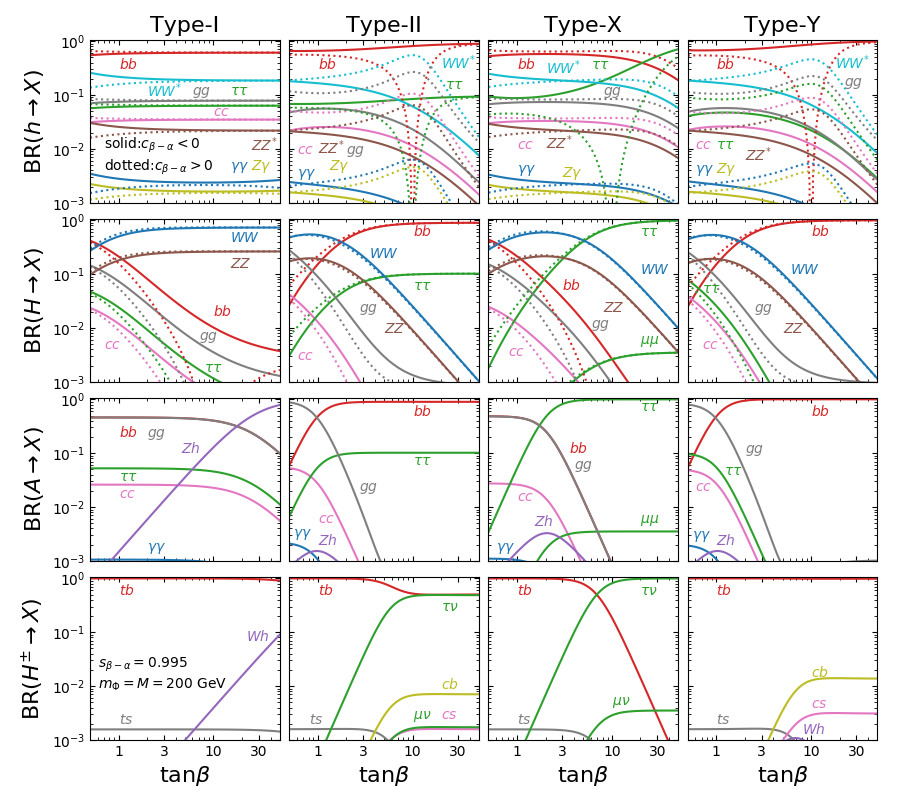}
\caption{
Decay branching ratios for $h$, $H$, $A$ and $H^\pm$ as a function of $\tan\beta$ in the case of $m_{\Phi}=M=200$ GeV and $\sba=0.995$. Solid lines show results of $c_{\beta-\alpha}<0$ and dotted lines are those of $c_{\beta-\alpha}>0$. 
Results for Type-I, Type-II, Type-X and Type-Y of the THDM are shown from the left panels to the right panels. 
}
\label{FIG:BR3}
\end{figure}
\begin{figure}[t]
\includegraphics[width=\linewidth]{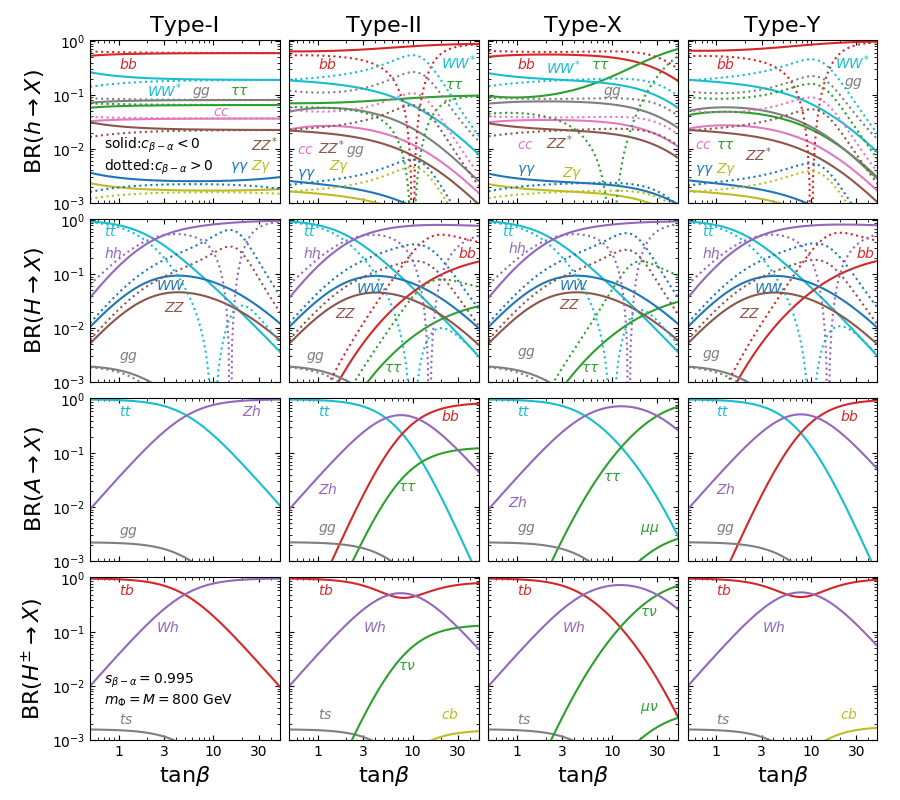}\hspace{0mm}
\caption{
Decay branching ratios for $h$, $H$, $A$ and $H^\pm$ as a function of $\tan\beta$ in the case of $m_{\Phi}=M=800$ GeV and $\sba=0.995$. Solid lines show results of $c_{\beta-\alpha}<0$ and dotted lines are those of $c_{\beta-\alpha}>0$. 
Results for Type-I, Type-II, Type-X and Type-Y of the THDM are shown from the left panels to the right panels. 
}
\label{FIG:BR4}
\end{figure}

We here discuss total widths and branching ratios for the neutral Higgs bosons and the charged Higgs bosons in four types of the THDMs in order for later discussion about direct searches of heavy Higgs bosons.   
We describe the behavior of the total widths and the branching ratios in cases with the alignment limit, $s_{\beta-\alpha}=1$ and without taking  the alignment limit, $\sba=0.995$. 
In the numerical computations, we use the beta version of {\tt H-COUP~v3}~\cite{H-COUPv3}, where the QCD corrections presented in previous subsections are included. 
In the QCD correction functions $C_{\phi}^{}$ and $E_{\phi}^\textrm{virt}$, polylog functions appear. 
We use {\tt CHAPLIN}~\cite{Buehler:2011ev} for the numerical evaluation of such polylog functions.  
We have confirmed that our numerical results for the total widths and the branching ratios are consistent with {\tt 2HDMC}~\cite{Eriksson:2009ws}. 

We here show the case that masses of the additional Higgs bosons as well as $M$ are degenerate; i.e., $m_{\Phi} \equiv m_{H}=m_{A}=m_{H^{\pm}}$ and $M = m_\Phi$. 
While the $m_{\Phi}$ is set to be $m_{\Phi}= 200~{\rm GeV}$ or $800~{\rm GeV}$, $\tan\beta$ is scanned in the following range, $0.5<\tan\beta<50$. 
We note that, without depending on $\tan\beta$, results with $m_{\Phi}=200~{\rm GeV}$ for Type-II and Type-X are already excluded by the constraint from the flavor physics  (also, for Type-I and Type-Y in lower $\tan\beta$ regions, $\tan\beta \lesssim 2$)~\cite{Misiak:2017bgg,Misiak:2020vlo}. 
Nevertheless, we show them in order to compare results among four types of the THDM. 
For the SM parameters, we use the following values of the $\overline{\rm MS}$ quark masses at a scale of each pole mass;
\begin{align}
\overline{m}_{b}(m_{b})=4.18~{\rm GeV}~\mbox{\cite{Zyla:2020zbs}},
\quad \overline{m}_{c}(m_{c})=1.28~{\rm GeV}~\mbox{\cite{Zyla:2020zbs}},
\quad \overline{m}_{t}(m_{t})=162.3~{\rm GeV}~\mbox{\cite{Alekhin:2013nda}}.
\end{align} 
 For the mass of the strange quark we use the running mass at $\mu=2~{\rm GeV}$~\cite{Garden:1999fg}, $\overline{m}_{s}(2~{\rm GeV})=0.097~{\rm GeV}$.
 Running quark masses at an arbitrary scale are derived by Eq.~\eqref{eq:rmq}. 
 In the derivation, the running strong coupling is evaluated with the following values of $\Lambda_{\rm QCD}^{N_f}$~\cite{Tanabashi:2018oca}, 
 \begin{align}
 \Lambda_{\rm QCD}^{6}= 89~{\rm MeV},
\quad  \Lambda_{\rm QCD}^{5}= 210~{\rm MeV}, 
\quad  \Lambda_{\rm QCD}^{4}= 292~{\rm MeV},
\quad  \Lambda_{\rm QCD}^{3}= 332~{\rm MeV}, 
 \end{align}
for $N_f$= 6,\ 5,\ 4,\ {\rm and}\ 3, respectively. 
The input value of the CKM matrix elements and the total width for the weak gauge bosons as well as the top quark are taken as \cite{Zyla:2020zbs},
 \begin{align}
 &V_{tb}=0.999172,
 \quad\quad V_{ts}=0.3978,
 \quad\quad V_{cb}=0.04053, \\
 &\Gamma_{W}=2.085~{\rm GeV}, 
 \quad\quad  \Gamma_{Z}=2.4952~{\rm GeV}, 
  \quad\quad \Gamma_{t}=1.42~{\rm GeV}.
 \end{align}
The former is relevant for the charged Higgs decays into quarks, $H^{\pm}\to tb$, $H^{\pm}\to ts$ and $H^{\pm}\to cb$. 
 The latter is used in computation of the Higgs boson decays into off-shell particles. 
 
Before we show numerical behaviors of the total widths and the branching ratios, we mention the loop induced decays of the charged Higgs bosons.  
The branching ratio of  $H^{\pm}\to W^{\pm}Z $ can be enhanced  when the mass difference between $H^{\pm }$ and $A$ is taken to some extent~\cite{Kanemura:1997ej,Abbas:2018pfp}. 
Whereas, in the following numerical results, where the additional Higgs bosons are degenerate, the branching ratio of $H^{\pm}\to W^{\pm}Z $ is at most $\mathcal{O}(10^{-4})$ in the present parameter choices. 
Furthermore, the branching ratio of $H^{\pm}\to W^{\pm}\gamma$ is smaller than that of $H^{\pm}\to W^{\pm}Z$. 
 
The following numerical results for the total widths and the branching ratios are similar to those given in Ref.~\cite{Kanemura:2014bqa}, where the systematic studies have been done. 
 Nevertheless, we here show them because there are some developments from the previous study. 
 Main difference from Ref.~\cite{Kanemura:2014bqa} is that we compute the decay processes including higher-order QCD corrections. 
 Also, we incorporate the above mentioned decay processes for the charged Higgs bosons, $H^{\pm}\to W^{\pm}Z$ and $H^{\pm}\to W^{\pm}\gamma$,  in the evaluation of the total width.

In Fig.~\ref{FIG:width}, we show the total decay widths for the neutral Higgs bosons and the charged Higgs bosons as a function of $\tan\beta$ in the cases of $m_{\Phi}=M=200$ GeV and  $m_{\Phi}=M=800$~GeV.  
Different values of $s_{\beta-\alpha}$ are taken in each panel, namely $\sba=1$  in the top panels, $\sba=0.995$ with $c_{\beta-\alpha}<0$ in the middle panels, and $s_{\beta-\alpha}=0.995$ with $c_{\beta-\alpha}>0$ in the bottom panels. 
For the $h$ decays, in the alignment limit $\sba=1$, the couplings with fermions and weak gauge bosons coincide with those in the SM at tree level, so that the total decay width does not depend on $\tb$. 
On the contrary,  when $\tan\beta$ increases at $\sba=0.995$ with $c_{\beta-\alpha}<0$, the total width also increases due to the effect of $\tb$ enhancement on $h\to b\bar{b}$ $(h\to \tau\bar{\tau})$ in Type-II and Type-Y (Type-X). 
For the heavy Higgs bosons $H, A$ and $H^{\pm}$, the total widths vary in the both cases of $\sba=1$ and $\sba\neq1$. 
While those in Type-I  monotonically decrease except for $H$ with a mass of 800 GeV, there appears the dip at a certain value of $\tan\beta$ for each additional Higgs boson in Type-II, X and Y. 

In Fig.~\ref{FIG:BR1}, we show $\tan\beta$ dependence of the decay branching ratios for the neutral Higgs bosons and the charged Higgs bosons in the alignment limit, $s_{\beta-\alpha}=1$, with $m_{\Phi}=M=200~{\rm GeV}$. 
For the SM-like Higgs boson decays,  there is no $\tan\beta$ dependence of all the decay modes, since all the scaling factors $\kappa_{X}^{h}$ are unity when $s_{\beta-\alpha}=1$. 
We note that, in addition, the squared scaling factors of the fermion couplings for $H$ and $A$ are common and simply expressed by $\zeta_{f}$ parameters; i.e., $|\kappa_{f}^{H}|^{2}=|\kappa_{f}^{A}|^{2}=\zeta_{f}^{2}$ at $s_{\beta-\alpha}=1$. 
In the case with $m_{\Phi}=M=200~{\rm GeV}$, the decay mode into a pair of the top quarks  does not open for the $H (A)$ decays.  
Hence, for $\tan\beta>1$, the main decay mode of $H$ is $ H\to b\bar{b}$  except for Type-X, as similar to the SM-like Higgs boson decays. 
For Type-X, the main decay mode is $ H\to \tau\bar{\tau}$ due to the $\tan\beta$ enhancement for the leptonic decays, which also causes $H\to \mu\bar{\mu}$ with about 0.3\% for $\tan\beta \gtrsim 4$. 

 For decays of $A$, one can see that the behavior of the branching ratios for decays into fermions is similar to those of $H$ for all types of the THDM because of $|\kappa_{f}^{H}|^{2}=|\kappa_{f}^{A}|^{2}$. 
 The difference from $H$ decays appears in the decay into $gg$.
Namely, BR$(A\to gg)$ is relatively larger than  BR$(H\to gg)$. 
This mainly comes from the fact that the NLO QCD correction is more significant than $H\to gg$, although the expressions at the LO are also different between $H$ and $A$. 

Apart from the neutral Higgs bosons, decays including a top quark exist for the charged Higgs bosons.  
While the decay into $tb$ is the main decay mode for Type-I and Type-Y, the decay into $\tau\nu $ can be dominant in high $\tan\beta$ regions for Type-X. 
For Type-II both of the bottom Yukawa and the tau Yukawa coupling are enhanced by $\tan\beta$. 
As a consequence, the branching ratio ${\rm BR}(H^{\pm}\to \tau\nu)$ approaches to ${\rm BR}(H^{\pm}\to tb)$ in high $\tan\beta$ regions, in which effect of the top Yukawa coupling is negligible. 

  In Fig.~\ref{FIG:BR2}, the  branching ratios in the case of $s_{\beta-\alpha}=1$ and $m_{\Phi}=M=800~{\rm GeV}$ are shown as a function of $\tan\beta$. 
For $h$ decays,  the behavior does not change much from the case with $m_{\Phi}=M=200~{\rm GeV}$ since the decay rates  do not depend on the mass of the additional Higgs bosons at tree level. 
 Main difference from Fig.~\ref{FIG:BR1} is appearance of the decays into $t\bar{t}$ in $H$ and $A$. 
 It dominates the branching ratios of $H$ and $A$ for Type-I with any value of $\tan\beta$. 
 On the other hands, for Type-II, Type-X  and Type-Y, $H \to t\bar{t}$ and $A \to t\bar{t}$ can be dominant only for $\tan\beta \lesssim$ 10 since the decay rate is proportional to $\cot^{2}\beta$. 
  
  Next, we move on cases without taking the alignment limit, $s_{\beta-\alpha}\neq 1$. 
  In these cases, the branching ratios of fermionic decay modes of $H$ and $h$ vary with a sign of $\cba$. 
  Furthermore, for decays of heavy Higgs bosons,  additional decay modes,  such as $H\to VV$ $(V=W, Z)$ $H\to hh$, $A\to Zh$ and $H^{\pm}\to W^{\pm} h$, shall appear. 
  Therefore, their decay patterns can drastically change from the case of the alignment limit. 
 
In Fig.~\ref{FIG:BR3},  we show $\tan\beta$ dependence of the branching ratios for $h$, $H$, $A$ and $H^{\pm}$  in the case with $s_{\beta-\alpha}=0.995$ and $m_{\Phi}=M=200~{\rm GeV}$. 
For decays of $h$ and $H$, predictions in the cases with $c_{\beta-\alpha}<0$ and $c_{\beta-\alpha}>0$ are separately plotted by solid lines and dotted lines, respectively.
Regarding the decay of $h$ one can see clear $\tan\beta$ dependence for all the decay modes.
In particular, the branching ratio for $h\to b\bar{b}$ remarkably increases by $\tan\beta$. 
For the CP-even Higgs boson $H$, the decays into the on-shell weak gauge bosons $H\to ZZ$ and $H\to W^{+}W^{-}$, which are proportional to $m_{H}^{3}$ as seen in Eq.~\eqref{eq:hvv} of Appendix~\ref{sec:app1},  can dominate.
Whereas, the decay into $b\bar{b}$ ($\tau\bar{\tau}$) overcomes them for large $\tan\beta$ in Type-II and Type-Y (Type-X). 
Similarly, decays into a scalar boson and an off-shell vector boson  $A\to h Z^{\ast}  $ and $H^{\pm}\to h W^{\pm \ast} $ can be sizable in Type-I for large $\tan{\beta}$. 

In Fig.~\ref{FIG:BR4}, the branching ratios in the case with $s_{\beta-\alpha}=0.995$ and $m_{\Phi}=M=800~{\rm GeV}$ are also shown.
While $H\to t\bar{t}$ and $A\to t\bar{t}$ can be the main decay mode as similar to Fig.~\ref{FIG:BR2}, for decays of $H$, $H\to hh$ can be dominant due to the large scalar coupling $\lambda_{Hhh}$. 
Apart from this, one can see that the branching ratio for $H\to t\bar{t}$ and $H\to hh$ are close to 0 at $\tan\beta \sim10$ and  $\tan\beta\sim16$, respectively,  when $c_{\beta-\alpha}>0$. 
This is because the scaling factor $\kappa_{t}^{H}$ and the scalar coupling $\lambda_{Hhh}$ vanish at those values of $\tan\beta$.
We note that the value of $\lambda_{Hhh}$ depends on the value of $M$ as we can see from Eq.~\eqref{lam_Hhh}.
Therefore, the decay width for $H\rightarrow hh$ can change if we consider the non-degenerate case; i.e., $m_{\Phi}\neq M$.

\ken{The branching ratios including QCD corrections are discussed in the above paragraphs. 
On the other hands, there are a lots of studies on EW corrections to decays of the SM-like Higgs boson~\cite{Arhrib:2003ph,Arhrib:2016snv,Kanemura:2004mg,Kanemura:2014dja,Kanemura:2015mxa,Kanemura:2017wtm,Kanemura:2018yai,Kanemura:2019kjg,Chen:2018shg,Gu:2017ckc,Castilla-Valdez:2015sng,Xie:2018yiv,Altenkamp:2017ldc,Altenkamp:2017kxk,Altenkamp:2017ldc}
and additional Higgs bosons~\cite{Krause:2016oke,Krause:2016xku,Krause:2019qwe,Denner:2018opp,Su:2019dsf}. 
 NLO EW corrections to $h\to f\bar{f}$  can be evaluated by utilizing {\tt H-COUP v2}~\cite{Kanemura:2019slf}. 
 Also, in the program, those to $h\to VV^{\ast}\to Vf\bar{f}$ are calculated. 
NLO EW corrections to on-shell two-body decays of $H, ~A$ and $H^{\pm}$ will be implemented in {\tt H-COUP v3}~\cite{H-COUPv3}. 
In {\tt 2HDECAY}~\cite{Krause:2018wmo}, NLO corrections to on-shell two-body decays of $h,~H,~A$, and $H^{\pm}$ are evaluated. 
In addition, NLO EW corrections to $h/H\to V^{(*)}V^{(*)} \to 4f$ are calculated in {\tt Prophecy4f}~\cite{Denner:2019fcr}.}

\section{Direct searches at the LHC}\label{sec:direct}
In this section, we present current constraints on the parameter space in the THDMs from direct searches 
for heavy Higgs bosons with the LHC Run-II data.

Let us briefly summarize the procedure how we obtain the constraints on the parameters in the THDMs
from model-independent analyses for heavy Higgs boson searches at the LHC.
First, we compute production cross sections of heavy neutral Higgs bosons, $\phi=H$ and $A$, in the THDMs
for the gluon-fusion process ($pp\to\phi$) and for the bottom-quark associated (or bottom-quark annihilate) process ($pp\to\phi(b\bar b)$)
at the NNLO in QCD 
by using {\tt Sushi-1.7.0}~\cite{Harlander:2012pb,Harlander:2016hcx}.
For the charged Higgs boson production $pp\to tH^\pm$, 
we use the values given at the NLO QCD by the Higgs cross section working group (HXSWG)~\cite{deFlorian:2016spz}, based on Refs.~\cite{Berger:2003sm,Dittmaier:2009np,Flechl:2014wfa,Degrande:2015vpa}.
Second, we calculate decay branching ratios of the Higgs bosons in the THDMs, including higher-order QCD corrections,
as described in Sec.~\ref{sec:decay}.
 \begin{table}[t]
\begin{center}
\begin{tabular}{llll}\hline\hline
Constrained quantity &Applicable mass region &&Reference  \\\hline\hline
$\sigma(\phi)\times {\rm BR}(\phi \to\tau\tau)$ & $200<m_\Phi<2000$ GeV 
&& Fig.~7(a) in \cite{Aaboud:2017sjh} \\\hline 
$\sigma(\phi(bb))\times {\rm BR}(\phi \to\tau\tau)$ & $200<m_\Phi<2000$ GeV 
&& Fig.~7(b) in \cite{Aaboud:2017sjh} \\\hline
$\sigma(\phi(bb))\times {\rm BR}(\phi \to b{b})$ & $450<m_\Phi<1400$ GeV 
&& Fig.~8 in \cite{Aad:2019zwb} \\\hline
$\sigma(\phi)\times {\rm BR}(\phi \to t{t})$ & $400<m_\Phi<5000$ GeV 
&& Fig.~14 in \cite{Aaboud:2018mjh} \\\hline
$\sigma(H)\times {\rm BR}(H\to hh)\times {\rm BR}(h\to b{b})^2$& $260<m_\Phi<2000$ GeV 
&& Fig.~9(a) in \cite{Aaboud:2018knk} \\\hline
$\sigma(H)\times {\rm BR}(H\to WW)$ & $200<m_\Phi<2000$ GeV 
&& Fig.~5 in \cite{Aaboud:2017gsl} \\\hline
$\sigma(H)\times {\rm BR}(H\to ZZ)$ & $200<m_\Phi<2000$ GeV 
&& Fig.~6 in \cite{Aaboud:2017rel} \\\hline
$\sigma(A)\times {\rm BR}(A\to Zh)\times {\rm BR}(h\to b{b})$ & $200<m_\Phi<2000$ GeV 
&& Fig.~6(a) in \cite{Aaboud:2017cxo} \\\hline  
$\sigma(A(bb))\times {\rm BR}(A\to Zh)\times {\rm BR}(h\to b{b})$& $200<m_\Phi<2000$ GeV 
&& Fig.~6(b) in \cite{Aaboud:2017cxo} \\\hline
$\sigma(tH^\pm)\times {\rm BR}(H^\pm\to tb)$ & $200<m_\Phi<2000$ GeV 
&& Fig.~8 in \cite{Aaboud:2018cwk}\\\hline
$\sigma(tH^\pm)\times {\rm BR}(H^\pm\to\tau\nu)$ & $200<m_\Phi<2000$ GeV 
&& Fig.~8(a) in \cite{Aaboud:2018gjj}\\\hline  
\hline  
\end{tabular}
\caption{
List of constraints used in this study from direct searches for heavy Higgs bosons at the 13~TeV LHC.
}
\label{tab:lhc}
\end{center}
\end{table}
Finally, we compute the production cross sections times the branching ratios for each parameter point 
for each search channel at the LHC listed in Table~\ref{tab:lhc}, and compared with
the upper limits at 95\% CL with 36~fb$^{-1}$ data to obtain the constraints.
Here, as in Sec.~\ref{sec:decay}, we assume the common heavy Higgs boson masses $m_H=m_A=m_{H^\pm}\,(\equiv m_\Phi)$ and also $M=m_\Phi$.
Because we are interested in the near alignment scenario, we consider 
the value of $s_{\beta-\alpha}$ as 1, 0.995, 0.99 and 0.98 both for $c_{\beta-\alpha}<0$ and $c_{\beta-\alpha}>0$. 
We note that we use the expected upper limits, not the observed ones, from the LHC analyses 
in order for the HL-LHC projection.   
Although we use the ATLAS data, listed in Table~\ref{tab:lhc}, the similar limits have been reported 
by the CMS experiment~\cite{Khachatryan:2016qkc,Sirunyan:2018taj,Sirunyan:2017uhk,Sirunyan:2017isc,Khachatryan:2016cfx,Sirunyan:2019xls}.
We also note that, although new analyses with full Run-II data (139~fb$^{-1}$) are available 
for some channels; e.g., $\phi\to\tau\bar\tau$~\cite{Aad:2020zxo}, 
we use the upper limit with 36~fb$^{-1}$ data for a fair comparison with the other channels. 
Similar phenomenological studies have been done earlier in; e.g., Refs.~\cite{Arbey:2017gmh,Arhrib:2018ewj}.

\subsection{Production cross sections for the additional Higgs bosons}

\begin{figure}
 \includegraphics[height=0.19\textwidth]{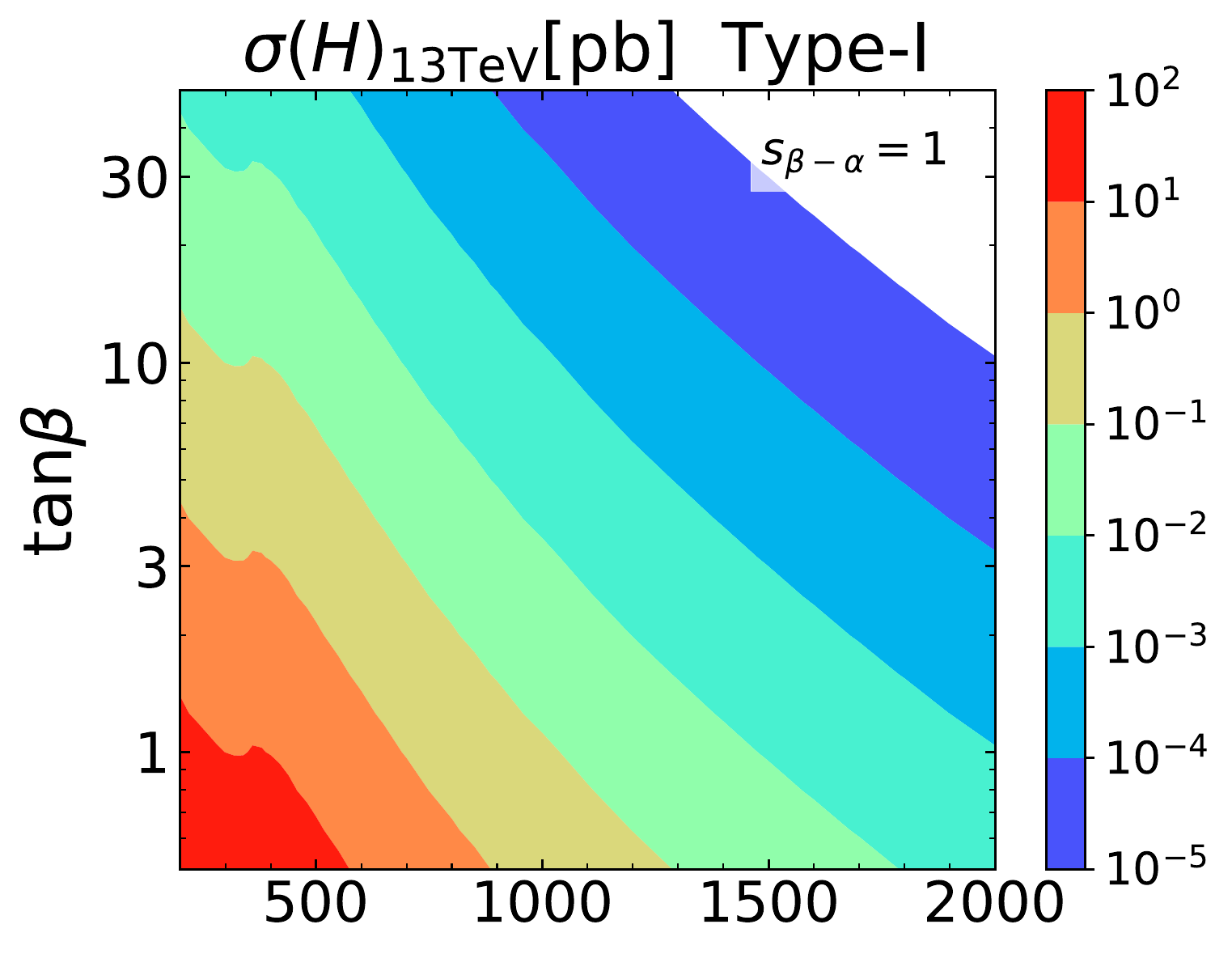}  
 \includegraphics[height=0.19\textwidth]{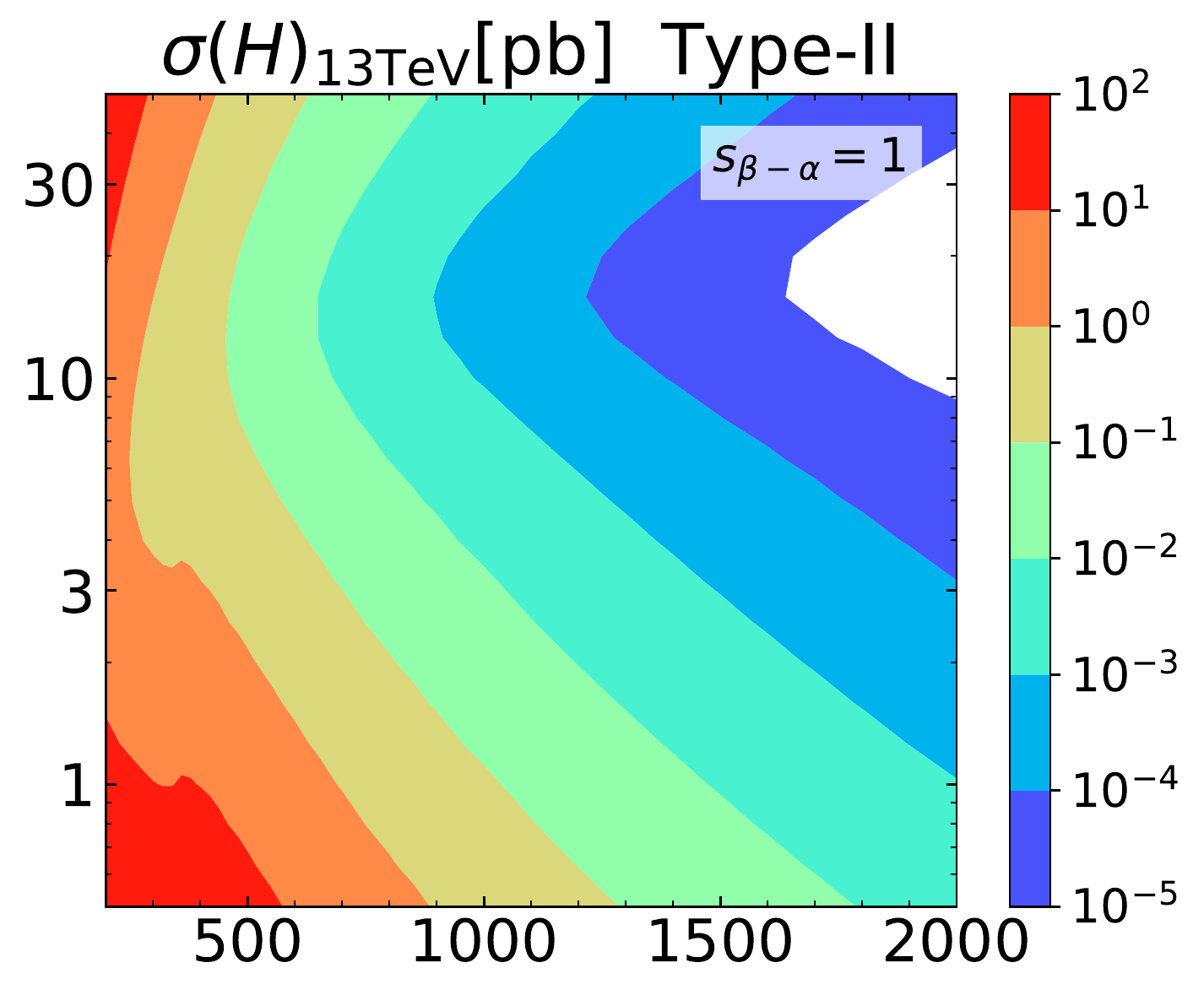}\quad  
 \includegraphics[height=0.19\textwidth]{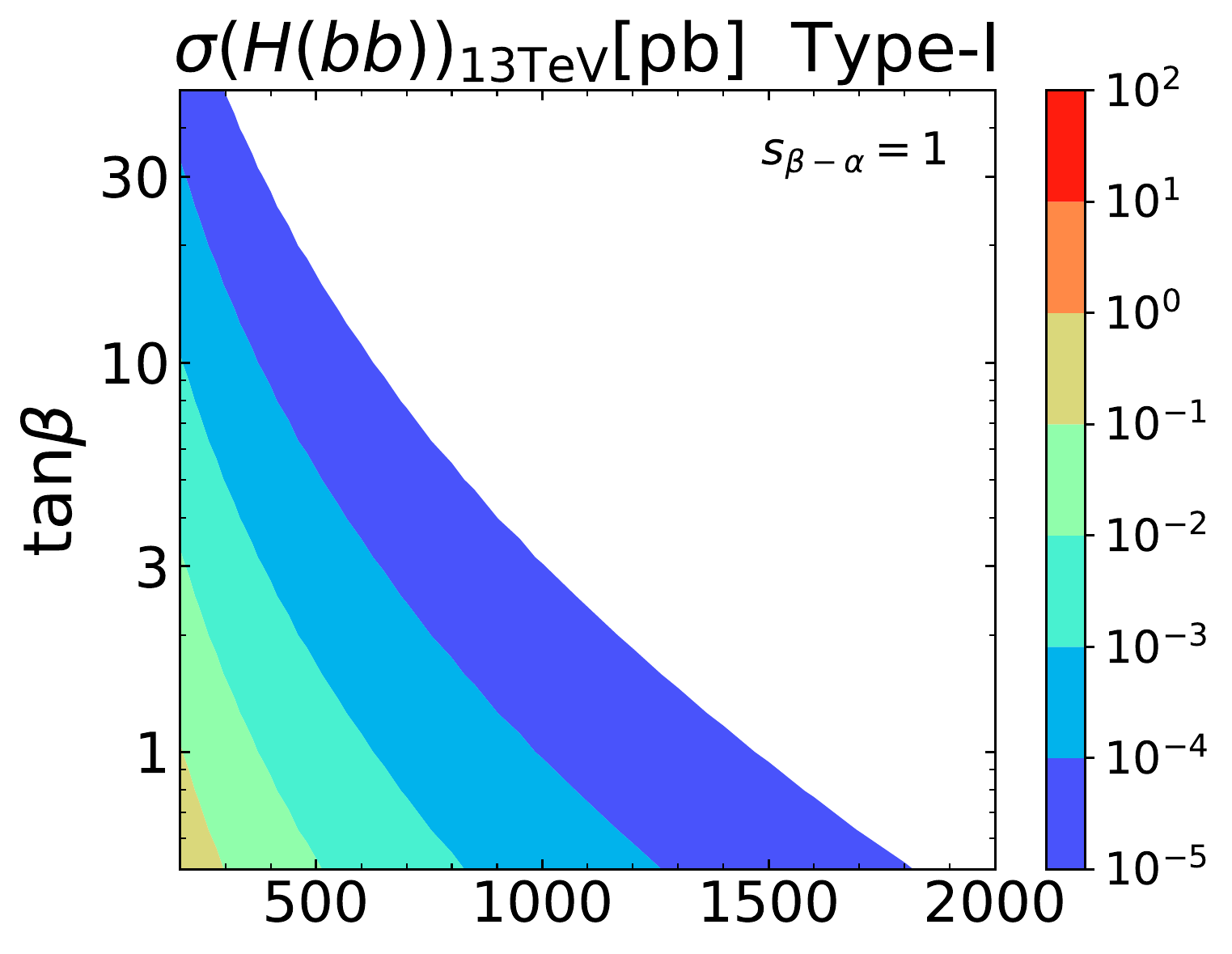}  
 \includegraphics[height=0.19\textwidth]{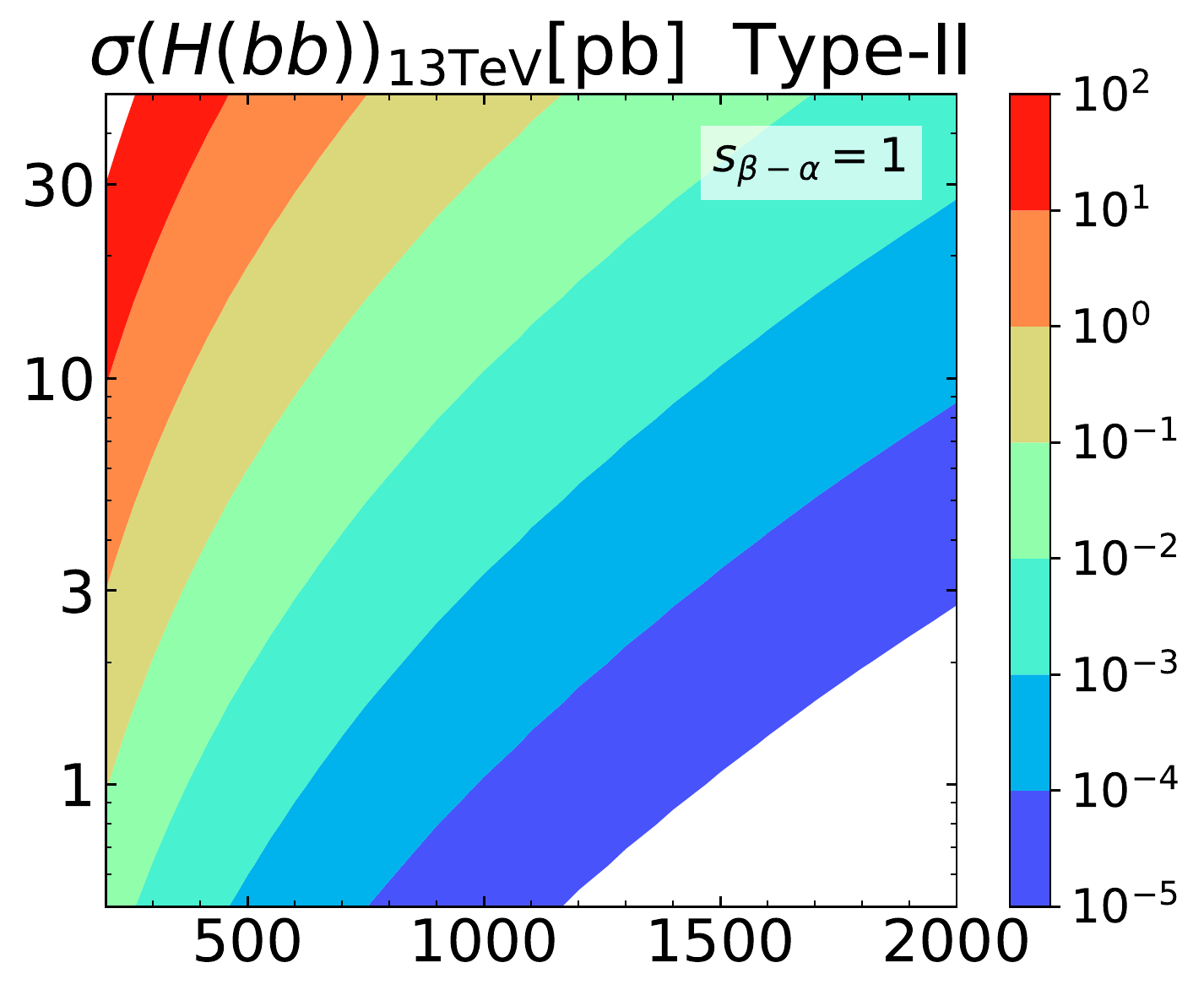}
 \includegraphics[height=0.18\textwidth]{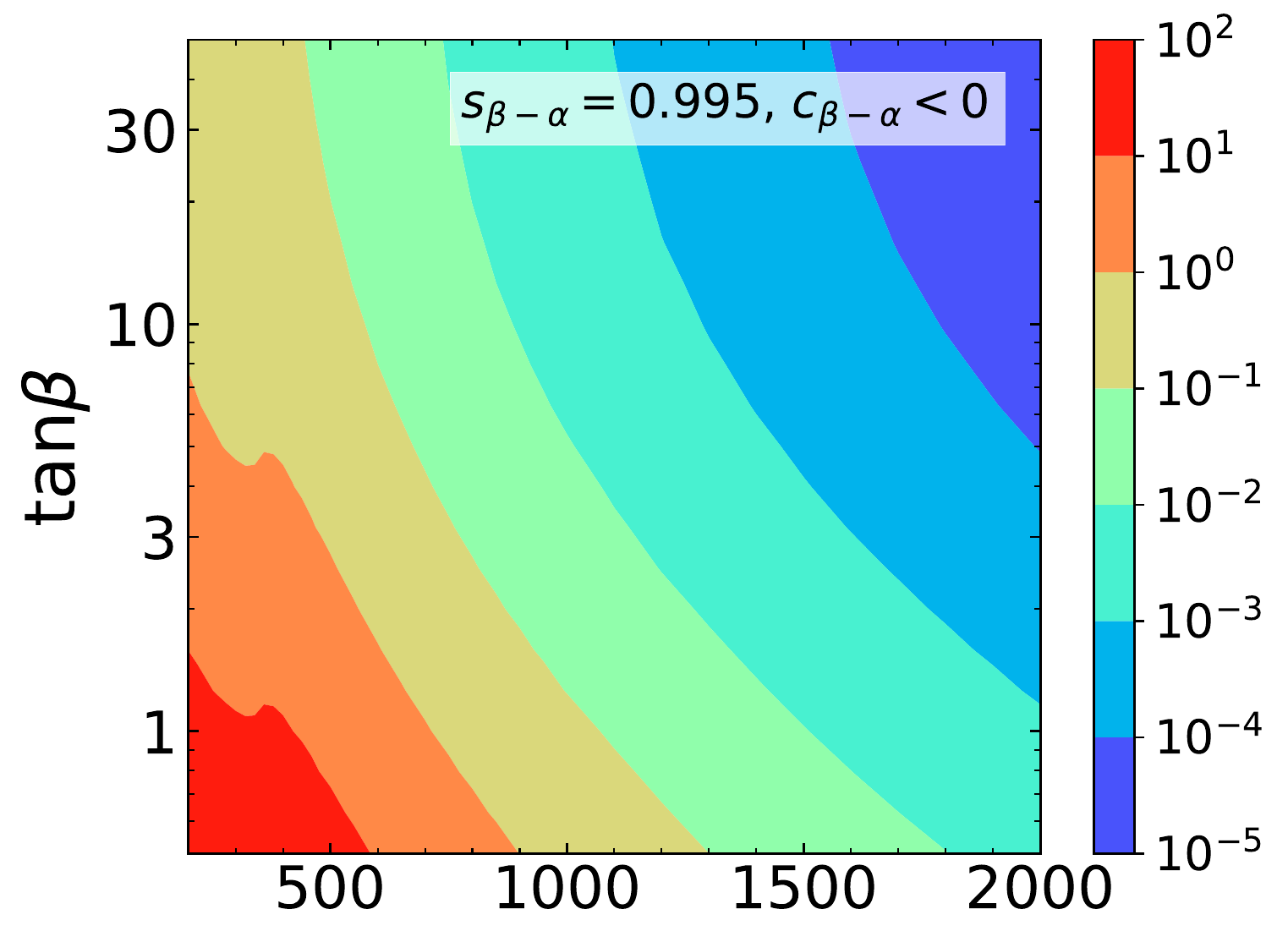}  
 \includegraphics[height=0.18\textwidth]{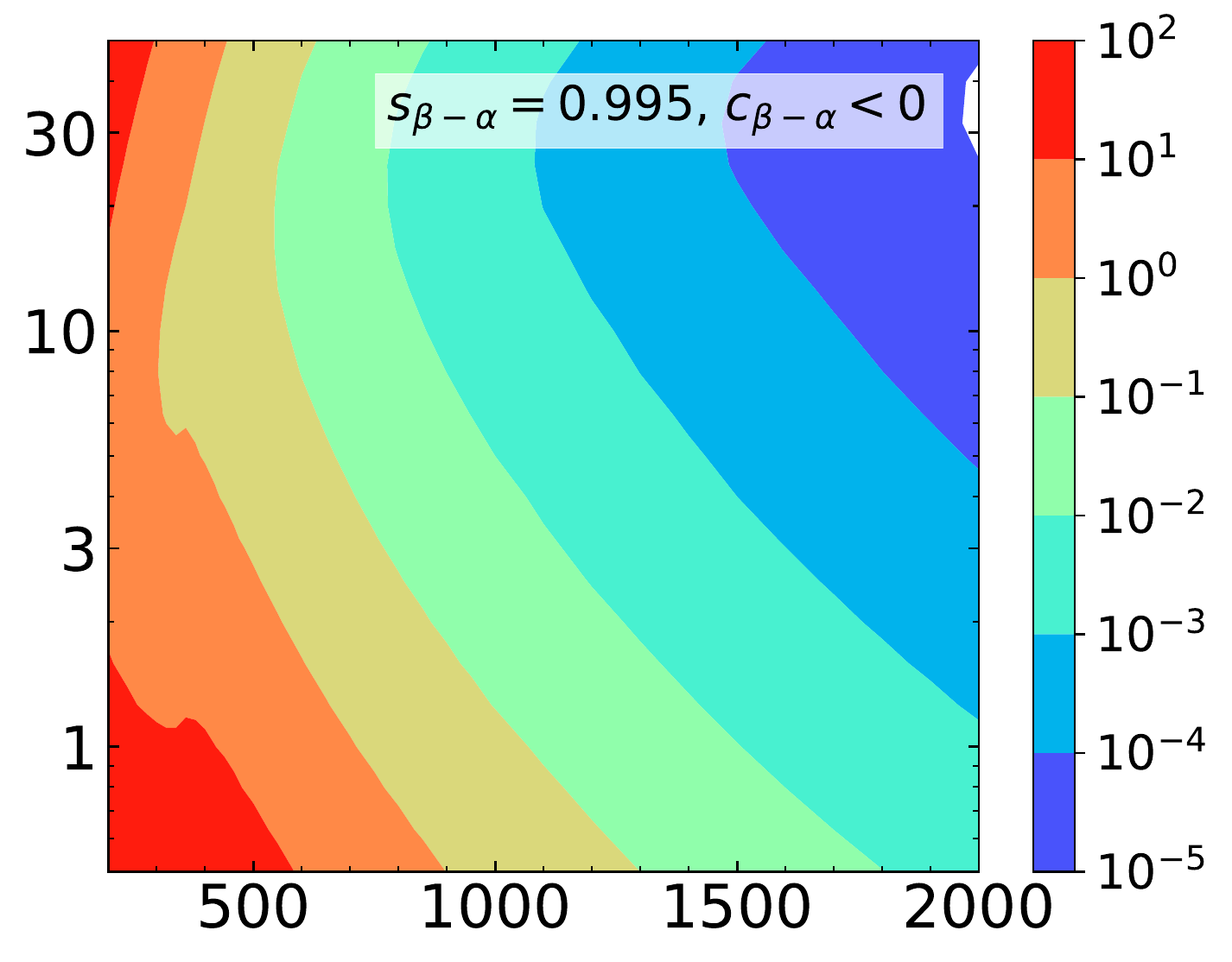}\quad  
 \includegraphics[height=0.18\textwidth]{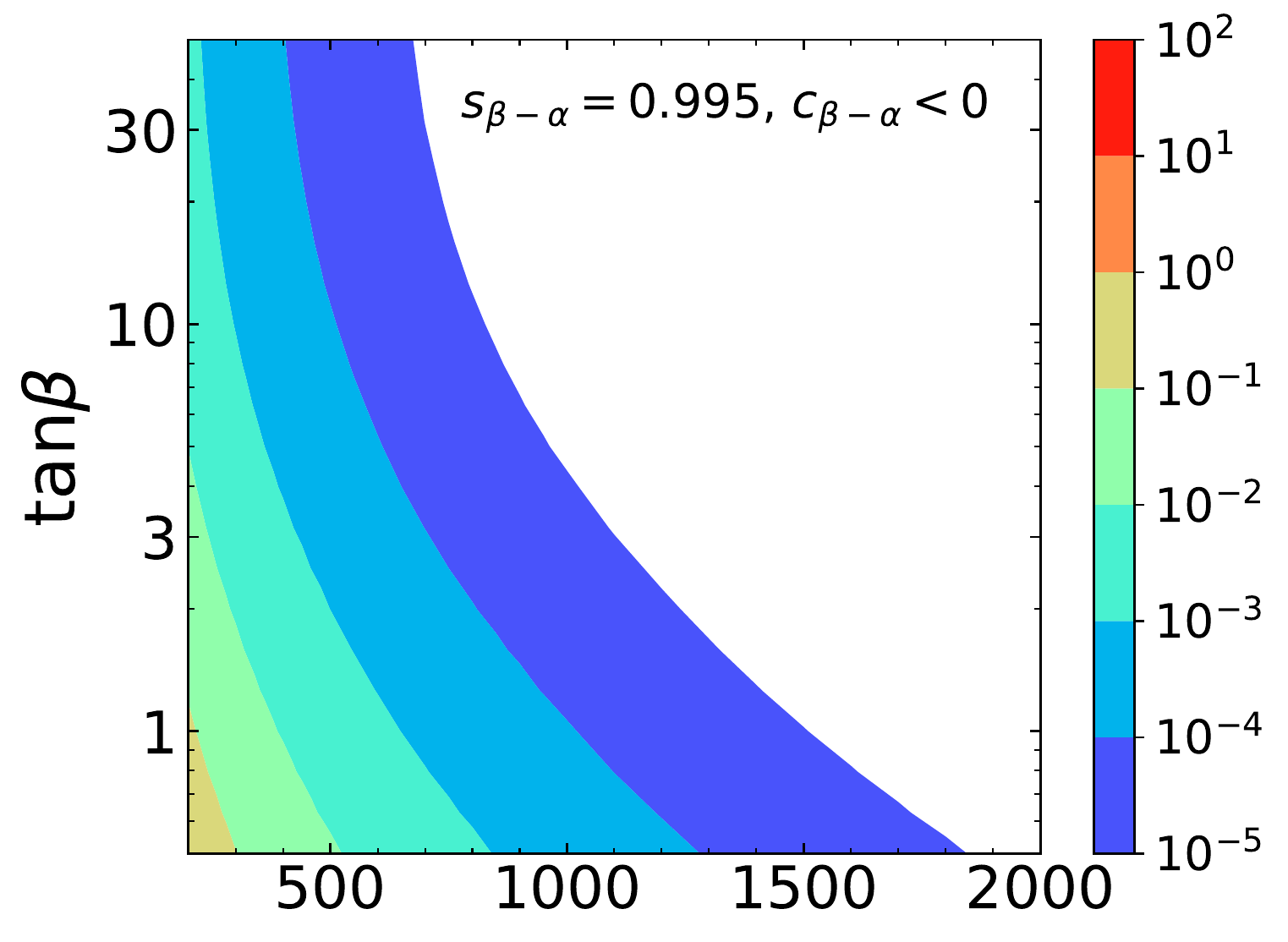}  
 \includegraphics[height=0.18\textwidth]{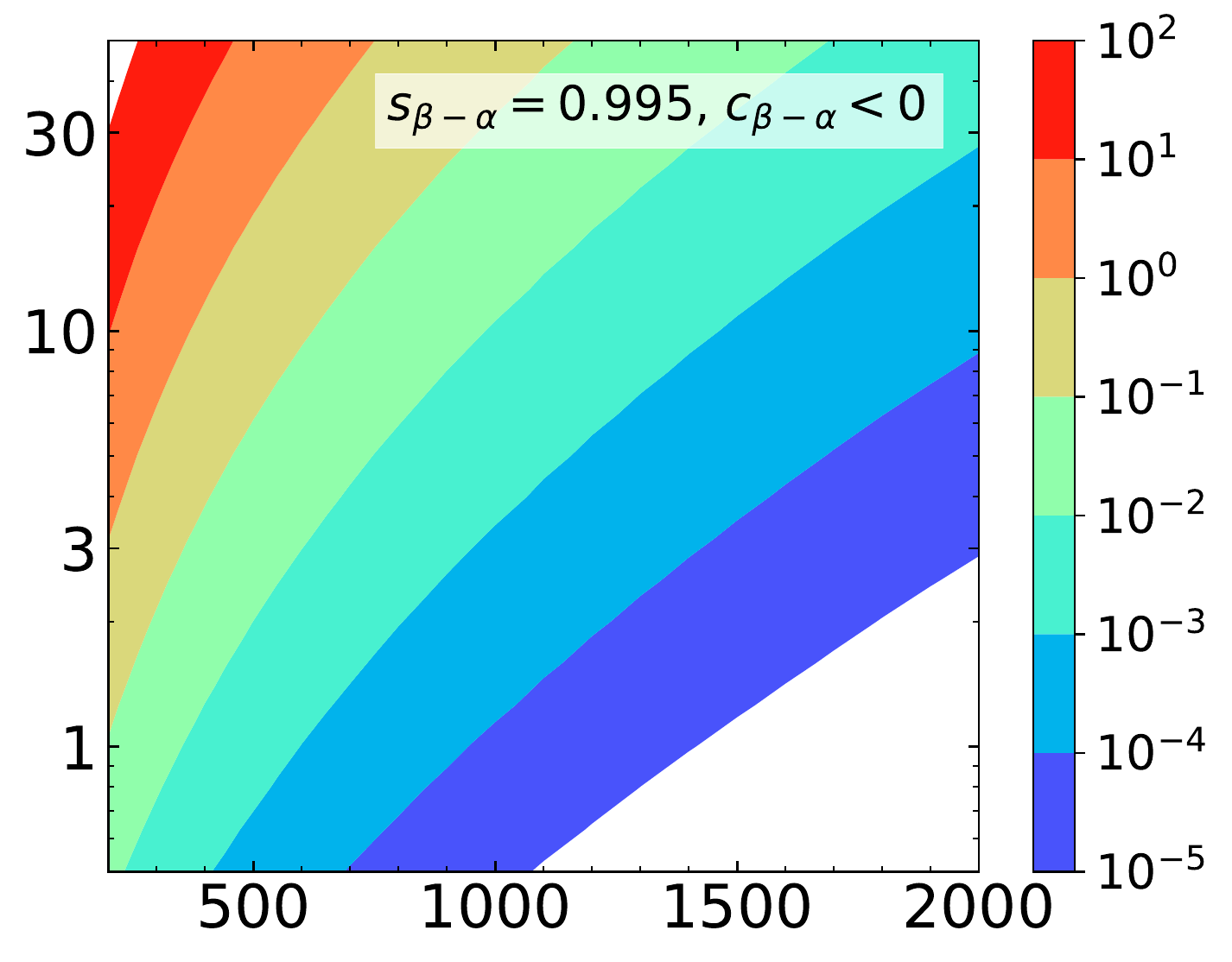}
 \includegraphics[height=0.18\textwidth]{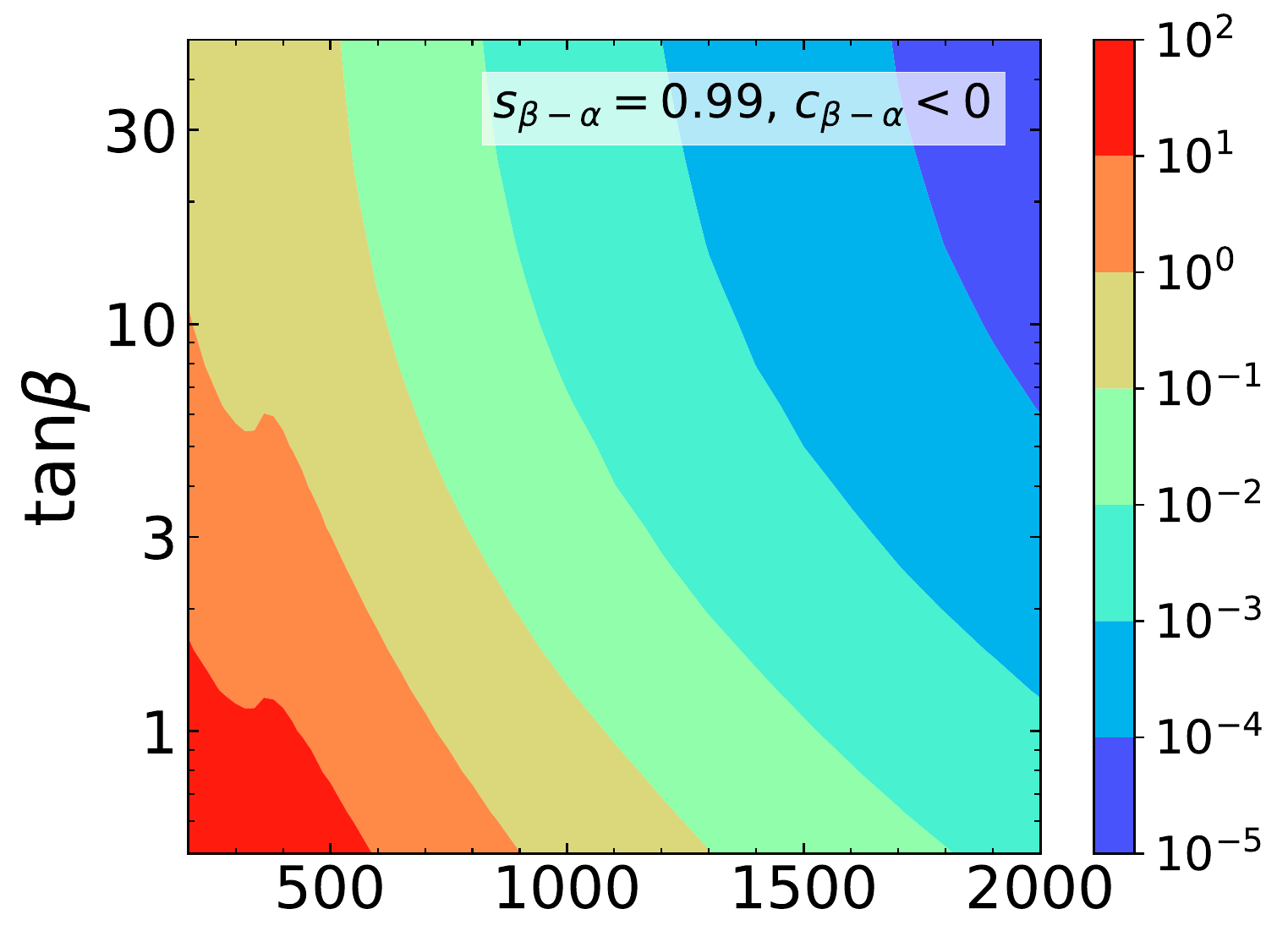}  
 \includegraphics[height=0.18\textwidth]{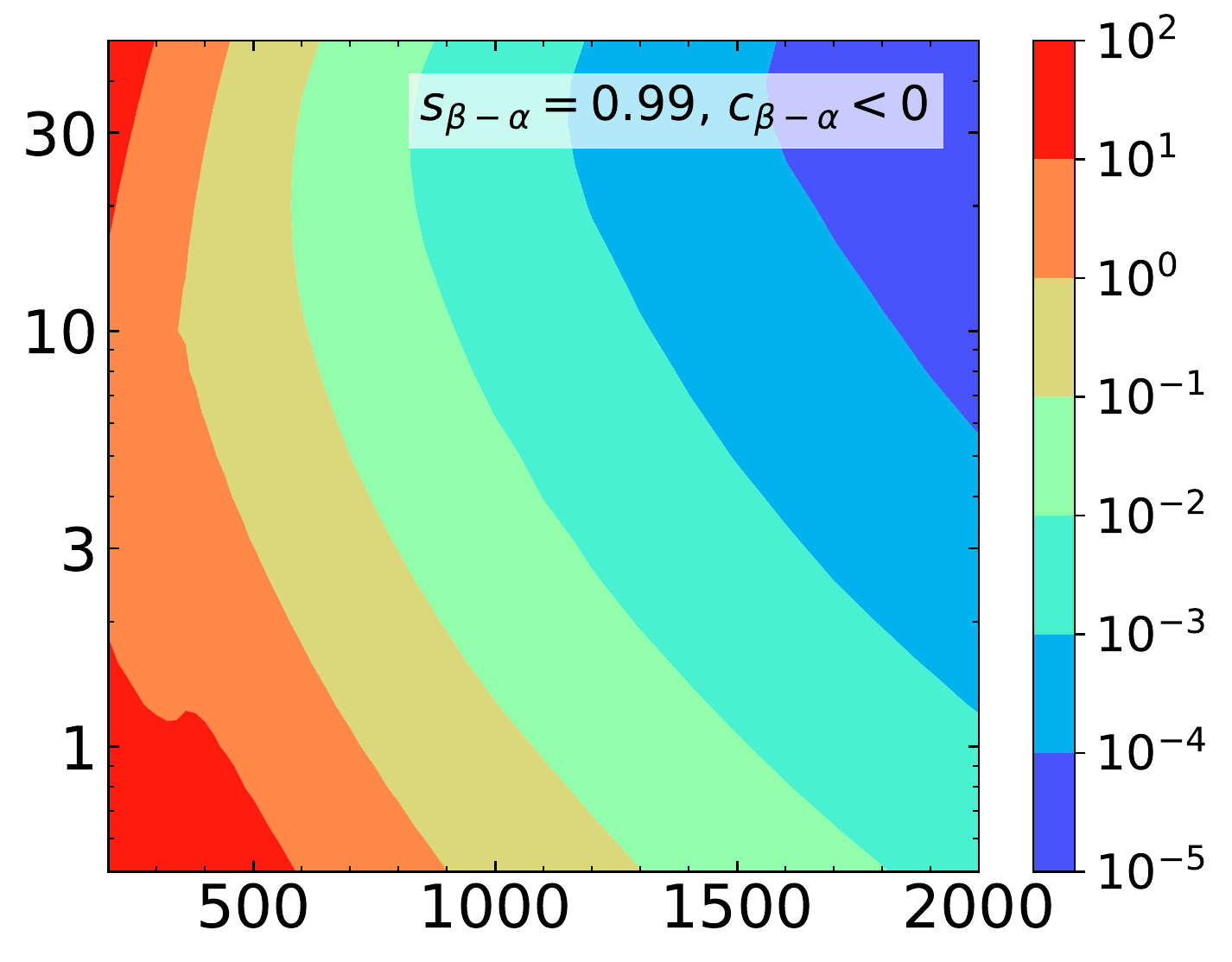}\quad  
 \includegraphics[height=0.18\textwidth]{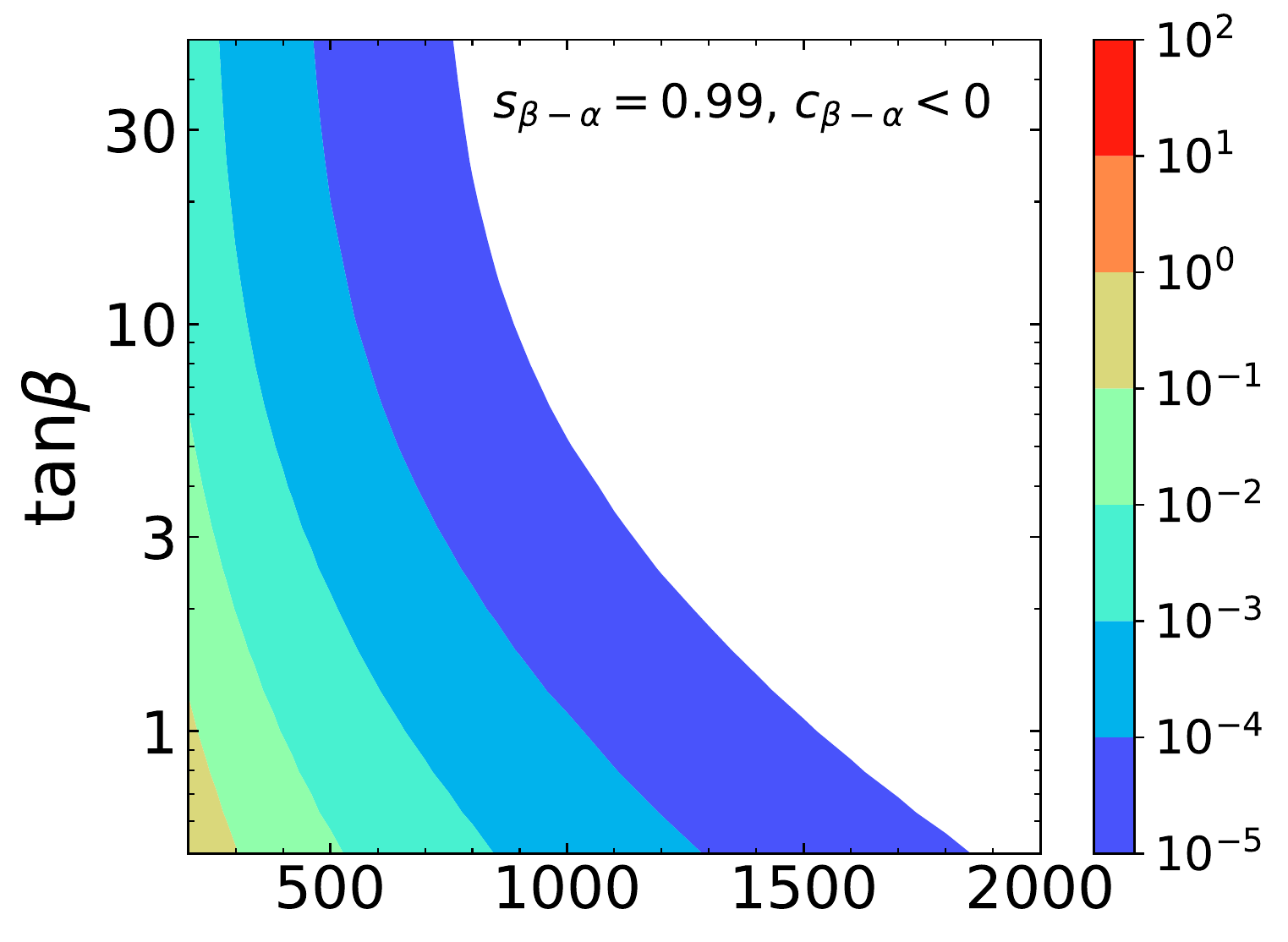}  
 \includegraphics[height=0.18\textwidth]{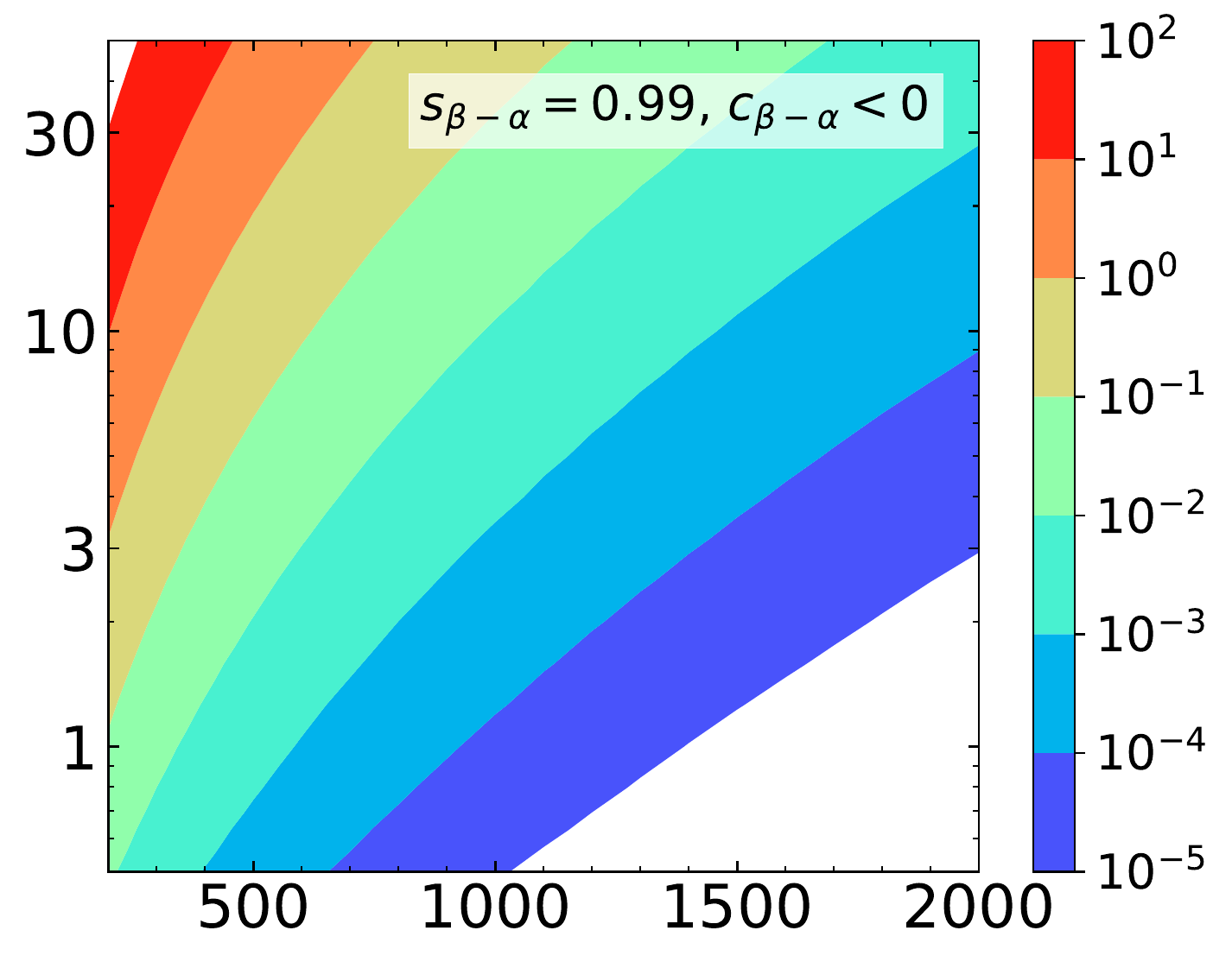}
 \includegraphics[height=0.194\textwidth]{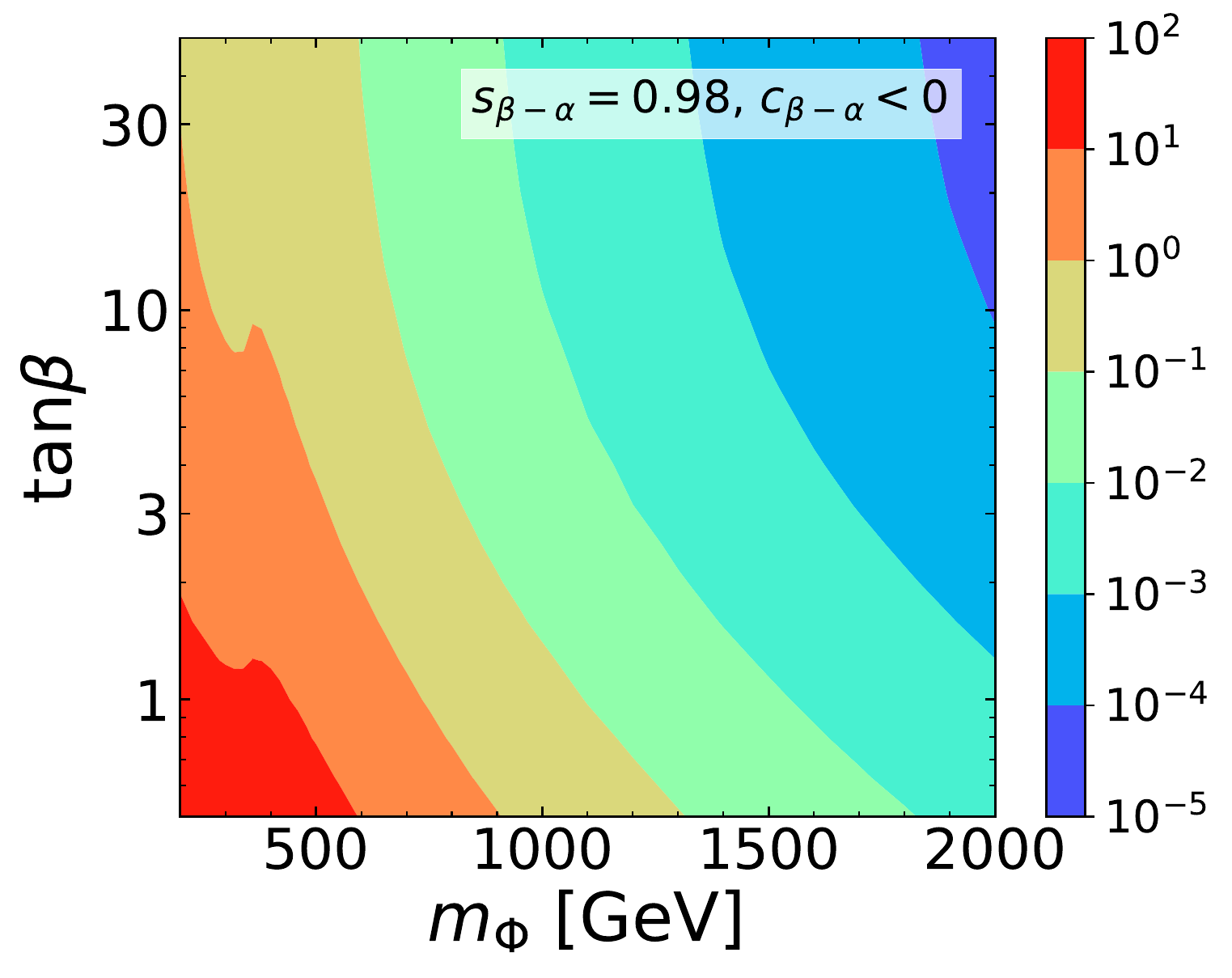}  
 \includegraphics[height=0.194\textwidth]{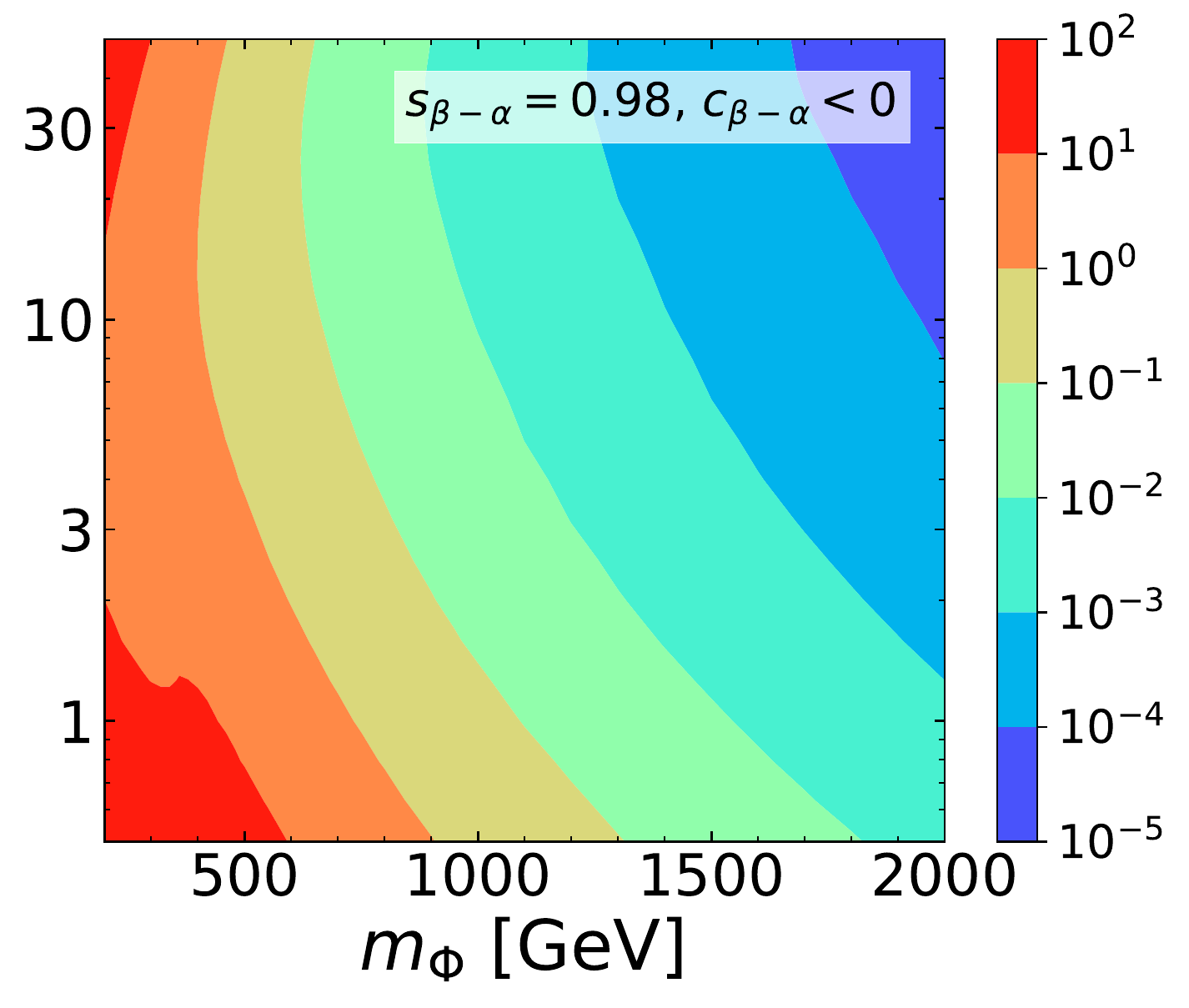}\quad  
 \includegraphics[height=0.194\textwidth]{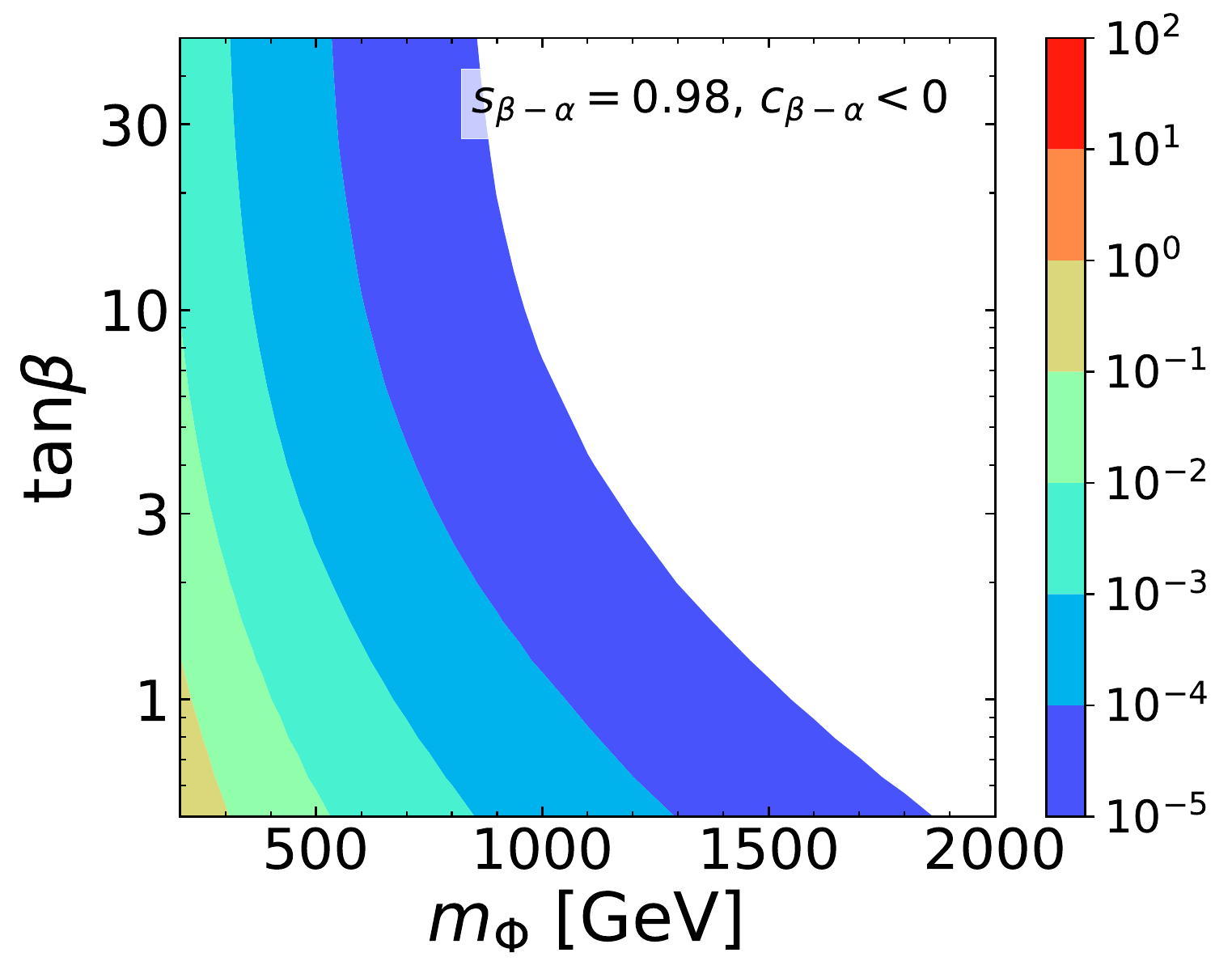}  
 \includegraphics[height=0.194\textwidth]{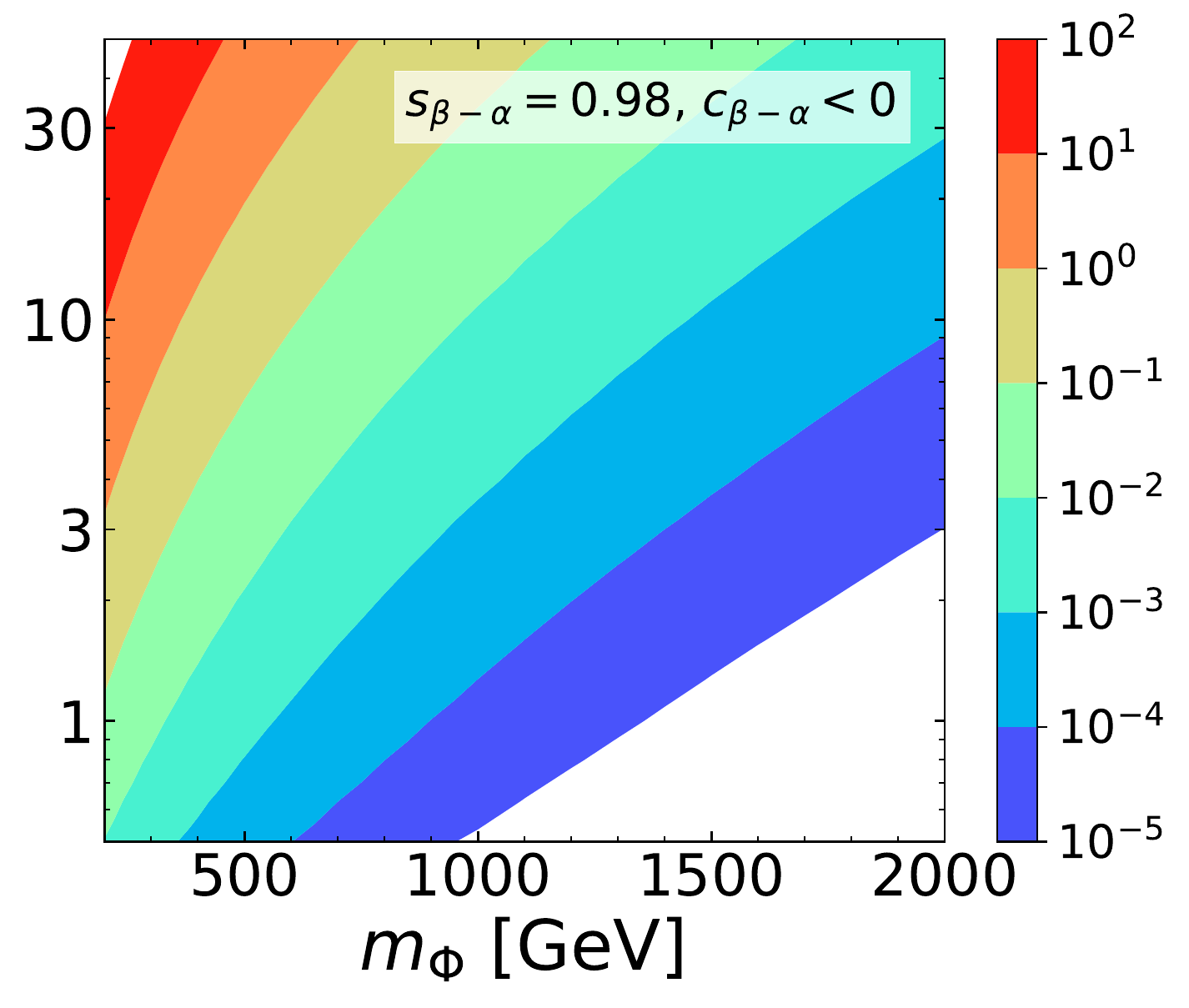}  
\caption{
Production cross sections for the CP-even heavy Higgs boson $H$ at the 13~TeV LHC 
on the $m_\Phi$--$\tan\beta$ plane.
Panels in two columns  from the left (right) show the production via the gluon fusion (the bottom-quark associated)
in the Type-I and Type-II THDMs, where
the value of $s_{\beta-\alpha}$ is set to be 1, 0.995, 0.99 and 0.98  
with $c_{\beta-\alpha} < 0$ from the top to the bottom panels.
The cross sections are shown with different colors from blue to red,
corresponding to from $10^{-5}$~pb to $10^{2}$~pb.     
}
\label{fig:xsec-H_cn}
\end{figure}     

\begin{figure}
 \includegraphics[height=0.19\textwidth]{Fig/H_t1_gg_sba1}  
 \includegraphics[height=0.19\textwidth]{Fig/H_t2_gg_sba1}\quad  
 \includegraphics[height=0.19\textwidth]{Fig/H_t1_bb_sba1}  
 \includegraphics[height=0.19\textwidth]{Fig/H_t2_bb_sba1}
 \includegraphics[height=0.18\textwidth]{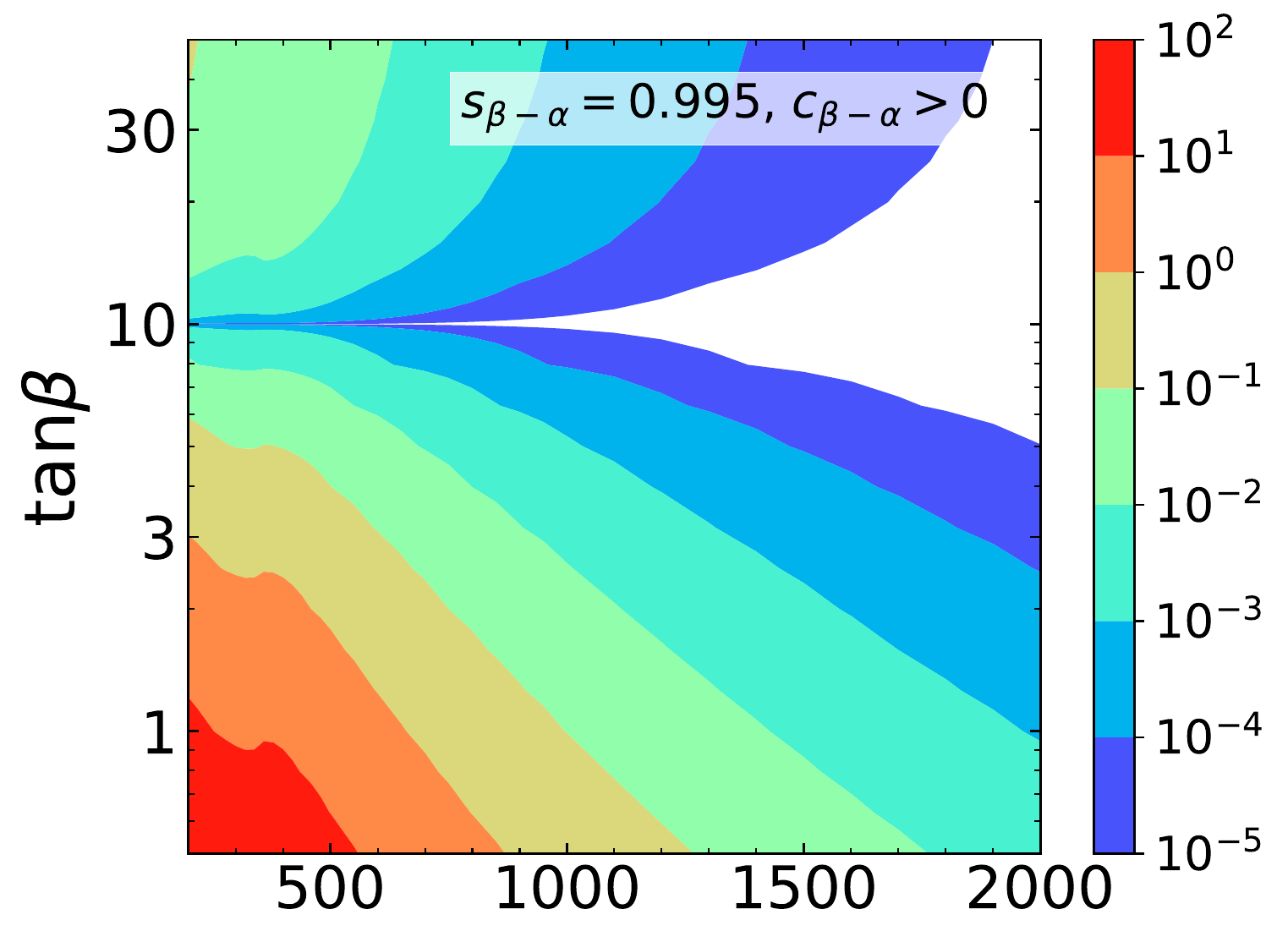}  
 \includegraphics[height=0.18\textwidth]{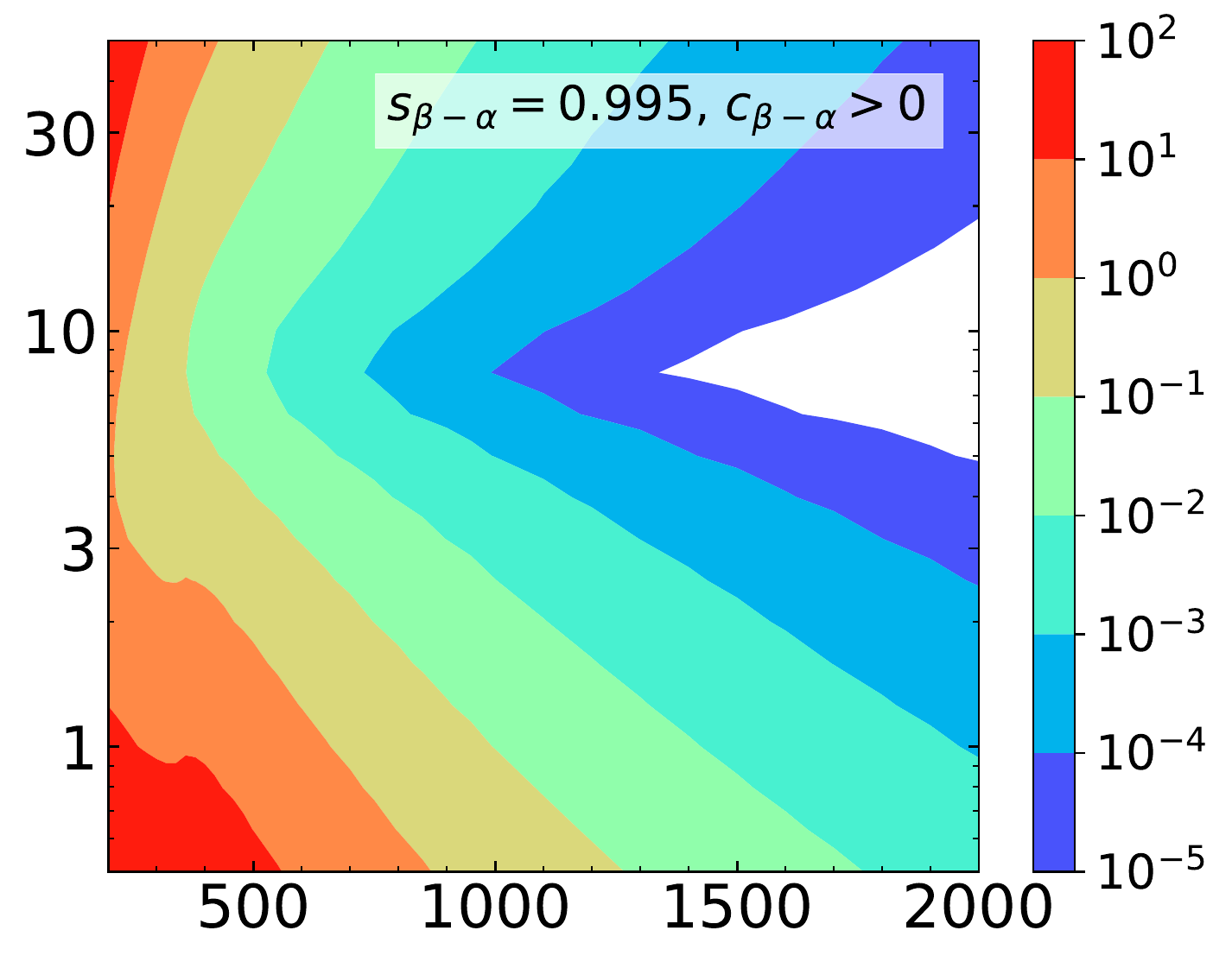}\quad  
 \includegraphics[height=0.18\textwidth]{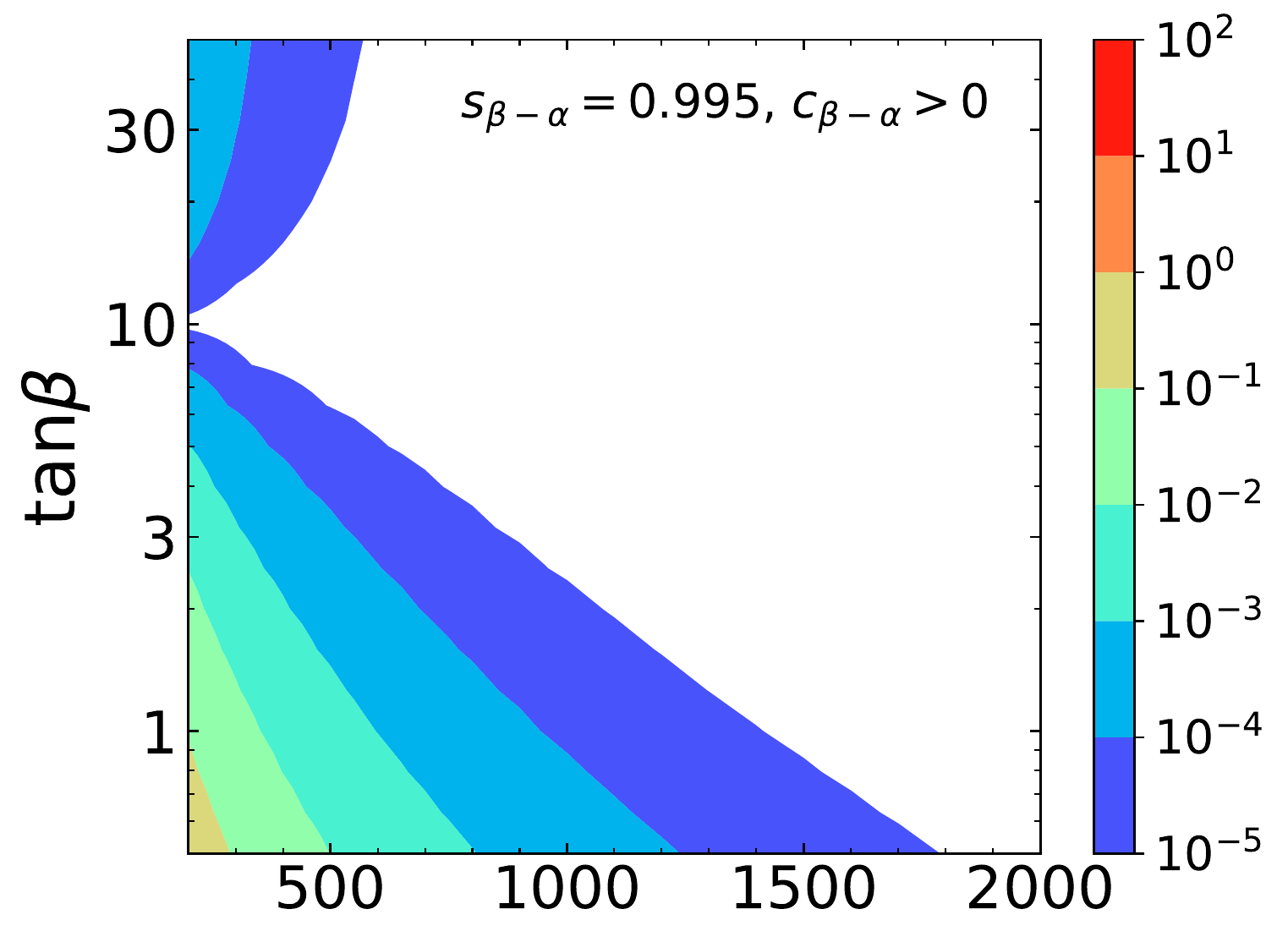}  
 \includegraphics[height=0.18\textwidth]{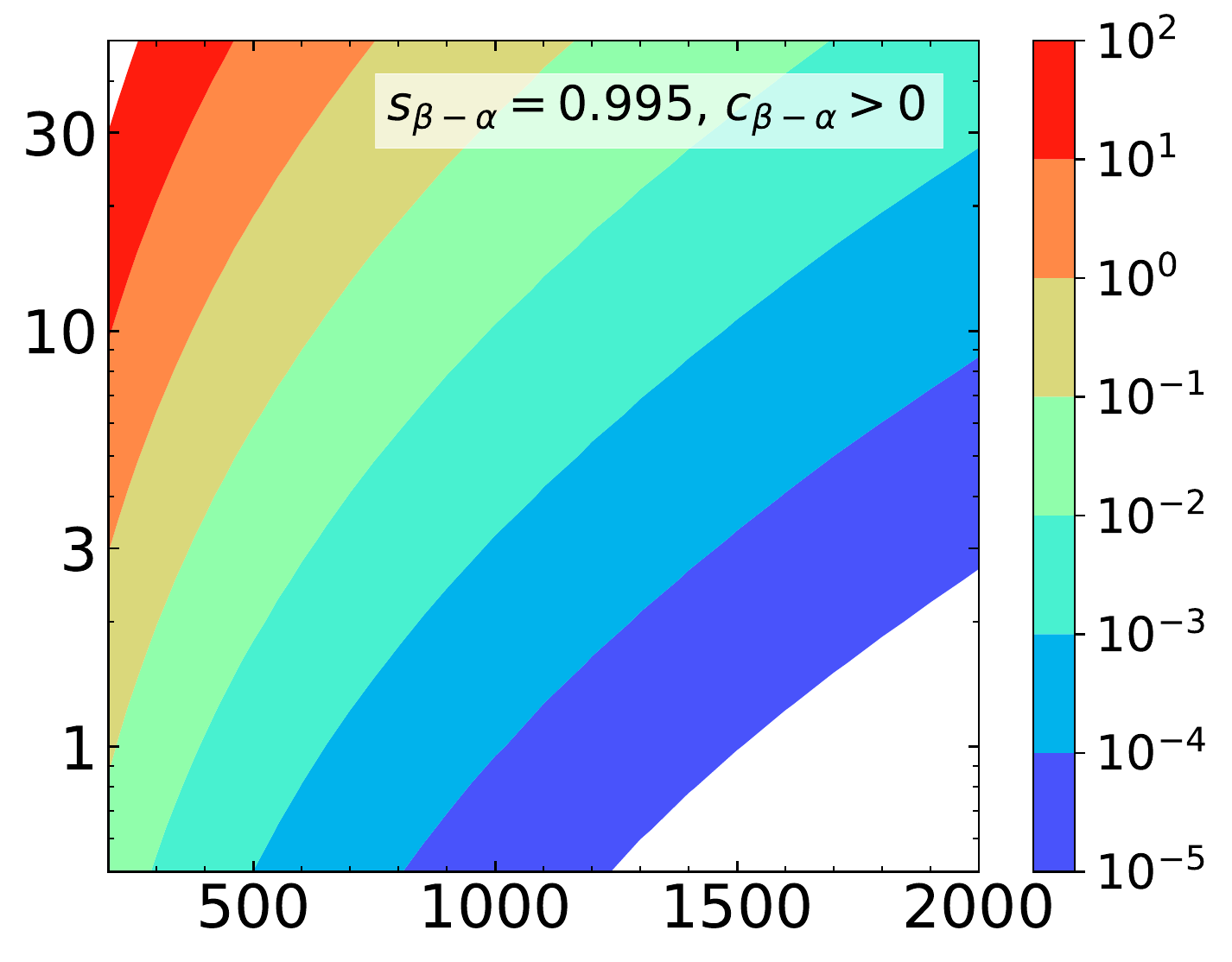}
 \includegraphics[height=0.18\textwidth]{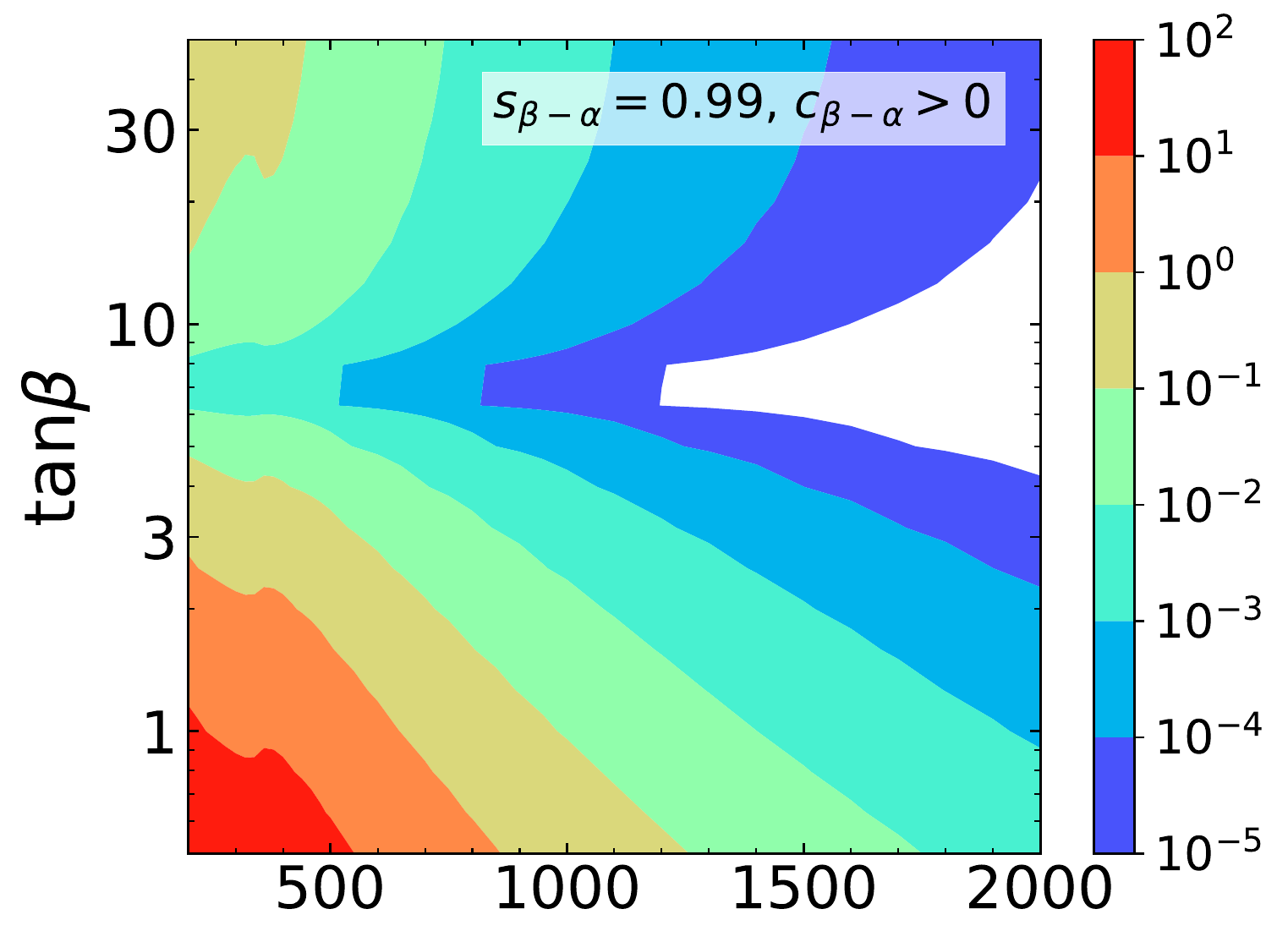}  
 \includegraphics[height=0.18\textwidth]{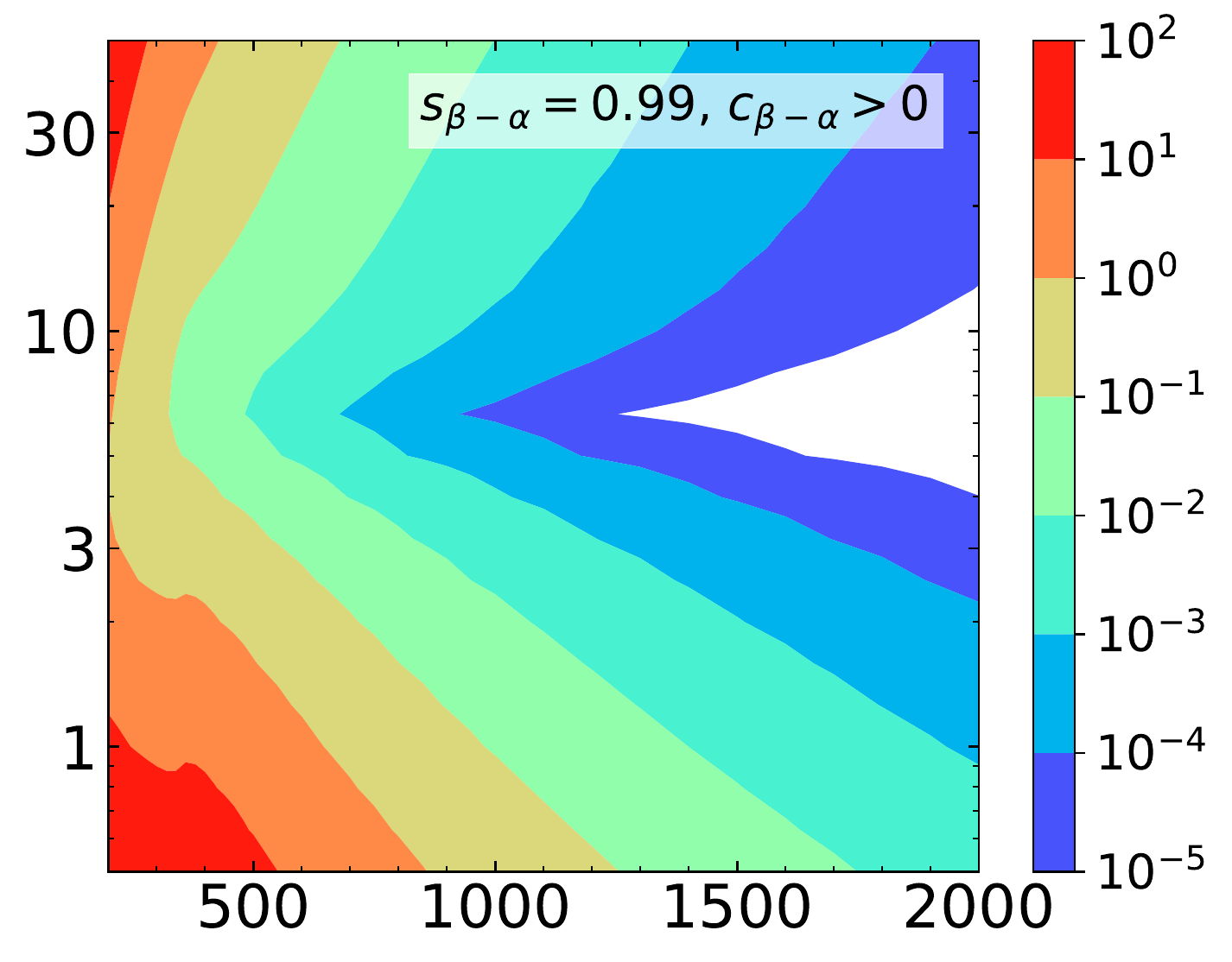}\quad  
 \includegraphics[height=0.18\textwidth]{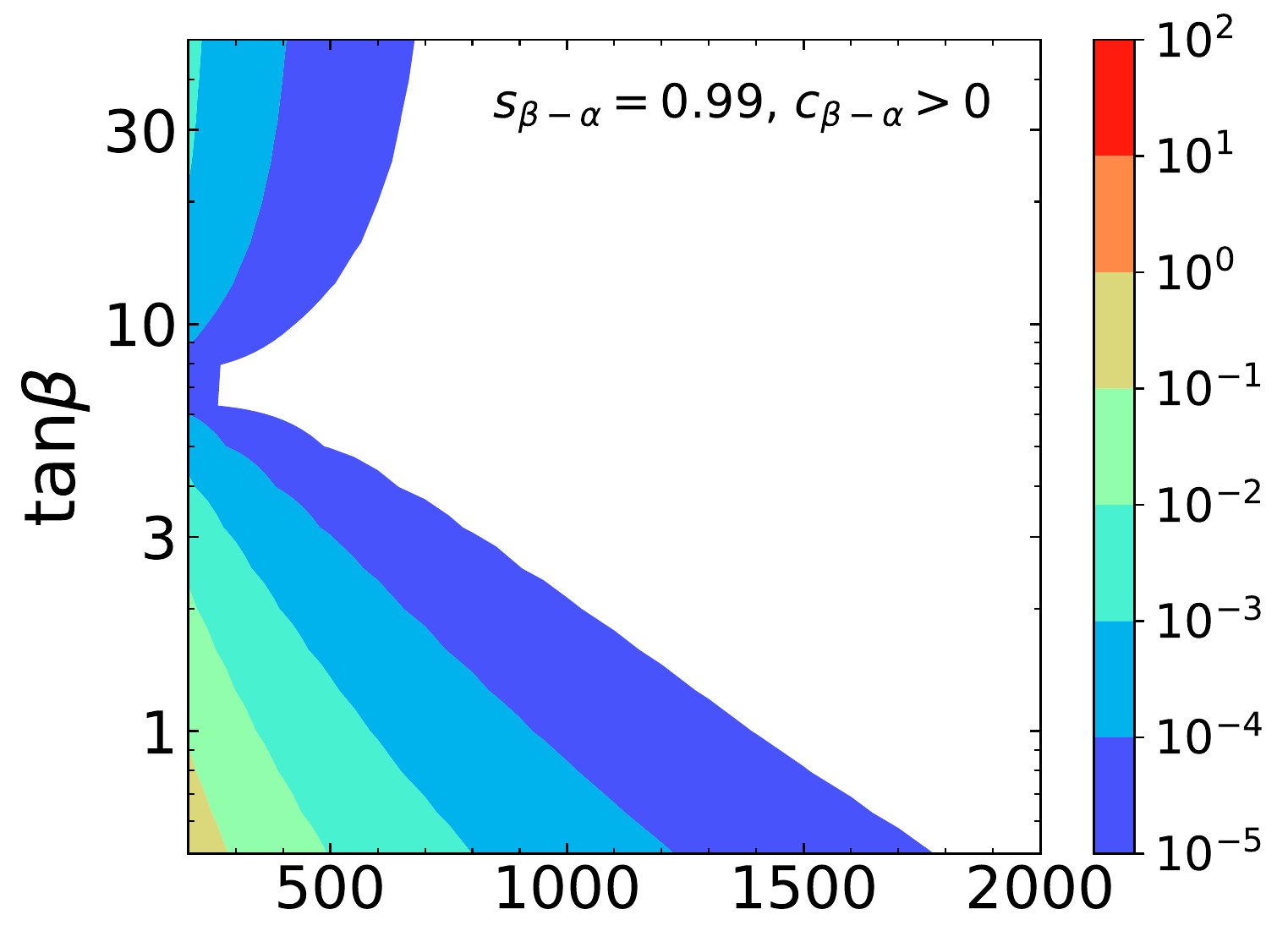}  
 \includegraphics[height=0.18\textwidth]{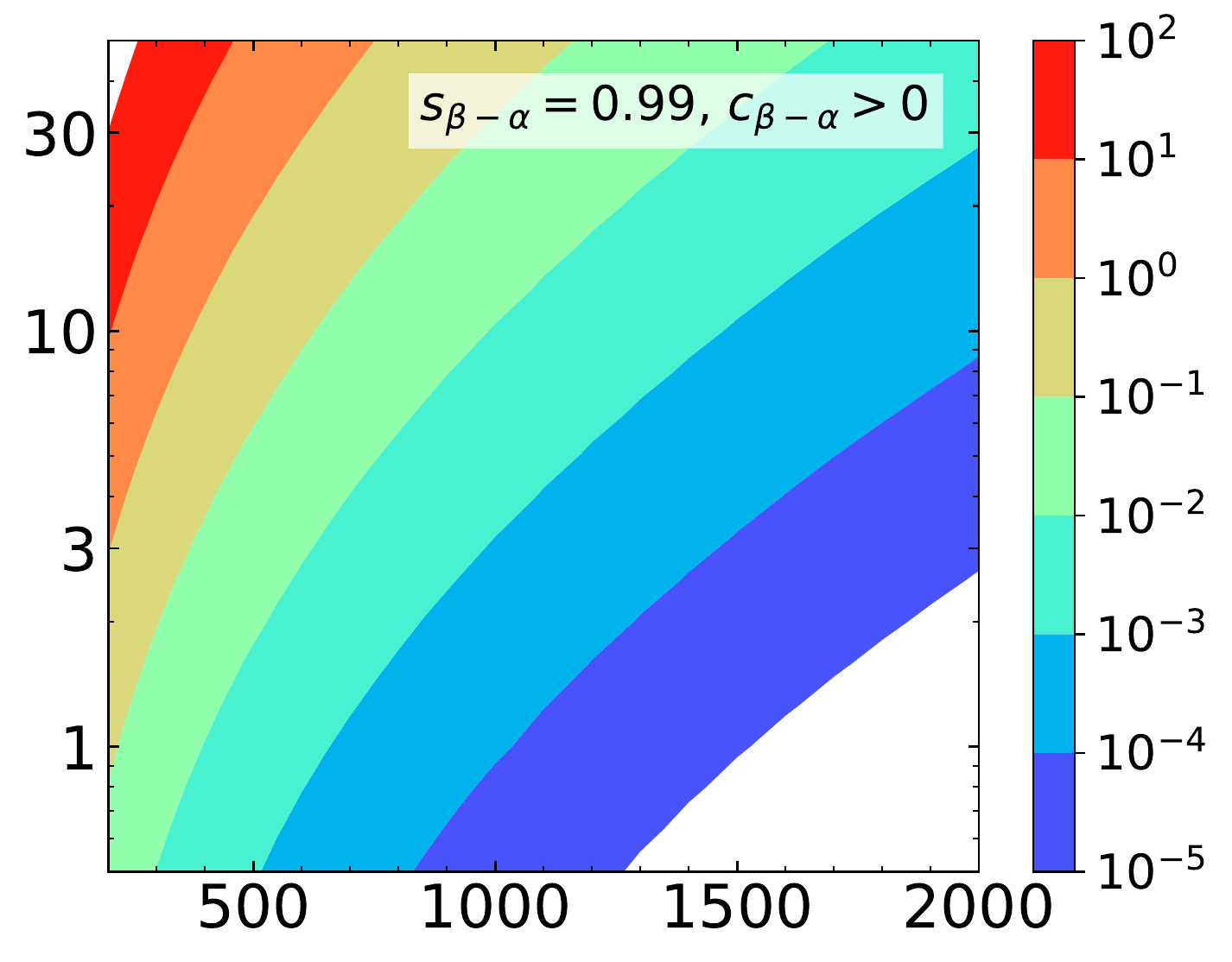}
 \includegraphics[height=0.194\textwidth]{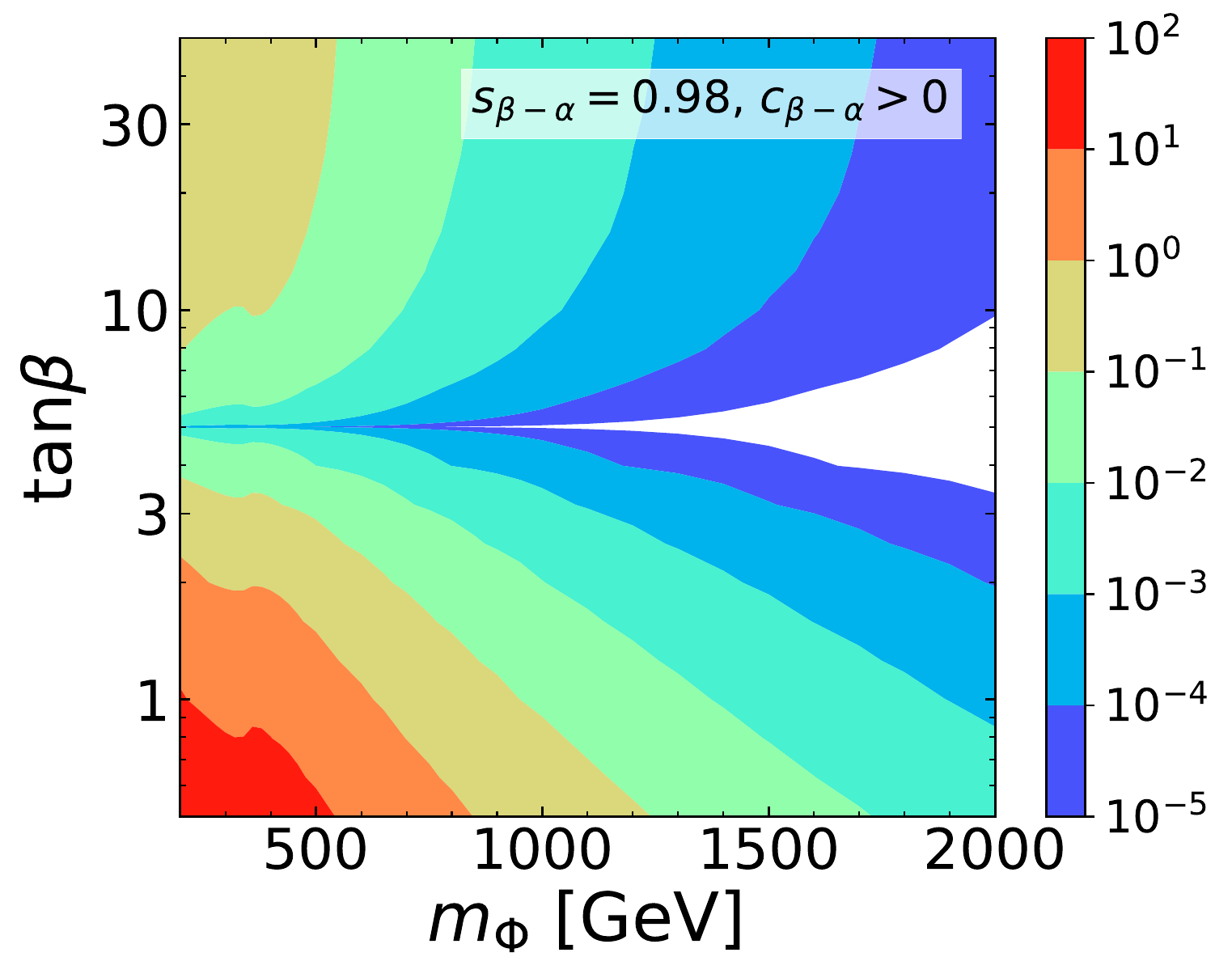}  
 \includegraphics[height=0.194\textwidth]{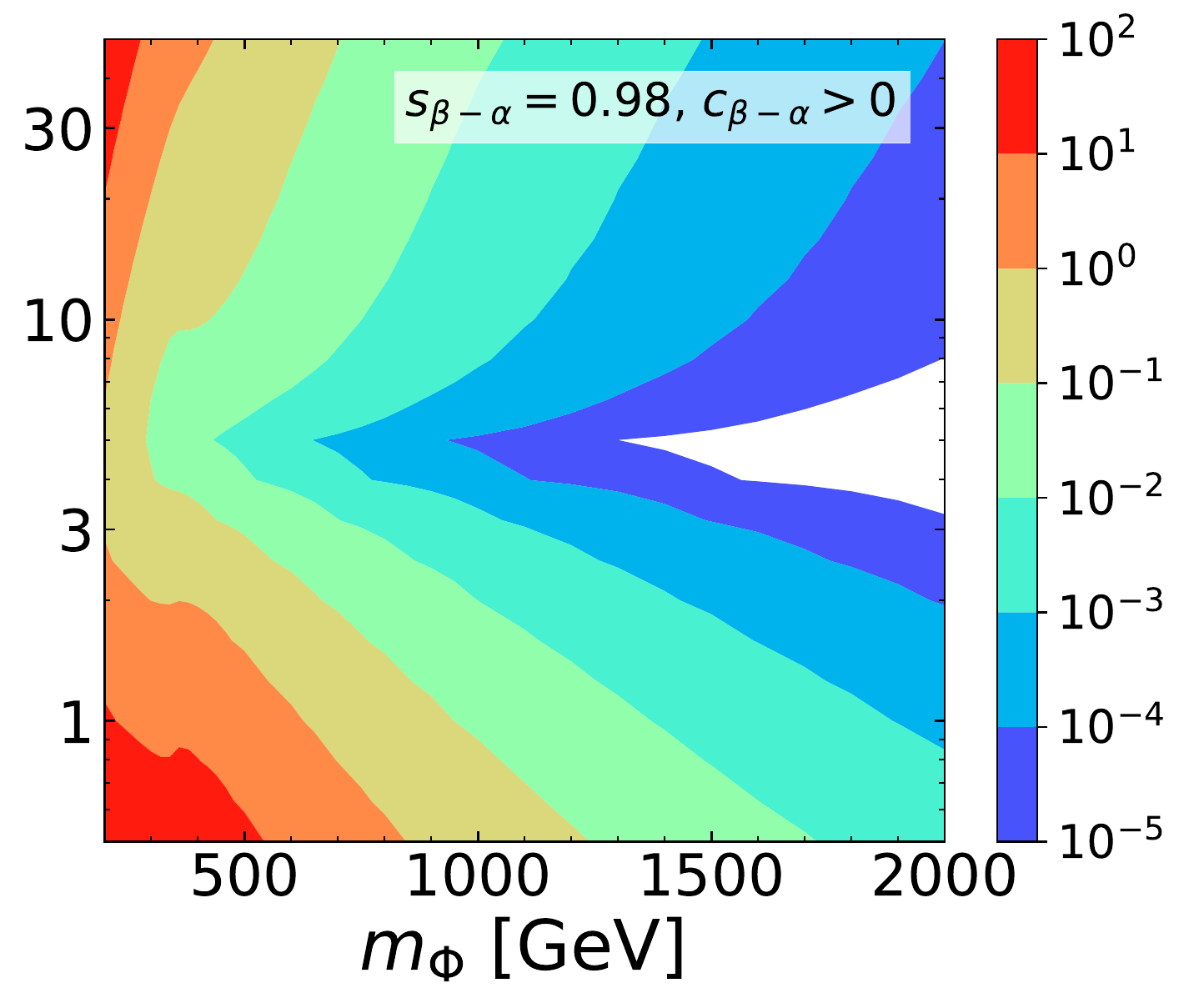}\quad  
 \includegraphics[height=0.194\textwidth]{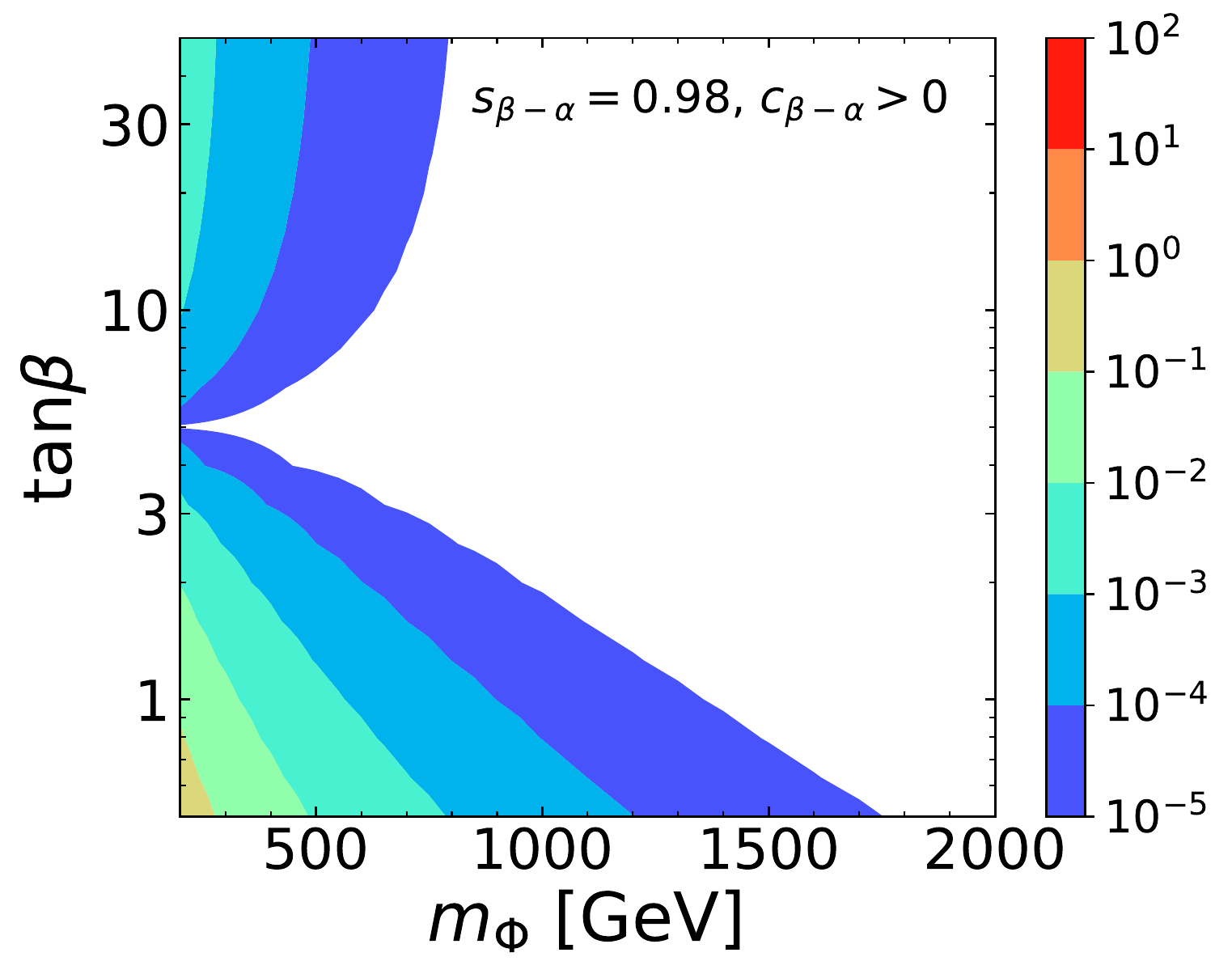}  
 \includegraphics[height=0.194\textwidth]{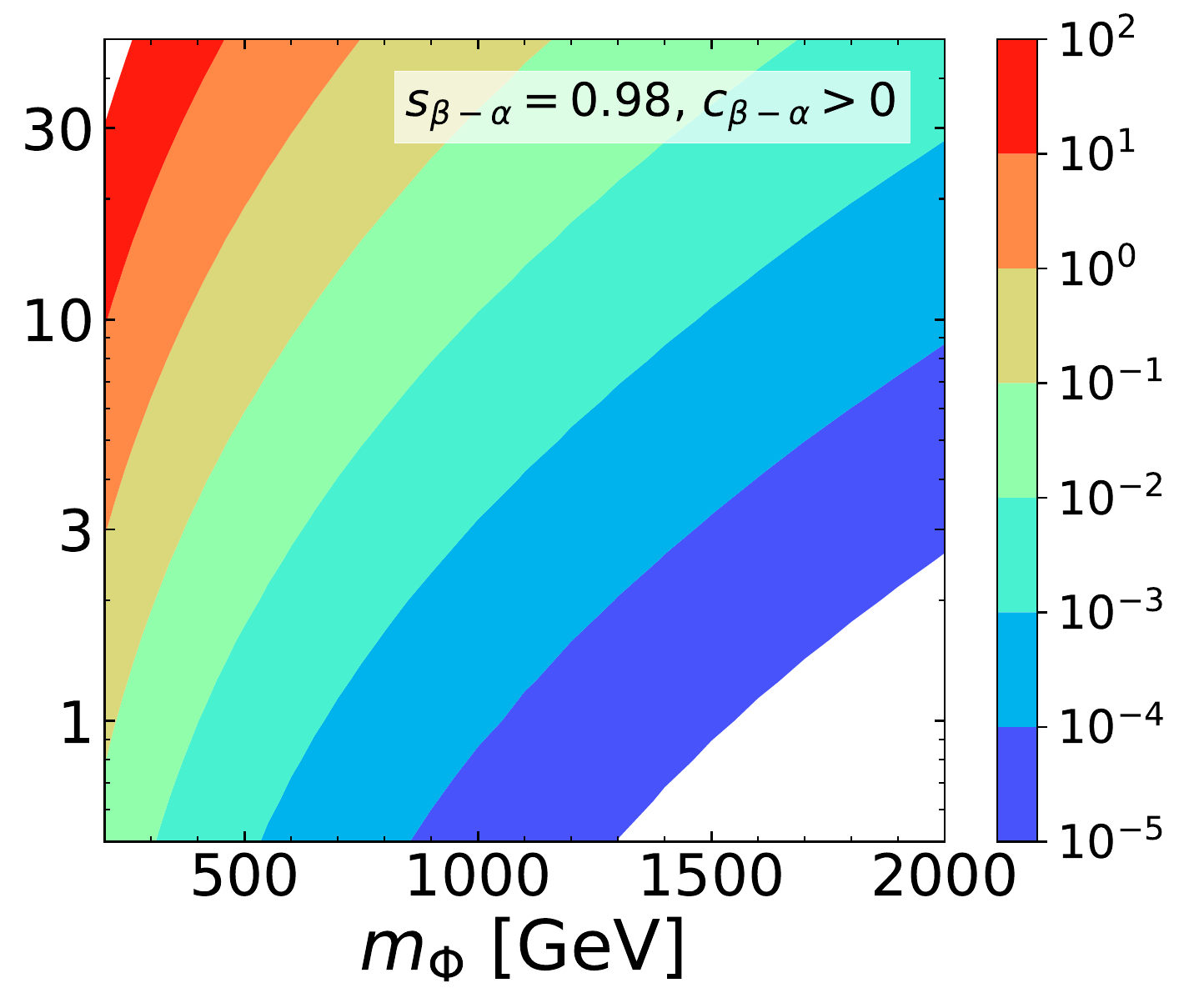}  
\caption{
Production cross sections for the CP-even heavy Higgs boson $H$ at the 13~TeV LHC 
on the $m_\Phi$--$\tan\beta$ plane.
Panels in two columns  from the left (right) show the production via the gluon fusion (the bottom-quark associated)
in the Type-I and Type-II THDMs, where
the value of $s_{\beta-\alpha}$ is set to be 1, 0.995, 0.99 and 0.98  
with $c_{\beta-\alpha} > 0$ from the top to the bottom panels.
The cross sections are shown with different colors from blue to red,
corresponding to from $10^{-5}$~pb to $10^{2}$~pb. 
}
\label{fig:xsec-H_cp}
\end{figure}    

Before we discuss current constraints on the parameter space from direct searches, 
we present production rates for the heavy Higgs bosons at the 13~TeV LHC.
Figure~\ref{fig:xsec-H_cn} shows cross sections for the CP-even heavy Higgs boson $H$
via the gluon fusion process (left two columns) and via the bottom-quark associated process (right two columns)
on the $m_\Phi$--$\tan\beta$ plane.
We only show the cases in the Type-I and Type-II THDMs
since the lepton sector is irrelevant for the productions, 
namely the productions in Type-X and Type-Y are same as in Type-I and Type-II, respectively.
The value of $s_{\beta-\alpha}$ is set to be 1, 0.995, 0.99, and 0.98 with $c_{\beta-\alpha}< 0$
from the top to the bottom panels.

For the gluon-fusion process, shown in the left two columns  in Fig.~\ref{fig:xsec-H_cn}, the Higgs bosons are produced via quark loops.
Therefore, the difference of the Yukawa sector between Type-I and Type-II in Eq.~\eqref{kappa_f}
leads to significantly different dependence on the model parameters.
In Type-I, where the top-quark loop is entirely dominant, 
the larger $\tb$ is, the smaller the cross section is for a fixed mass.
One can also see the threshold enhancement of the top-quark loop at $m_\Phi\sim 2m_t$. 
In Type-II, the top-quark loop is dominant for small $\tb$,
while the production via the bottom-quark loop becomes dominant for large $\tb$
because of the bottom-Yukawa enhancement.
The $\sba$ dependence of the cross sections is very small for small $\tb$.
In the large $\tb$ region, on the other hand,
the cross sections for a fixed mass tend to be larger as $\sba$ deviates from the alignment limit.
The production via the bottom-quark associated process, shown in the right two columns  in Fig.~\ref{fig:xsec-H_cn}, is entirely subdominant in Type-I, 
while that becomes dominant for large $\tb$ in Type-II. 

In Fig.~\ref{fig:xsec-H_cp}, similar to Fig.~\ref{fig:xsec-H_cn}, but for $c_{\beta-\alpha}>0$, 
we show the production rates.
In this case, except for the $b$-associate process in Type-II, 
the cross sections show a peculiar $\tb$ dependence since the top and the bottom Yukawa in Type-I
and the top Yukawa in Type-II, given in Eq.~\eqref{kappa_f}, vanishes for a certain $\sba$ and $\tb$;
 e.g., $\tb\sim10$ for the $\sba=0.995$ case.
 
\begin{figure}
 \includegraphics[height=0.205\textwidth]{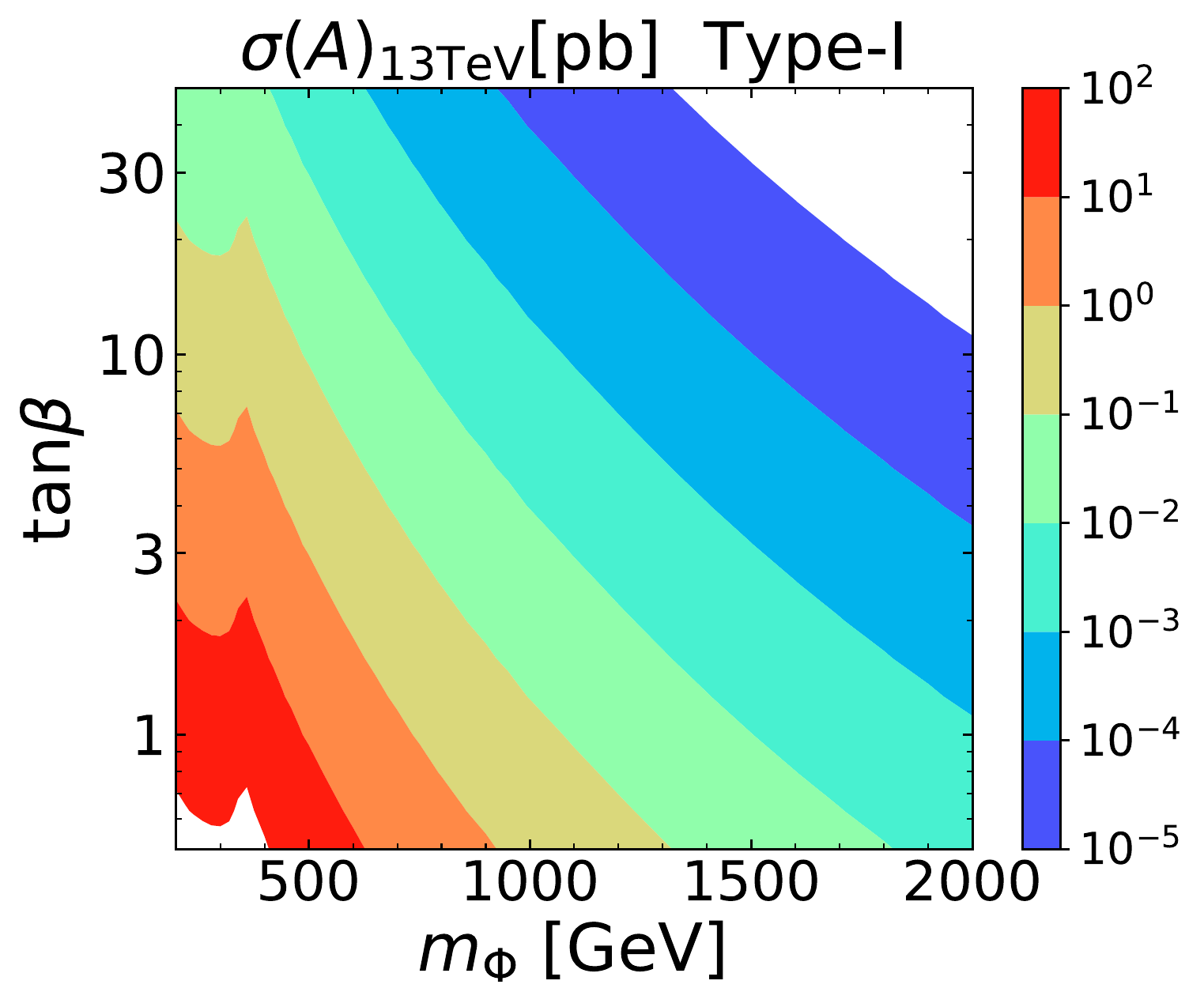}  
 \includegraphics[height=0.205\textwidth]{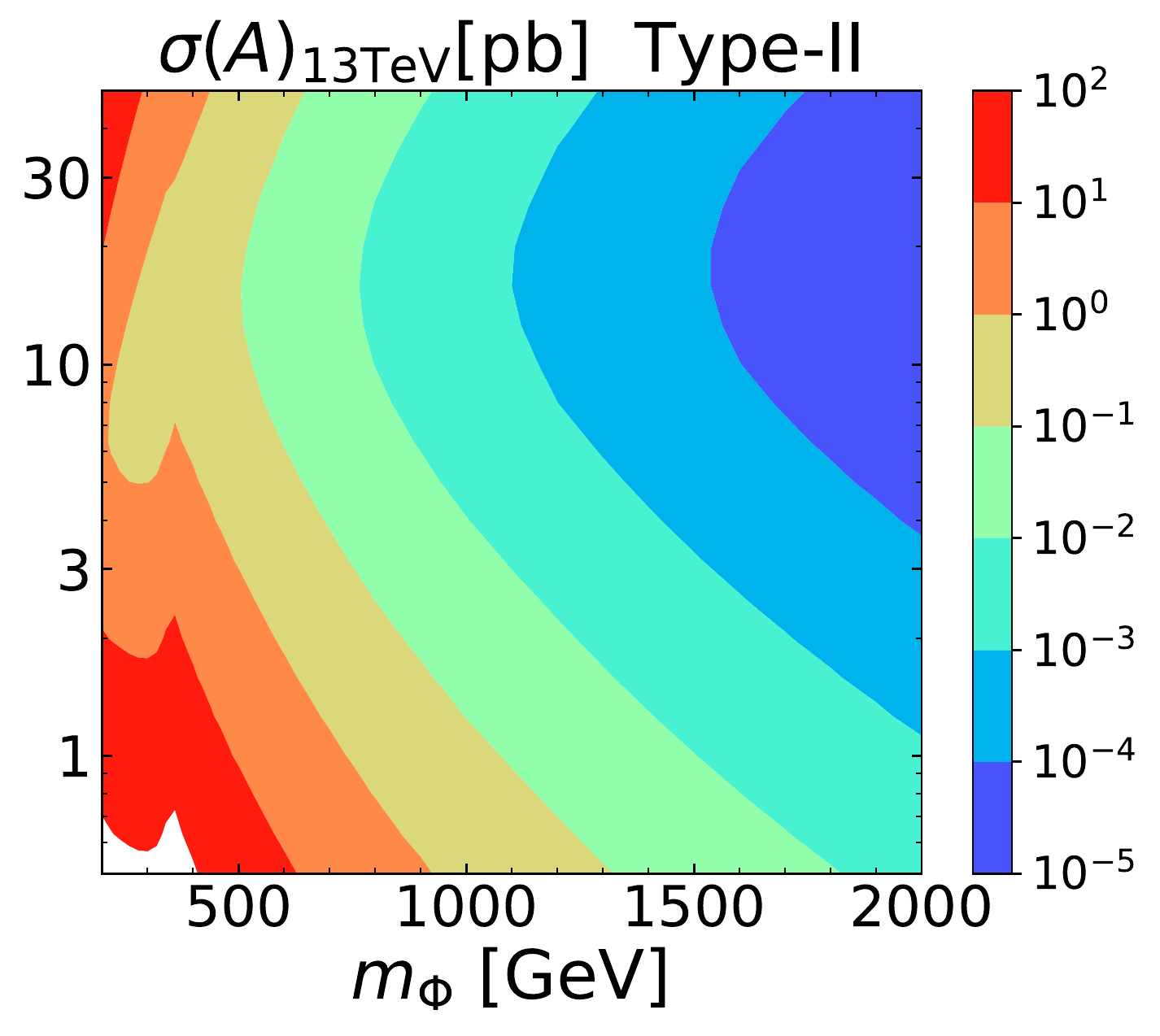}\quad  
 \includegraphics[height=0.205\textwidth]{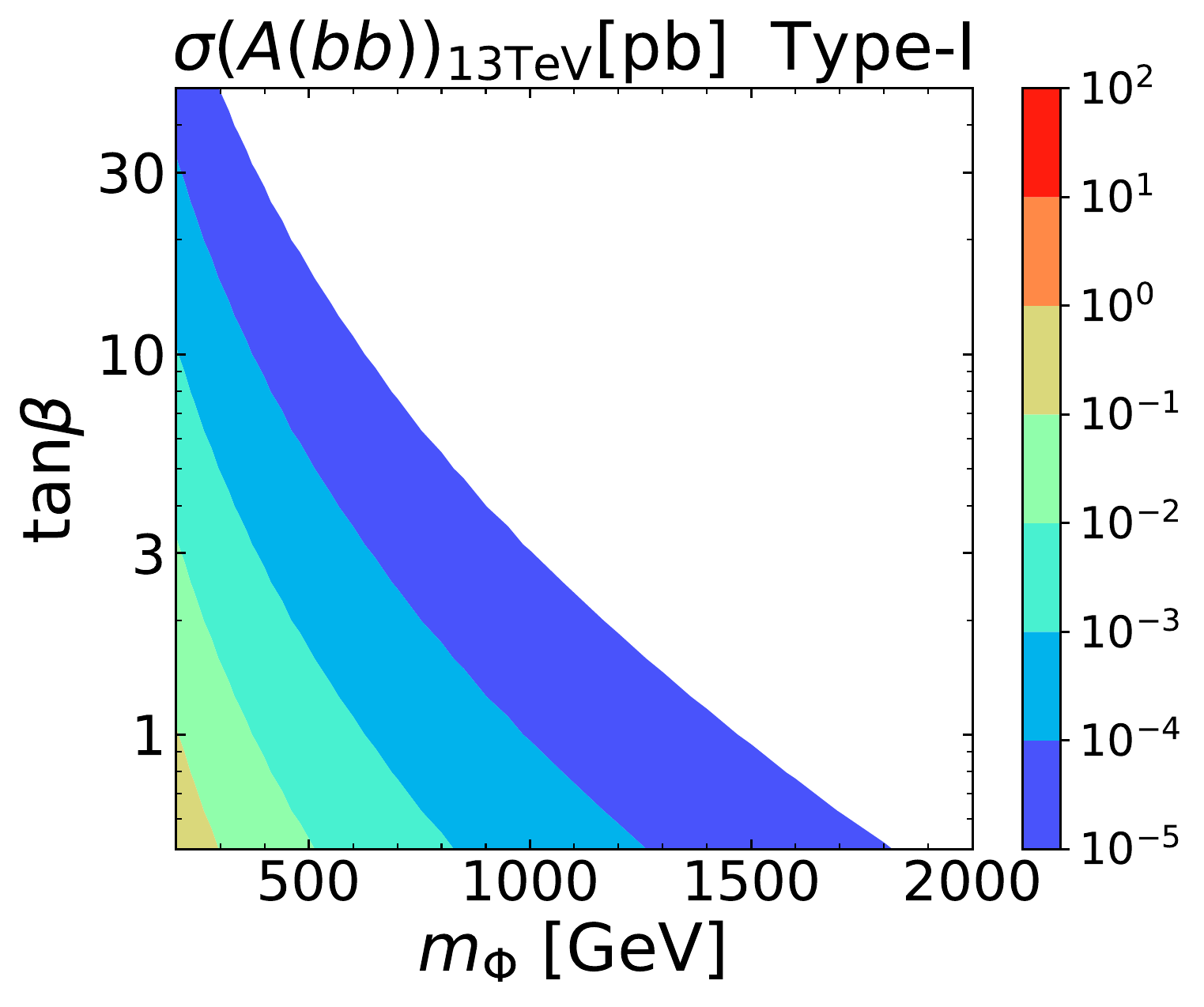}  
 \includegraphics[height=0.205\textwidth]{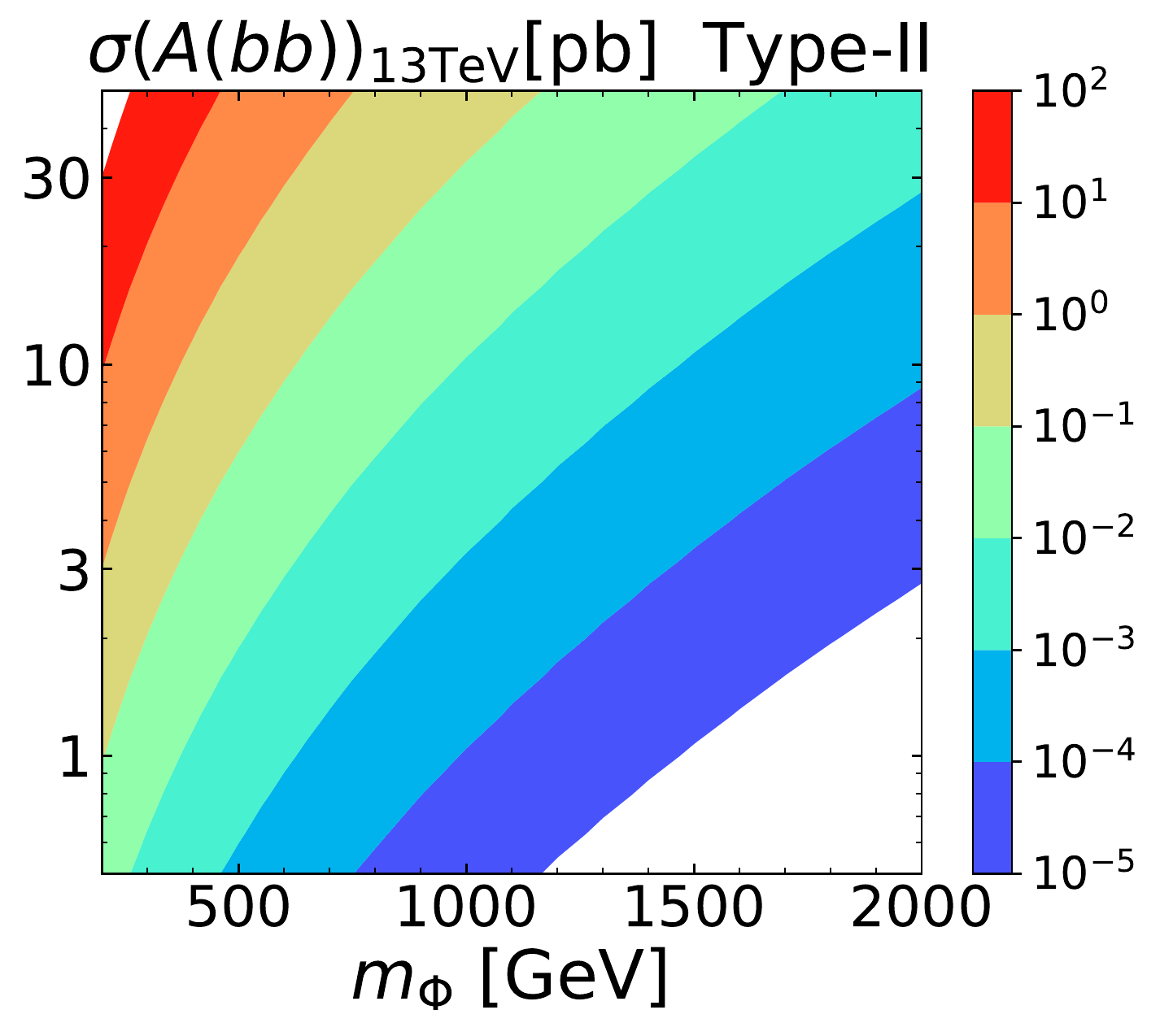}  
\caption{
Production cross sections for the CP-odd Higgs boson $A$ at the 13~TeV LHC 
on the $m_\Phi$--$\tan\beta$ plane.
Panels from the left to the right show the production via the gluon fusion 
in the Type-I and Type-II THDMs, and via the bottom-quark associated process
in the Type-I and Type-II THDMs, respectively.
}
\label{fig:xsec-A}
\end{figure} 

\begin{figure}
 \includegraphics[height=0.205\textwidth]{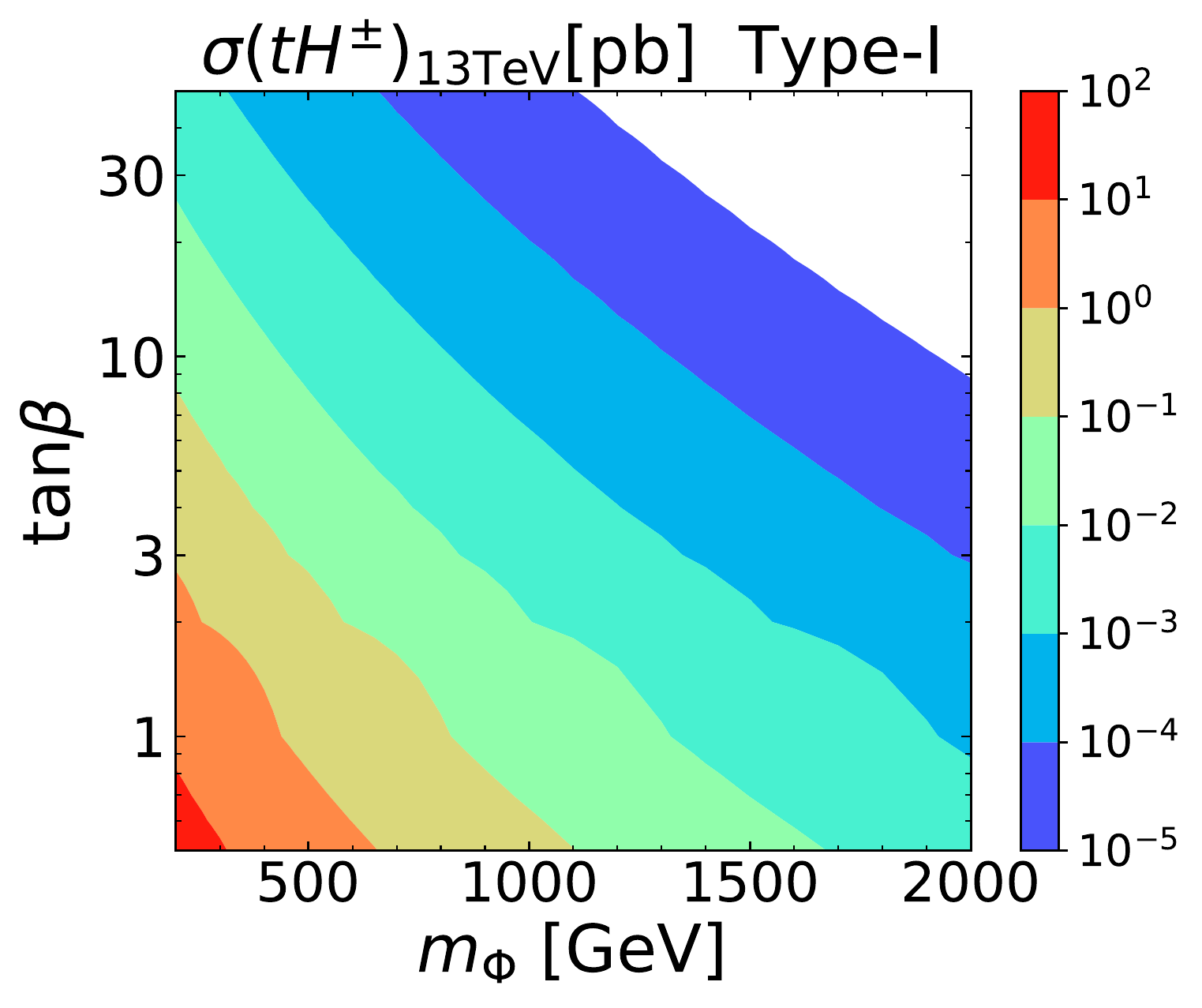}  
 \includegraphics[height=0.205\textwidth]{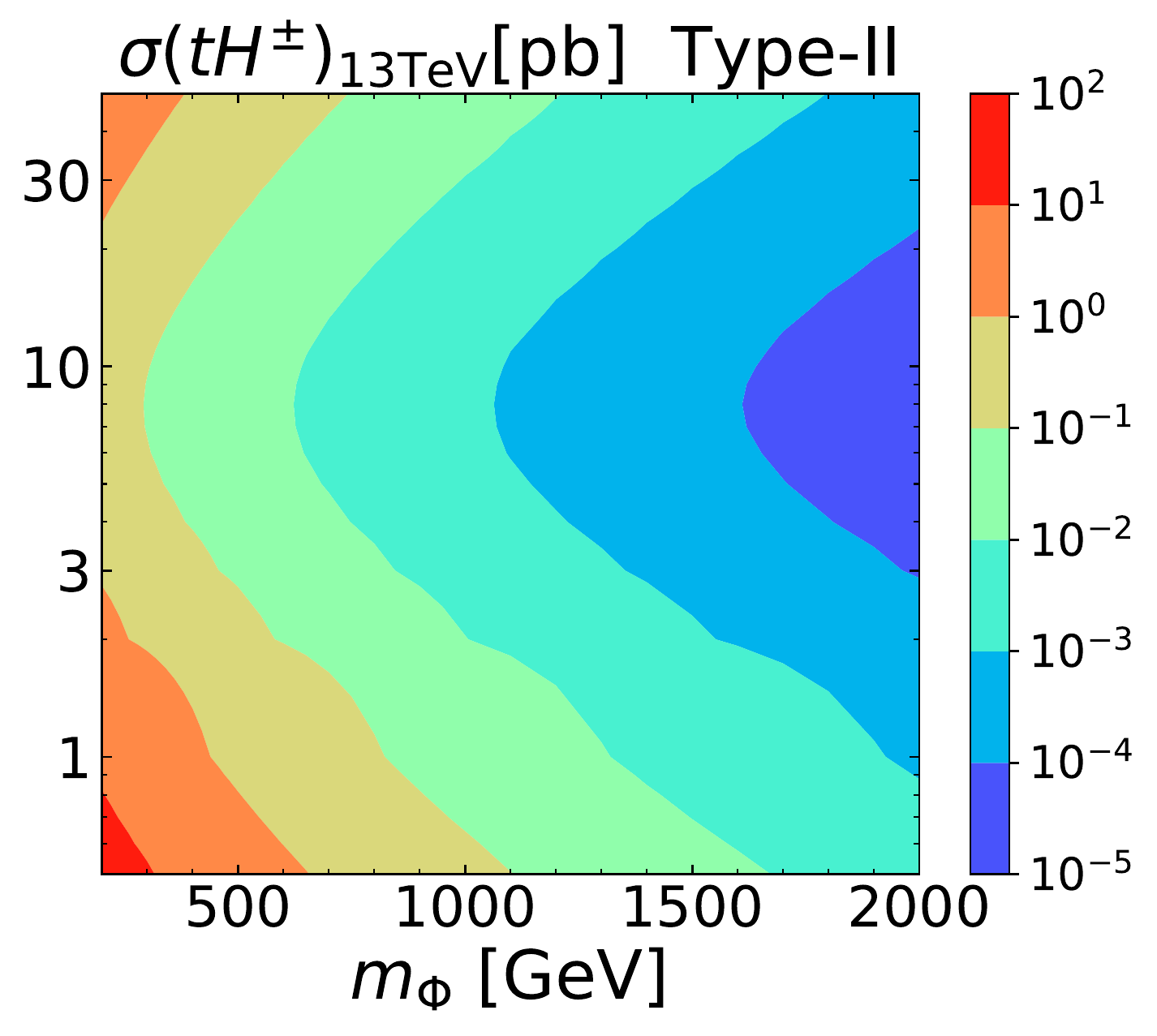}    
\caption{
Production cross sections for the charged Higgs boson $\hpm$ at the 13~TeV LHC 
on the $m_\Phi$--$\tan\beta$ plane in the Type-I (left) and Type-II (right) THDMs.
}
\label{fig:xsec-C}
\end{figure} 

Figure~\ref{fig:xsec-A} presents production rates for the CP-odd Higgs boson $A$.
The production processes are same as those for $H$, shown in Figs.~\ref{fig:xsec-H_cn} and \ref{fig:xsec-H_cp},
namely the gluon fusion process (left two columns)
and the bottom-quark associated process (right two columns).
Different from the CP-even Higgs bosons, the production rates only depend on $\tb$  
because of the Yukawa structure in Eq.~\eqref{kappa_f}. 
The global parameter dependence of the cross sections via the gluon fusion
is similar to that for $H$ with $\sba=1$, but 
the production rate for $A$ is slightly larger than that for $H$ 
at each point on the $m_\Phi$--$\tan\beta$ plane.
The parameter dependence of the cross sections via the bottom-quark annihilation
is as same as for $H$ with $\sba=1$.

In Fig.~\ref{fig:xsec-C}, at the LHC charged Higgs bosons $H^\pm$ are mainly produced in association with a top quark via $gb\to tH^\pm$ for $m_{H^{\pm}}>m_{t}$,
whose cross sections are shown.
Similar to the productions for $A$, the cross section only depends on $\tb$.
For a fixed mass, in Type-I, the larger $\tb$ is, the smaller the production rate is.
 In Type-II, on the other hand, up to $\tb\sim7$, the larger $\tb$ is, the smaller the production rate is,
 similar to the Type-I case. 
 However, for $\tb\gtrsim7$, the production rate becomes larger for larger $\tb$ due to $\tb$ enhancement of the bottom-Yukawa coupling.

We here mention other heavy Higgs boson productions. 
Although we assume $m_{\hpm}=m_H$ in this study, if the $H/A\to\hpm W^{\mp}$ decay is kinematically allowed,
the production via $gg\to H\to\hpm W^{\mp}$ can be comparable with that via $gb\to tH^\pm$~\cite{Dicus:1989vf,BarrientosBendezu:1998gd,Moretti:1998xq,Akeroyd:2016ymd}. 
\ken{
Heavy Higgs bosons are also produced in electroweak processes such as $HA$, $\hpm h/H/A$, and $H^+H^-$~\cite{Moretti:2001pp,Alves:2005kr}, as well as in loop induced processes such as $H^{\pm}W^{\mp}$~\cite{Brein:2000cv} and $H^+H^-$~\cite{Brein:1999sy}.
}

\subsection{Constraints from the direct searches}

\begin{figure}
 \includegraphics[height=0.235\textwidth]{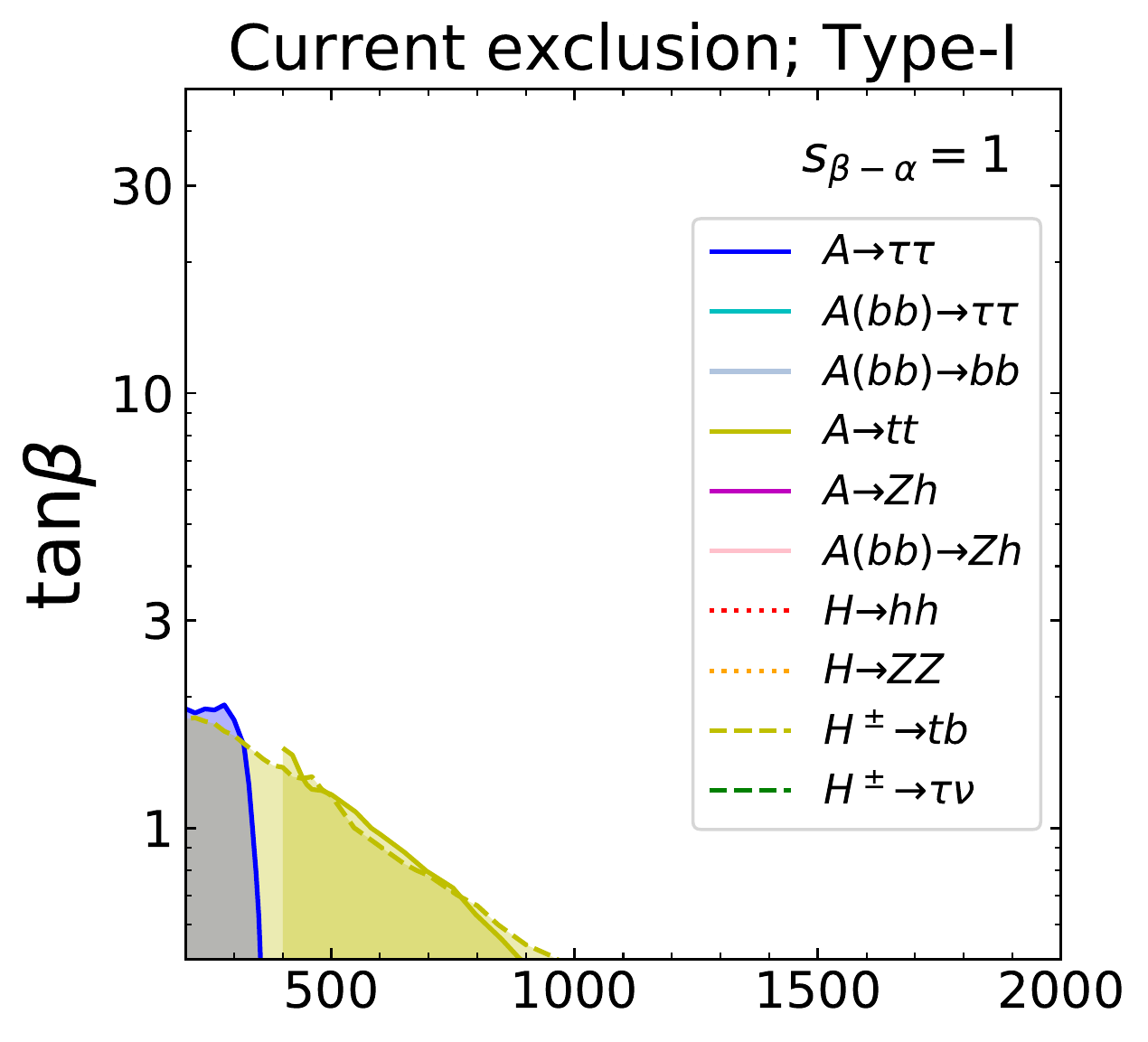}
 \includegraphics[height=0.235\textwidth]{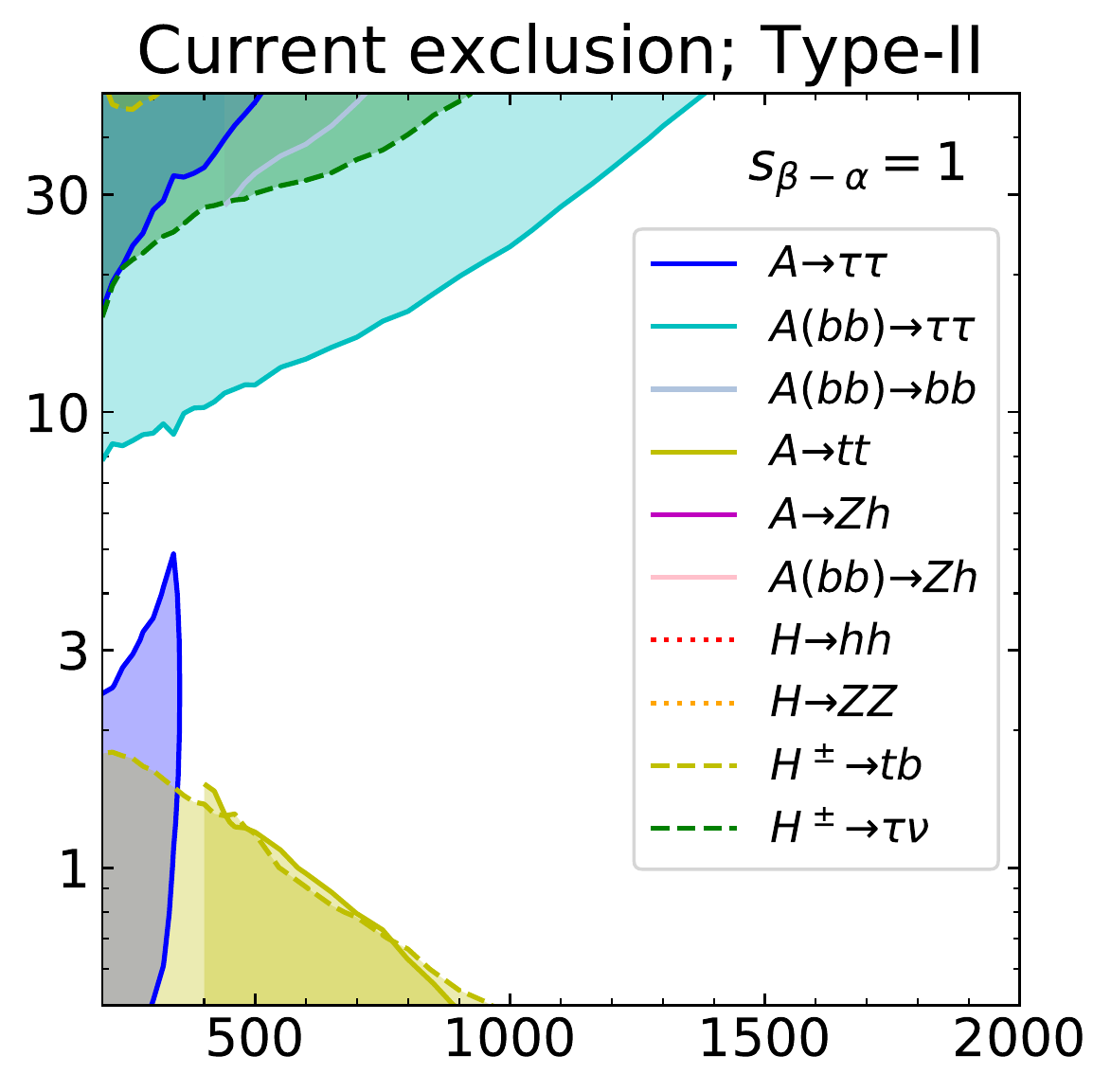}
 \includegraphics[height=0.235\textwidth]{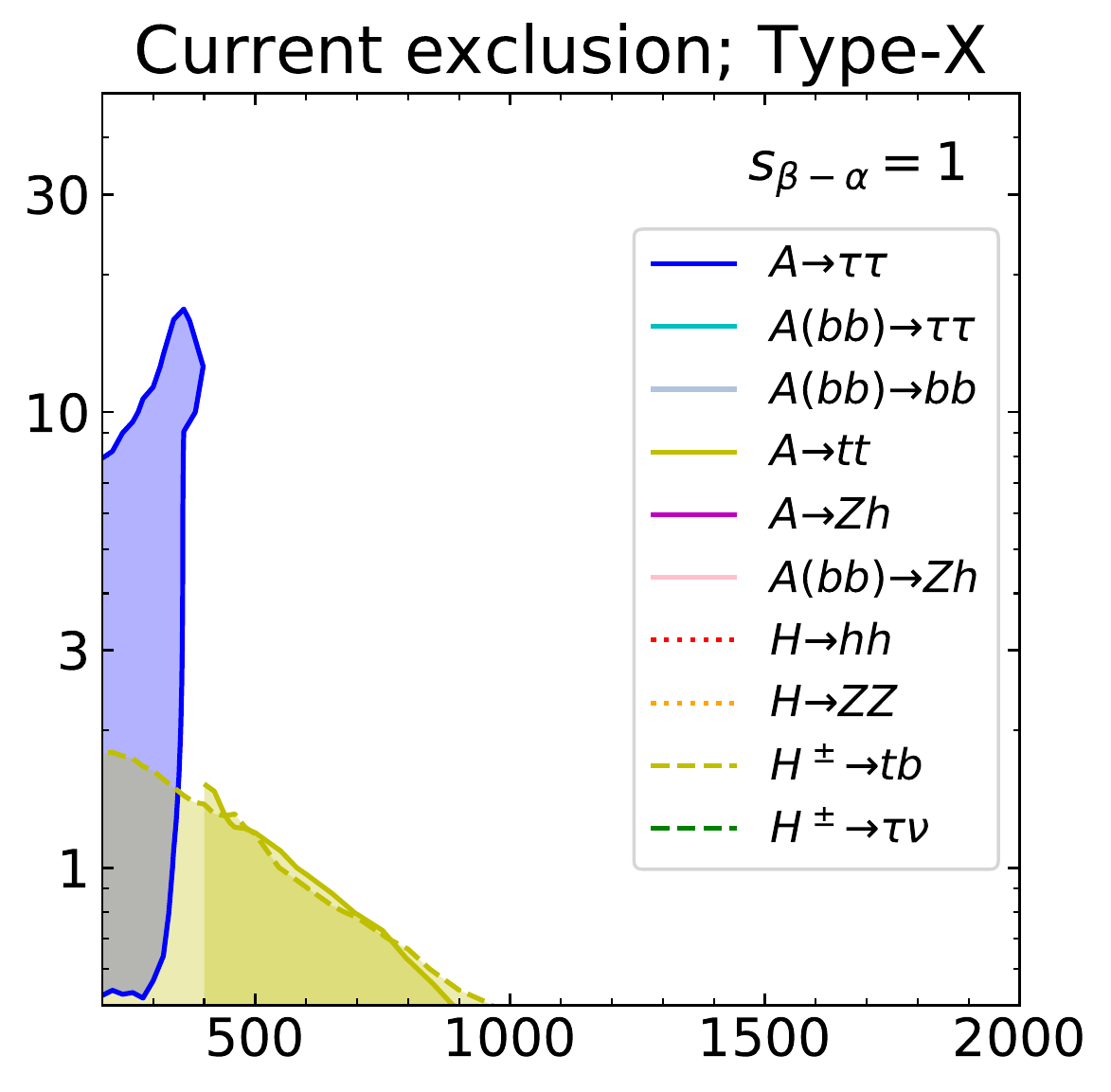}
 \includegraphics[height=0.235\textwidth]{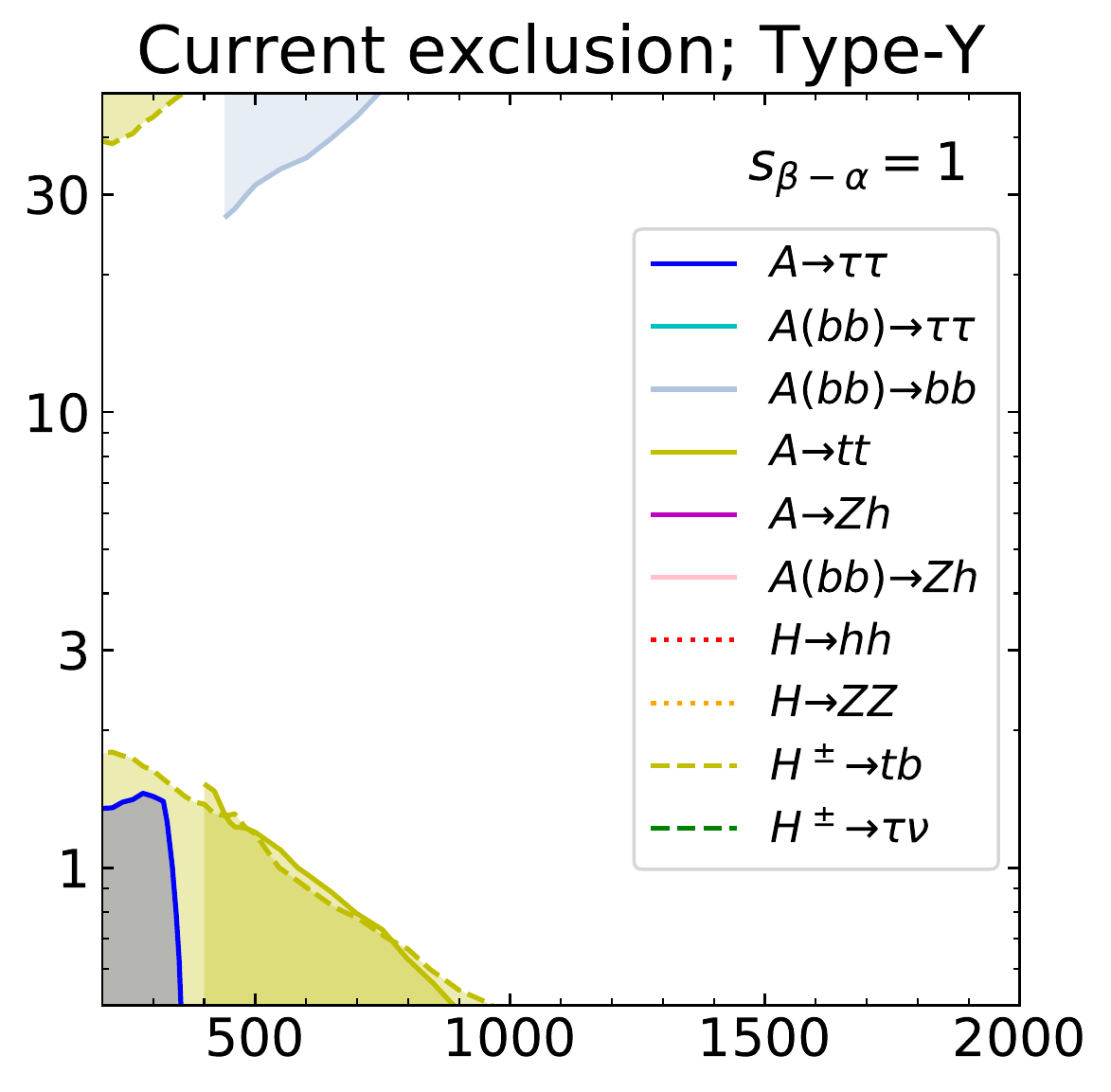}
 \includegraphics[height=0.22\textwidth]{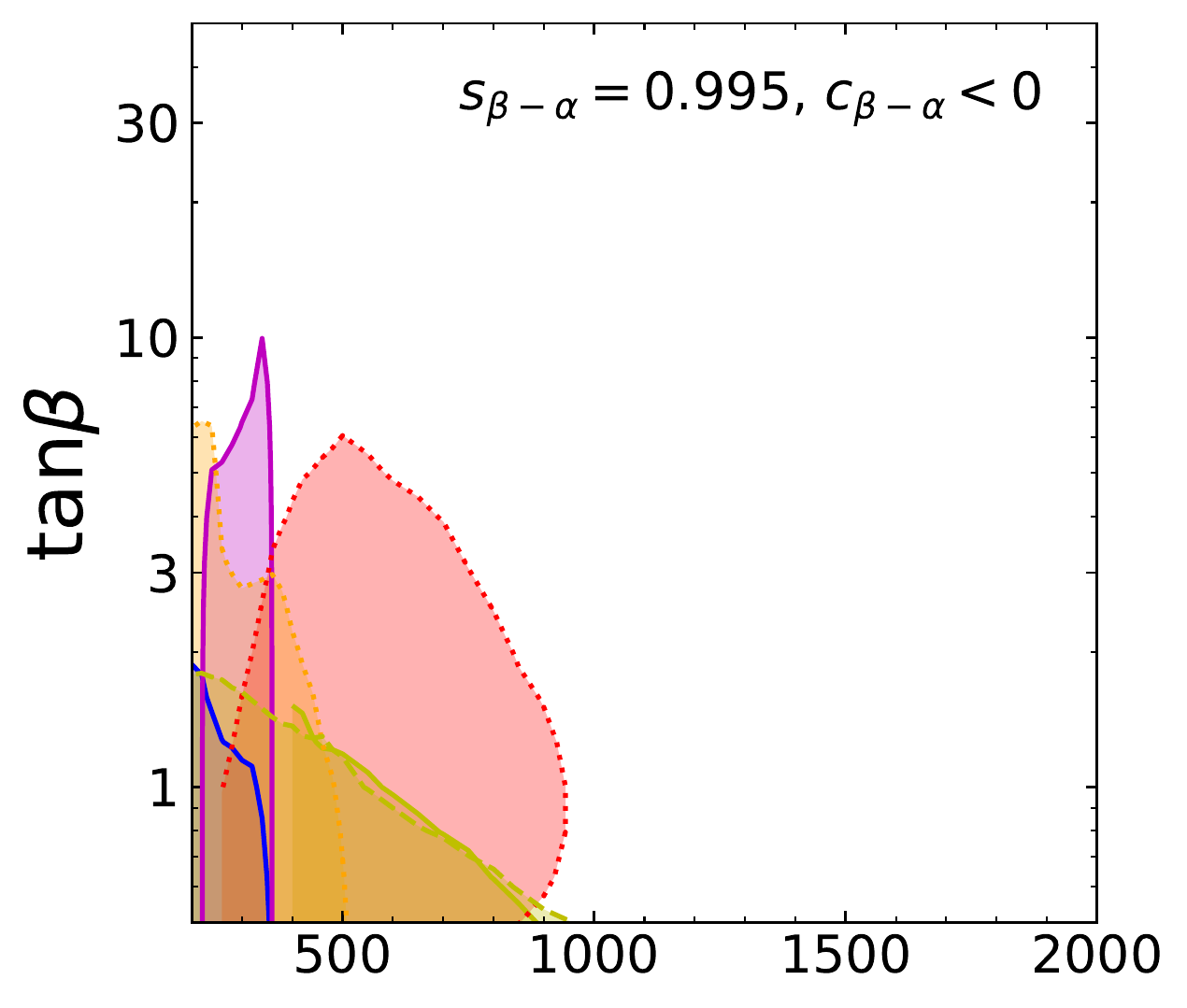}
 \includegraphics[height=0.22\textwidth]{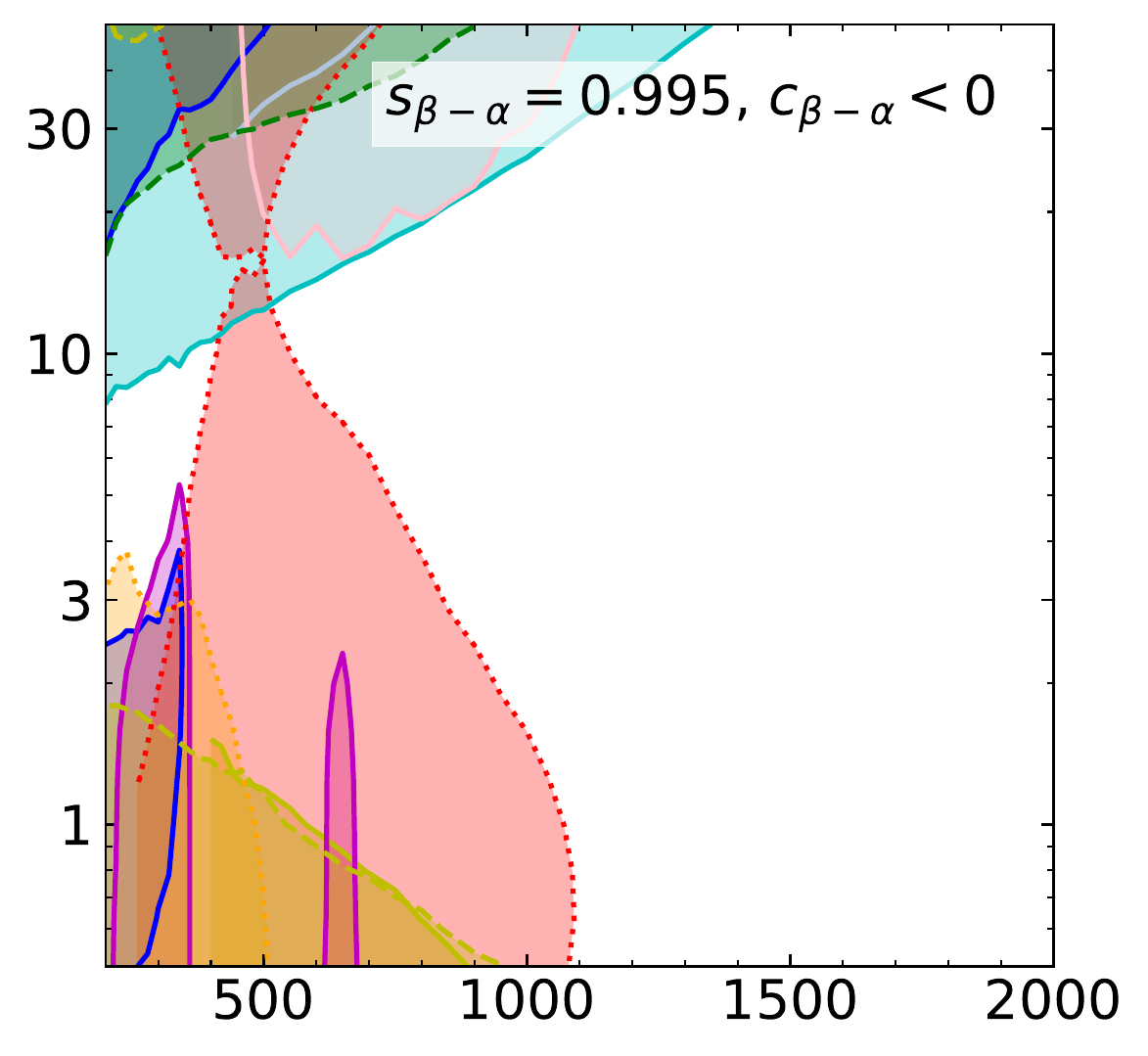}
 \includegraphics[height=0.22\textwidth]{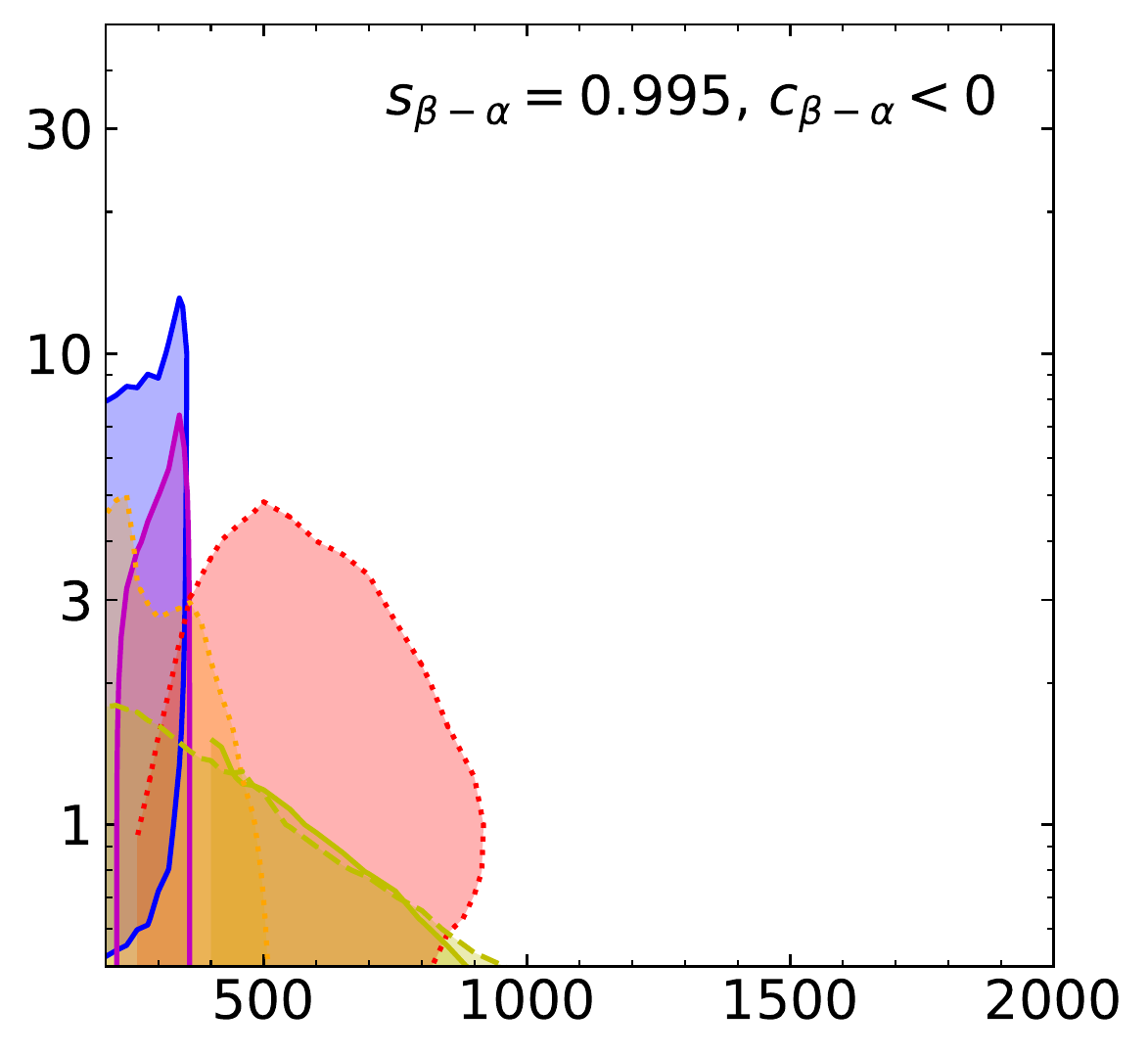}
 \includegraphics[height=0.22\textwidth]{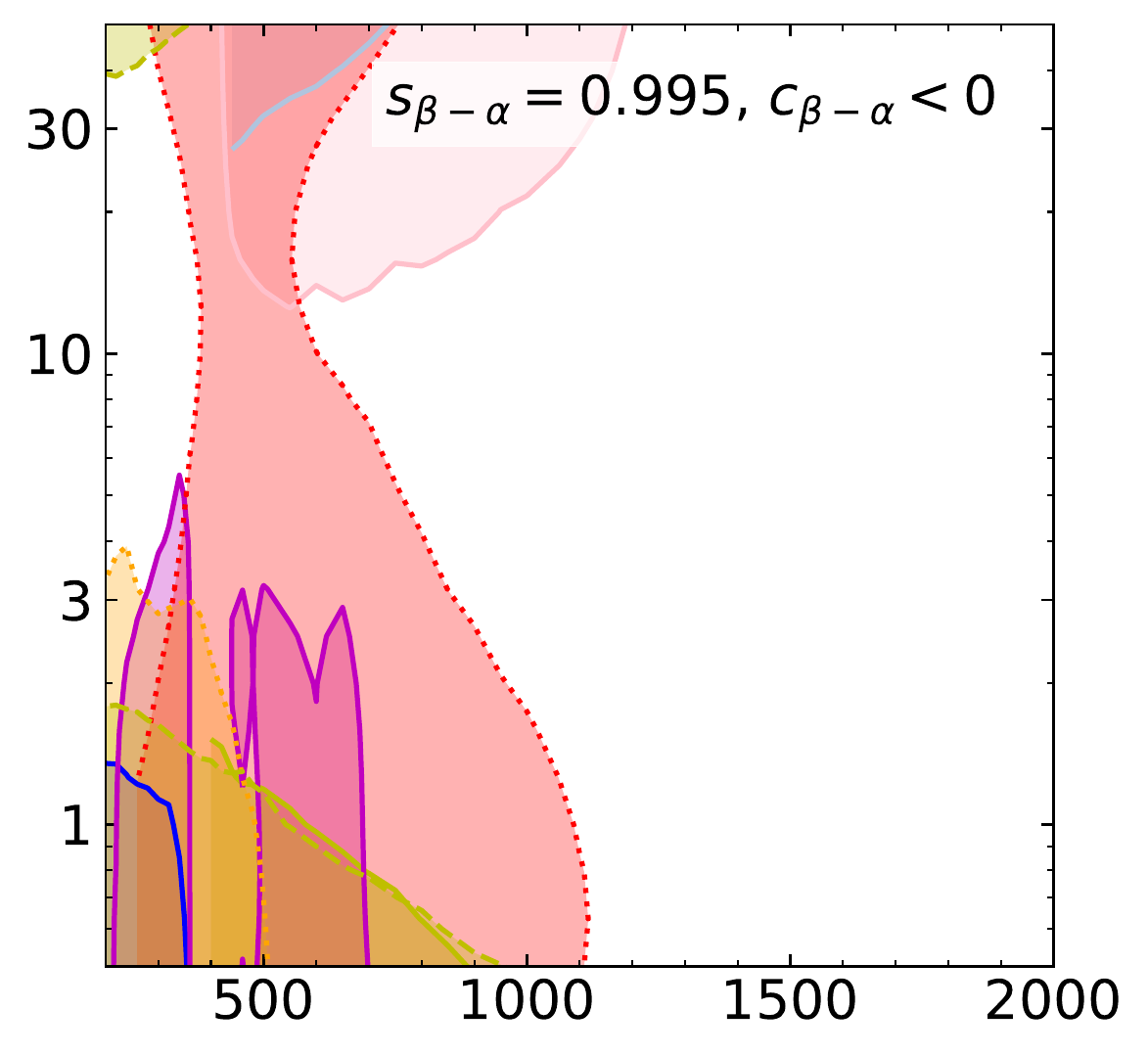}   
 \includegraphics[height=0.22\textwidth]{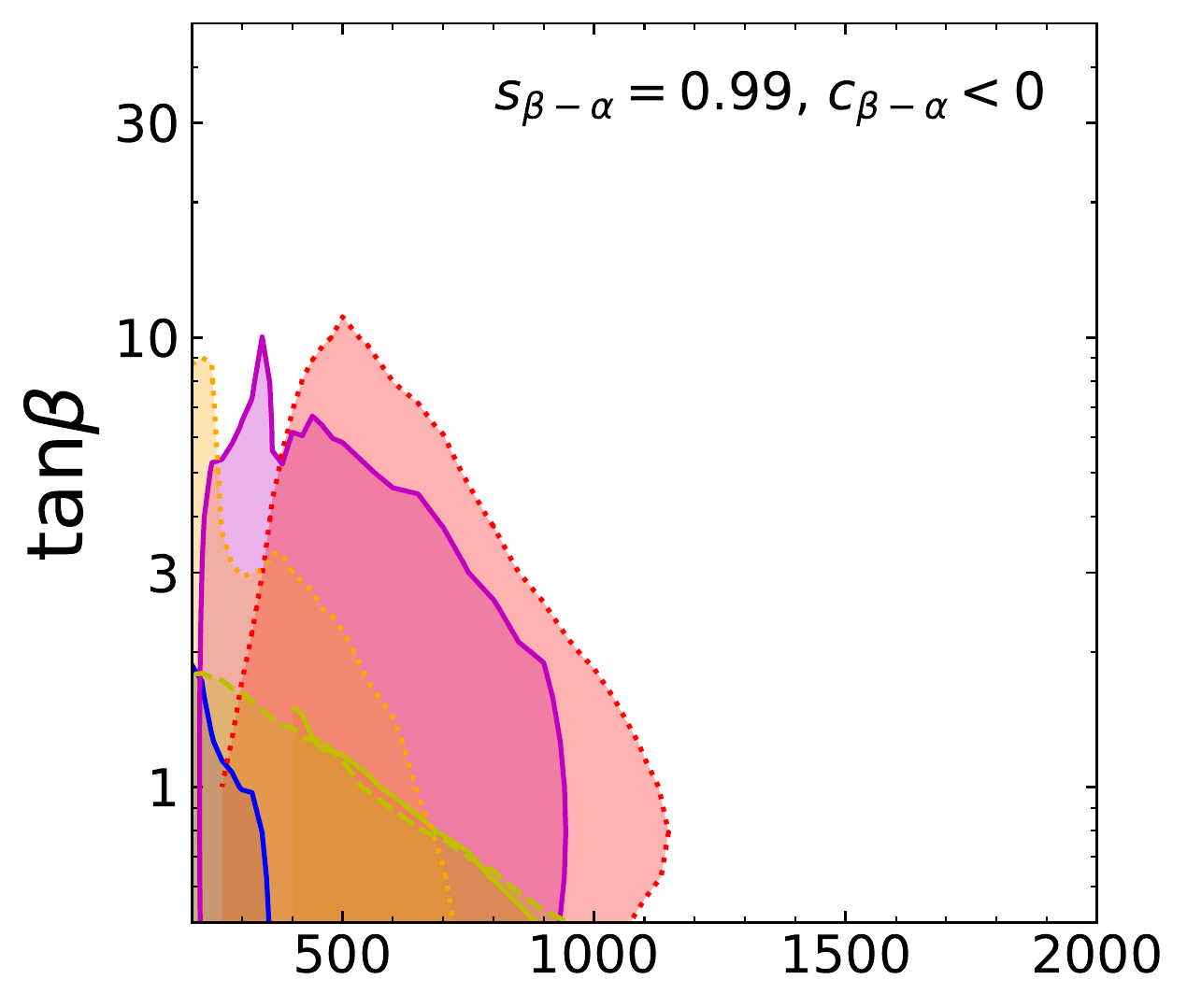}
 \includegraphics[height=0.22\textwidth]{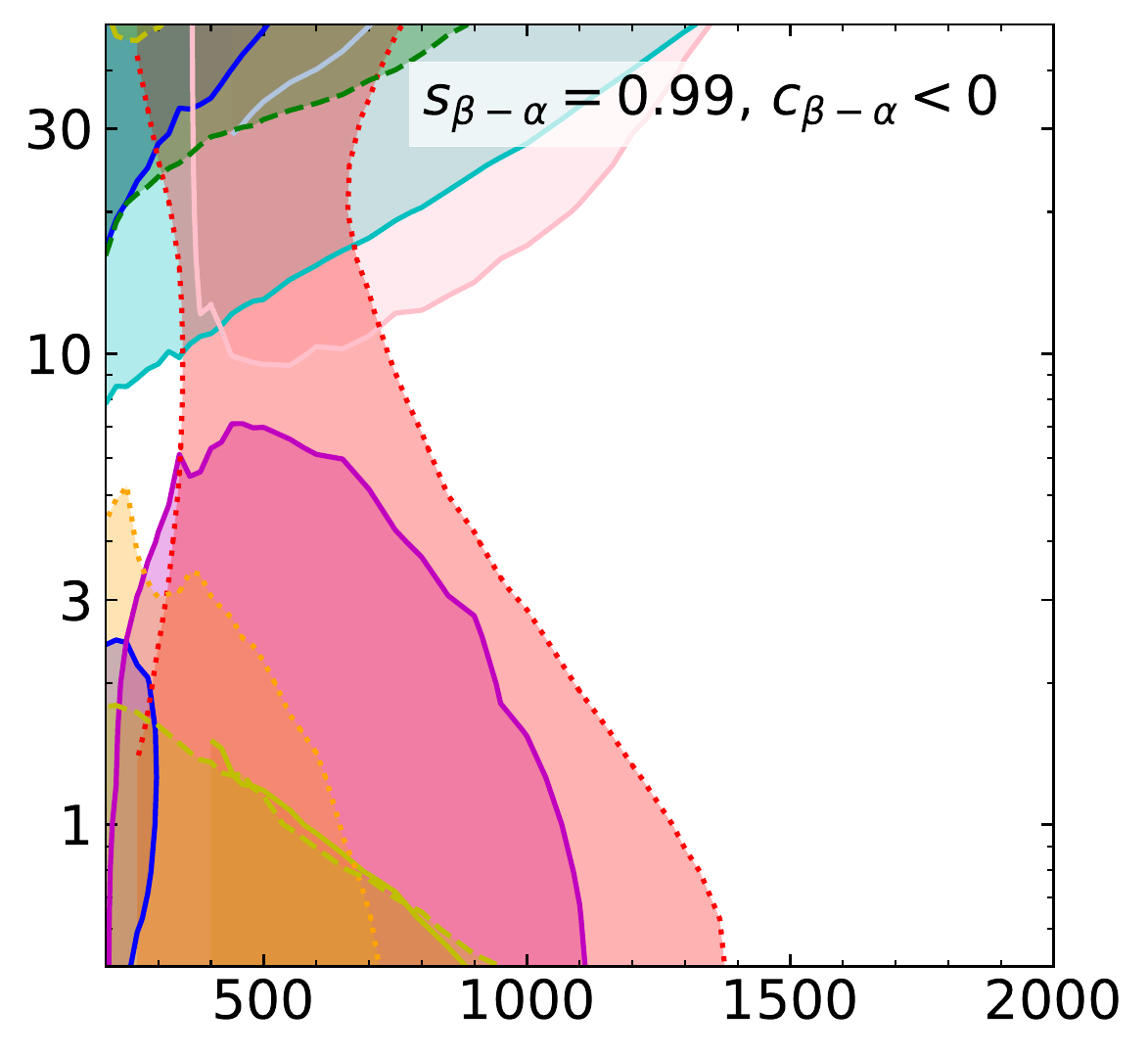}
 \includegraphics[height=0.22\textwidth]{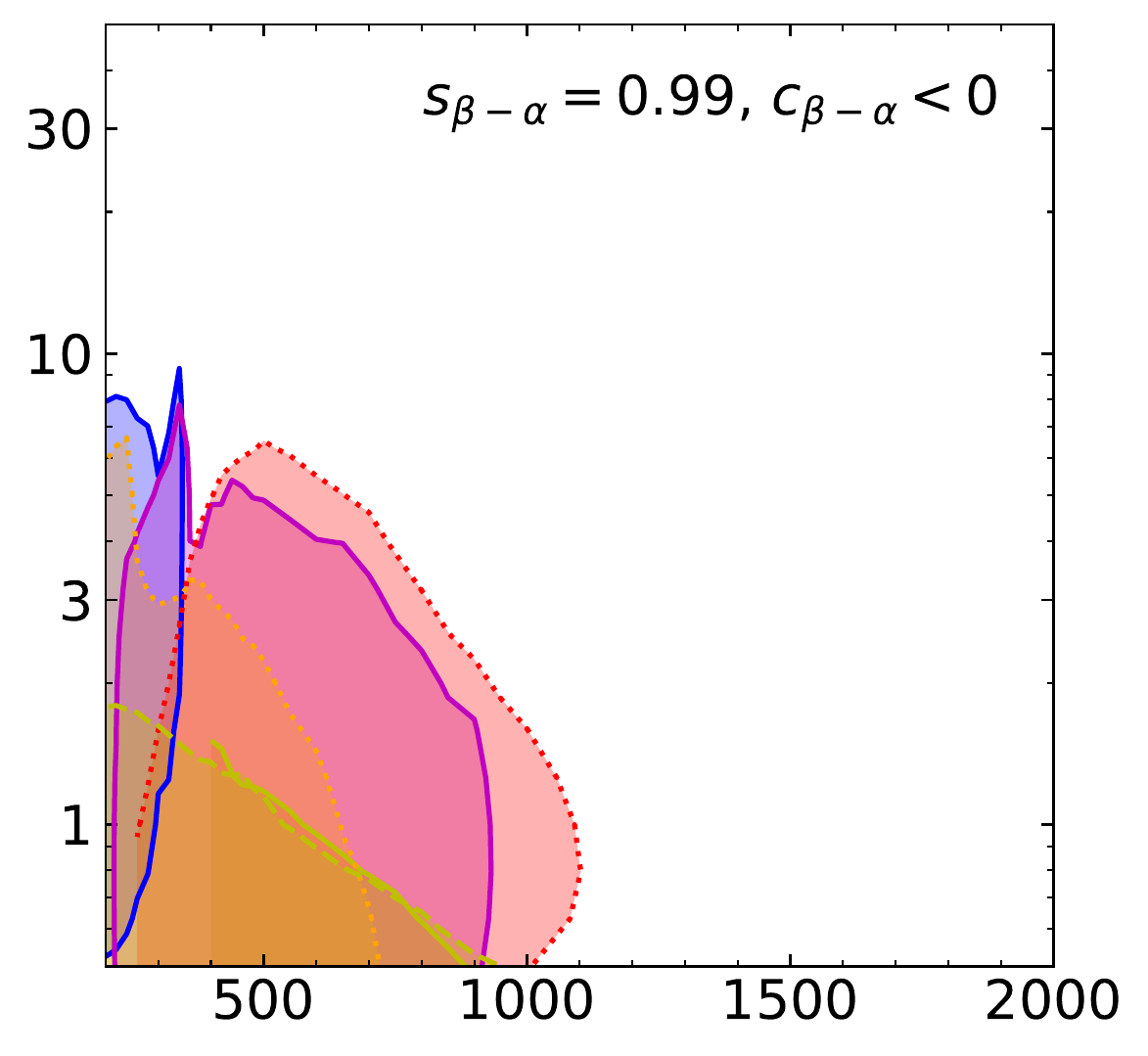}
 \includegraphics[height=0.22\textwidth]{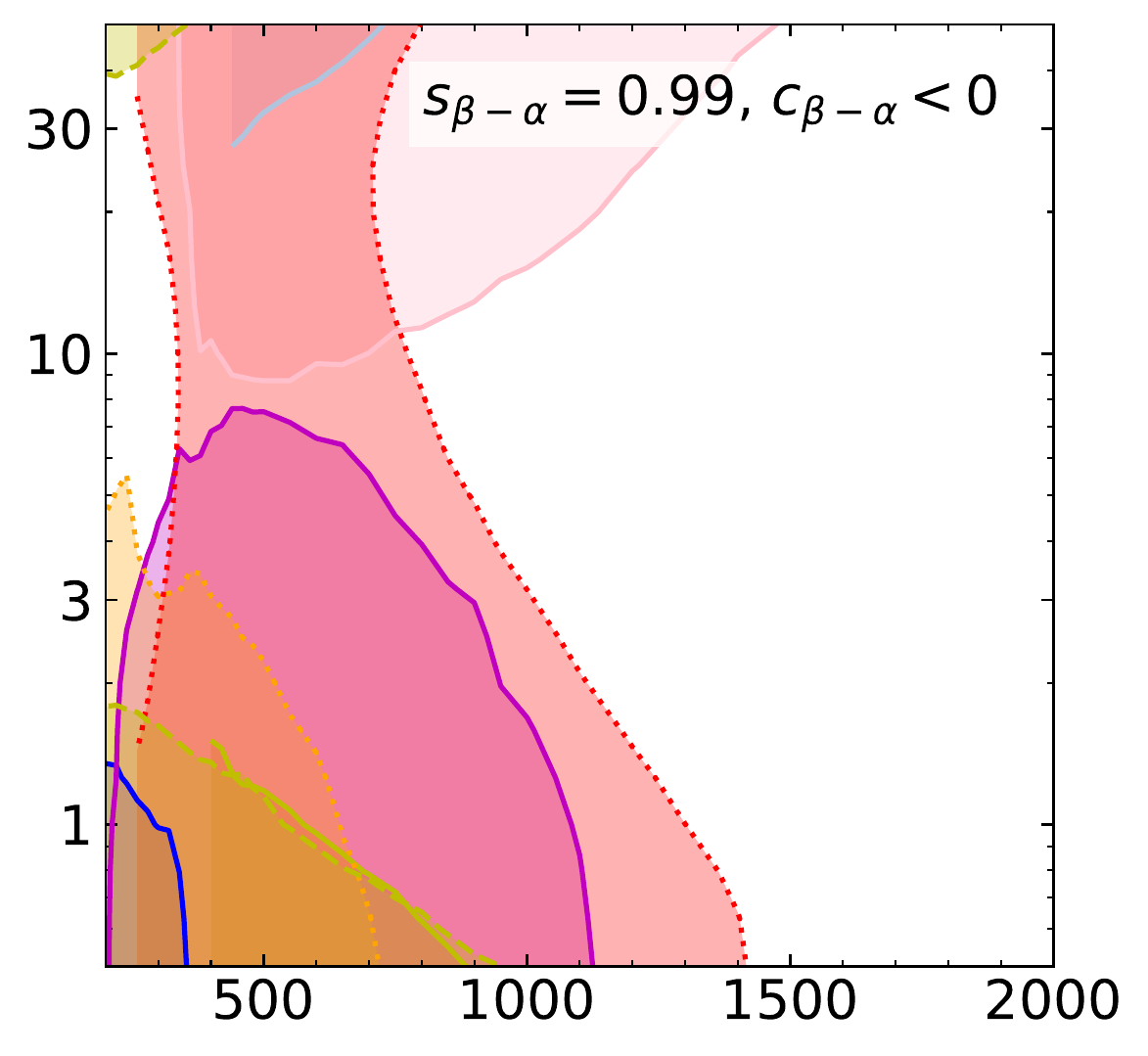} 
 \includegraphics[height=0.237\textwidth]{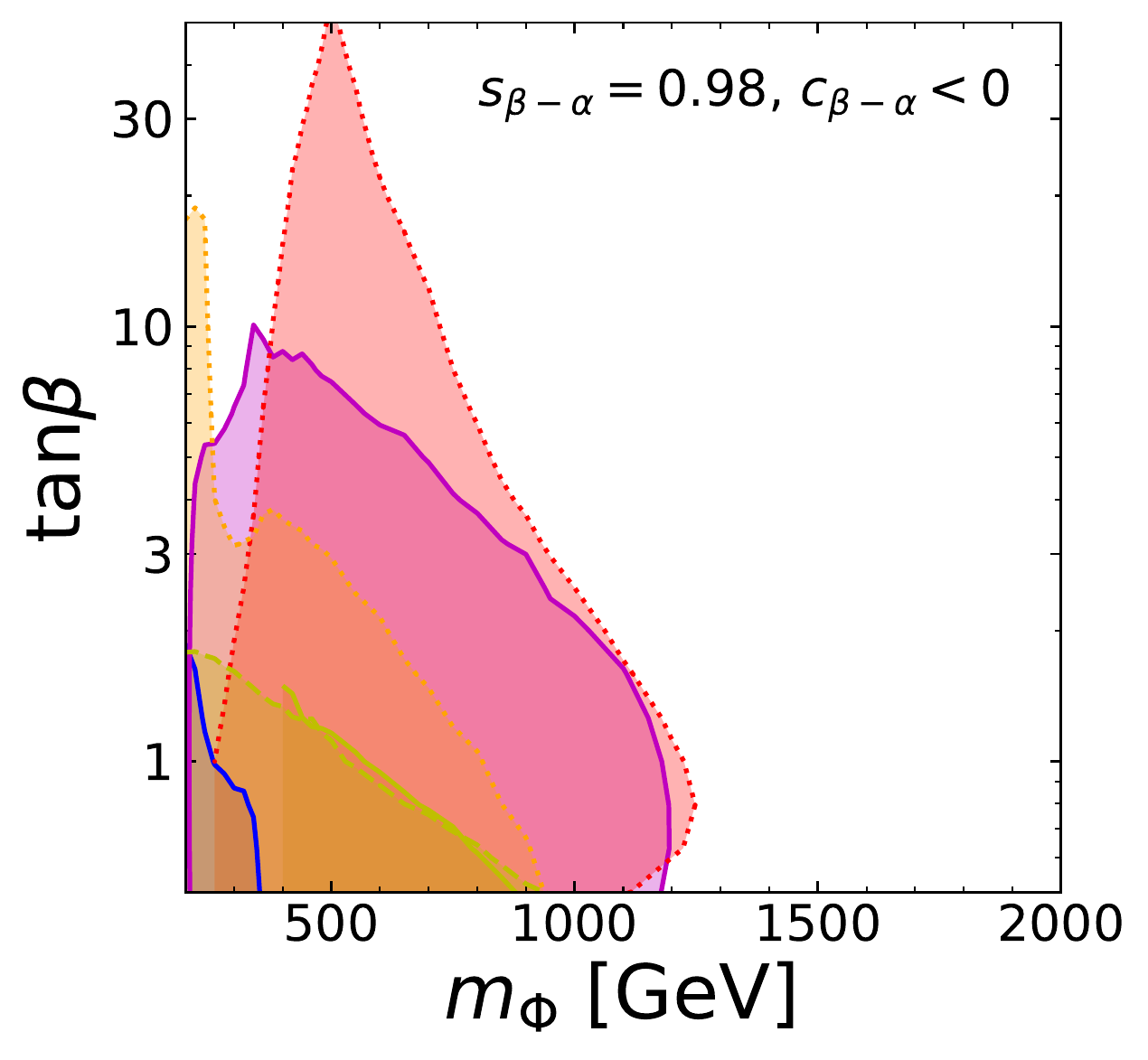}
 \includegraphics[height=0.237\textwidth]{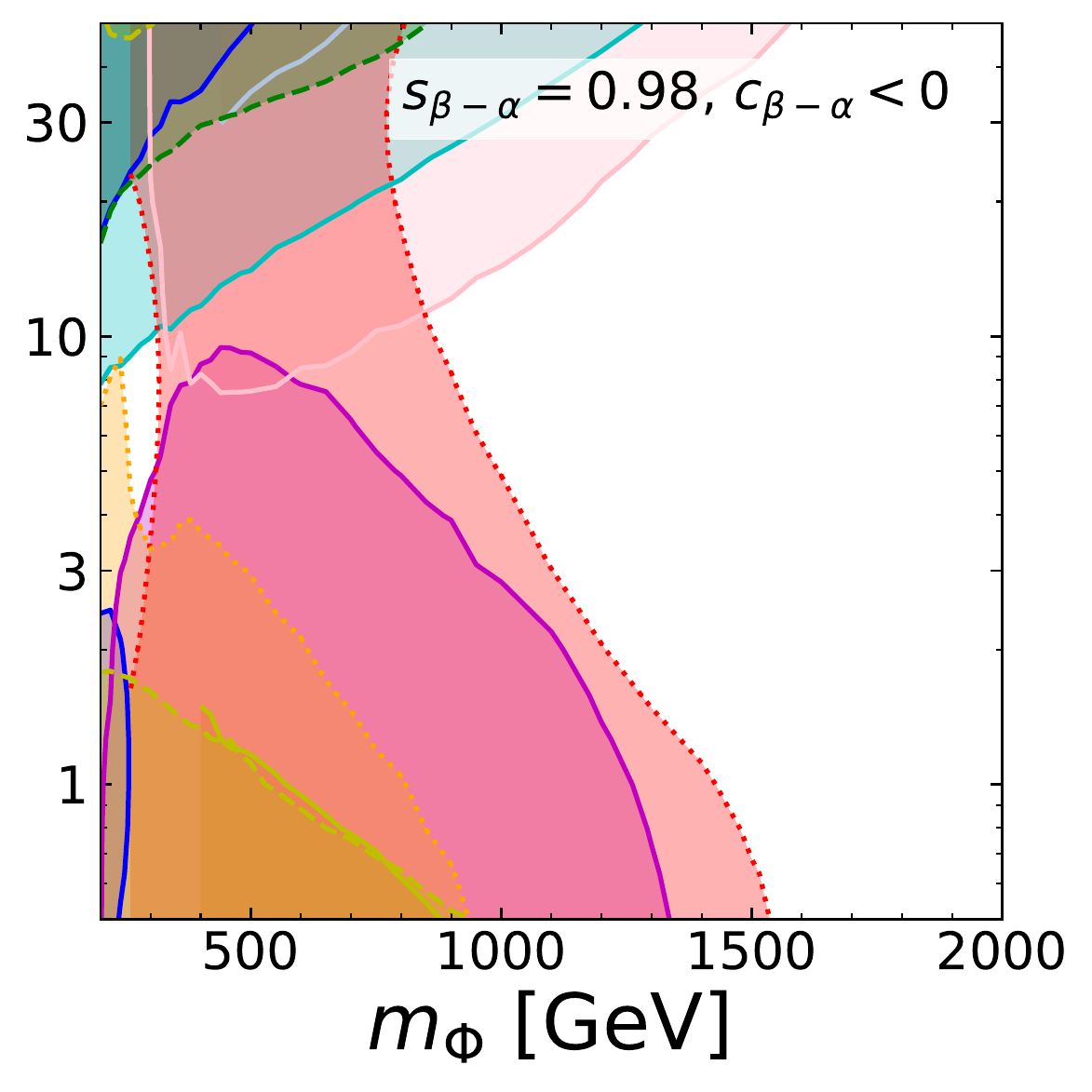}
 \includegraphics[height=0.237\textwidth]{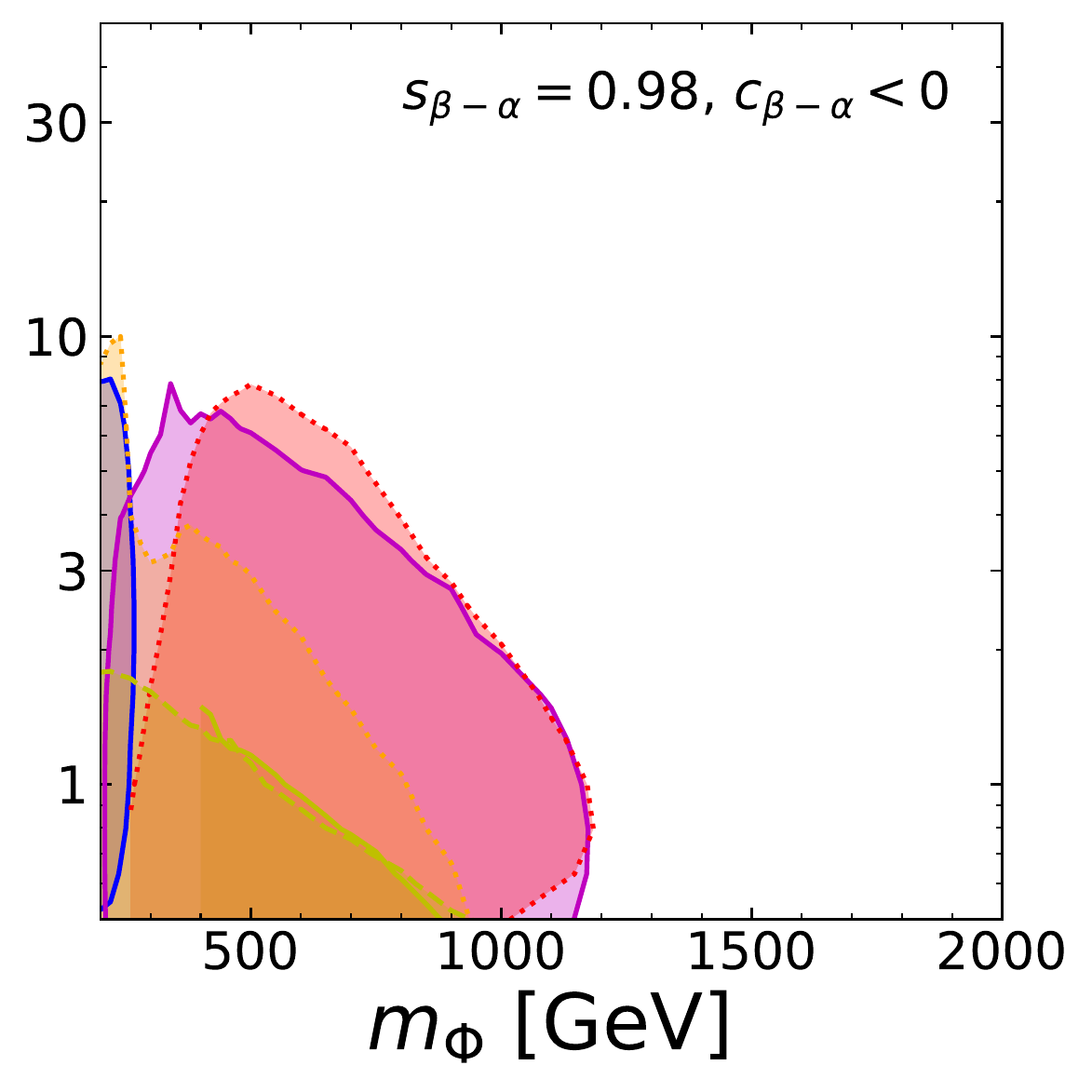}
 \includegraphics[height=0.237\textwidth]{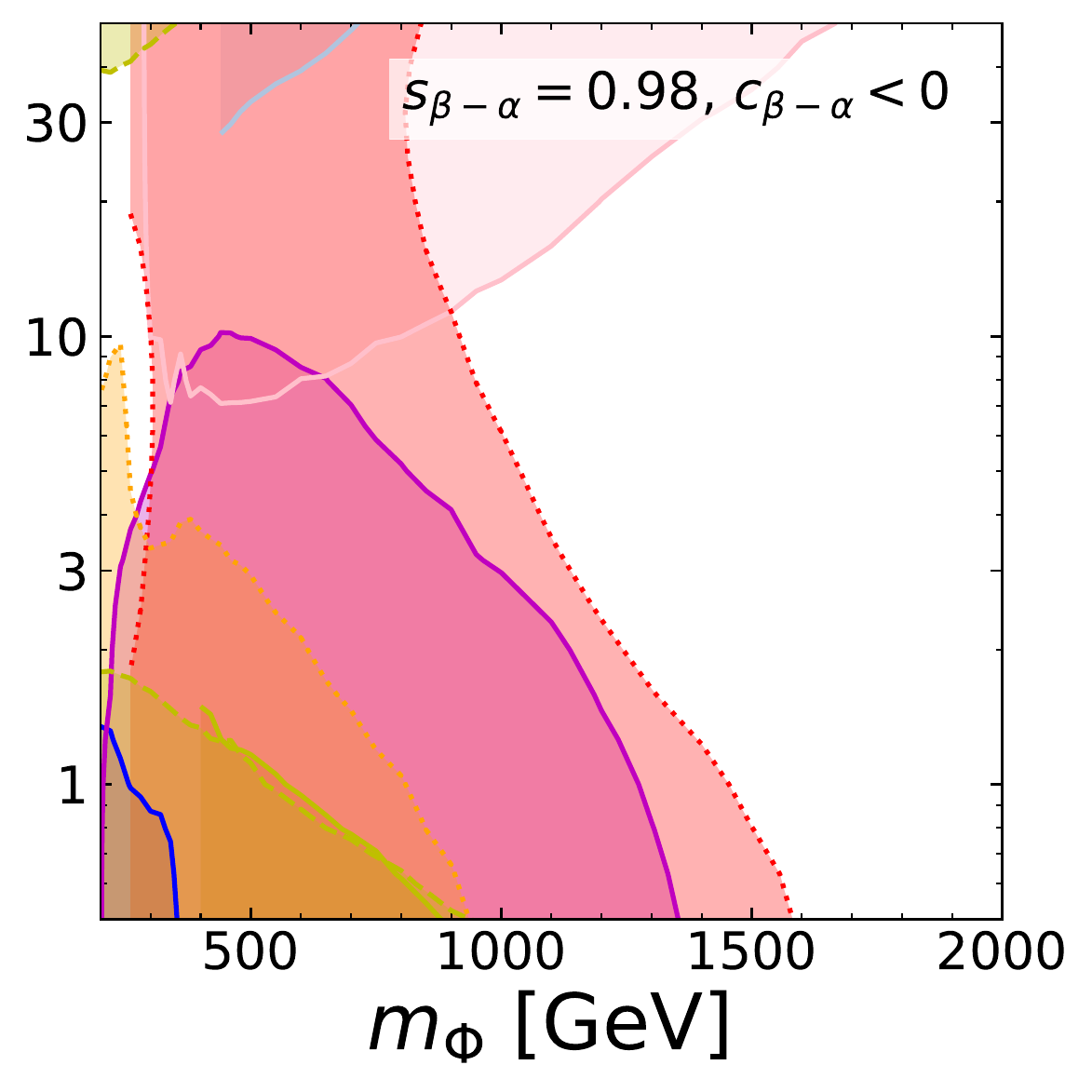}    
\caption{
Regions on the $m_\Phi$--$\tan\beta$ plane excluded at 95\% CL 
in the Type-I, Type-II, Type-X and Type-Y THDMs (from the left to the right panels) 
via direct searches for heavy Higgs bosons with the 36~fb$^{-1}$ LHC Run-II data.
The value of $s_{\beta-\alpha}$ is set to be 1, 0.995, 0.99 and 0.98 with $c_{\beta-\alpha}<0$ from the top to the bottom panels.
}
\label{fig:current-n}
\end{figure} 

\begin{figure}
 \includegraphics[height=0.235\textwidth]{Fig/ex_rs13_t1_sba1.pdf}
 \includegraphics[height=0.235\textwidth]{Fig/ex_rs13_t2_sba1.pdf}
 \includegraphics[height=0.235\textwidth]{Fig/ex_rs13_t3_sba1.pdf}
 \includegraphics[height=0.235\textwidth]{Fig/ex_rs13_t4_sba1.pdf}
 \includegraphics[height=0.22\textwidth]{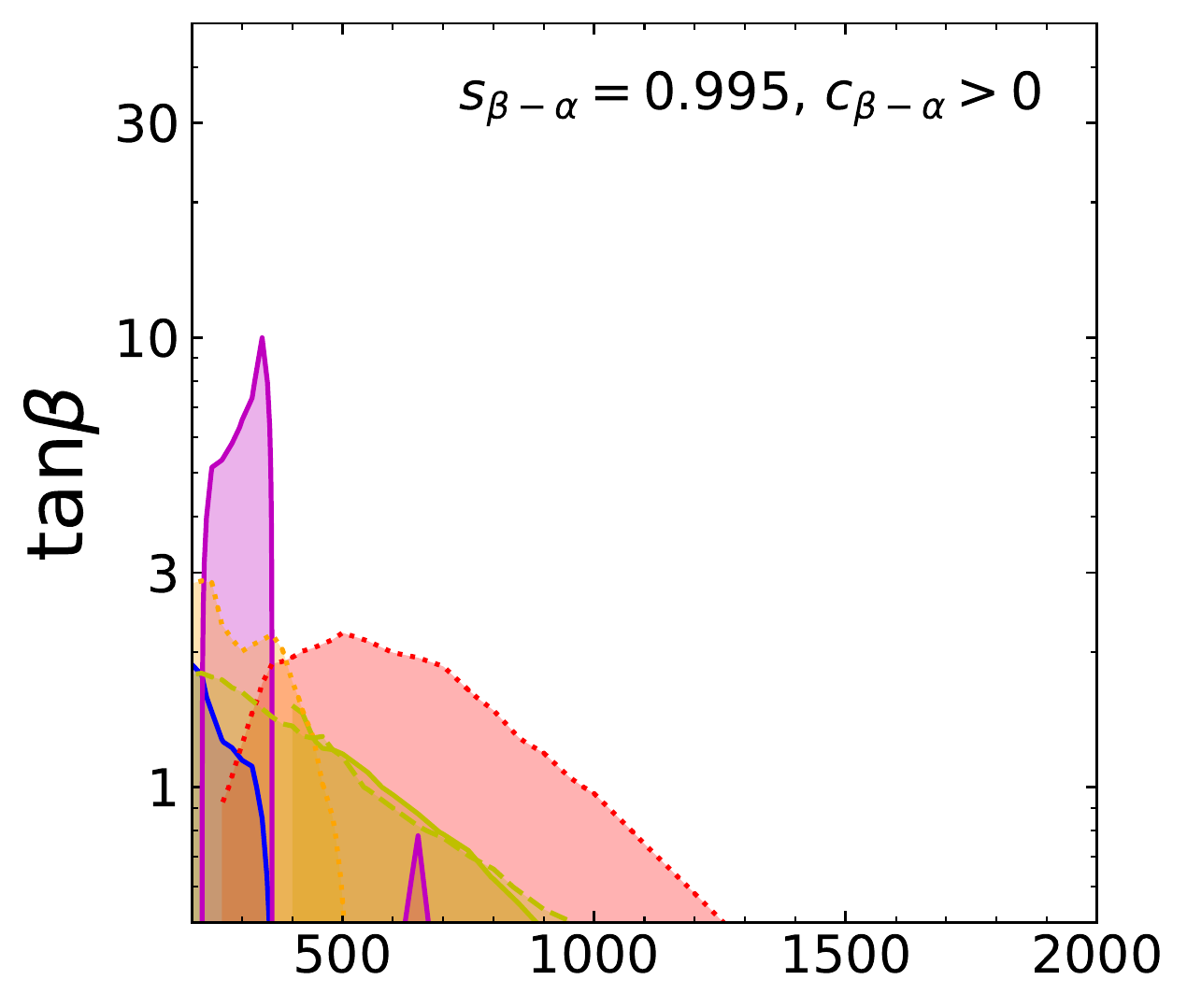}
 \includegraphics[height=0.22\textwidth]{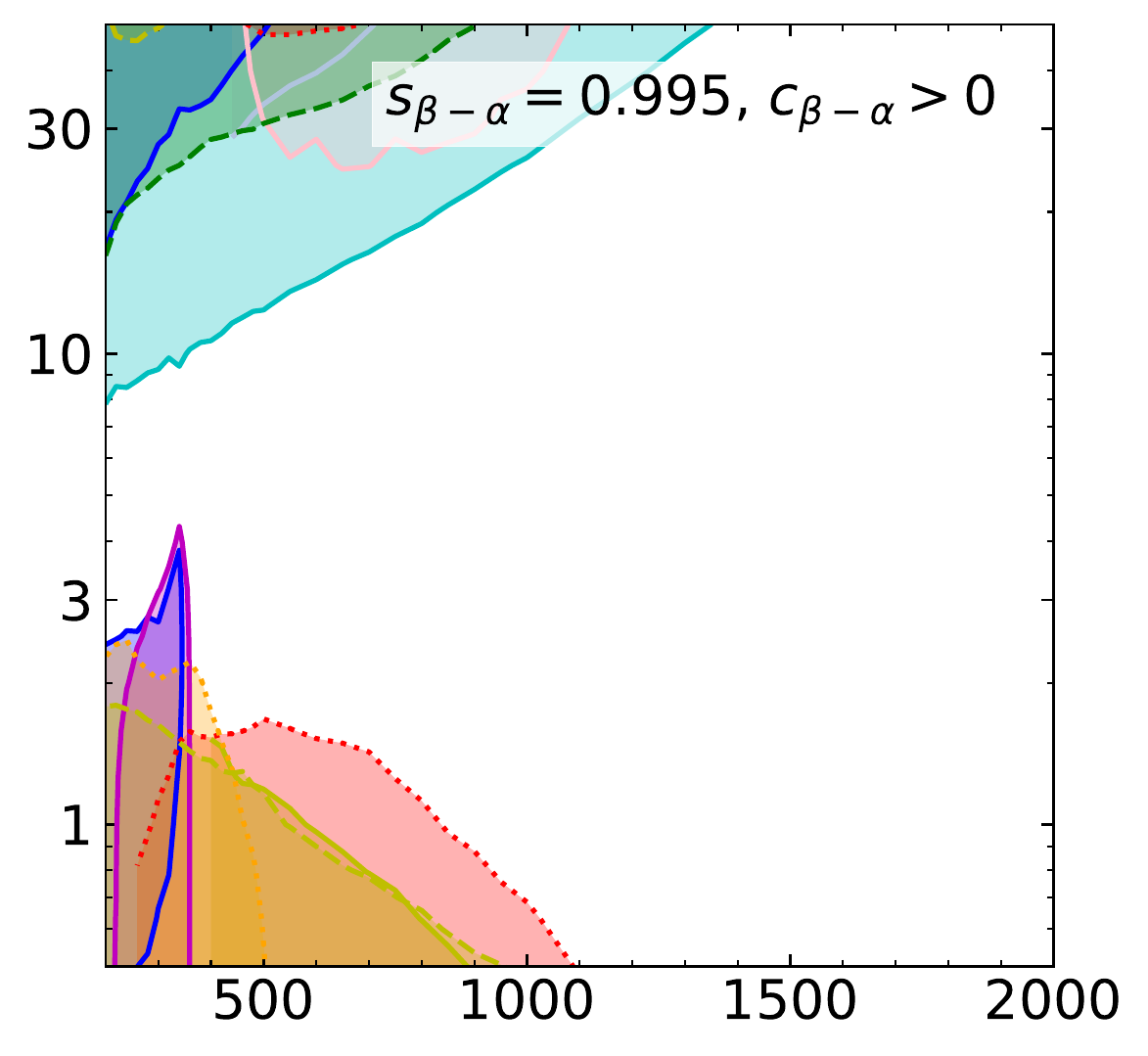}
 \includegraphics[height=0.22\textwidth]{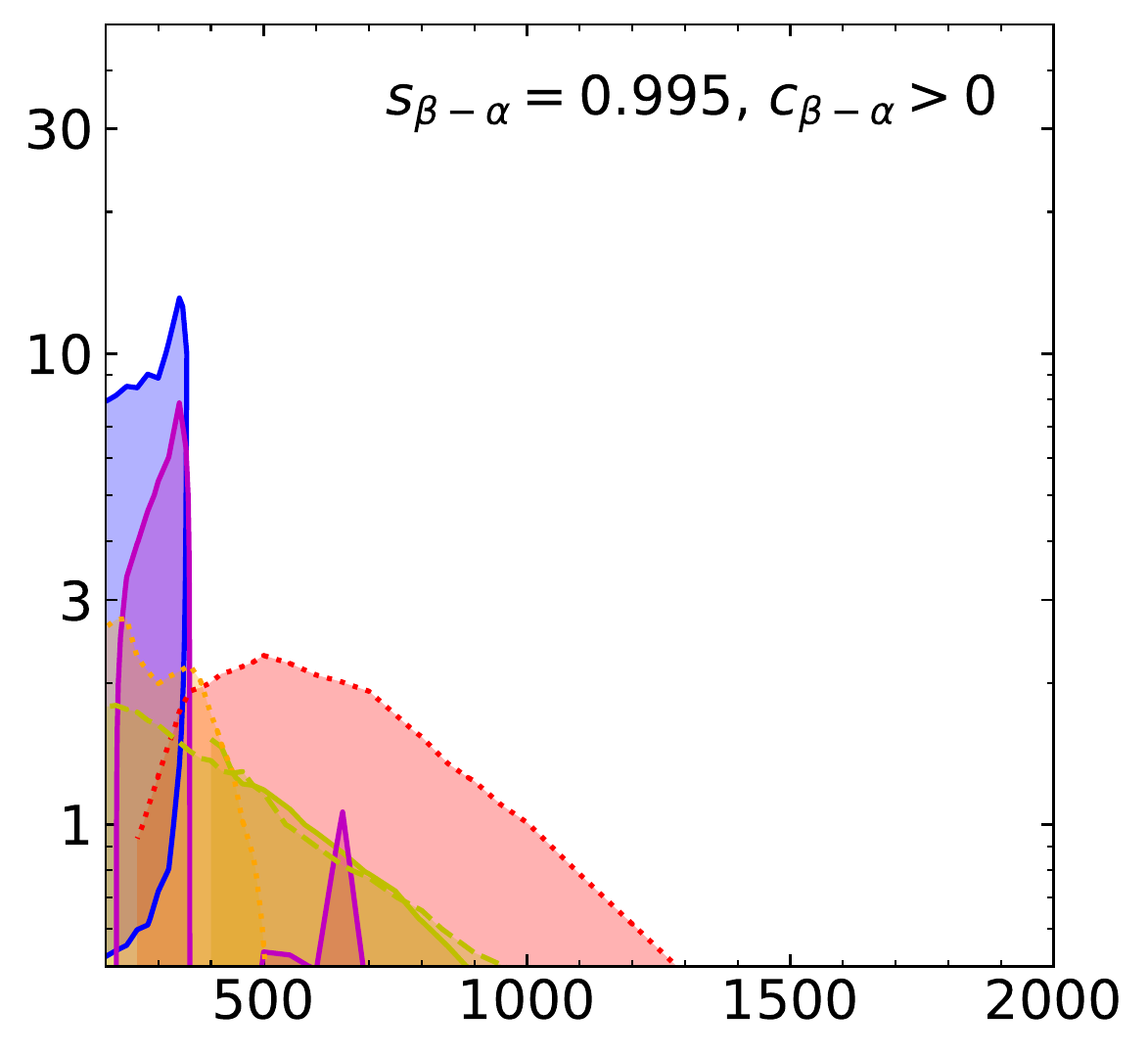}
 \includegraphics[height=0.22\textwidth]{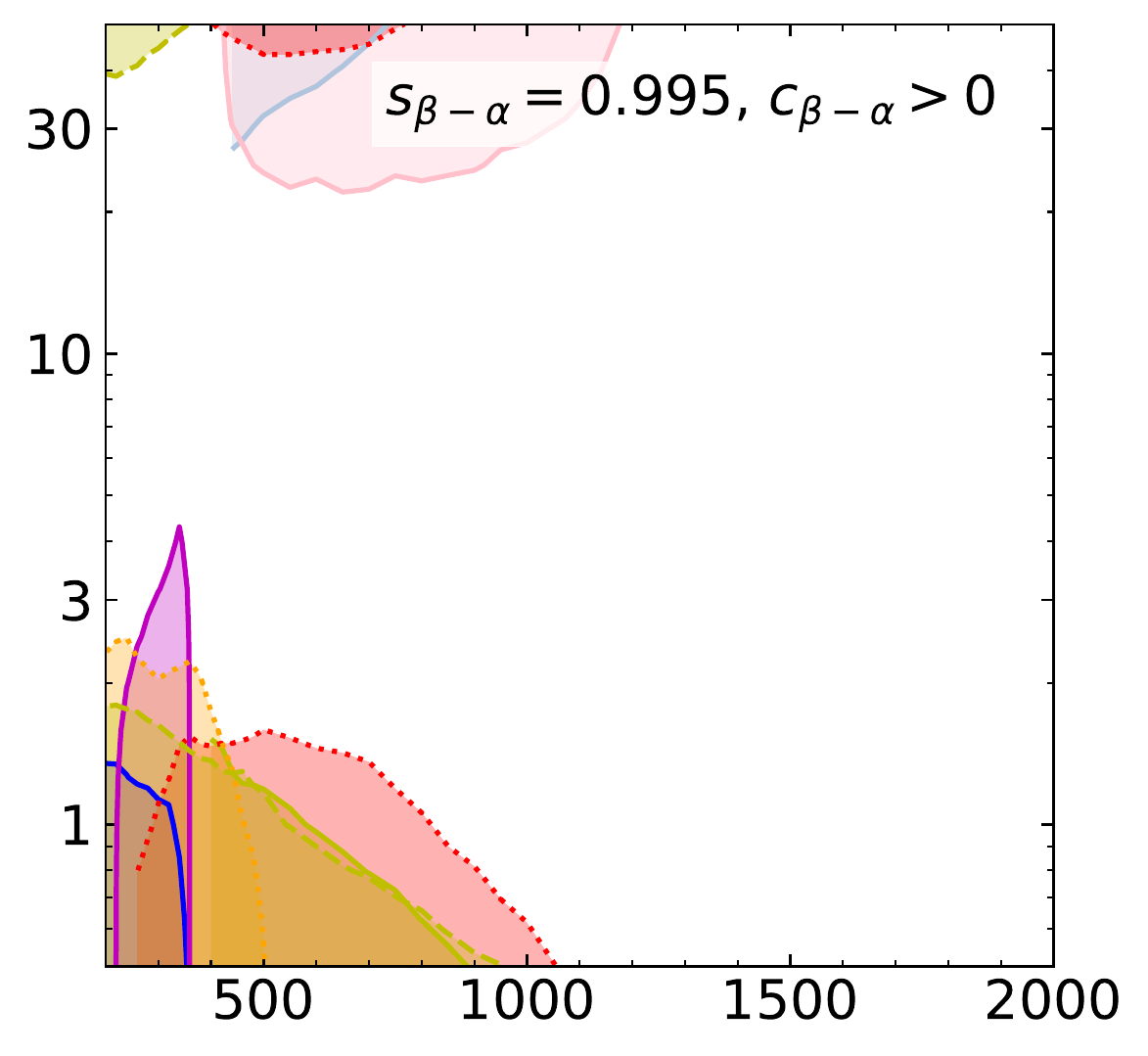}   
 \includegraphics[height=0.22\textwidth]{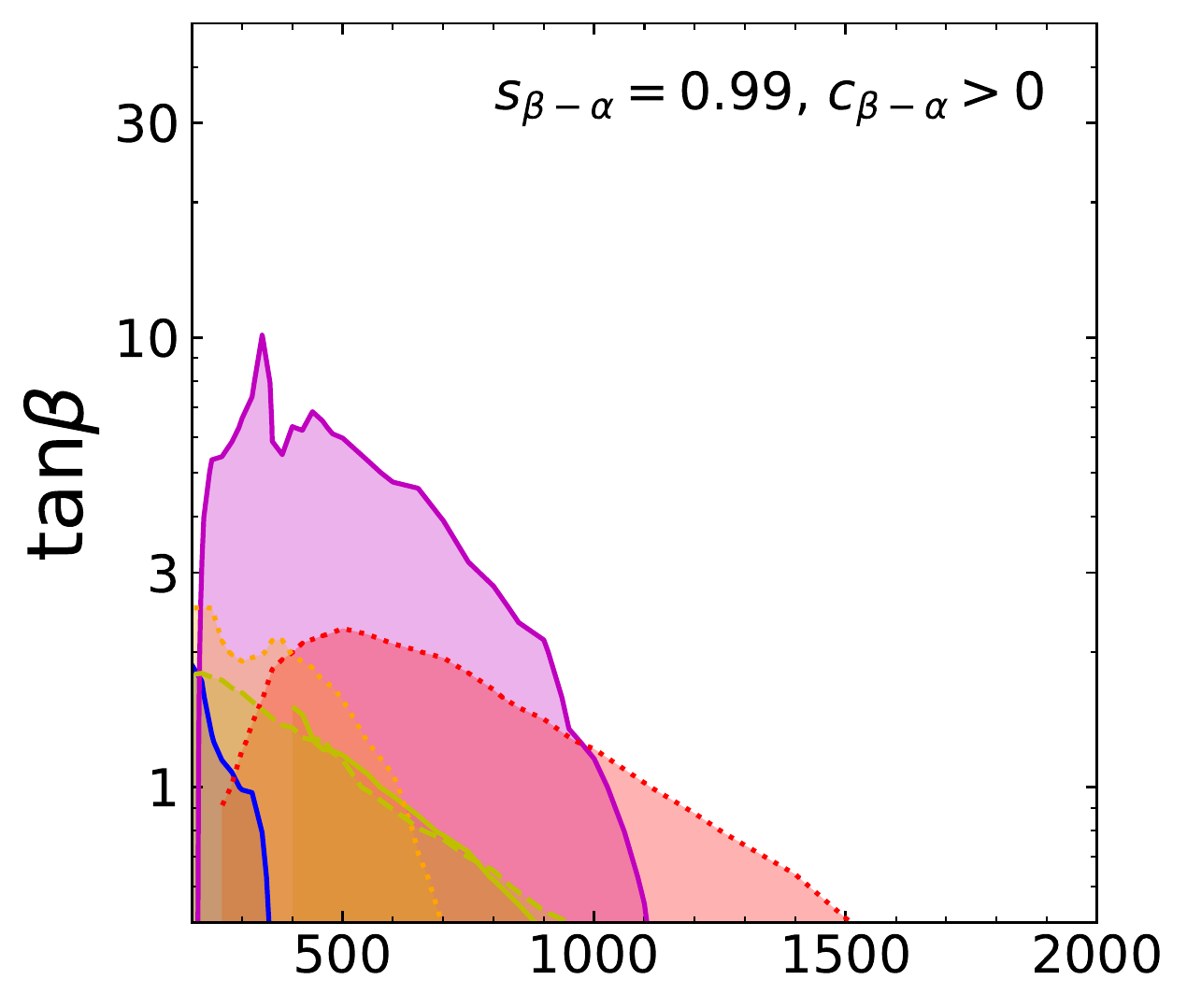}
 \includegraphics[height=0.22\textwidth]{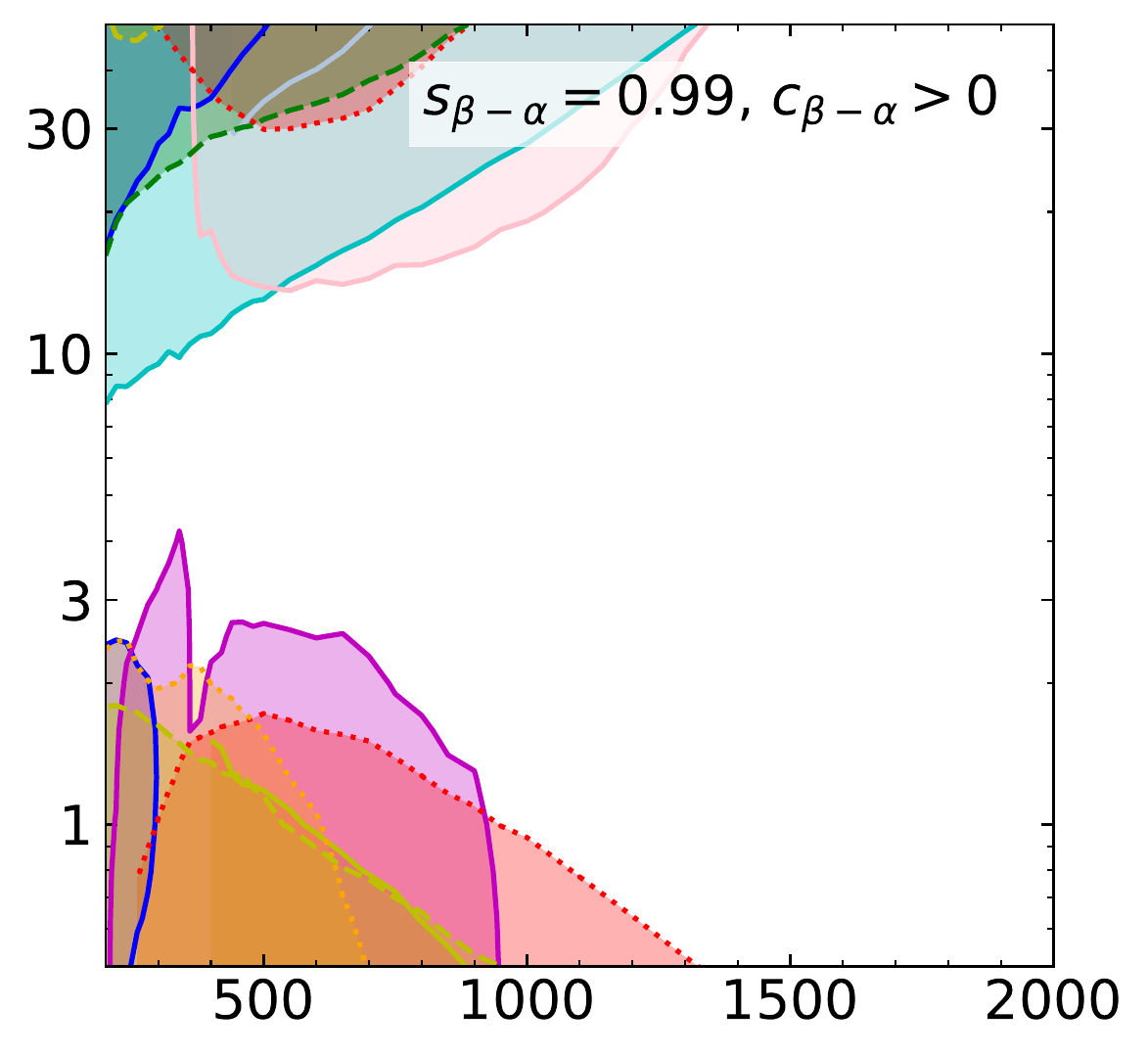}
 \includegraphics[height=0.22\textwidth]{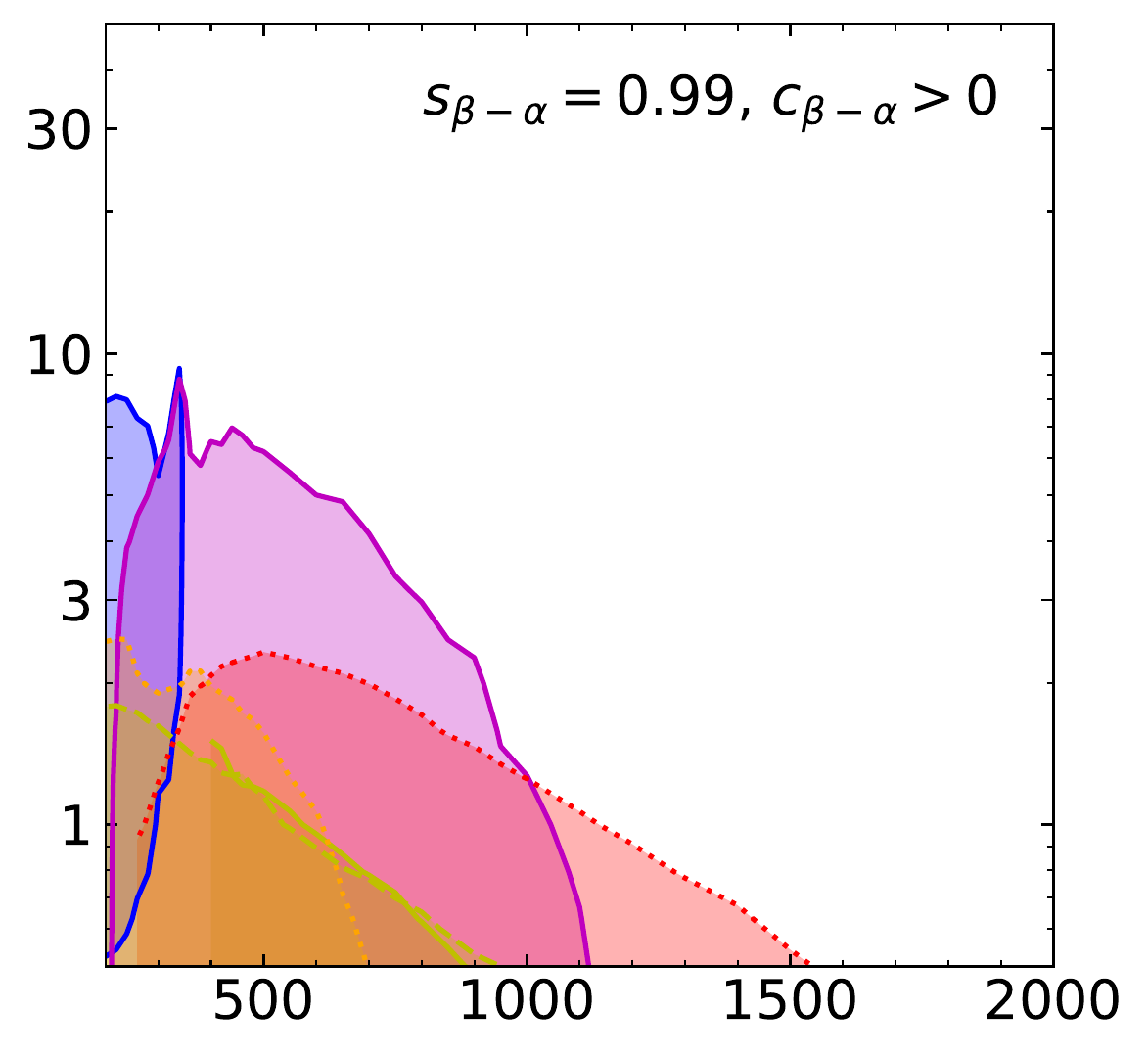}
 \includegraphics[height=0.22\textwidth]{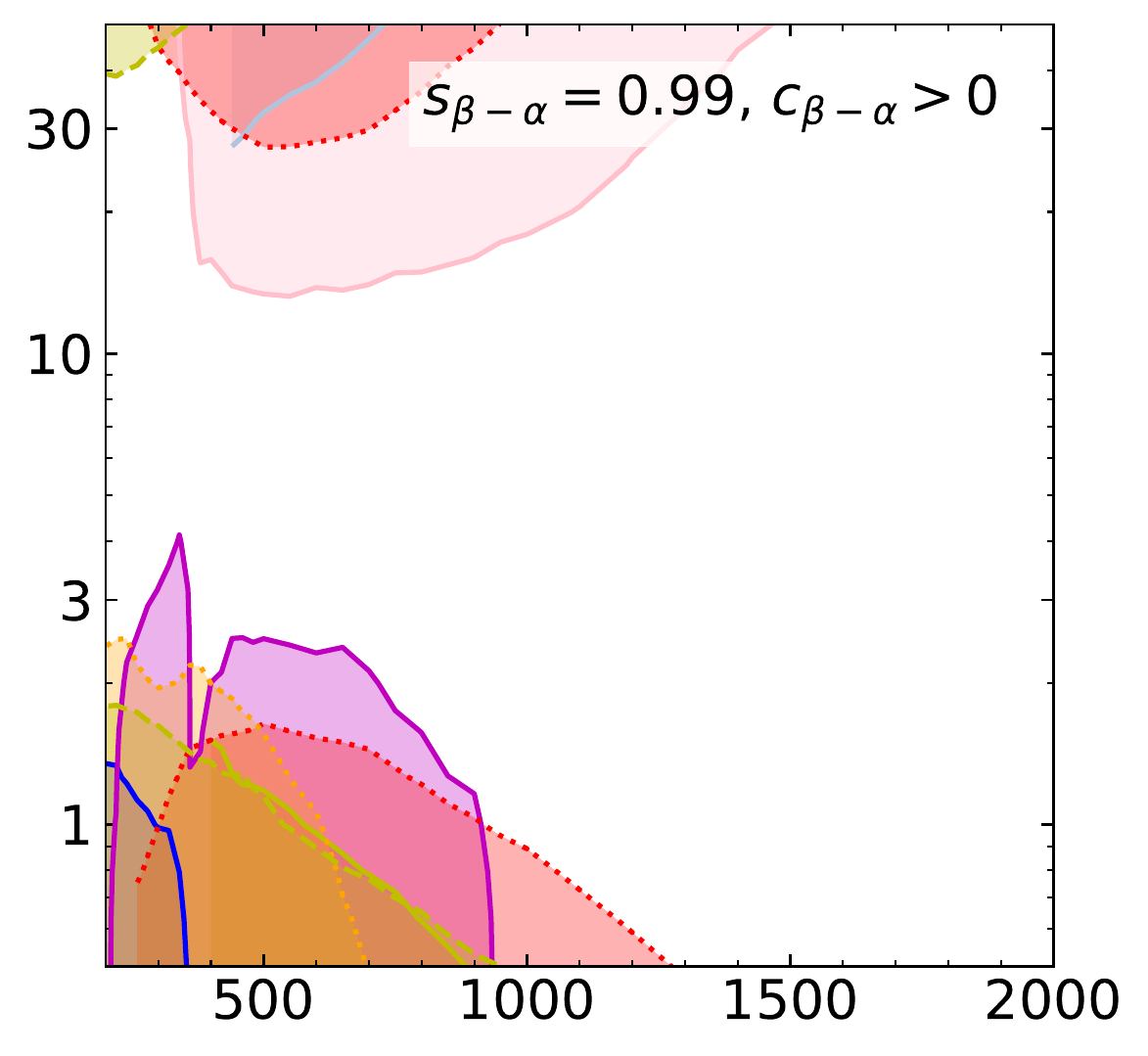} 
 \includegraphics[height=0.237\textwidth]{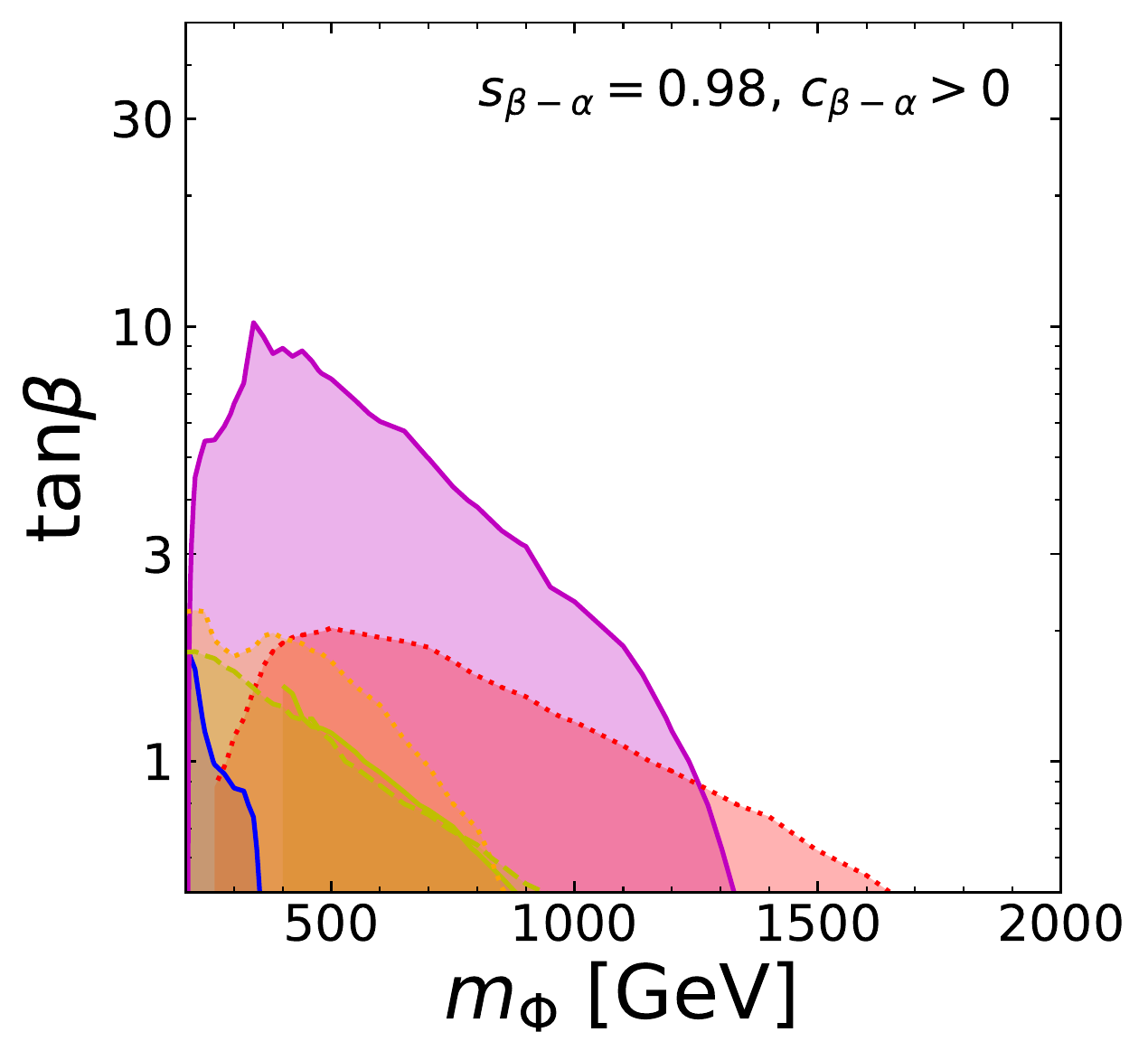}
 \includegraphics[height=0.237\textwidth]{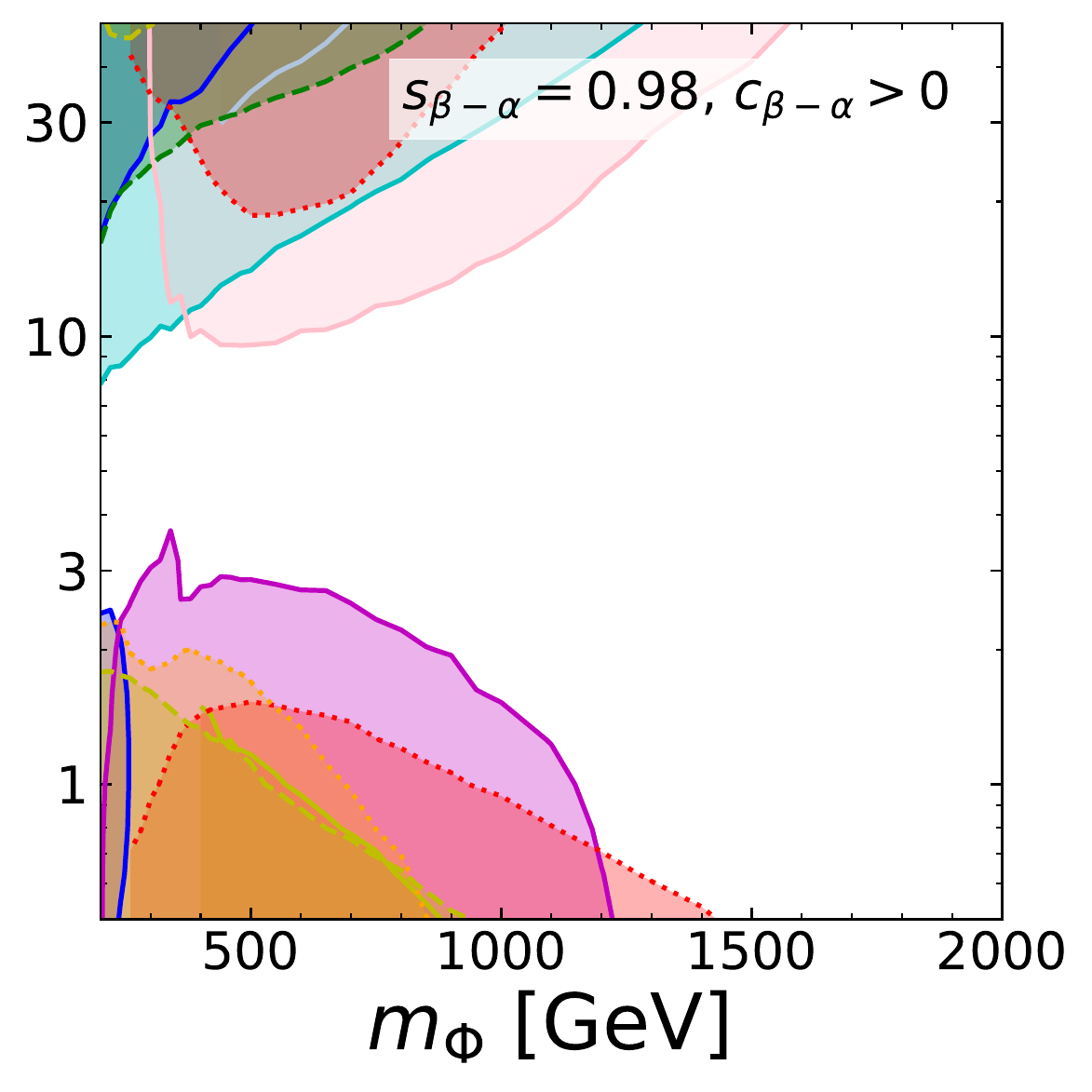}
 \includegraphics[height=0.237\textwidth]{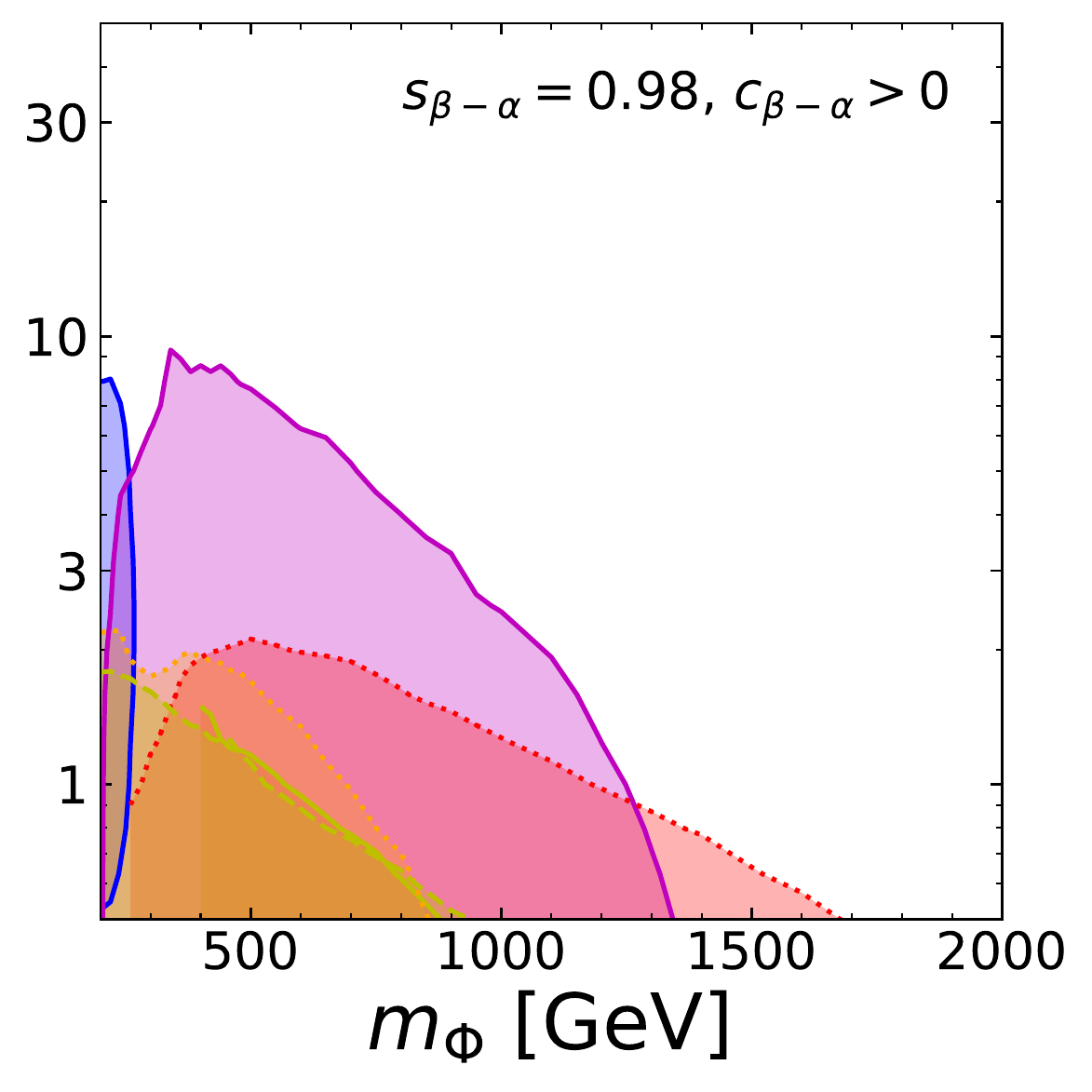}
 \includegraphics[height=0.237\textwidth]{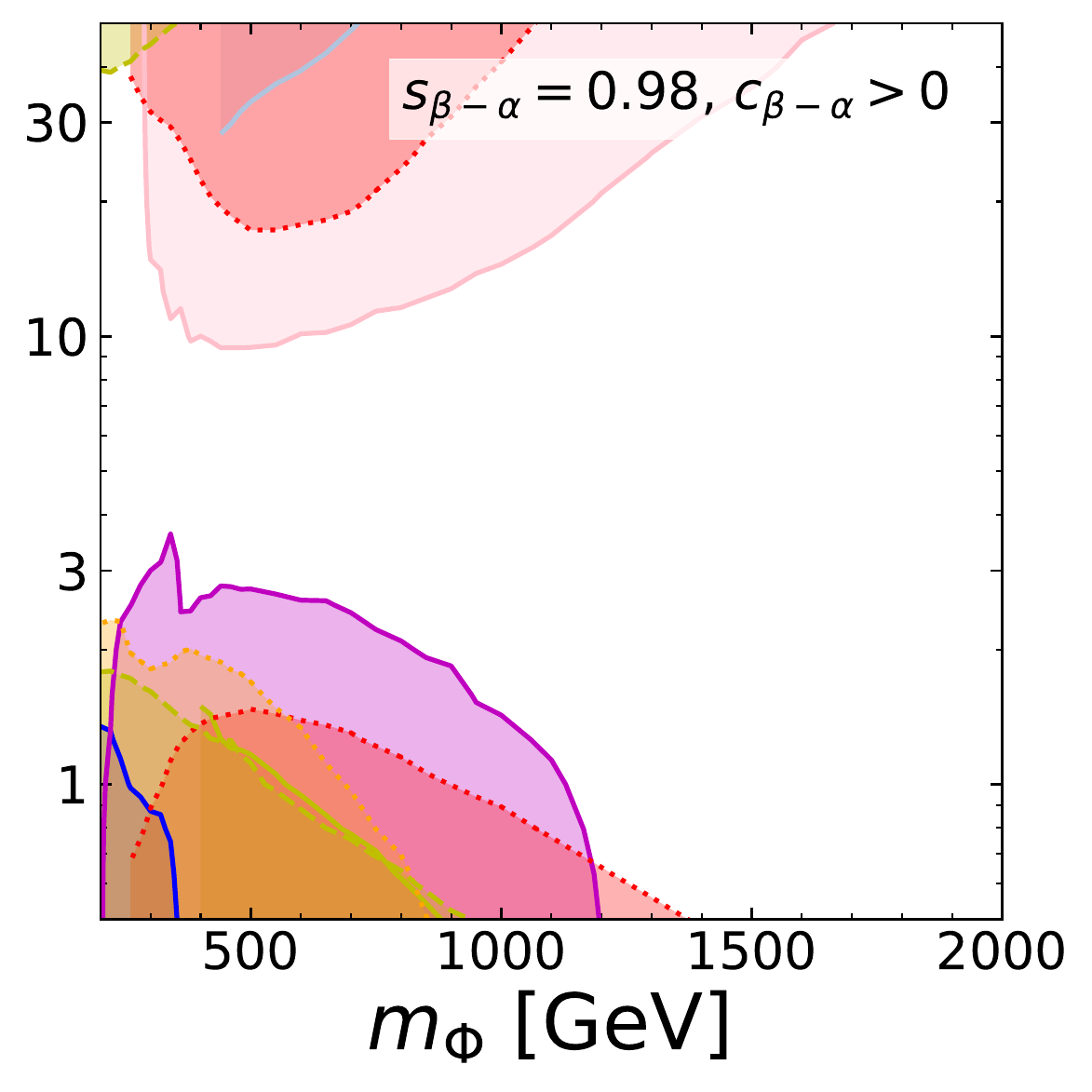} 
\caption{
Regions on the $m_\Phi$--$\tan\beta$ plane excluded at 95\% CL 
in the Type-I, Type-II, Type-X and Type-Y THDMs (from the left to the right panels)
via direct searches for heavy Higgs bosons with the 36~fb$^{-1}$ LHC Run-II data.
The value of $s_{\beta-\alpha}$ is set to be 1, 0.995, 0.99 and 0.98 with $c_{\beta-\alpha} > 0$ from the top to the bottom panels.
}
\label{fig:current-p}
\end{figure} 

Now, let us turn to discuss constraints on the parameter space in each THDM 
from direct searches for heavy Higgs bosons with the LHC Run-II data.

In Fig.~\ref{fig:current-n}, we show exclusion
regions at 95\% CL on the $m_\Phi$--$\tan\beta$ plane
in the Type-I, Type-II, Type-X and Type-Y THDMs (from the left to the right panels)
via various direct searches for heavy Higgs bosons with the 36~fb$^{-1}$ LHC Run-II data listed in Table~\ref{tab:lhc}.
The value of $s_{\beta-\alpha}$ is set to be 1, 0.995, 0.99 and 0.98 with $c_{\beta-\alpha}<0$ from the top to the bottom panels.
The shaded regions with dotted, solid, and dashed border lines denote the exclusion regions for $H$, $A$, and $H^\pm$, respectively. 

Each exclusion region is understood by each production rate, shown in Figs.~\ref{fig:xsec-H_cn}--\ref{fig:xsec-C},
times each branching ratio, depicted in Figs.~\ref{FIG:BR1}--\ref{FIG:BR4}.
We highlight several points for $A$, $H$ and $\hpm$ in order.

Regarding to the CP-odd Higgs boson $A$;
\begin{itemize}
 \item 
 For large $\tb$, exclusion regions only appear in the Type-II and the Type-Y THDMs, 
 in which the production via the bottom-quark loop as well as the bottom-quark associated production becomes dominant.  
 \item 
 The $A\to\tau\bar\tau$ channel is significant only for $m_A<2m_t$ or for large $\tb$ in Type-II. 
 We note that, although the branching ratio of the $A\to\tau\bar\tau$ decay is even dominant for large $\tb$ in Type-X, 
 the production rate is too small to be constrained.%
 \footnote{Four-$\tau$ final states from the $pp\to HA$ process in Type-X can be relevant~\cite{Kanemura:2011kx}.} 
 For $m_A>2m_t$, the $A\to t\bar t$ channel becomes relevant in the small $\tb$ region in all the types.%
\footnote{Because there is no specific analysis for the spin-0 resonance in the $t\bar t$ final state in the LHC Run-II,
we use the limit for $Z'$~\cite{Aaboud:2018mjh}, which is valid from the Run-I 8~TeV analysis~\cite{Aad:2015fna}.}
 \item 
 Since the $A\to Zh$ decay only occurs for the non-alignment case,
 the exclusion regions are remarkably different between for the alignment case and for the non-alignment case.
 The region of the exclusion from the $A\to Zh$ channel becomes larger from $\sba=0.995$ to $0.98$, since
 the decay rate for $A\to Zh$ is proportional to $\cba^2$.
\end{itemize}

Regarding to the CP-even heavier Higgs boson $H$;
\begin{itemize}
 \item The production rate via the gluon fusion for the heavier CP-even Higgs boson $H$ is smaller than that for the $A$ production, as mentioned above.
Moreover, in the non-alignment case, the fermionic branching ratios of $H$ for low $\tb$ is smaller than those for $A$ due to the decays into a pair of the weak gauge bosons, which are forbidden for $A$. 
 Therefore, the constraints are slightly weaker than the $A$ case, and we do not present the exclusions explicitly 
 for the $H\to\tau\bar\tau$, $H(b\bar b)\to\tau\bar\tau$, $H(b\bar b)\to b\bar b$ and $H\to t\bar t$ channels. 
 \item For $m_H>2 m_{W,Z}$ and/or $m_H>2 m_h$,
 the peculiar decay modes for $H$ are $H\to WW$, $H\to ZZ$ and $H\to hh$ for the non-alignment case
 and give rise to the relatively large exclusions. 
 The region of the constraint from $H\to WW$ is similar to $H\to ZZ$, but smaller, so we do not show it explicitly. 
 \item We note that, as mentioned in Sec.~\ref{sec:decay}, the $H\to hh$ decay depends on $M^2$.
 For a non-degenerate case $M\ne m_\Phi$, the exclusion region from the $H\to hh$ channel can be different that for the degenerate case.
\end{itemize}

Regarding to the charged Higgs boson $\hpm$;
\begin{itemize}
 \item For the near alignment scenario, in the low $\tb$ region ($\tb\lesssim5$), the $H^\pm\to tb$ decay is dominant for all the types, 
 therefore the exclusions of the low-mass and low-$\tb$ region from the $\hpm\to tb$ channel are almost same for all the panels.
 \item In the large $\tb$ region, the constraint from the $\hpm\to\tau\nu$ channel can be significant only in Type-II. 
Although the branching ratio of the $\hpm\to\tau\nu$ is even dominant for large $\tb$ in Type-X, 
the constraint is insignificant due to the small production rate. 
 \item We note that, as mentioned in Sec.~\ref{sec:model}, in Type-II and Type-Y there is an independent constrain from flavor observables on the mass of charged Higgs bosons, $m_{\hpm}\gtrsim800$~GeV. 
\end{itemize}

Figure~\ref{fig:current-p} shows the same as in Fig.~\ref{fig:current-n}, but for the $c_{\beta-\alpha}>0$ case.
The global picture of the exclusion regions is same as for the $c_{\beta-\alpha}<0$ case. 
A remarkable difference is that the constraints for $H$ in the non-alignment case are much weaker for around $\tb\sim 7-10$
due to the strong suppression of the production rates.
Although $\sigma(A\to Zh)$ does not depend on the sign of $\cba$, the exclusion regions for $\cba>0$ in Type-II and Y are smaller
than those for $\cba<0$. 
This is because the analysis includes the $h\to b\bar b$ decay, whose branching ratio has a singular behavior for $\cba>0$;
see Figs.~\ref{FIG:BR3} and \ref{FIG:BR4}.

\begin{table}
\begin{center}
\begin{tabular}{c||c|c|c|c}\hline\hline
$\sba$               & Type-I & Type-II   & Type-X  & Type-Y \\\hline\hline
$0.995$        & $t_\beta \geq  0.54$ ($t_\beta \geq  0.54$)   &  -- ($0.57 \leq t_\beta \leq 1.6$)   & $0.43\leq t_\beta \leq 4.1$  ($0.42 \leq t_\beta \leq 4.2$)   & -- (--)  \\\hline
$0.990$        & $t_\beta \geq 0.86$ ($t_\beta \geq 0.86$)      &  -- (--)                               & $0.71\leq t_\beta \leq 2.0$  ($0.72 \leq t_\beta \leq 2.5$)   & -- (--)  \\\hline
$0.980$        & $t_\beta \geq 1.3$ ($t_\beta \geq 1.3$)        &  -- (--)                               &  -- (--)         &         -- (--)  \\\hline\hline
\end{tabular}
\caption{95\% CL allowed range of $\tan\beta$ for the case with $c_{\beta-\alpha} < 0$ ($c_{\beta-\alpha} > 0$)
from the signal strength of the discovered Higgs boson at the LHC~\cite{Aad:2019mbh}. 
The hyphen denotes no allowed region. 
}
\label{tab:tb}
\end{center}
\end{table}

Before closing this section, we briefly discuss the signal strength for the discovered Higgs boson measured at the LHC Run-II experiment, 
which provides independent constraints on the parameter space from those given by the direct searches discussed in this section. 
Measurements of the signal strength set constraints on the Higgs boson couplings; i.e., the $\kappa$ values defined in Sec.~\ref{sec:model}, which 
can be translated into those on $s_{\beta-\alpha}$ and $\tan\beta$. 
In Table~\ref{tab:tb}, we summarize the 95\% CL allowed range of $\tan\beta$ in the THDMs with fixed values of $s_{\beta-\alpha}$.
The $\kappa$ values are extracted from Ref.~\cite{Aad:2019mbh}, which are presented in Table~\ref{tab:kappa} as a reference. 
We see that except for the Type-I THDM it gives severe constraints on $\tan\beta$, because $\kappa_b$ and/or $\kappa_\tau$ can significantly differ from unity in the Type-II, Type-X and Type-Y THDMs even 
for the approximate alignment case.

\section{Combined results of direct searches at the HL-LHC and precision tests at the ILC}\label{sec:synergy}
Now, let us turn to investigate how the current parameter space in the THDMs discussed in the previous section
can be explored further in future experiments,  
especially by
direct searches for heavy Higgs bosons at the HL-LHC
as well as by precision measurements of the Higgs boson couplings at the ILC. 
We note that complementarity for direct searches for heavy Higgs bosons between at the LHC and the ILC500 
was discussed for the THDMs in Ref.~\cite{Kanemura:2014dea}.  

In order to obtain the sensitivity projection to the HL-LHC with 3000~fb$^{-1}$ of integrated luminosity, 
we rescale the current expected sensitivity by $\sqrt{3000/36}\sim9.1$.
We also perform a further rescaling of the sensitivity from $\sqrt{s}=13$~TeV to $\sqrt{s}=14$~TeV
by taking into account the ratio of the signal cross sections, $\sigma(m_\Phi)_{\rm 14TeV}/\sigma(m_\Phi)_{\rm 13TeV}$.
Here, we assume that signal and background increase by the same amount from 13~TeV to 14~TeV, 
which can be conservative particularly for the high-mass region. 
Detailed projection with systematic uncertainties for the $\phi\to\tau\bar\tau$ channel was performed in the report for the HL-LHC~\cite{Cepeda:2019klc},
where one can see the higher sensitivity for $m_\Phi\gtrsim 1200$~GeV.

\begin{table}
\begin{center}
\begin{tabular}{c||c|ccccc}\hline\hline
               & Current (ATLAS, CMS)  & HL-LHC (ATLAS, CMS) & ILC250 & ILC500 & (1$\sigma$ [\%]) \\\hline\hline
$\kappa_Z^{}$ & $(1.11 \pm 0.08, 1.00\pm 0.11)$ &  (2.6, 2.4)  & 0.38 & 0.30\\\hline
$\kappa_W^{}$ & $(1.05 \pm 0.09, -1.13^{+0.16}_{-0.13}              )$ &  (3.1, 2.6) & 1.8 & 0.40\\\hline
$\kappa_b^{}$ & $(1.03^{+0.19}_{-0.17}, 1.17^{+0.27}_{-0.31})$ & (6.2, 6.0)&  1.8 & 0.60\\\hline
$\kappa_t^{}$ & $(1.09^{+0.15}_{-0.14}, 0.98\pm0.14)$        & (6.3, 5.5) &  -- & 6\\\hline
$\kappa_c^{}$ & (--, --)        & (--, --) &   2.4 & 1.2\\\hline
$\kappa_\tau^{}$ & $(1.05^{+0.16}_{-0.15}, 1.02\pm0.17)$ & (3.7, 2.8)& 1.9 & 0.80\\\hline
$\kappa_\mu^{}$  & (--, $0.80^{+0.59}_{-0.80}$)      & (7.7, 6.7)& 5.6 & 5.1\\\hline
$\kappa_g^{}$    & $(0.99^{+0.11}_{-0.10}, 1.18^{+0.16}_{-0.14})$ & (4.2, 4.0)& 2.2 & 0.97\\\hline
$\kappa_\gamma^{}$ & $(1.05\pm 0.09, 1.07^{+0.14}_{-0.15} )$ & (3.7, 2.9)&1.1 & 1.0\\\hline
$\kappa_{Z\gamma}^{}$ & (--, --)                 & (12.7, --)& 16 & 16\\\hline
$\kappa_{h}^{}$ & (--, --)                 & (--, --)& -- & 27\\\hline\hline
\end{tabular}
\caption{Summary for the current measurements and expected $1\sigma$ accuracies of the $\kappa$ values. 
For the current measurements, we refer to the values, assuming that the branching ratio of the decay into BSM particles is zero, which are given by the ATLAS experiments with 80 fb$^{-1}$~\cite{Aad:2019mbh} and the CMS experiments with 35.9 fb$^{-1}$~\cite{Sirunyan:2018koj}. 
For the HL-LHC, we refer to the expected accuracies given in Ref.~\cite{Cepeda:2019klc} using systematic uncertainties at the Run-II experiment. 
For the ILC250, we refer to the expected accuracies given by the ILC with 250 GeV and 2000 fb$^{-1}$~\cite{Fujii:2017vwa}.
For the ILC500, the expected accuracies are based on the results of the ILC250 combining the simulations at $\sqrt{s} = 350$ GeV with 200 fb$^{-1}$ and those at $\sqrt{s} = 500$ GeV with 4000 fb$^{-1}$~\cite{Fujii:2017vwa}.}
\label{tab:kappa}
\end{center}
\end{table}

\begin{figure}
 \includegraphics[height=0.235\textwidth]{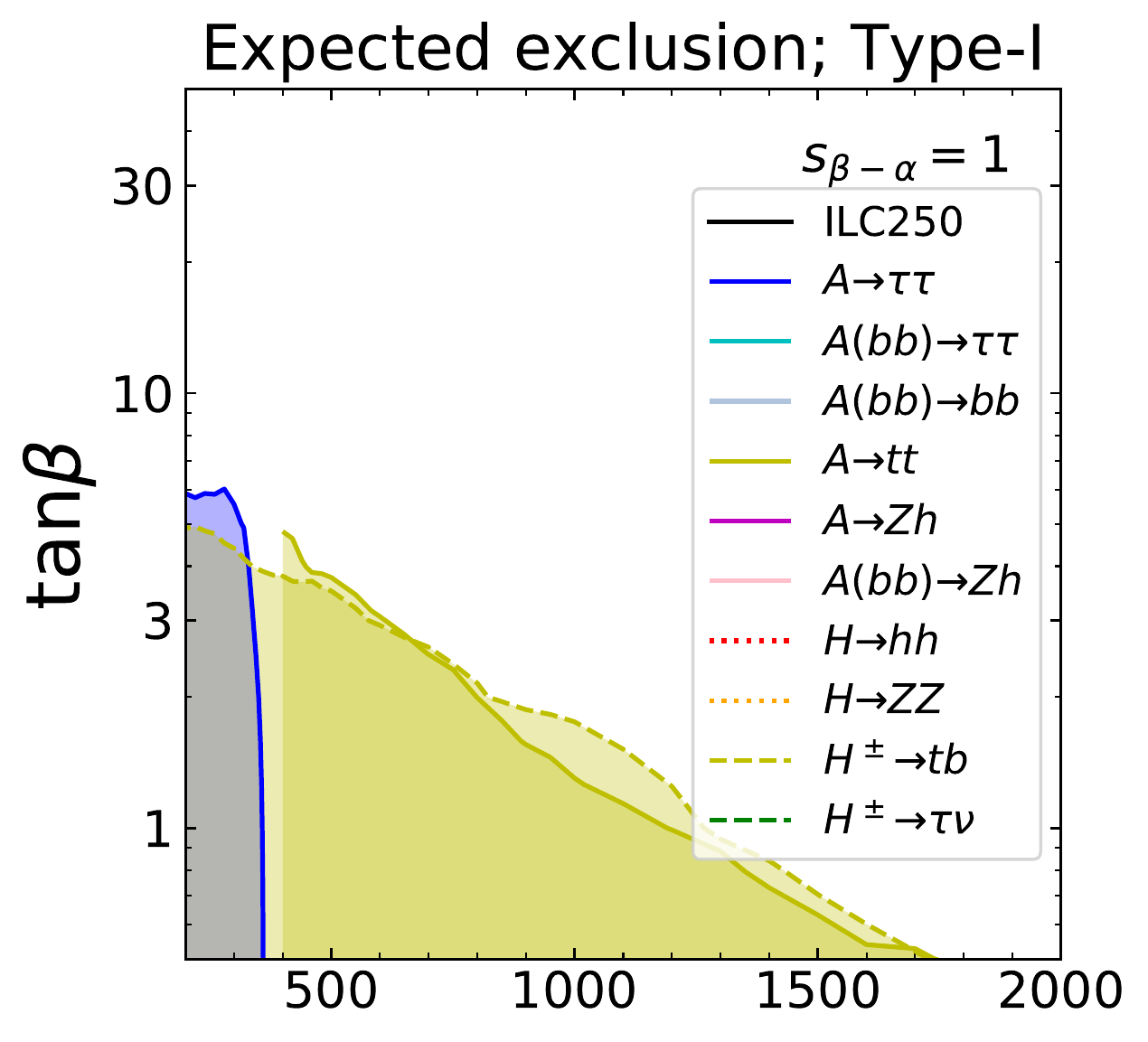}
 \includegraphics[height=0.235\textwidth]{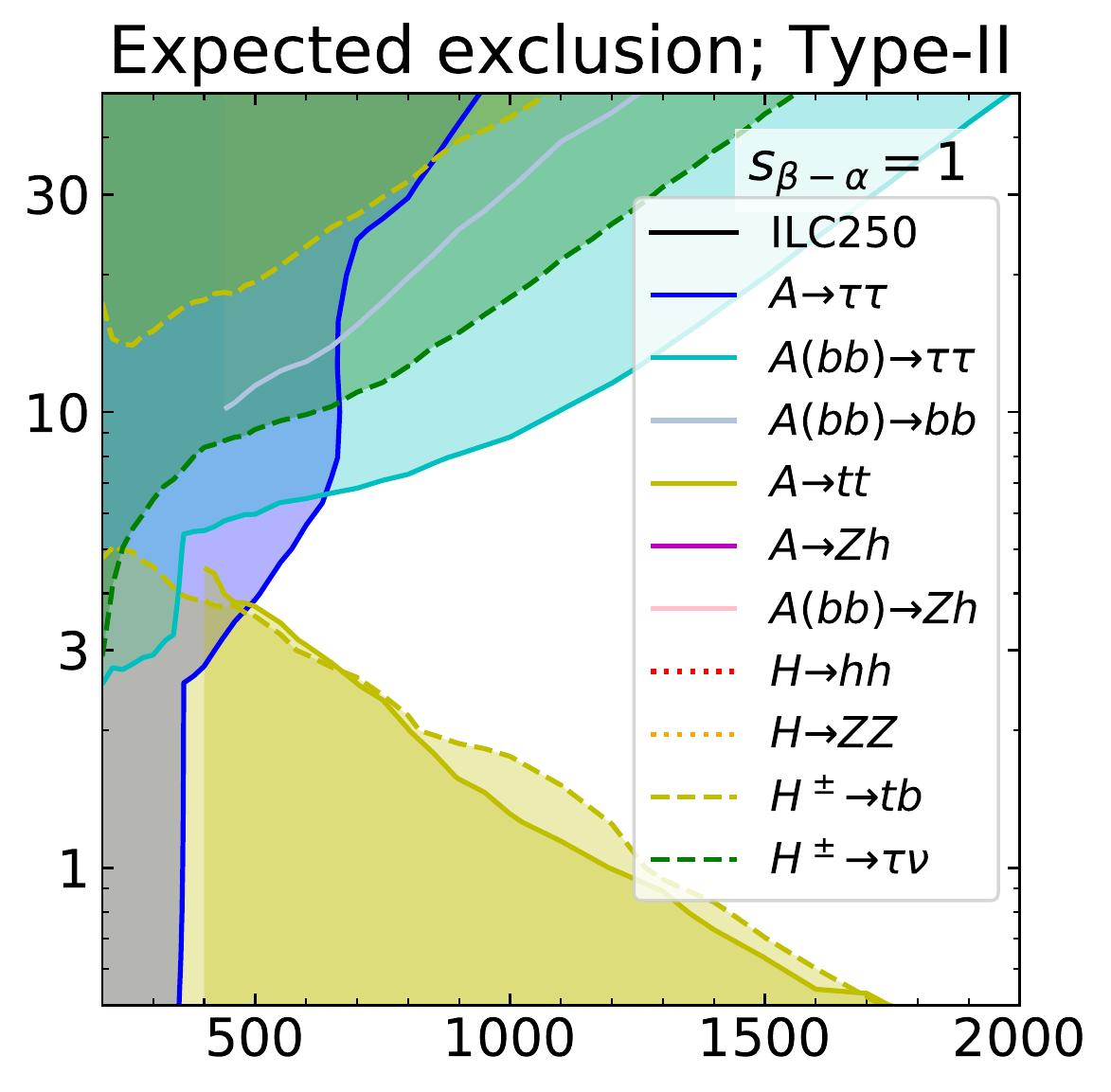}
 \includegraphics[height=0.235\textwidth]{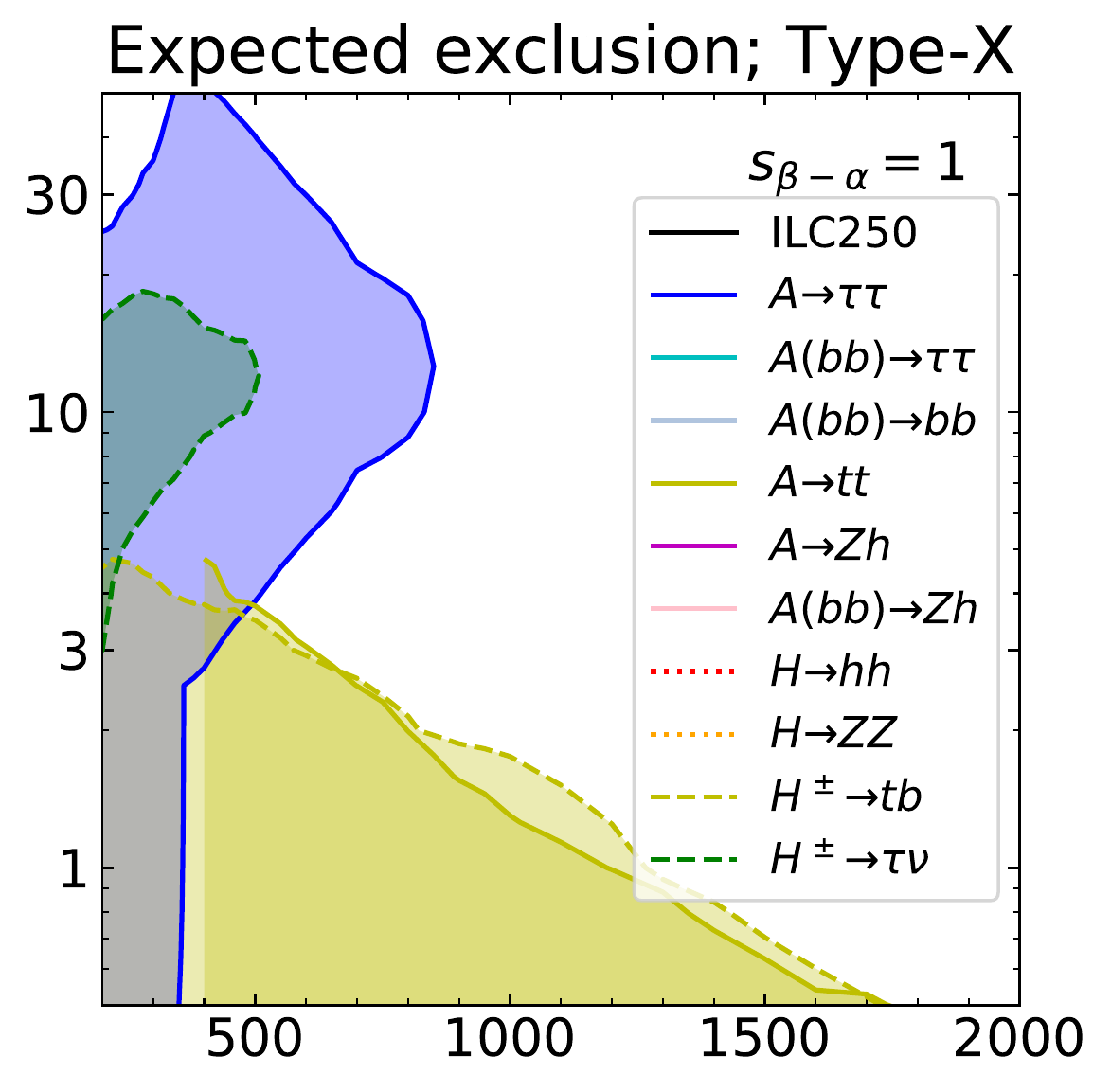}
 \includegraphics[height=0.235\textwidth]{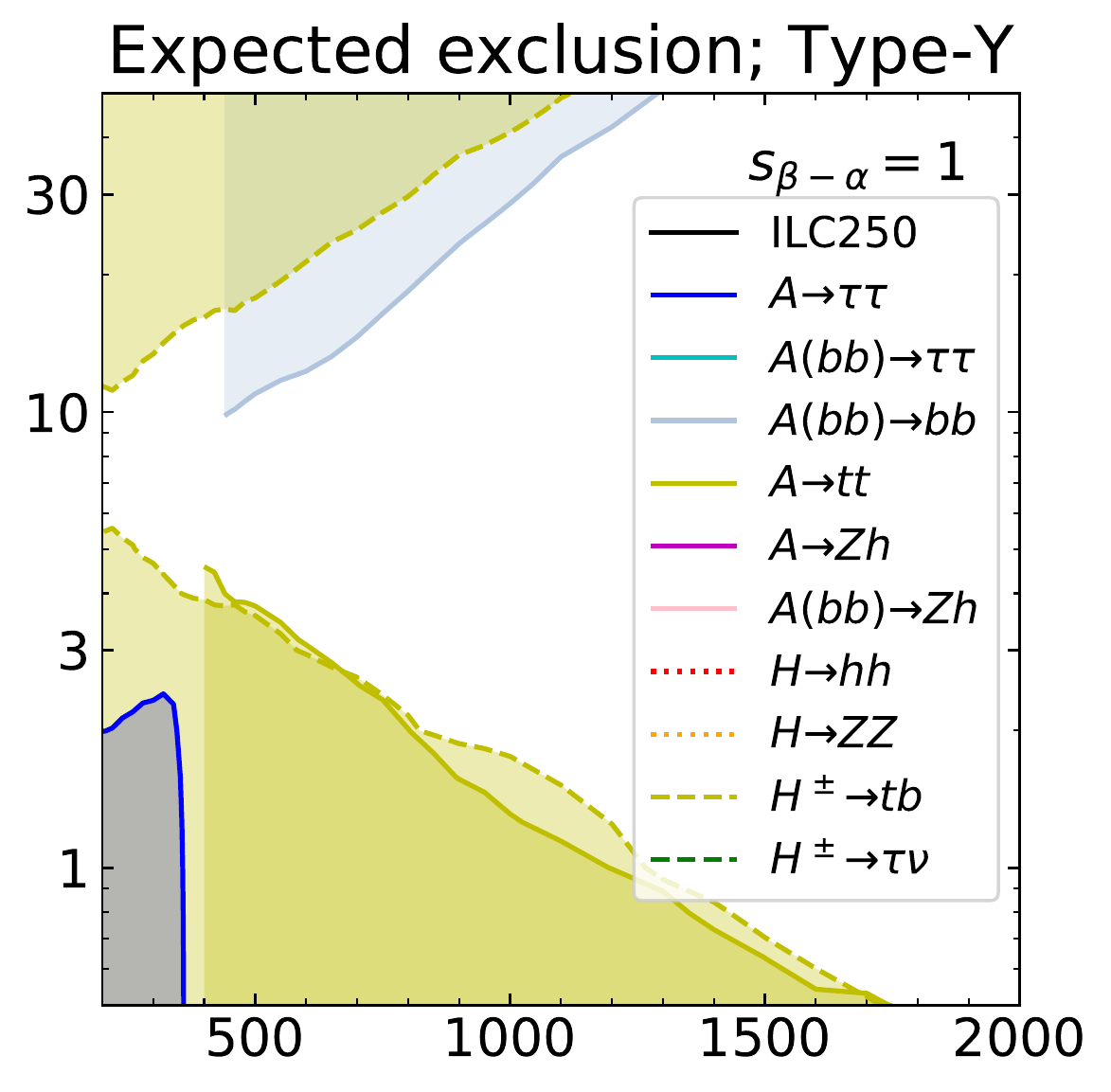}
 \includegraphics[height=0.220\textwidth]{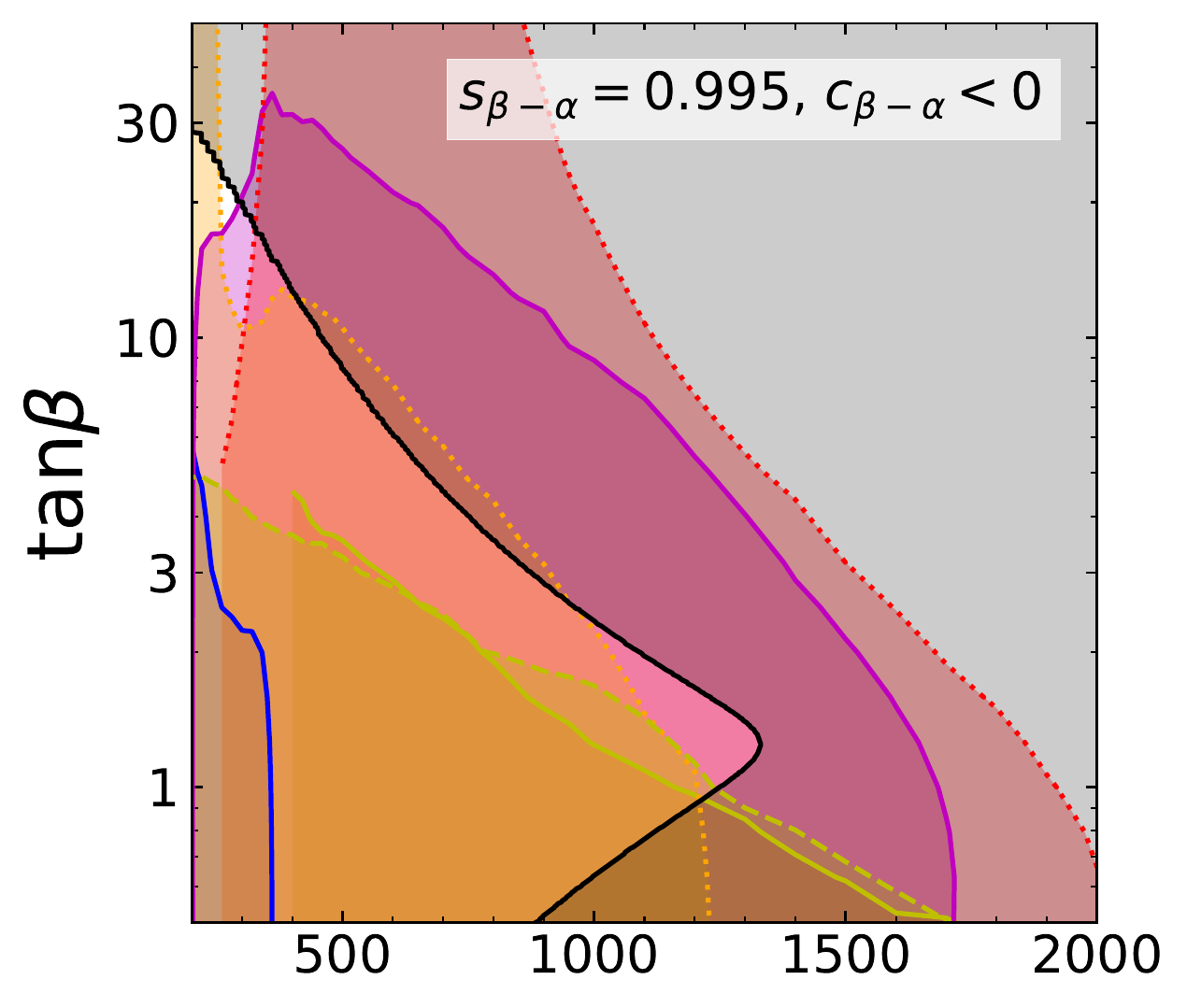}
 \includegraphics[height=0.220\textwidth]{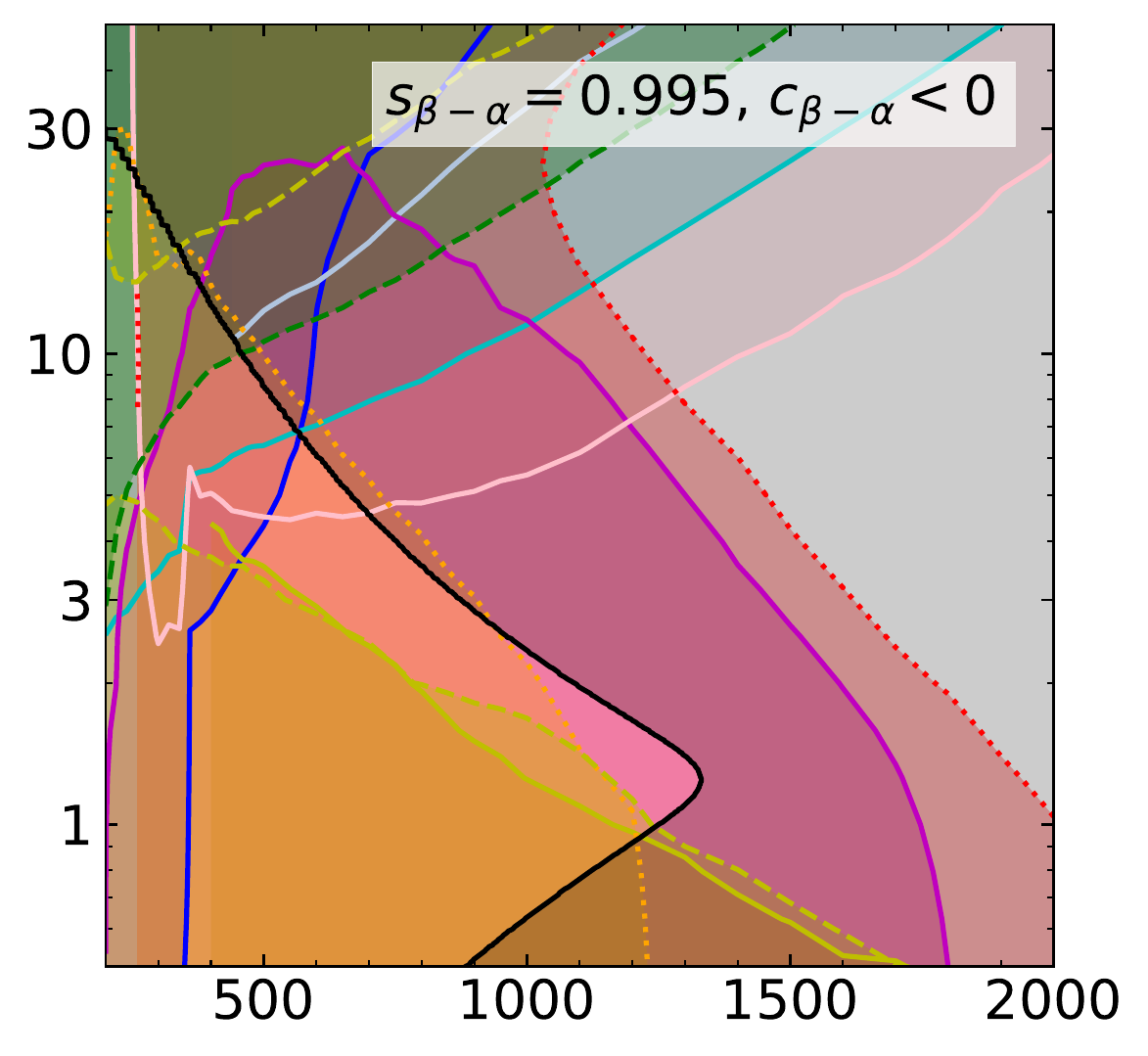}
 \includegraphics[height=0.220\textwidth]{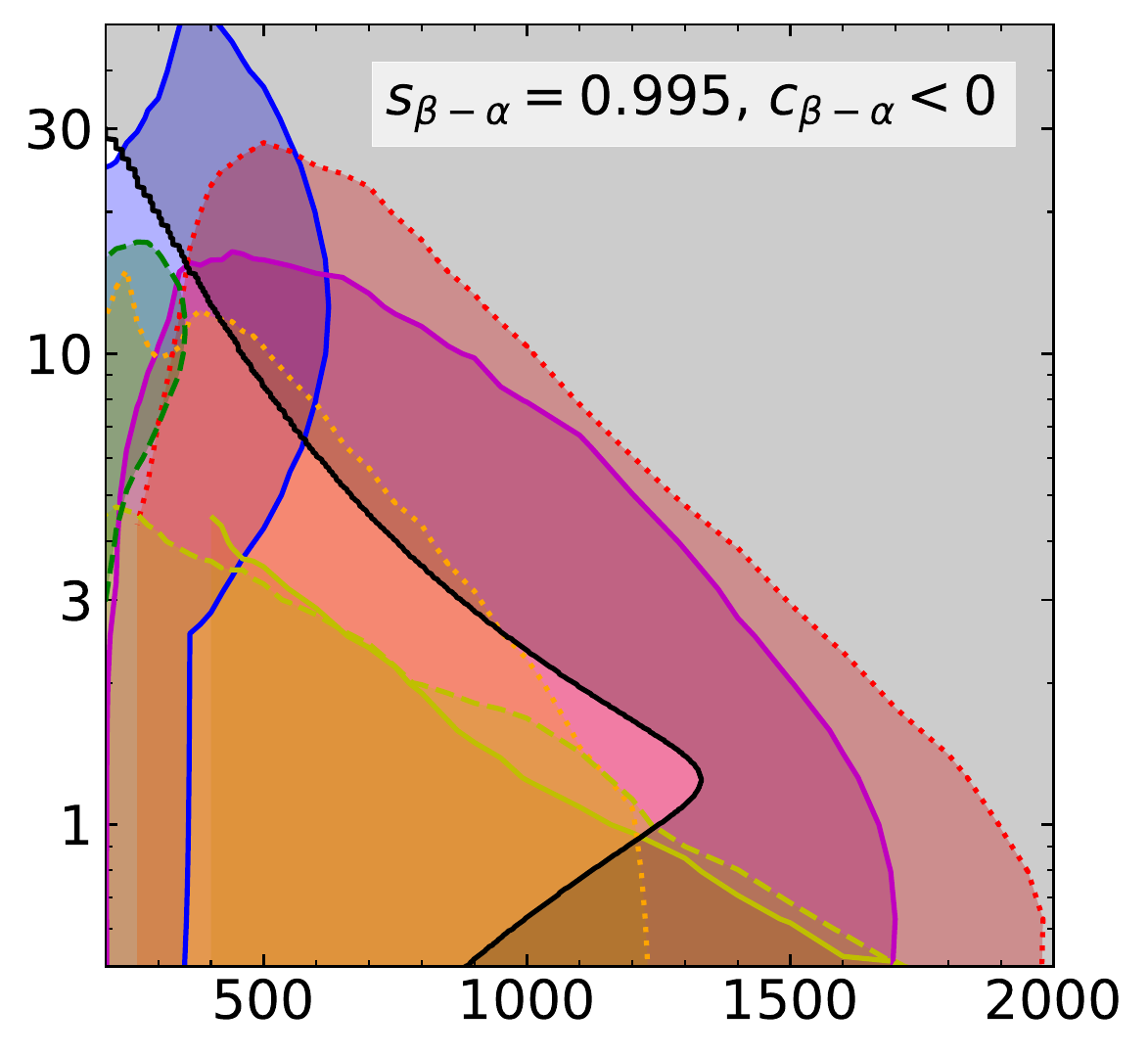}
 \includegraphics[height=0.220\textwidth]{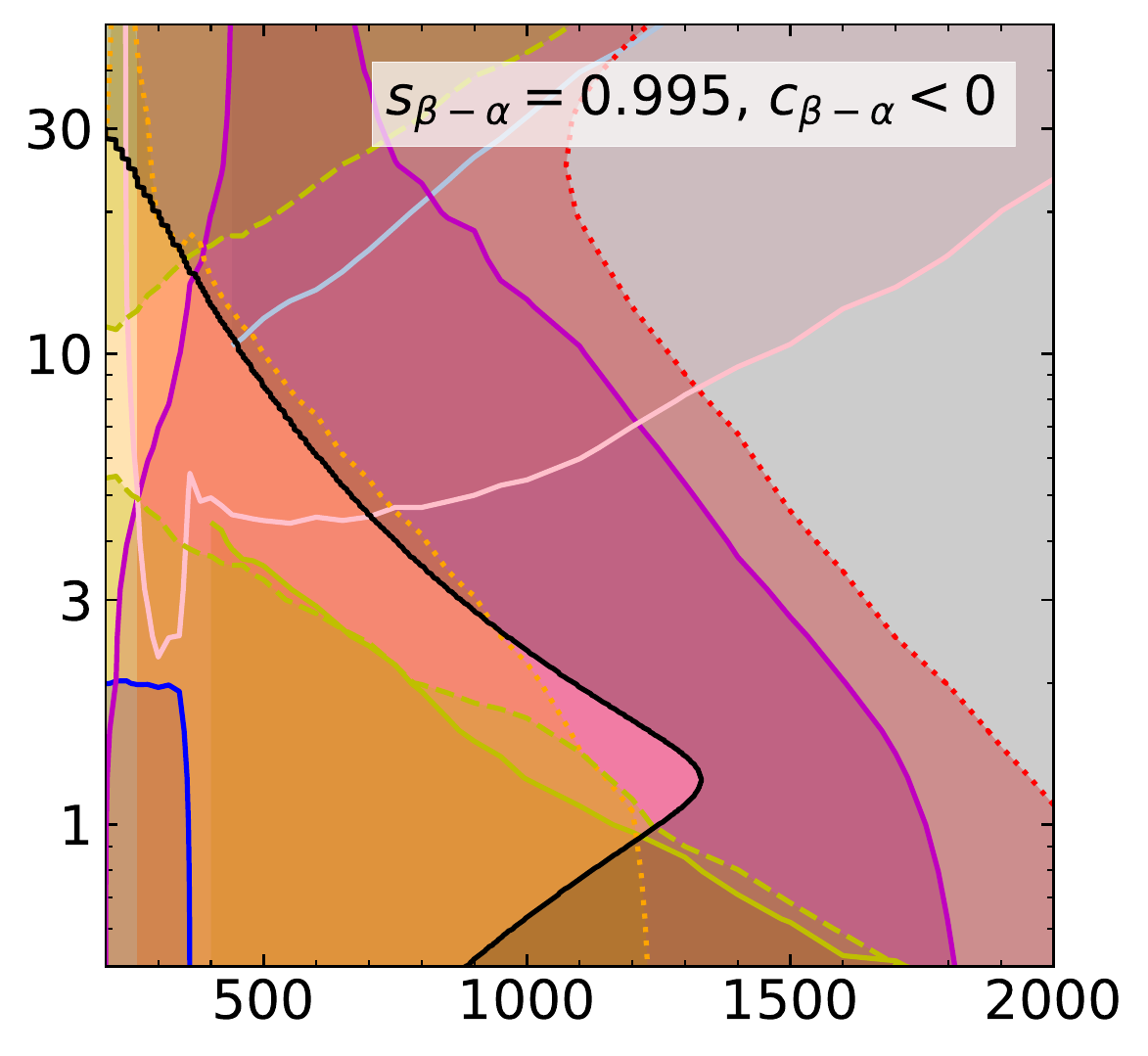}   
 \includegraphics[height=0.220\textwidth]{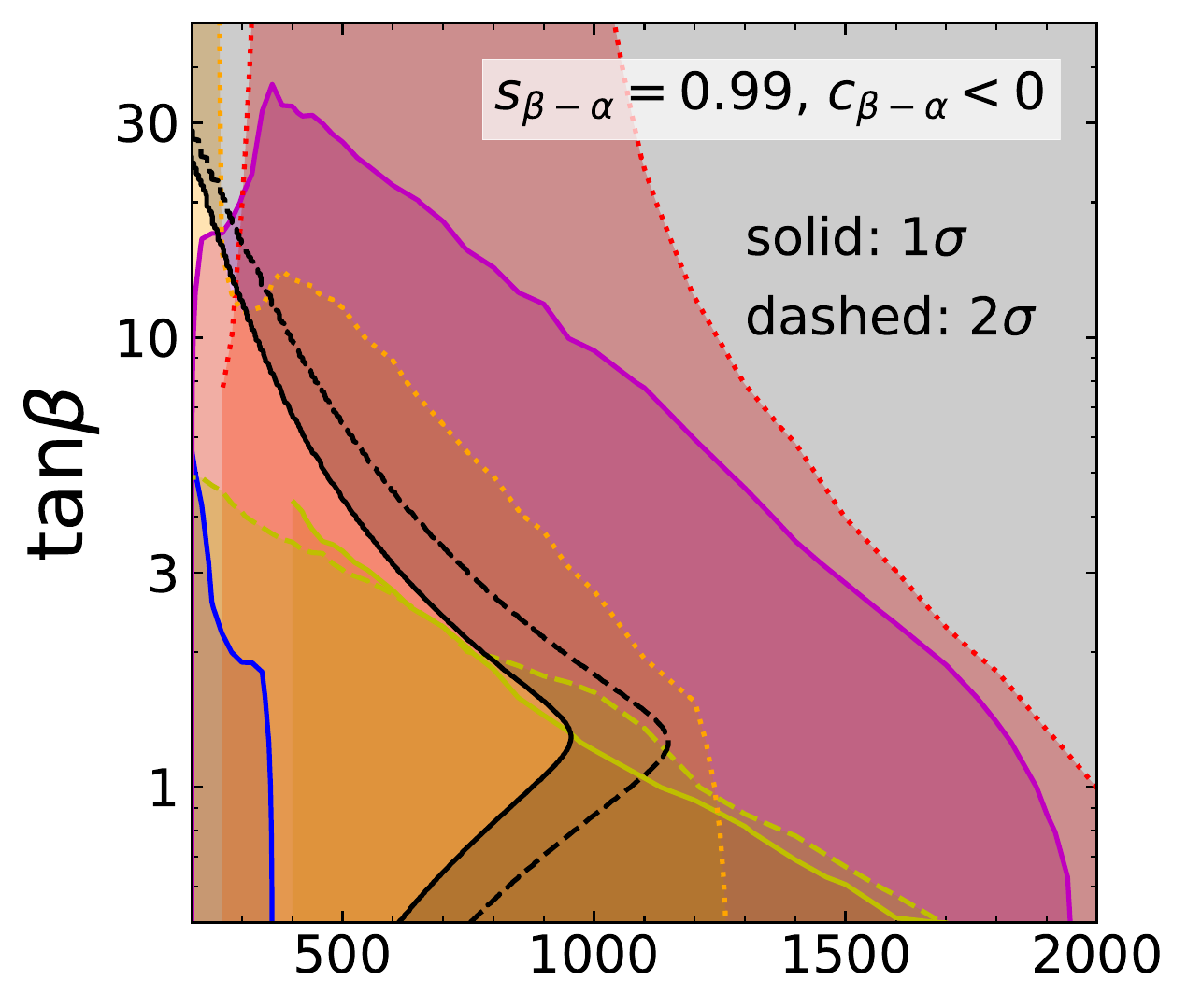}
 \includegraphics[height=0.220\textwidth]{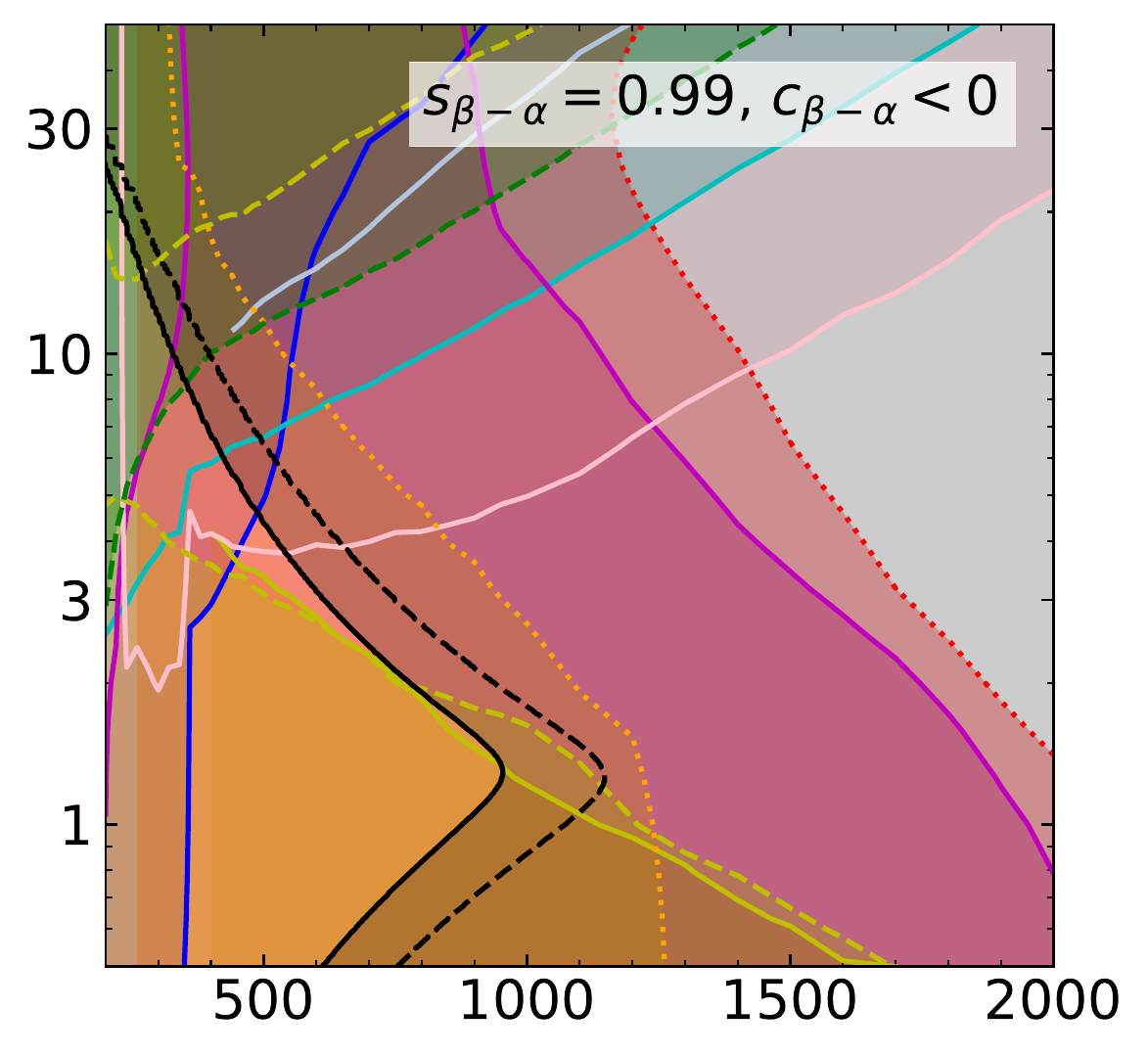}
 \includegraphics[height=0.220\textwidth]{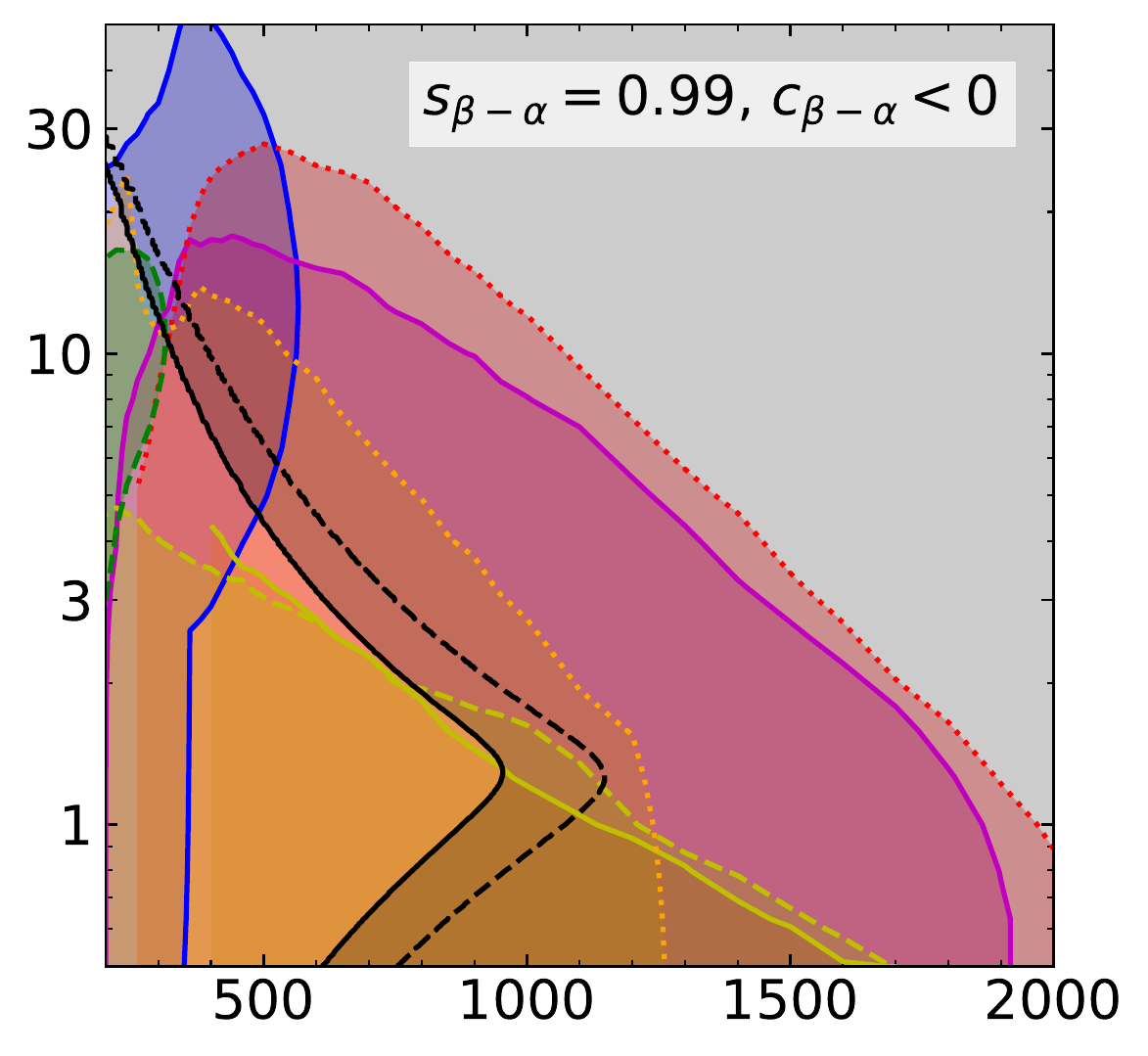}
 \includegraphics[height=0.220\textwidth]{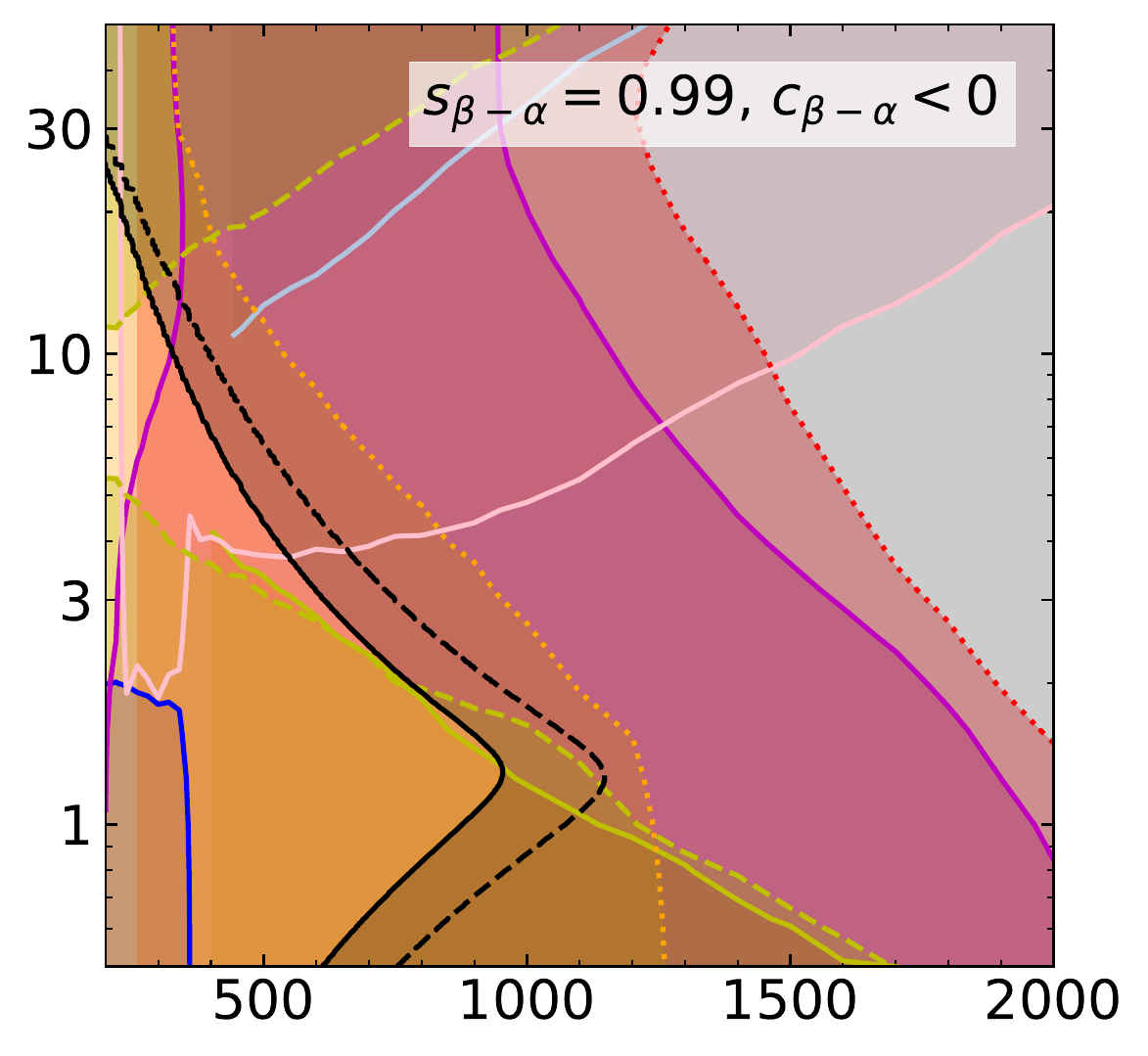} 
 \includegraphics[height=0.237\textwidth]{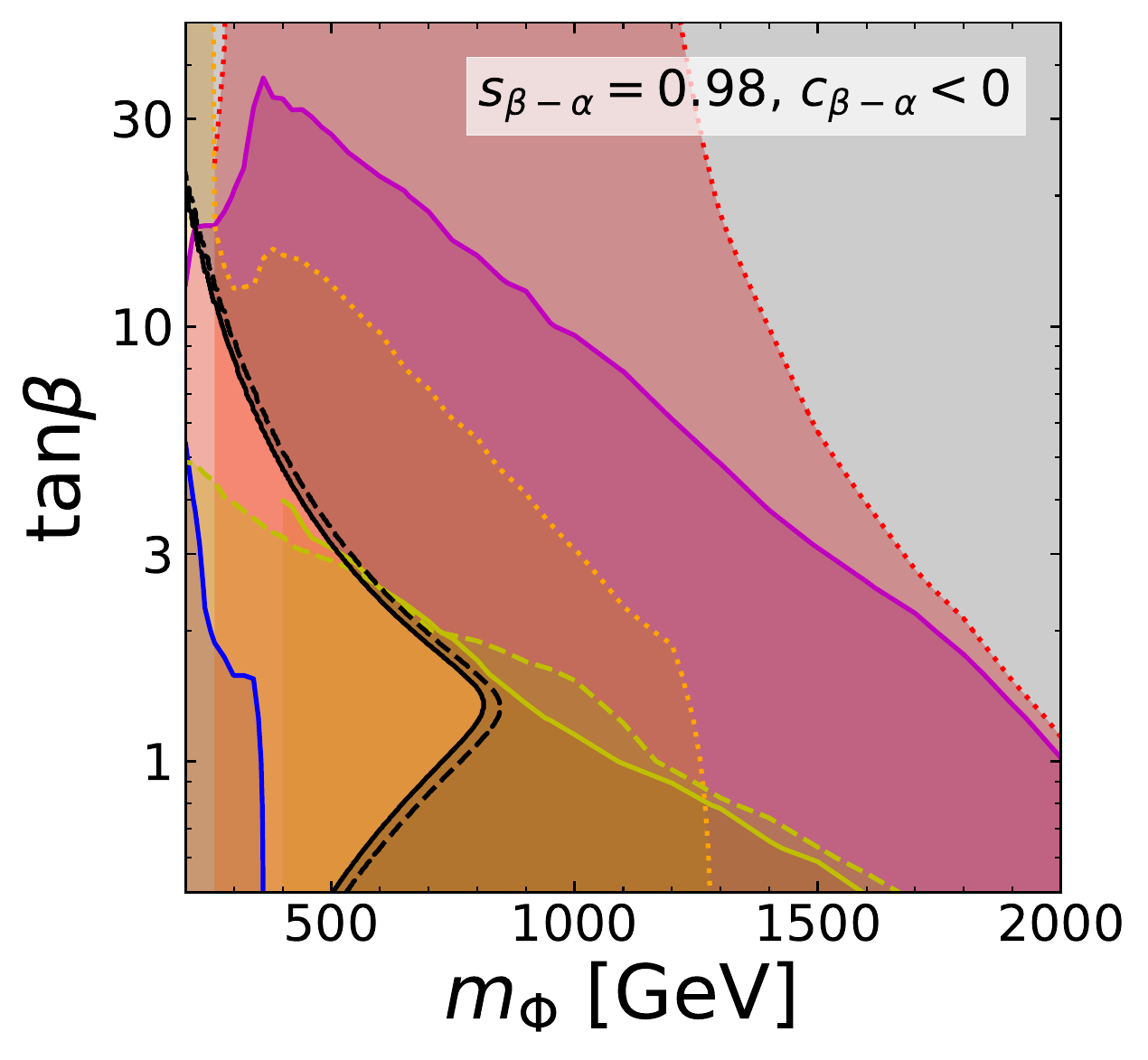}
 \includegraphics[height=0.237\textwidth]{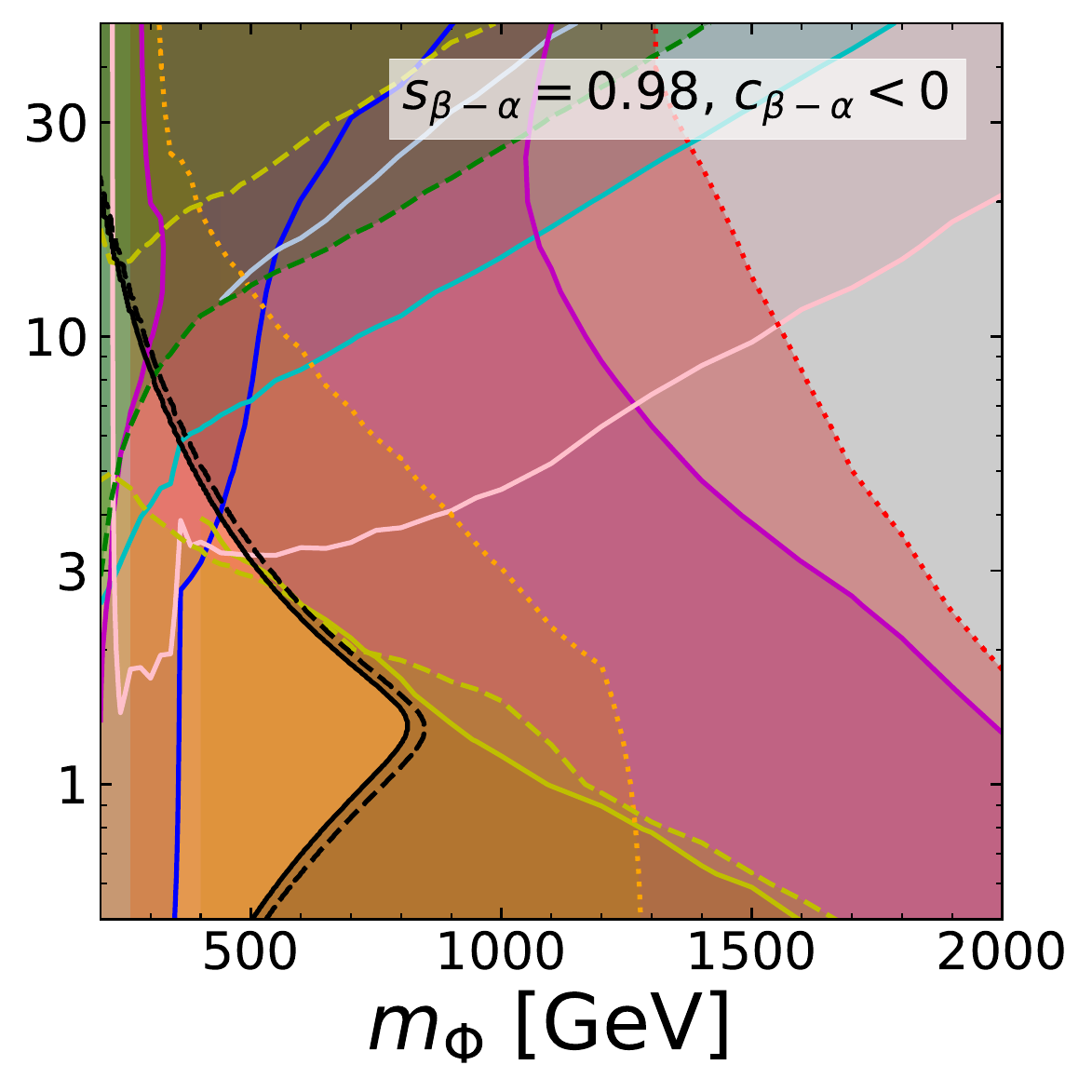}
 \includegraphics[height=0.237\textwidth]{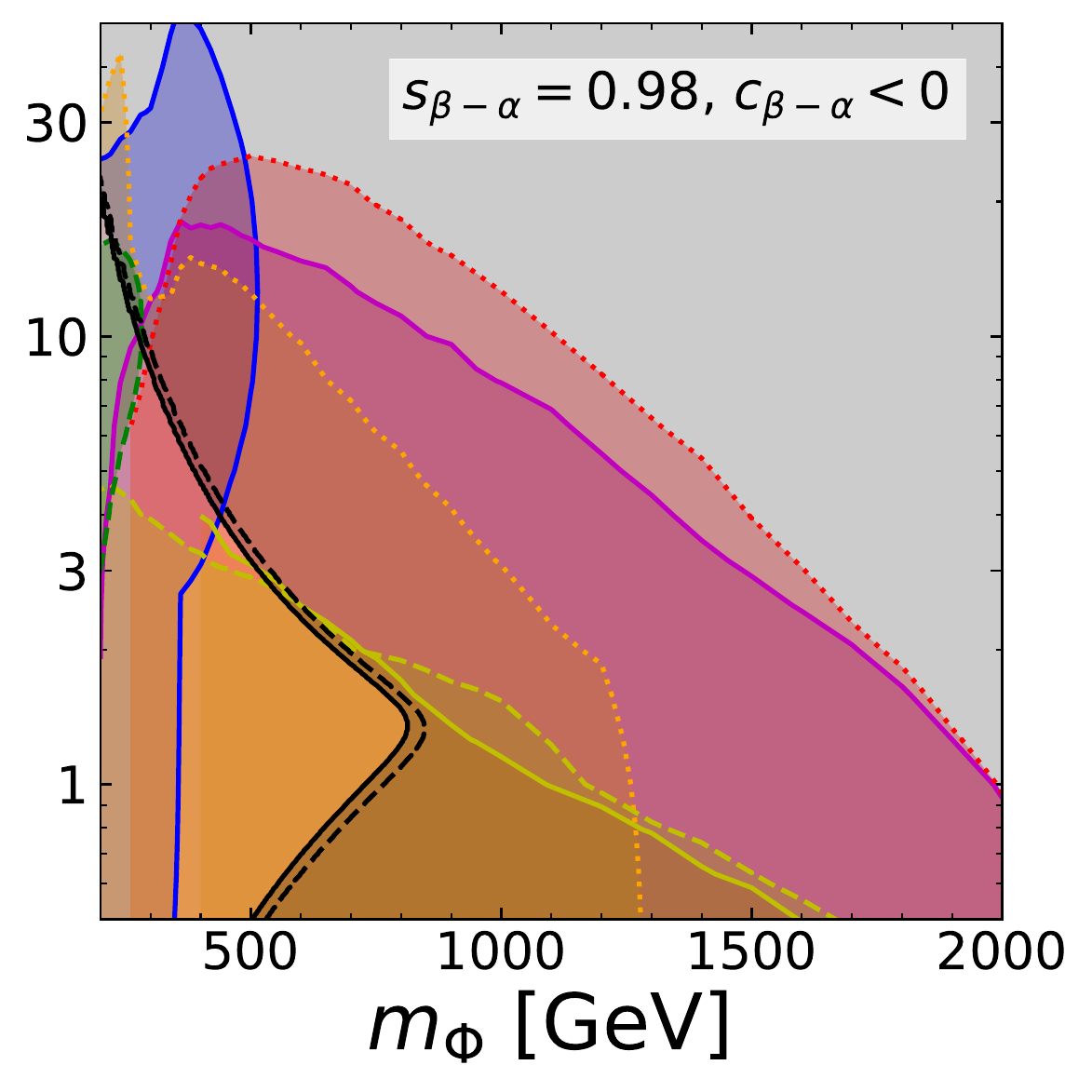}
 \includegraphics[height=0.237\textwidth]{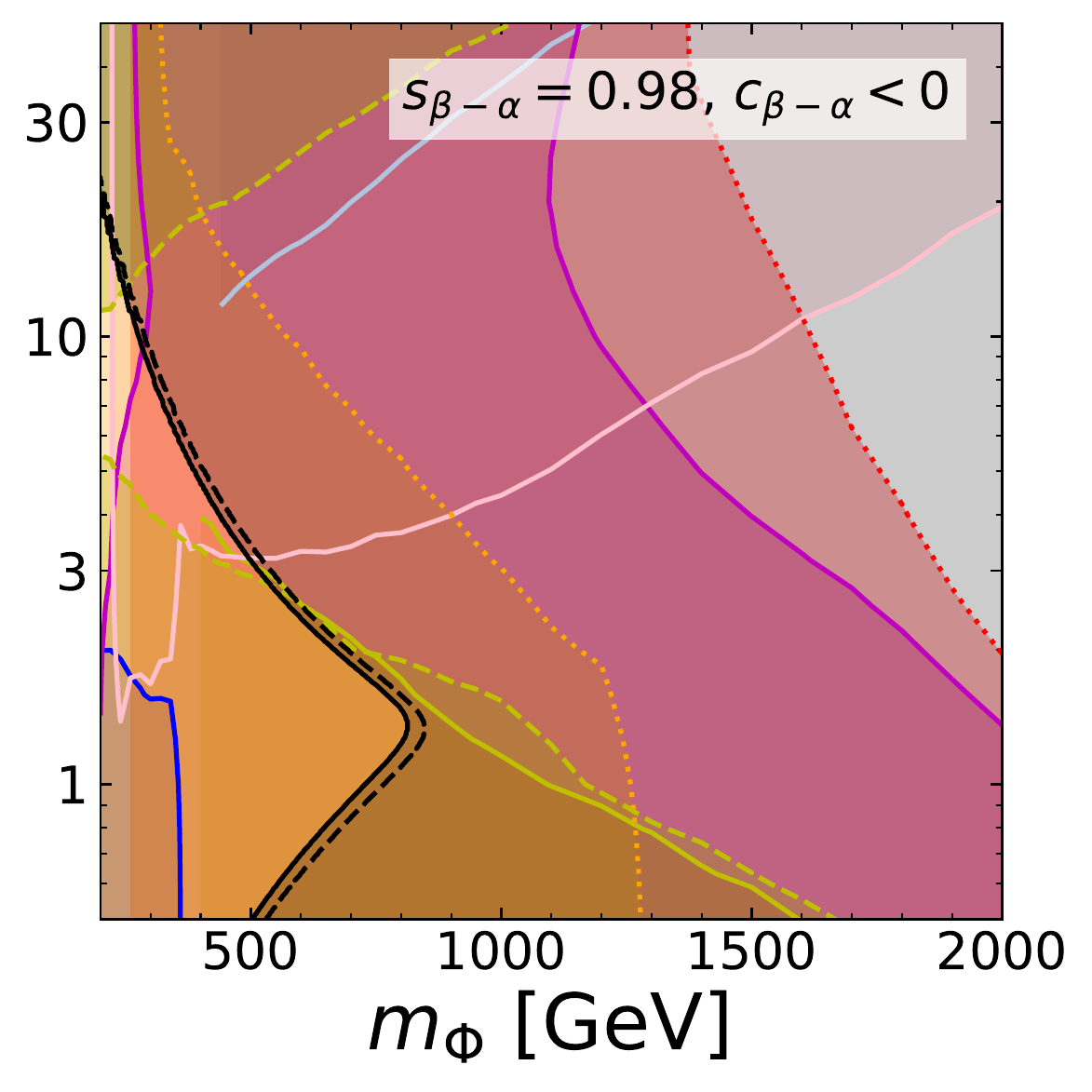}  
\caption{
Regions on the $m_\Phi$--$\tan\beta$ plane expected to be excluded at 95\% CL 
in the Type-I, Type-II, Type-X and Type-Y THDMs (from the left to the right panels)
via direct searches for heavy Higgs bosons at the HL-LHC and via precision measurements of the Higgs boson couplings at the ILC.
The value of $s_{\beta-\alpha}$ is set to be 1, 0.995, 0.99 and 0.98 with $c_{\beta-\alpha} < 0$ from the top to the bottom panels.
}
\label{fig:combine-n}
\end{figure}  

\begin{figure}
 \includegraphics[height=0.235\textwidth]{Fig/ex_rs14_t1_sba1_ilc.pdf}
 \includegraphics[height=0.235\textwidth]{Fig/ex_rs14_t2_sba1_ilc.pdf}
 \includegraphics[height=0.235\textwidth]{Fig/ex_rs14_t3_sba1_ilc.pdf}
 \includegraphics[height=0.235\textwidth]{Fig/ex_rs14_t4_sba1_ilc.pdf}
 \includegraphics[height=0.220\textwidth]{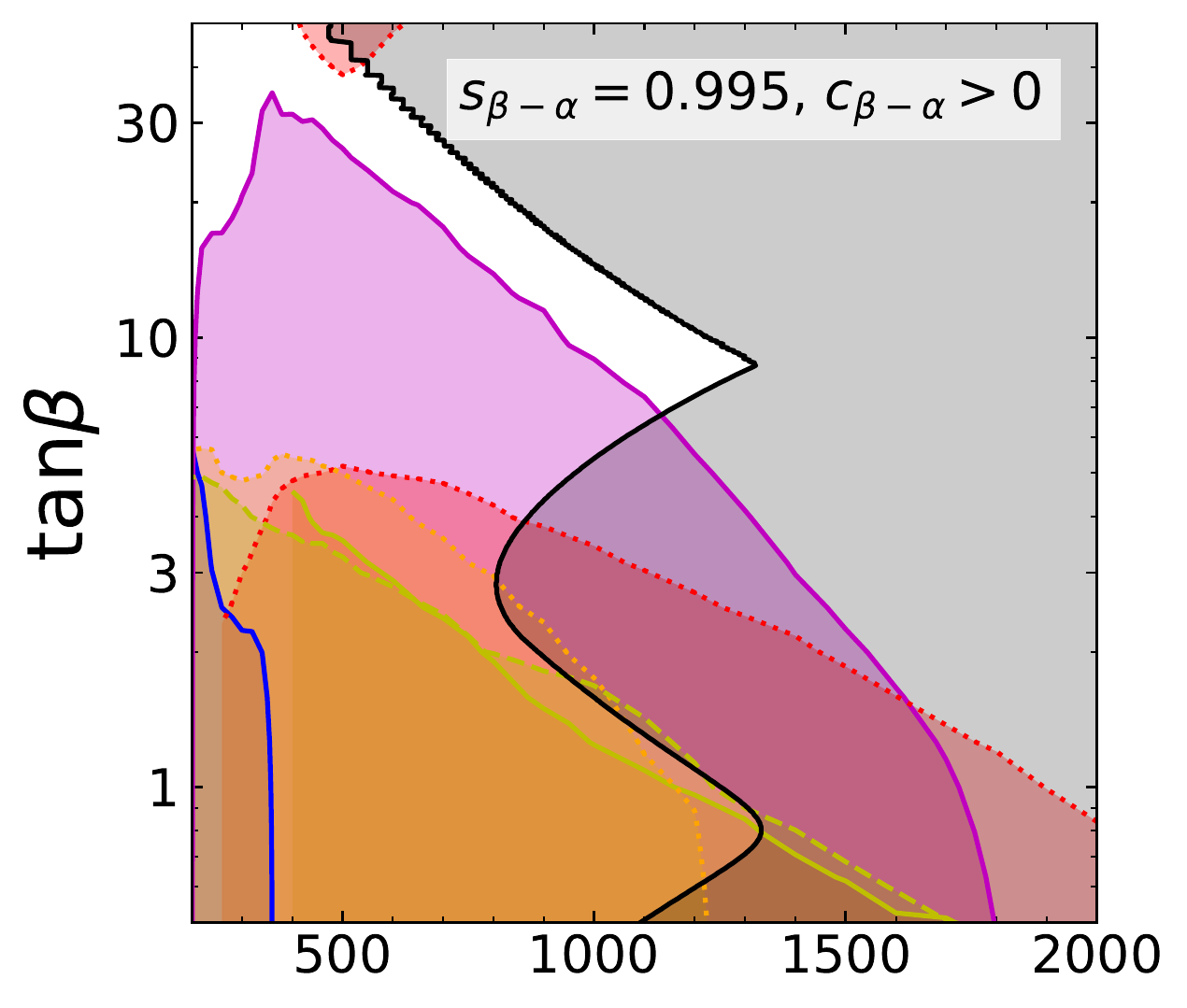}
 \includegraphics[height=0.220\textwidth]{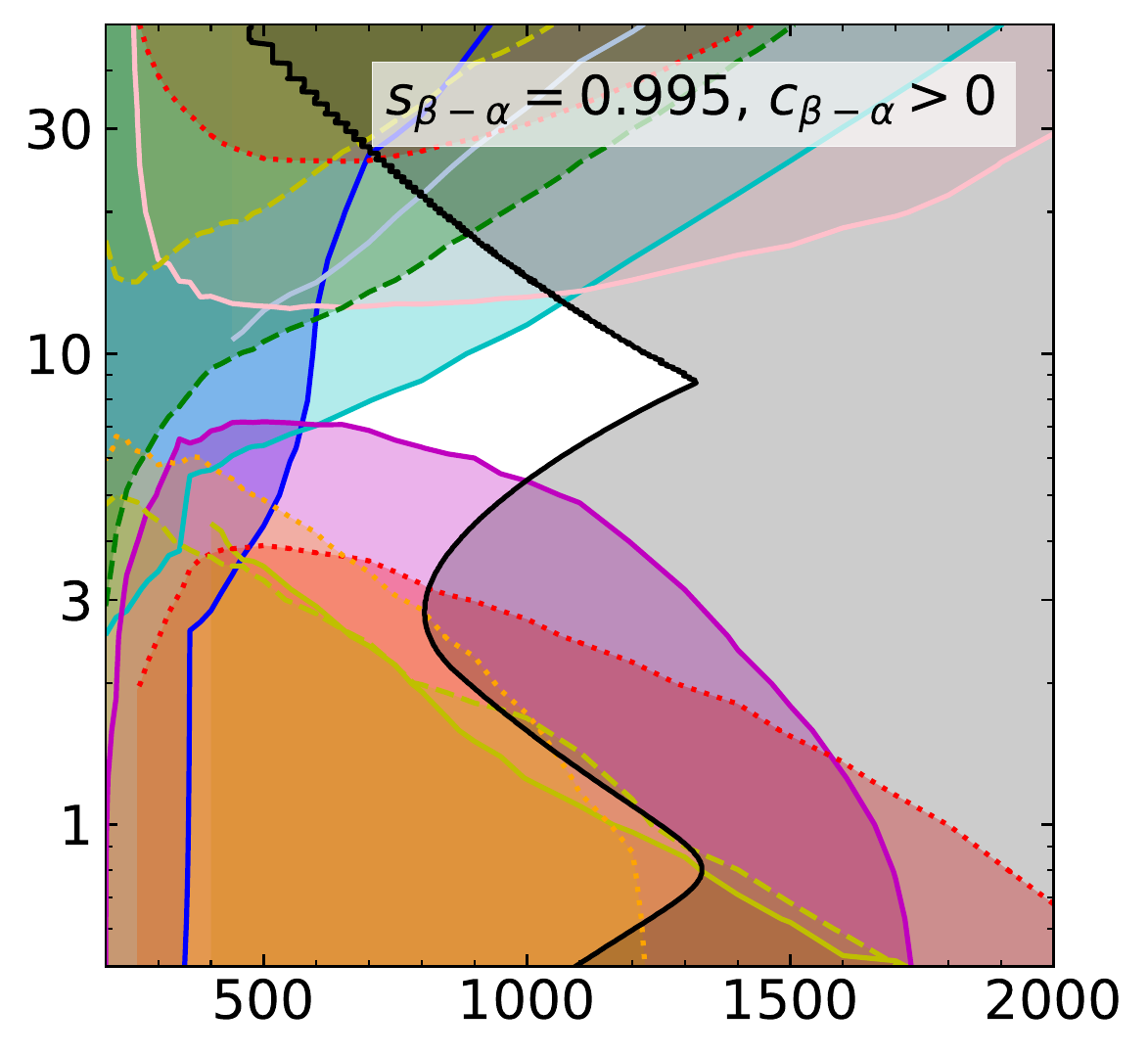}
 \includegraphics[height=0.220\textwidth]{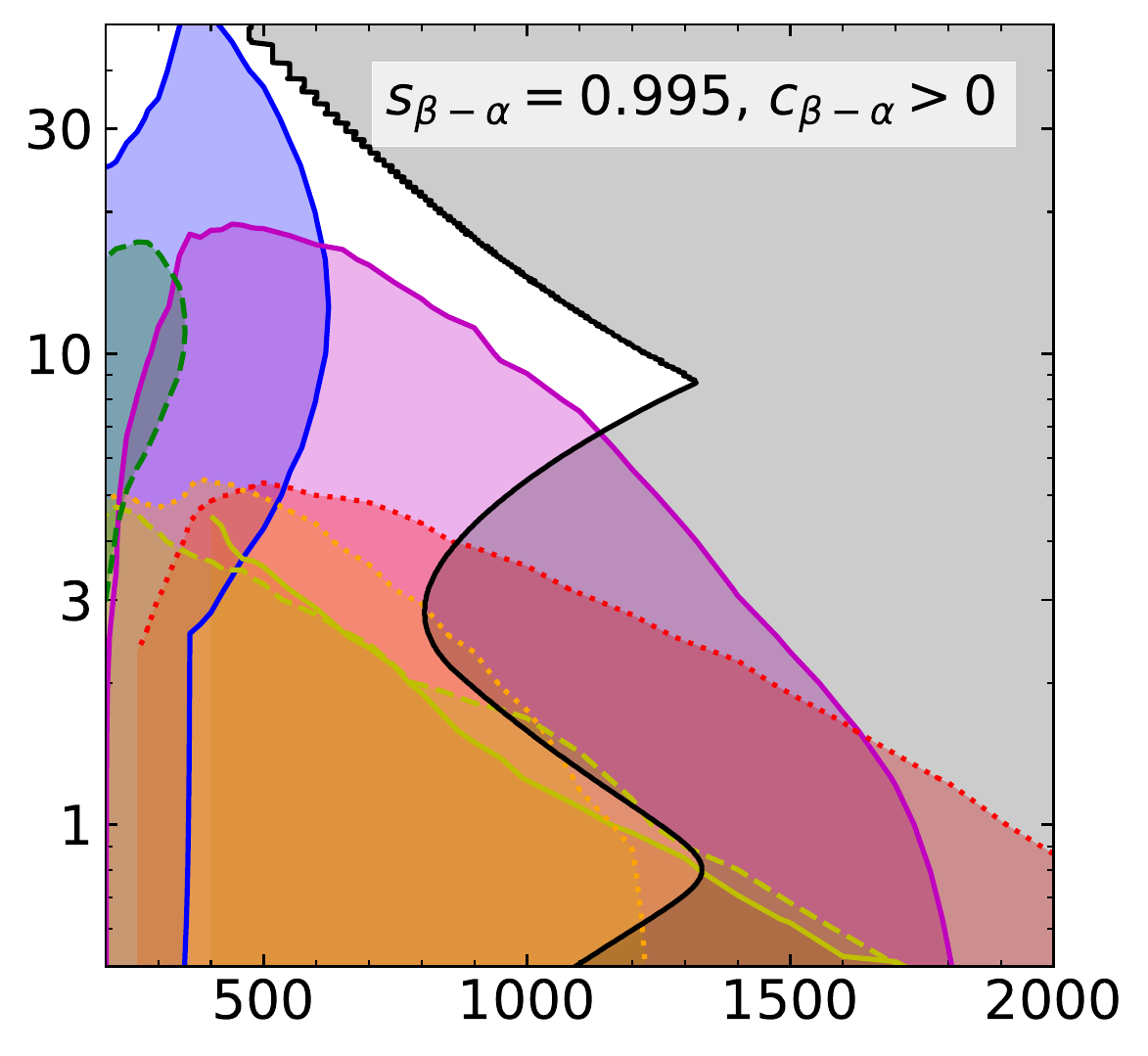}
 \includegraphics[height=0.220\textwidth]{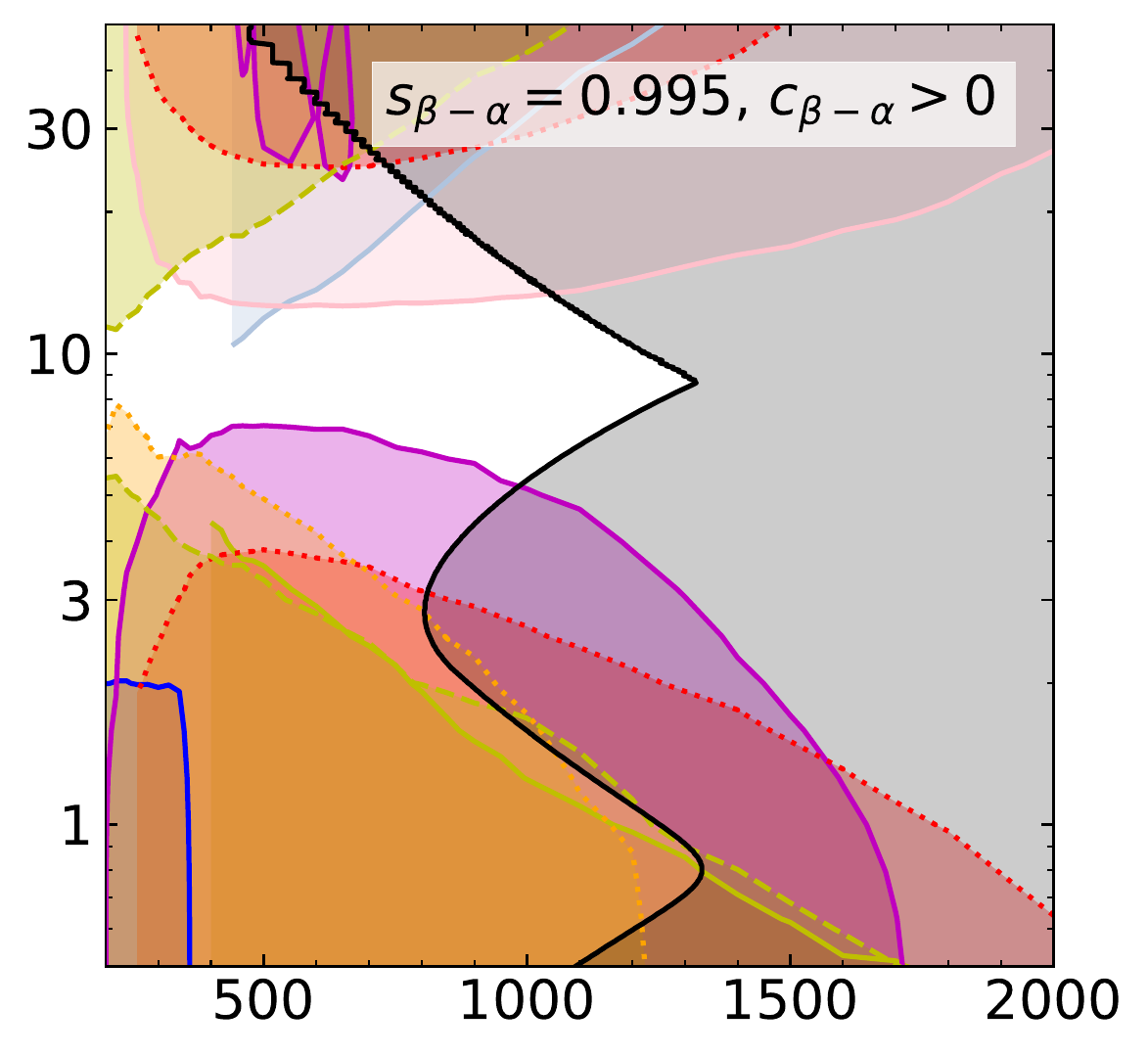}   
 \includegraphics[height=0.220\textwidth]{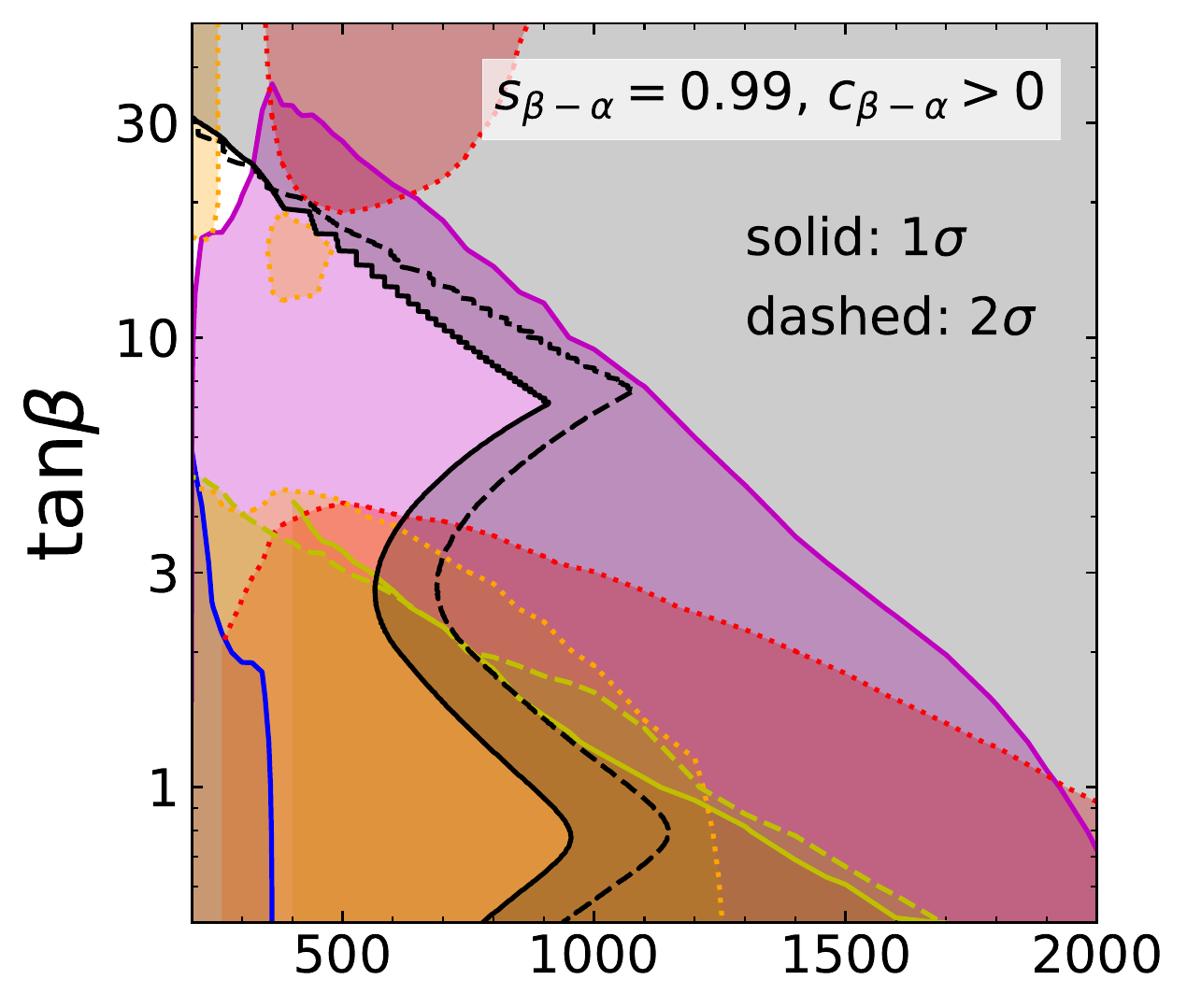}
 \includegraphics[height=0.220\textwidth]{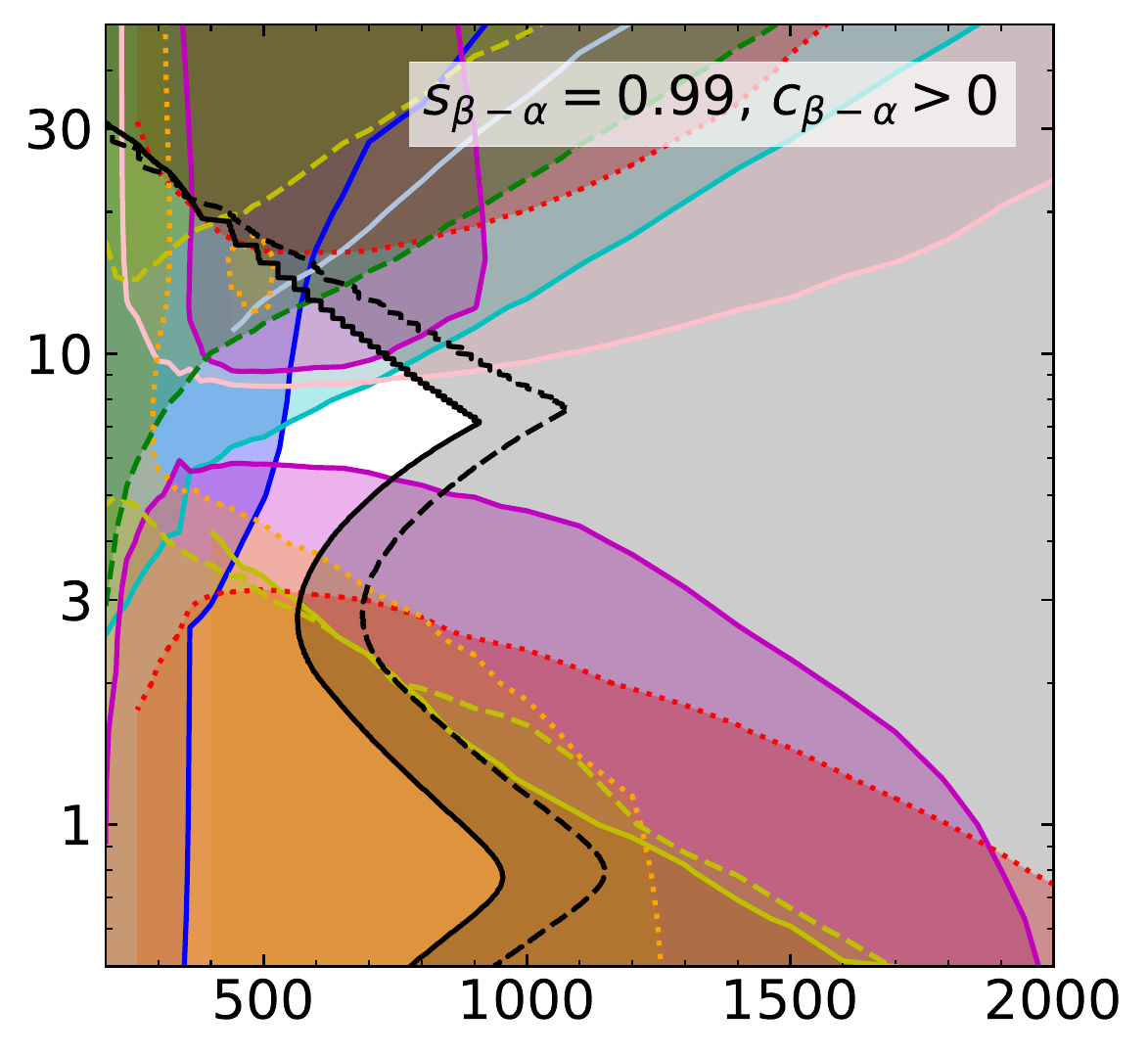}
 \includegraphics[height=0.220\textwidth]{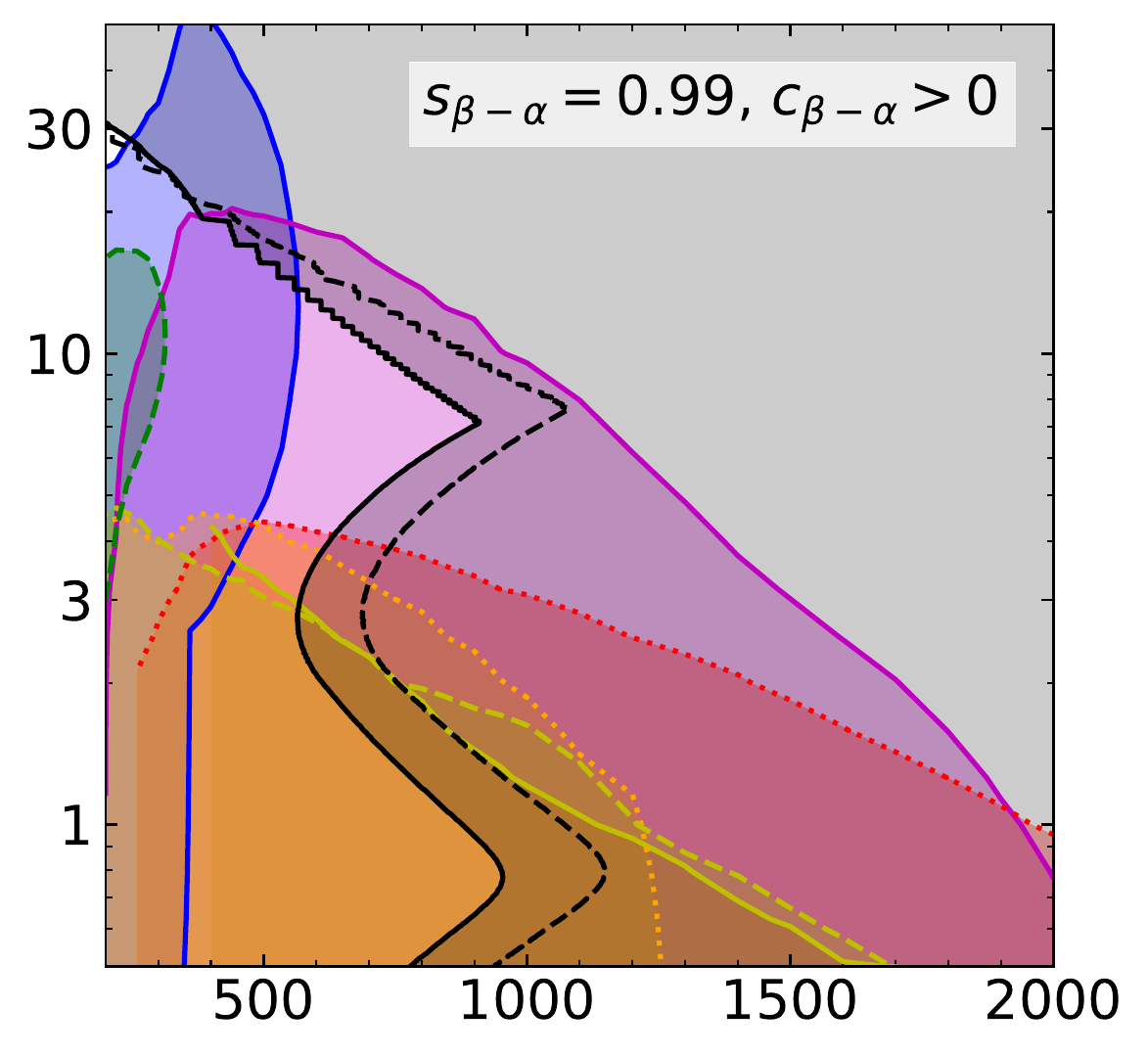}
 \includegraphics[height=0.220\textwidth]{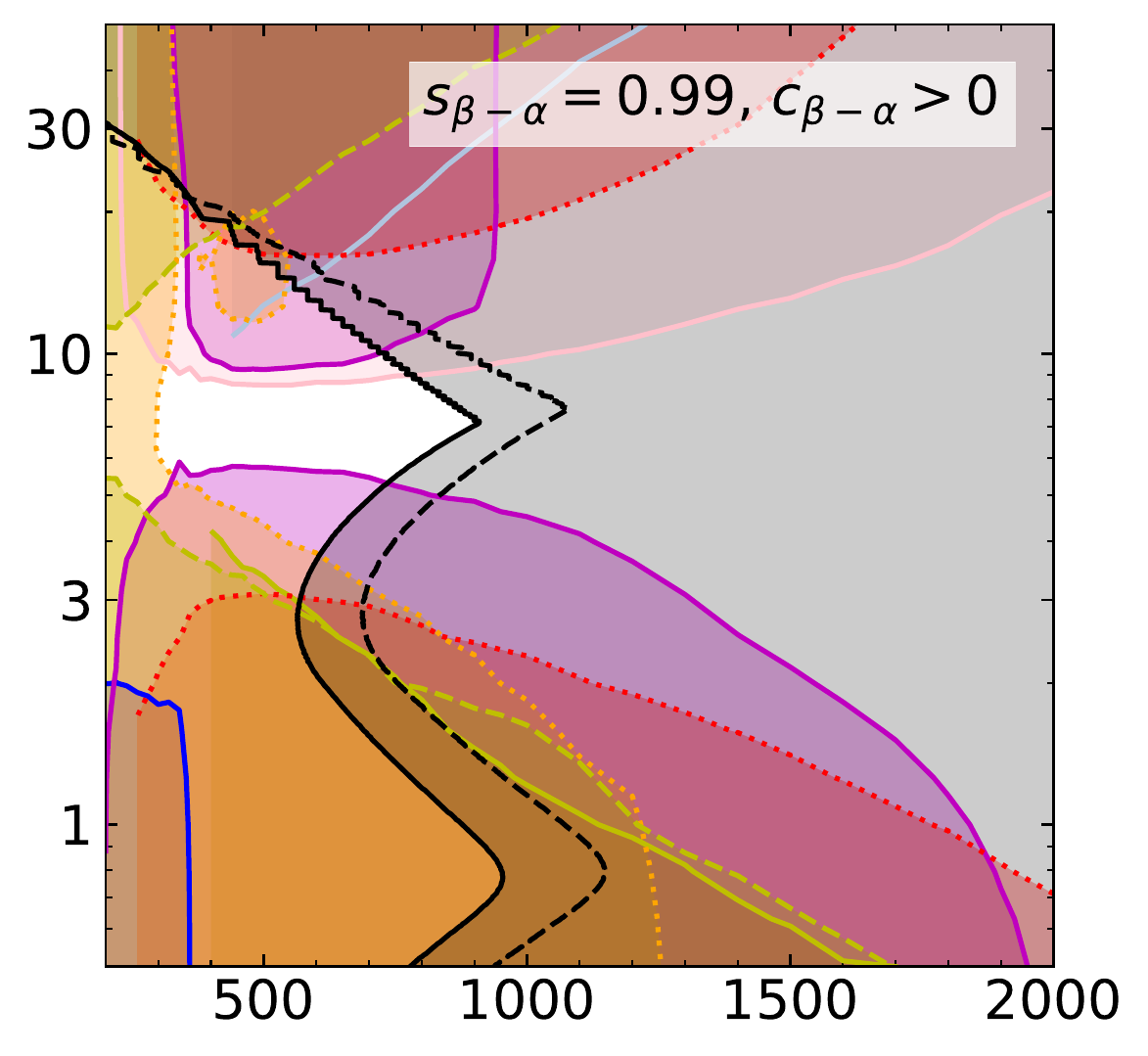} 
 \includegraphics[height=0.237\textwidth]{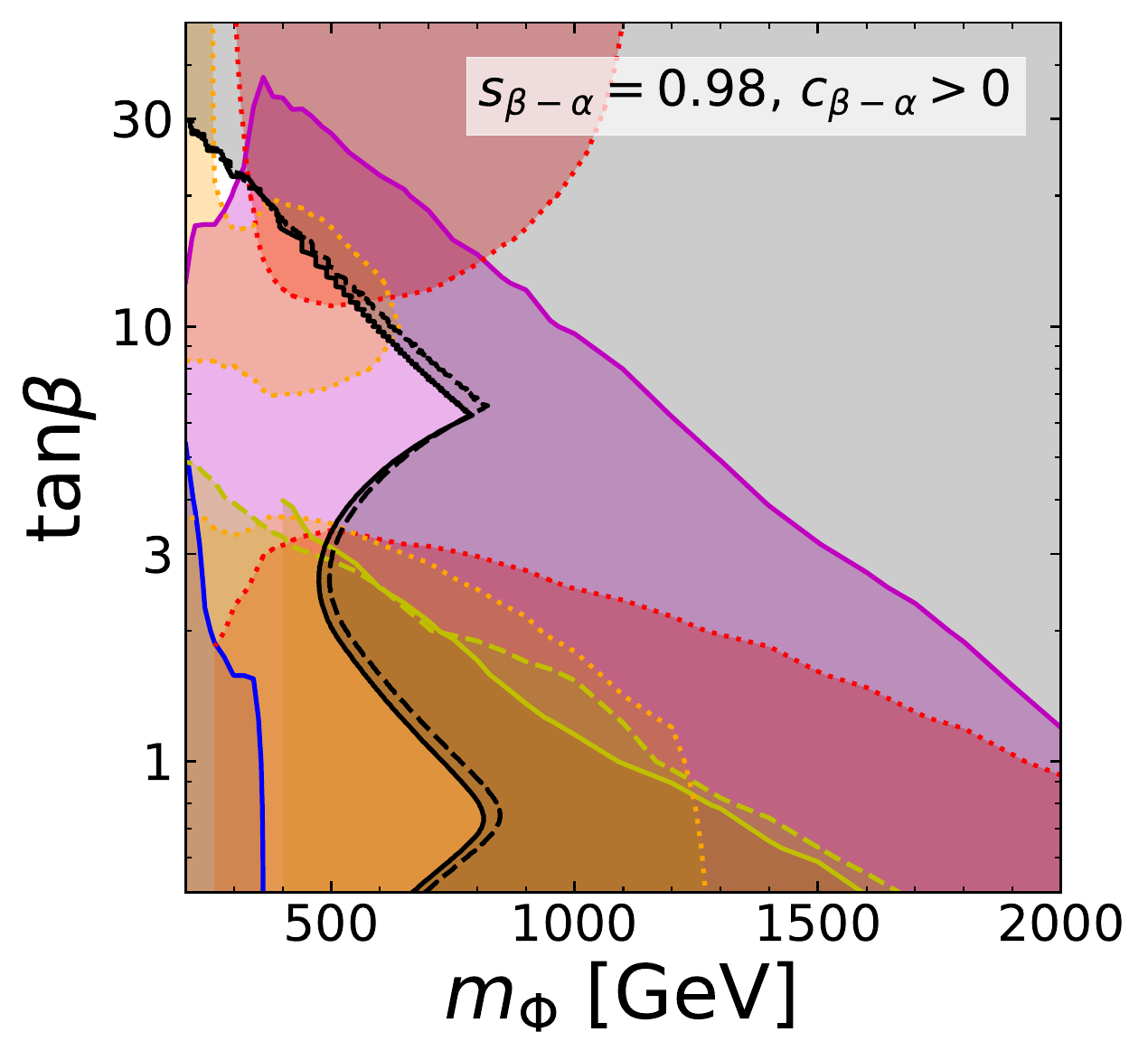}
 \includegraphics[height=0.237\textwidth]{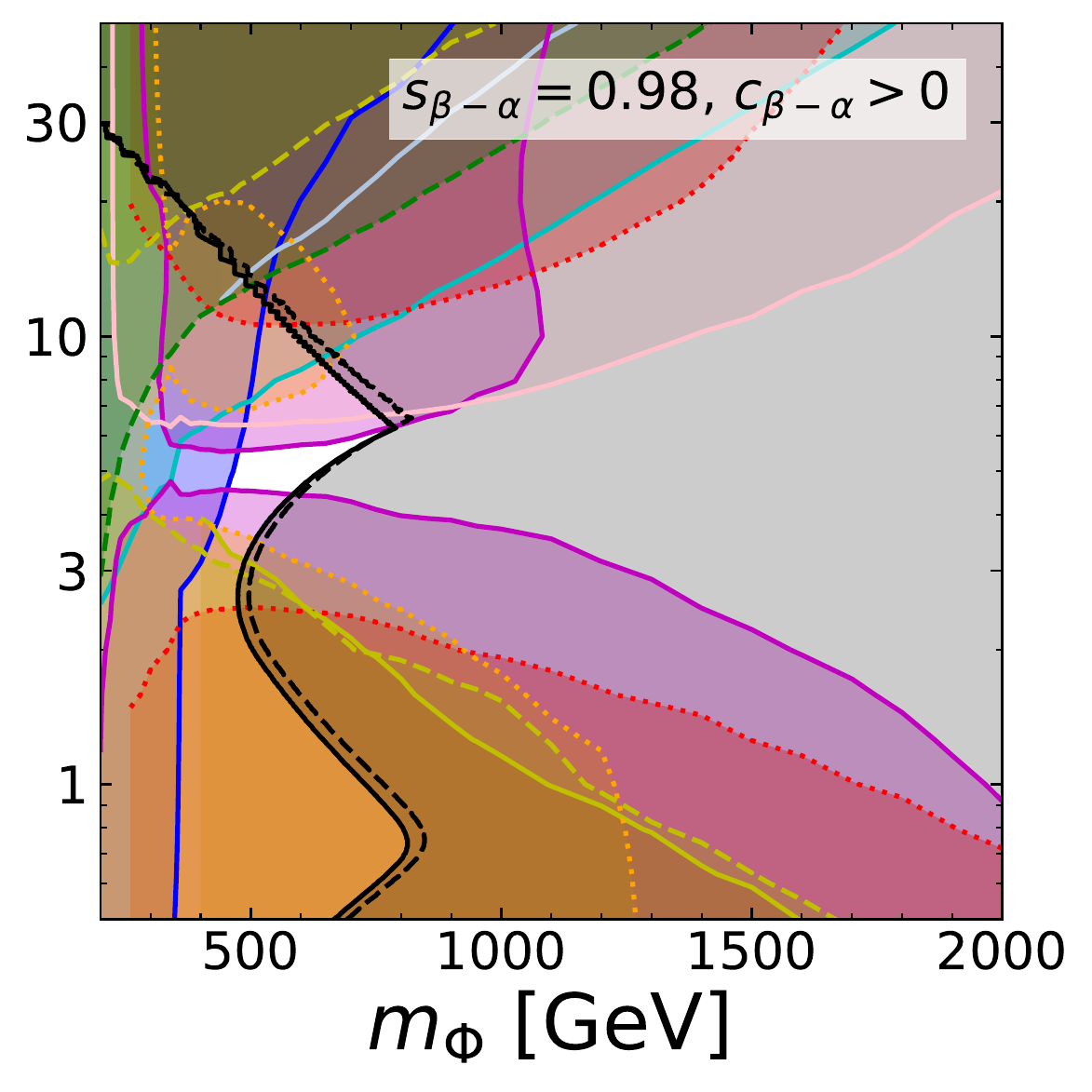}
 \includegraphics[height=0.237\textwidth]{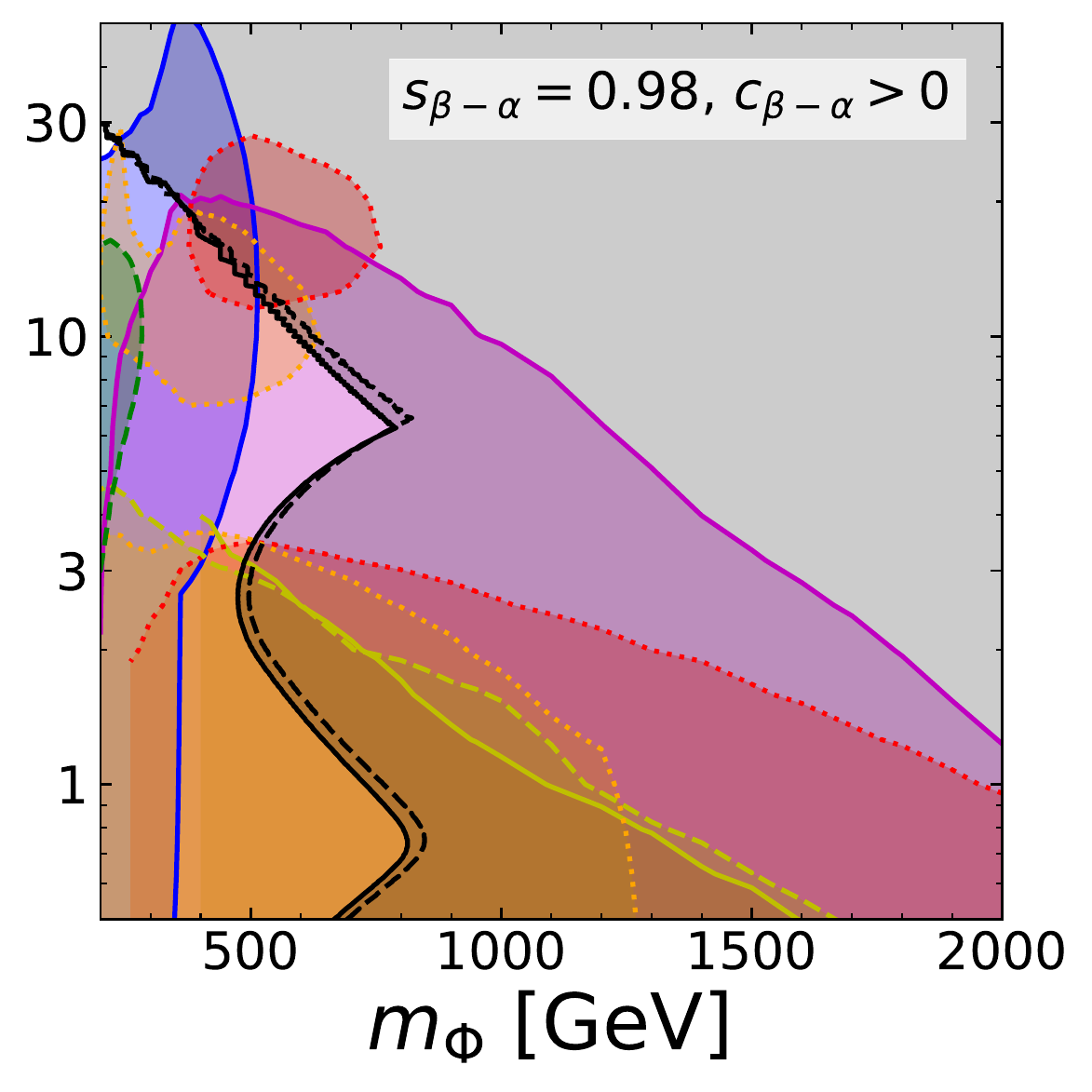}
 \includegraphics[height=0.237\textwidth]{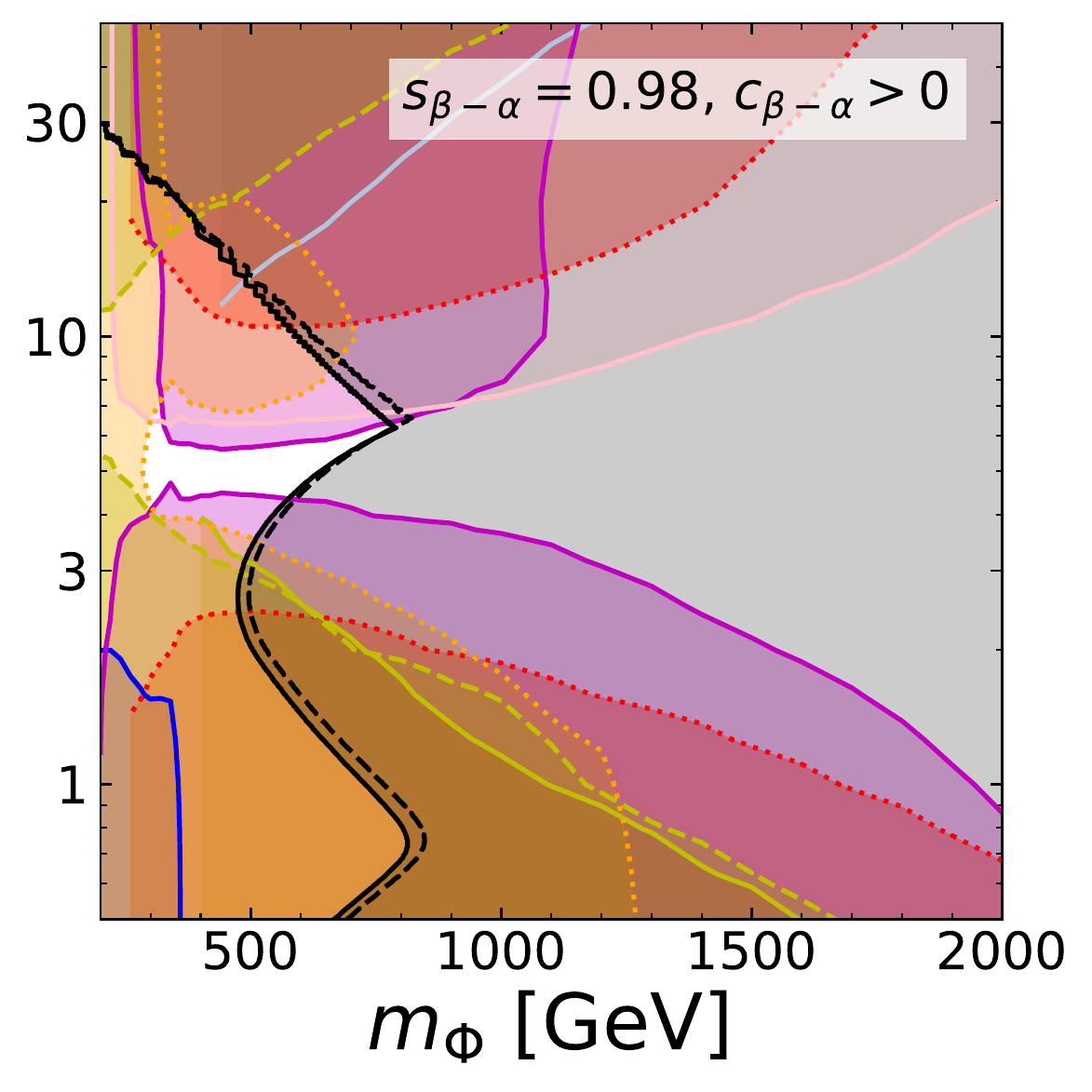} 
\caption{
Regions on the $m_\Phi$--$\tan\beta$ plane expected to be excluded at 95\% CL 
in the Type-I, Type-II, Type-X and Type-Y THDMs (from the left to the right panels)
via direct searches for heavy Higgs bosons at the HL-LHC and via precision measurements of the Higgs boson couplings at the ILC.
The value of $s_{\beta-\alpha}$ is set to be 1, 0.995, 0.99 and 0.98 with $c_{\beta-\alpha} > 0$ from the top to the bottom panels.
}
\label{fig:combine-p}
\end{figure}  

In addition, from precision measurements of the 125~GeV Higgs boson couplings, 
we can further constrain the parameter space in the THDMs. 
In Table~\ref{tab:kappa}, we summarize the current measurements of the $\kappa$ values at the LHC Run-II and the expected $1\sigma$ accuracies of their measurements at the HL-LHC and at the ILC. 
As we can see, the current uncertainties of the measured $\kappa$ values are not small, $10\%$ and $10$--$20\%$ level for $\kappa_V$ and $\kappa_f$, respectively. 
However, these uncertainties can be reduced significantly at those future collider experiments; e.g., $\kappa_Z^{}$ is expected to be measured with a few percent at the HL-LHC and less than $1\%$ at the ILC. 
As we explained in Sec.~\ref{sec:model}, if a nonzero deviation in a Higgs boson coupling is confirmed, 
an upper limit on the mass of the additional Higgs bosons can be given because the decoupling limit is no longer realized. 
In the following discussion, we numerically derive the upper limit on the common mass of the additional Higgs bosons $m_\Phi$ by imposing the bounds from perturbative unitarity and vacuum stability, which are discussed in Sec.~\ref{sec:model}. 
We will see that the upper limit appears for the non-alignment case $s_{\beta-\alpha} \neq 1$, depending on the value of $\tan\beta$. 

In Fig.~\ref{fig:combine-n}, we show
regions on the $m_\Phi$--$\tan\beta$ plane expected to be excluded at 95\% CL 
in the Type-I, Type-II, Type-X and Type-Y THDMs (from the left to the right panels)
via direct searches for heavy Higgs bosons at the HL-LHC and via precision measurements of the Higgs boson couplings at the ILC.
The search channels we consider are same as for the current constraints in Figs.~\ref{fig:current-n} and \ref{fig:current-p}.
The value of $s_{\beta-\alpha}$ is set to be 1, 0.995, 0.99 and 0.98 with $c_{\beta-\alpha}<0$ from the top to the bottom panels.
The shaded regions with solid, dotted, and dashed border lines denote the exclusion regions for $A$, $H$, and $H^\pm$, respectively. 

The global picture of the exclusion regions from the direct searches is similar to the current exclusions, but much wider parameter regions are excluded. 
Especially, for the non-alignment case $\sba\ne1$, the large portion in 
this parameter space is excluded via the $A\to Zh$ and $H\to hh$ channels, which set the lower-mass limit with a given $\tb$.
We note again that the exclusion region from the $H\to hh$ channel can be different for the $M\ne m_\Phi$ case.

Black shaded regions are the regions excluded from the constraints of perturbative unitarity and/or vacuum stability. 
Here, we assume the precision at the ILC250, and the Higgs boson couplings with weak bosons deviate with 1$\sigma$ (2$\sigma$) level\ken{, which corresponds to black solid (dashed) curves, }as
\begin{align}
 \kappa_V^h=[0.995,0.99,0.98] \pm 0.0038\ (0.0076). 
\end{align} 
For these constraints, we scan the value of $M^{2}$ with $M^{2}>0$, so that the black shaded region indicates that there is no value of 
$M^{2}$ which simultaneously satisfies the unitarity and the vacuum stability bounds. 
In the above sense, the black region can be regarded as a conservative excluded region. 
Interestingly, it is seen that a non-zero deviation for the 125~GeV Higgs couplings from the SM prediction sets an upper limit of the heavy Higgs masses.  
For $\sba=0.995$, the alignment limit is included by the $2\sigma$ error, so that the dashed curve does not appear. 

Details of the behavior of the upper limit from precision measurements on $m_\Phi$, shown in Fig.~\ref{fig:combine-n}, are following,
where explicit formulae of the constraints are given in Appendix~\ref{sec:app0}.
For $c_{\beta-\alpha} < 0$, the third condition of the vacuum stability bound given in Eq.~(\ref{eq:vs})
sets an upper limit on $M$ which is slightly smaller than $m_\Phi$ almost without depending on the value of $\tan\beta$; e.g.,  
$M \gtrsim 680$, 730 and 780 GeV being excluded for $x = -0.1$ and $m_\Phi = 800$, 900 and 1000 GeV, respectively,
where $x\equiv \pi/2-(\beta-\alpha)$.
The important point here is that the required value of $m_\Phi^{2} - M^{2} (> 0)$ gets larger for a larger value of $m_\Phi$. 
Whereas, the unitarity bound excludes a larger difference between $M^{2}$ and $m_\Phi^{2}$, which makes magnitudes of the $\lambda$ parameters larger, 
as seen in Eqs.~(\ref{eq:lam1})--(\ref{eq:lam45}).  
Therefore, for a fixed value of $s_{\beta-\alpha}$ and $\tan\beta$  we can find a critical value of $m^{2}_\Phi$, above which the solution of the value of $M^{2}$ to satisfy the both unitarity and vacuum stability bounds vanishes. 
Such an upper limit on $m_\Phi$ becomes stronger when the value of $\tan\beta$ differs from unity because the $\lambda_1$ or $\lambda_2$ parameter becomes significant so that the unitarity bound sets more severe 
constraint on $|M^{2}-m^{2}_\Phi|$.
We here emphasize that
the entire parameter space we consider is explored by combining the constraints from the direct searches at the HL-LHC and from the precision measurements of the 125~GeV Higgs boson couplings at the ILC.

Figure~\ref{fig:combine-p} shows the same as in Fig.~\ref{fig:combine-n}, but for the $c_{\beta-\alpha}>0$ case.
Because of the singular behaviors of the production cross section for $H$ and of the branching ratios for $h$
around $\tb\sim7-10$, shown in Figs.~\ref{fig:xsec-H_cp} and \ref{FIG:BR4}, 
a narrow parameter region in the Type-II and the Type-Y models remains without any constraints from the direct searches
even for low $m_\Phi$.
Similar to Fig.~\ref{fig:combine-n}, there appears an upper limit on $m_\Phi$ by the constraints of unitarity and vacuum stability in Fig.~\ref{fig:combine-p}. 
A remarkable difference, however, arises from the vacuum stability bound as compared with the case for $c_{\beta-\alpha} < 0$. 
In this case with a low $\tan\beta$ region, 
the condition $\lambda_2 > 0$ sets an upper limit on $M^{2}$ for a fixed value of $m^{2}_\Phi$ with $M^{2} \lesssim m_\Phi^{2}$.
This upper limit on $M^{2}$ gets milder when $\tan\beta$ becomes larger.   
When $\tan\beta$ exceeds a certain value, 
the upper limit on $M^{2}$ is almost fixed to be $m_\Phi^{2}$ due to the condition $\lambda_1 > 0$ instead of $\lambda_2 > 0$. 
Such a non-trivial $\tan\beta$ dependence on the vacuum stability bound provides two peaks of the upper limit on $m_\Phi$ as seen in Fig.~\ref{fig:combine-p}. 
As a result, some small parameter regions remain uncovered by both the HL-LHC and the ILC250.    

We here give a comment on the case, where the degeneracy between the common mass of the additional Higgs bosons $m_{\Phi}$ and $M$ is relaxed. 
In the above analysis, we have set $M=m_{\Phi}$ in the analysis of the exclusion region by the direct searches for simplicity. 
As we have mentioned in Sec.~\ref{sec:decay}, 
the decay width for $H\rightarrow hh$ depends on the value of $M$, and the exclusion region for $H$ might change if we consider the case of $M\neq m_{\Phi}$.
We note, however, that most of the parameter regions excluded by $H\rightarrow hh$ are also excluded by the $A\rightarrow Zh$ decay mode, which does not depend on the value of $M$.
Therefore, our main conclusion 
does not change even if we relax the degeneracy among $m_{\Phi}$ and $M$.

To summarize, the entire parameter space in the THDMs can be explored by the synergy between the direct searches at the HL-LHC and the precision measurements of the 125~GeV Higgs boson couplings at the ILC.
In other words, if we observed any deviations for the Higgs boson couplings at the ILC, we would be able to find 
additional Higgs bosons at the HL-LHC, or reject a certain type of new physics models. 
In order to quantify the above statement, we have also checked the 5$\sigma$ discovery sensitivity by naive rescaling. 
We find that the discovery regions are certainly smaller than the 95\% CL excluded region shown in Figs.~\ref{fig:combine-n} and \ref{fig:combine-p}.
Consequently, 
for $\cba<0$, we find that most of the parameter space is covered by the direct searches at the HL-LHC and the precision tests at the ILC250. 
For $\cba>0$, on the other hand, some parameter regions appear, which requires more data and/or more precision to be explored.



\section{Conclusions}\label{sec:conclusion}

We have discussed the possibility that a wide region of the parameter space in the four types of the THDMs can be explored by the 
combination of the direct searches for the additional Higgs bosons at the LHC and 
precision measurements of the discovered Higgs boson couplings at future lepton colliders. 
The direct searches give lower limits on the masses of the additional Higgs bosons, while the 
precision measurements set upper limits by using the perturbative unitarity and the vacuum stability bounds. 
Thus, these two searches play an complementary role to explore the parameter space. 
We first have shown that the parameter region excluded by the direct search at the LHC Run-II, and then 
shown that the exclusion expected by using the synergy between the direct searches at the HL-LHC and the precision tests
assuming the accuracy expected for the measurements of the Higgs boson couplings at the ILC with a collision energy of 250 GeV. 
It has been found that in the nearly alignment scenario 
most of the parameter space is explored by the direct searches of extra Higgs bosons and the precision tests. 
In the alignment limit where all the Higgs boson couplings take the SM-like values, there are parameter regions which cannot be excluded due to the suppression of  
the Higgs to Higgs decays, $H\to hh$ and $A\to Zh$, and no upper limit on the masses from the theoretical arguments. 

\begin{acknowledgments}
This work is supported in part by the Grant-in-Aid on Innovative Areas, the Ministry of Education, Culture, 
Sports, Science and Technology, No.~16H06492, No.~18F18321, No.~18F18022 and No.~20H00160 [S.K.],  
Early-Career Scientists, No.~20K14474 [M.K.], 
JSPS KAKENHI Grant No.~18K03648 [K.M.], 
and Early-Career Scientists, No.~19K14714 [K.Y.]. 
M. A. was supported in part by the Sasakawa Scientific Research Grant from The Japan Science Society. 
\end{acknowledgments}

\begin{appendix}

\section{Bounds from unitarity and vacuum stability}\label{sec:app0}

The unitarity bound is defined by $|a_i| \leq 1/2$ as we discuss in Sec.~\ref{sec:model}, 
where the independent eigenvalues of the $s$-wave amplitude matrix are given by~\cite{Kanemura:1993hm,Akeroyd:2000wc,Ginzburg:2005dt,Kanemura:2015ska} 
\begin{align}
a_1^\pm &=  \frac{1}{32\pi}
\left[3(\lambda_1+\lambda_2)\pm\sqrt{9(\lambda_1-\lambda_2)^2+4(2\lambda_3+\lambda_4)^2}\right],\\
a_2^\pm &=
\frac{1}{32\pi}\left[(\lambda_1+\lambda_2)\pm\sqrt{(\lambda_1-\lambda_2)^2+4\lambda_4^2}\right],\\
a_3^\pm &= \frac{1}{32\pi}\left[(\lambda_1+\lambda_2)\pm\sqrt{(\lambda_1-\lambda_2)^2+4\lambda_5^2}
\right],\\
a_4^\pm &= \frac{1}{16\pi}(\lambda_3+2\lambda_4\pm 3\lambda_5),\\
a_5^\pm &= \frac{1}{16\pi}(\lambda_3\pm\lambda_4),\\
a_6^\pm &= \frac{1}{16\pi}(\lambda_3\pm\lambda_5).
\end{align}
The vacuum stability bound is given by~\cite{Deshpande:1977rw,Klimenko:1984qx,Sher:1988mj,Nie:1998yn}
\begin{align}
\lambda_1 > 0,~~ \lambda_2 > 0,~~ \sqrt{\lambda_1\lambda_2}+\lambda_3+ \text{MIN}(0,\lambda_4+\lambda_5,\lambda_4-\lambda_5) > 0. \label{eq:vs}
\end{align}

As we see the above expression, the unitarity and the vacuum stability bounds constrain the value of the $\lambda$ parameters. 
Thus, it would be convenient to express these parameters in terms of the physical parameters as follows: 
\begin{align}
\lambda_1&=\frac{1}{v^2}\left[(m_\Phi^2-M^2)\tan^2\beta + m_h^2 + 2\tan\beta(m_\Phi^2  -m_h^2)x \right] + {\cal O}(x^2), \label{eq:lam1}\\
\lambda_2&=\frac{1}{v^2}\left[(m_\Phi^2-M^2)\cot^2\beta + m_h^2 - 2\cot\beta(m_\Phi^2  -m_h^2)x \right] + {\cal O}(x^2), \label{eq:lam2}\\
\lambda_3&=\frac{1}{v^2}\left[m_\Phi^2+ m_h^2-M^2 -2\cot2\beta(m_\Phi^2 - m_h^2)x\right] + {\cal O}(x^2), \label{eq:lam3}\\
\lambda_4&=\lambda_5=\frac{M^2-m_\Phi^2}{v^2}, \label{eq:lam45}
\end{align}
where $m_{\Phi}=m_{H}=m_{A}=m_{H^{\pm}}$, and $x \equiv \pi/2 -(\beta -\alpha)$ such that $x= 0$ corresponds to the alignment limit $s_{\beta-\alpha} = 1$.

\section{Decay rates at the leading order}\label{sec:app1}

We present the analytic expressions of the decay rates of the Higgs boson at the LO. 
In order to specify the LO formula, the subscript $0$ is put in the decay rate, $\Gamma_0$.

\subsection{Decays of the neutral Higgs bosons}

We define $\phi = h,~H$ or $A$ and ${\cal H}=h$ or $H$.
The decay rates into a fermion pair are given by 
\begin{align}
\Gamma_0({\phi} \to f\bar{f}) &= \sqrt{2}G_F\frac{m_\phi m_f^2}{8\pi}|\kappa^\phi_f|^2N_c^f\lambda_\phi\left(\frac{m_f^2}{m_\phi^2},\frac{m_f^2}{m_\phi^2}\right), 
\end{align}
where $N_c^f = 1\ (3)$ for $f$ being a lepton (quark) and 
\begin{align}
\lambda_{h,H}(x,y) = \lambda^{3/2}(x,y),\quad \lambda_A(x,y) = \lambda^{1/2}(x,y), 
\end{align}
with 
\begin{align}
\lambda(x,y) = (1-x-y)^2-2xy. 
\end{align}

The decay rates into a pair of on-shell weak bosons ($V=W,Z$) are given by 
\begin{align}
\Gamma_0(\phi \to VV)&=\sqrt{2}G_F\frac{m_\phi^3}{4\pi c_V}(\kappa_V^\phi)^2\left(\frac{3m_V^4}{m_\phi^4}-\frac{m_V^2}{m_\phi^2}+\frac{1}{4}\right)\lambda^{1/2}\left(\frac{m_V^2}{m_\phi^2},\frac{m_V^2}{m_\phi^2}\right),
\label{eq:hvv}
\end{align}
where  $c_V = 1\ (2)$ for $V=W\ (Z)$. 
When one of the weak bosons is off-shell, we obtain 
\begin{align}
\Gamma_0(\phi \to WW^*)
&=\frac{3m_W^4G_F^2m_\phi}{16\pi^3 } (\kappa_V^\phi)^2 F\left(\frac{m_W}{m_\phi}\right),\\
\Gamma_0(\phi \to ZZ^*)&=\frac{m_Z^4G_F^2m_\phi}{64\pi^3}(\kappa_V^\phi)^2F\left(\frac{m_Z}{m_\phi}\right)\left(7-\frac{40}{3}s_W^2+\frac{160}{9}s_W^4\right),
\end{align}
where 
\begin{align}
F(x)&=-|1-x^2|\left(\frac{47}{2}x^2-\frac{13}{2}+\frac{1}{x^2}\right)+3(1-6x^2+4x^4)|\log x|\notag\\
&+\frac{3(1-8x^2+20x^4)}{\sqrt{4x^2-1}}\cos^{-1}\left(\frac{3x^2-1}{2x^3}\right).
\end{align}

The loop induced decay rates are given by 
\begin{align}
\Gamma_0(\phi \to\gamma \gamma)&=\frac{\sqrt{2}G_F\alpha_{\text{em}}^2m_\phi^3}{256\pi^3 }\left|\kappa_V^{\phi} I_W^\phi + \sum_f \kappa_f^\phi Q_f^2N_c^fI_F^\phi -\frac{\lambda_{H^+H^-\phi}}{v}I_S^\phi \right|^2, \\
\Gamma_0(\phi \to Z\gamma)&=\frac{\sqrt{2}G_F\alpha_{\text{em}}^2m_\phi^3}{128\pi^3}\left(1-\frac{m_Z^2}{m_\phi^2}\right)^3\left|\kappa_V^{\phi}J_W^\phi + \sum_f \kappa_f^{\phi}Q_fN_c^fv_fJ_F^\phi -\frac{\lambda_{H^+H^-\phi}}{v}
\frac{g_Z^{}c_{2W}^{}}{2}J_S^\phi \right|^2, \\
\Gamma_0(\phi \to gg)&=\frac{\sqrt{2}G_F\alpha_s^2 m_\phi^3}{128\pi^3 }\left|\sum_q \kappa_q^{\phi} I_q^\phi \right|^2, 
\end{align}
where 
\begin{align}
I_W^\phi & = \frac{2m_W^2}{m_\phi^2}\left[6+\frac{m_\phi^2}{m_W^2}+(12m_W^2-6m_\phi^2)C_0(0,0,m_\phi^2;m_W,m_W,m_W)\right], \\ 
I_F^{\cal H} & = -\frac{8m_f^2}{m_{\cal H}^2}\left[1+\left(2m_f^2-\frac{m_{\cal H}^2}{2}\right)C_0(0,0,m_{\cal H}^2;m_f,m_f,m_f)\right], \\
I_F^A & = -4m_f^2C_0(0,0,m_A^2;m_f,m_f,m_f), \\
I_S^\phi &= \frac{2v^2}{m_\phi^2}[1+2m_{H^\pm}^2C_0(0,0,m_\phi^2;m_{H^\pm},m_{H^\pm},m_{H^\pm})],\\
J_W^\phi &=\frac{2m_W^2}{s_Wc_W(m_\phi^2-m_Z^2)}\Bigg\{\left[c_W^2\left(5+\frac{m_\phi^2}{2m_W^2}\right)-s_W^2\left(1+\frac{m_\phi^2}{2m_W^2}\right)\right]\notag\\
&\times\Bigg[1+2m_W^2C_0(0,m_Z^2,m_{\phi}^2;m_f,m_f,m_f) \notag \\
&+\frac{m_Z^2}{m_\phi^2-m_Z^2}[B_0(m_\phi^2;m_W,m_W)-B_0(m_Z^2;m_W,m_W)]\Bigg]\notag\\
&-6c_W^2(m_\phi^2-m_Z^2)C_0(0,m_Z^2,m_{\phi}^2;m_f,m_f,m_f) \notag\\
&+2s_W^2(m_\phi^2-m_Z^2)C_0(0,m_Z^2,m_{\phi}^2;m_f,m_f,m_f)\Bigg\}, \\
J_F^{\cal H} &=-\frac{8m_f^2 }{s_Wc_W(m_{\cal H}^2-m_Z^2)}\Big[1+\frac{1}{2}(4m_f^2-m_{\cal H}^2+m_Z^2)C_0(0,m_Z^2,m_{\cal H}^2;m_f,m_f,m_f)\notag\\
&+\frac{m_Z^2}{m_{\cal H}^2-m_Z^2}[B_0(m_{\cal H}^2;m_f,m_f)-B_0(m_Z^2;m_f,m_f)]\Big],\\
J_F^A &=-\frac{4m_f^2}{s_Wc_W}C_0(0,m_Z^2,m_A^2;m_f,m_f,m_f),\\
J_S^\phi &=\frac{2v^2}{e(m_\phi^2-m_Z^2)}
\Bigg\{1+2m_{H^\pm}^2C_0(0,m_Z^2,m_\phi^2;m_{H^\pm},m_{H^\pm},m_{H^\pm}) \notag\\
&+\frac{m_Z^2}{m_\phi^2-m_Z^2}\left[B_0(m_\phi^2;m_{H^\pm},m_{H^\pm})-B_0(m_Z^2;m_{H^\pm},m_{H^\pm})\right]\Bigg\}, 
\end{align}
where $C_{0}$ and $B_{0}$ are the Passarino-Veltman functions~\cite{Passarino:1978jh}. 

The decay rates into a scalar and an on-shell weak boson is given by 
\begin{align}
\Gamma_0(\phi\to \phi'V)&=\frac{|g_{\phi \phi' V}|^2}{16\pi}\frac{m_\phi^3}{m_V^2}\lambda^{3/2}\left(\frac{m_V^2}{m_\phi^2},\frac{m_{\phi'}^2}{m_\phi^2}\right), 
\end{align}
where the scalar-scalar-gauge couplings are given in Appendix of Ref.~\cite{Kanemura:2015mxa}. 
When the weak boson is off-shell, the decay rate is given by 
\begin{align}
\Gamma_0(\phi\to \phi' W^{*}) &= 9g^2|g_{\phi\phi' W}|^2\frac{m_\phi}{128 \pi^3}G\left(\frac{m_{\phi'}^2}{m_\phi^2},\frac{m_W^2}{m_\phi^2}\right), \\
\Gamma_0(\phi\to \phi' Z^*)& =3g_Z^2|g_{\phi\phi' Z}|^2\frac{m_\phi}{256 \pi^3}G\left(\frac{m_{\phi'}^2}{m_\phi^2},\frac{m_Z^2}{m_\phi^2}\right)\left(7-\frac{40}{3}s_W^2+\frac{160}{9}s_W^4\right), 
\end{align}
where the function $G(x,y)$ is given as 
\begin{align}
G(x,y)&=\frac{1}{12y}\Bigg\{2\left(-1+x\right)^3-9\left(-1+x^2\right)y+6\left(-1+x\right)y^2\notag\\
&+6\left(1+x-y\right)y\sqrt{-\lambda(x,y)}
\left[\tan^{-1}\left(\frac{-1+x-y}{\sqrt{-\lambda(x,y)}}\right)+\tan^{-1}\left(\frac{-1+x+y}{\sqrt{-\lambda(x,y)}}\right)\right]\notag\\
&-3\left[1+\left(x-y\right)^2-2y\right]y\log x\Bigg\}. 
\end{align}

Finally, the decay rates into two lighter scalar bosons are given by 
\begin{align}
\Gamma_0 (\phi \to \phi'\phi'') &=
(1+\delta_{\phi'\phi''})\frac{|\lambda_{\phi\phi'\phi''}|^2}{16\pi m_\phi}
\lambda^{1/2}\left(\frac{m_{\phi'}^2}{m_\phi^2},\frac{m_{\phi''}^2}{m_\phi^2}\right). 
\end{align}
An example of this type of the decay is $H \to hh$. 

\subsection{Decays of the charged Higgs bosons}
Decays of the charged Higgs bosons  into two on-shell fermions are given by
\begin{align}
\Gamma_{0}(H^{\pm}\to f f^{\prime})&=
\frac{ G_{F}m_{H^{\pm}} }{4\sqrt{2}\pi }N_{c}^{f}C_{f}
\lambda^{1/2}\left(\frac{m^{2}_{f}}{m^{2}_{H^{\pm}}}, \frac{m^{2}_{f^{\prime}}}{m^{2}_{H^{\pm}}} \right) \notag \\
&\times \Bigg[ \left(1-\frac{m^{2}_{f}}{m^{2}_{H^{\pm}}}-\frac{m^{2}_{f^{\prime}}}{m^{2}_{H^{\pm}}} \right)
\Big\{ m_{f}^{2}\zeta_{f}^{2} +m_{f^{\prime}}^{2}\zeta_{f^{\prime}}^{2} \Big\}
+4\frac{m^{2}_{f} m^{2}_{f^{\prime}} }{m_{H^{\pm}}^{2}}\zeta_{f}\zeta_{f^{\prime}}\Bigg],
\end{align}
where a factor $C_{f}$ is $C_{f}=|V_{ud}|^{2}~(1)$  for the decay into the up-type quark and the down-type quark (the lepton and the neutrino).
The decay into an off-shell top quark and an on-shell down-type quark, $H^{\pm}\to  t^{\ast} q_{d}\to W^{\pm} b q_{d} \ (q_{d}=d,s,b)$, is expressed by
\begin{align}
\Gamma_{0}(H^{\pm}\to W^{\pm} b q_{d})=\frac{3 V_{tb}V_{tq_{d}} }{256\pi^{3} m_{H^{\pm}}}g^{2}\frac{m_{t}^{4}}{v^{2}}\zeta^{2}_{u}
H\left(\frac{m_{t}^{2}}{m^{2}_{H^{\pm}}}, \frac{m_{W}^{2}}{m^{2}_{H^{\pm}}}\right),
\end{align} 
with
\begin{align}
\notag
H(x,y)&=
\frac{1}{{4 {x}^3
   {y}}}
   \Big[
   2{y}^2 \left\{4 ({x}-1) {y}+3
   {x}\right\} \log {y} \\ \notag
&+2 ({x}-1) \left\{(3 {x}-1) {x}^3-3
   ({x}+1) {x} {y}^2+4 {y}^3\right\} \log \frac{{x}-1}{{x}-{y }} \\ 
   &-{x} ({y}-1) \left\{(6 {x}-5)
   {x}^2-4 ({x}-2) {y}^2+(3
   {x}-2) {x} {y}\right\} 
   \Big],  
\end{align}
where the mass of down-type quark is neglected. 

The on-shell decays into a neutral Higgs boson and a W boson are expressed by
\begin{align}
\Gamma_{0}(H^{\pm}\to  \phi W^{\pm}  )=\frac{m_{H^{\pm}}^{3}}{16\pi m^{2}_{W}}
|g_{H^{\pm}\phi W}|^{2} \lambda^{\frac{3}{2}}\left(\frac{m_{\phi}^{2}}{m_{H^{\pm}}^{2}},\frac{m_{W}^{2}}{m_{H^{\pm}}^{2}}\right),
\end{align}
where the coupling $g_{H^{\pm}\phi W}$ for each neutral Higgs boson is given in Appendix of Ref.~\cite{Kanemura:2015mxa}.
The decays into a neutral Higgs boson and an off-shell W boson is given by
\begin{align}
\Gamma_{0}(H^{\pm}\to \phi W^{\pm \ast})=9\frac{m_{H^{\pm}}}{128\pi^{3}}g^{2}|g_{H^{\pm}\phi W}|^{2}
G\left(\frac{ m_{\phi}^{2} }{m^{2}_{H^{\pm}} }, \frac{ m_{W}^{2} }{m^{2}_{H^{\pm}} } \right).
\end{align}

For the loop induced decay rates, $H^{\pm}\to W^{\pm} Z$ and $H^{\pm}\to W^{\pm} \gamma $, the concrete expressions of fermion loop contributions and  boson loop contributions are given in Refs.~\cite{CapdequiPeyranere:1990qk,Kanemura:1999tg} and Refs.~\cite{Kanemura:1997ej,Kanemura:1999tg}, respectively.

\end{appendix}

\bibliography{references}
\bibliographystyle{JHEP}

\end{document}